%% file: thesis.tex
\documentclass[12pt, a4paper,openany,oldfontcommands]{memoir}

\usepackage{graphicx}
\usepackage{epsfig}
\usepackage{amsmath}
\usepackage{amssymb}
\usepackage{amsthm}
\usepackage{booktabs}
\usepackage{longtable}
\usepackage[figuresright]{rotating}
\usepackage{latexsym}
\usepackage{textcomp}
\usepackage[vcentermath]{youngtab}
\usepackage{type1cm}
\usepackage{eso-pic}
\usepackage{mathrsfs}
\DisemulatePackage{setspace}
\usepackage{setspace}
\usepackage{graphics}
\usepackage{titlesec}
\usepackage{times}
\usepackage[]{geometry}
\usepackage{lipsum}
\usepackage{cite}
\usepackage{rotating}
\usepackage{float}
\usepackage{ctable}
\usepackage{slashed}
\usepackage[pdftex,colorlinks=true,linktocpage=true,urlcolor=magenta,linkcolor=blue,citecolor=red,backref=page]{hyperref}

\usepackage[normalem]{ulem}  

\def\be {\begin{equation}}
\def\ee {\end{equation}}
\def\bea {\begin{eqnarray}}
\def\eea {\end{eqnarray}}
\def\bc {\begin{center}}
\def\ec {\end{center}}
\def\nn {\nonumber}
\def\Tr{\operatorname{\textrm{Tr}}}
\def\sumint{\sum\!\!\!\!\!\!\!\!\!\!\!\!\!\!\!\!\!\int\limits}
\def\sumintb{\sum\!\!\!\!\!\!\!\!\!\int\limits}
\def\sumintf{\sum\!\!\!\!\!\!\!\!\!\!\int\limits}

\def\gm {\gamma}
\def\de {\delta}
\def\eps {\epsilon}
\def\wt {\widetilde}
\def\la {\langle}
\def\ra {\rangle}
\def\lrw {\leftarrow}
\def\mn {\mu\nu}
\def\th {\theta}
\def\ov {\overline}
\def\sg {\sigma}
\def\om {\omega}

\def\pp {\perp}
\def\pl {\shortparallel}
\DeclareMathOperator{\sech}{sech}
\newcommand\abcom[2]{\genfrac{}{}{0pt}{}{#1}{#2}}

\setlrmargins{*}{*}{1.2}
\setulmargins{*}{*}{1}
\setmarginnotes{5mm}{45.23mm}{\onelineskip}
\checkandfixthelayout

\makechapterstyle{mychapterstyle}{%
    \renewcommand{\chapnamefont}{\color{black}\LARGE\rmfamily\bfseries}%
    \renewcommand{\chapnumfont}{\LARGE\rmfamily\bfseries}%
    \renewcommand{\printchapternum}{%
        \chapnumfont\thechapter%
        }%
}

\makeatletter
\newcommand\thickhrulefill{\leavevmode \leaders \hrule height 1ex \hfill \kern \z@}
\makechapterstyle{VZ14}{

\renewcommand\chapnamefont{\Large\scshape}
\renewcommand\printchapternum{%
\chapnamefont\null\thickhrulefill\quad
\@chapapp\space\thechapter\quad\thickhrulefill}

}
\makeatother
\chapterstyle{VZ14}

\setsecheadstyle{\color{black}\Large \bfseries}
\setsubsecheadstyle{\color{black}\Large\bfseries}
\setsubsubsecheadstyle{ \color{black}\normalfont\bfseries}
\setparaheadstyle{\normalfont\sffamily}
\makeevenhead{headings}{\thepage}{}{\footnotesize\bfseries\leftmark}
\makeoddhead{headings}{\footnotesize\bfseries\rightmark}{}{\thepage}

\setsecnumdepth{subsection}
 \maxsecnumdepth{subsection}
\settocdepth{subsection}
\maxtocdepth{subsection}

\DeclareGraphicsExtensions{.jpg,.pdf,.eps}

\begin{document}
\doublespacing
\pagenumbering{gobble}
\noindent
\begin{center}
\Large
   \textsc{\bfseries Non-perturbative study of spectral function\\ and its application in Quark Gluon Plasma}
\end{center} 
\begin{center}
\vskip 0.5cm
{\bf {\em By}} 
\vskip 0.2cm
{\bf {\large ARITRA BANDYOPADHYAY}}
\vskip 0.0cm
{\bf {\large  PHYS05201204004 }}
\vskip 0.2cm
{\bf {\large Saha Institute of Nuclear Physics, Kolkata}}
\vskip 0.7cm
{ {\em {\large A thesis submitted to the
\vskip 0.05cm
Board of Studies in Physical Sciences
\vskip 0.05cm
In partial fulfillment of requirements
\vskip 0.05cm
For the Degree of 
}}}
\vskip 0.05cm
{\bf {\large DOCTOR OF PHILOSOPHY}}
\vskip 0.1cm
{ {\em of}}
\vskip 0.1cm
{\bf {\large HOMI BHABHA NATIONAL INSTITUTE}}
\vfill
\begin{figure}[hbt]
\begin{center}
\includegraphics[scale=0.25]{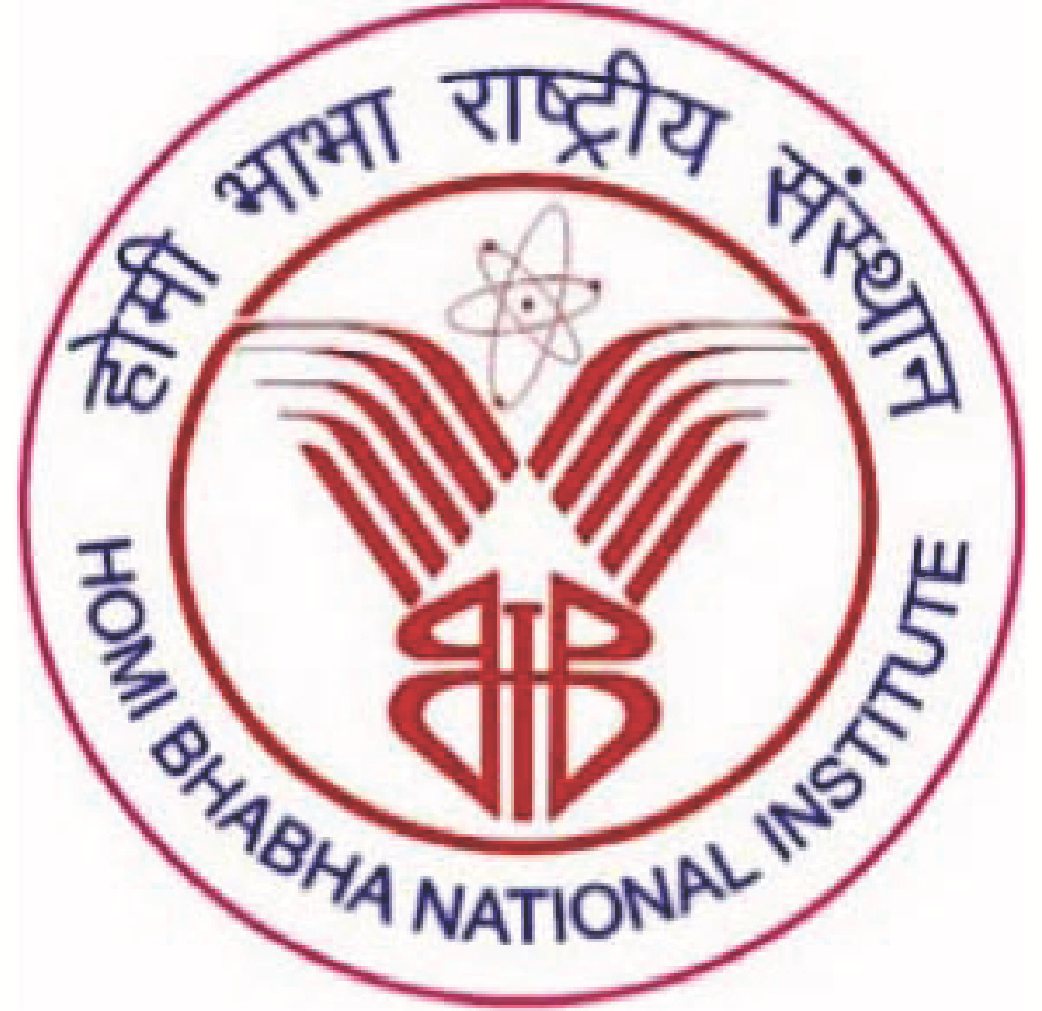}
\end{center}
\end{figure}
%
%
{\bf {\large July, 2017}}
\vfill
\end{center}

\newpage

%
%
\centerline{{\bf{\LARGE Homi Bhabha National Institute}}}
\vskip 0.2cm
\centerline{{\bf {\large Recommendations of the Viva Voce Committee}}}
\vskip 0.2cm
As members of the Viva Voce Committee, we certify that we have read the
dissertation prepared by {\bf{Aritra Bandyopadhyay}} entitled {\bf{``Non-perturbative study of spectral function and its application in Quark Gluon Plasma"}} and recommend that it maybe accepted as fulfilling the thesis requirement for the award of Degree of Doctor of Philosophy.
\vskip 0.1cm
\underline{\hspace{12.0cm}} Date:
\vskip -0.1cm 
Chairman - Professor Asit K De
\vskip 0.1cm
\underline{\hspace{12.0cm}} Date:
\vskip -0.1cm 
Guide / Convener - Professor Munshi G Mustafa
\vskip 0.1cm
\underline{\hspace{12.0cm}} Date:
\vskip -0.1cm 
Member 1 - Professor Pradip K Roy 
\vskip 0.1cm
\underline{\hspace{12.0cm}} Date:
\vskip -0.1cm 
Member 2 - Professor Debades Bandyopadhyay 
\vskip 0.1cm
\underline{\hspace{12.0cm}} Date:
\vskip -0.1cm 
Member 3 - Professor Sourav Sarkar 
\vskip 0.1cm
\underline{\hspace{12.0cm}} Date:
\vskip -0.1cm 
Examiner - Professor Hiranmaya Mishra
\vskip -0.1cm
\rule{14cm}{1pt}
\vskip -0.1cm
\hspace{0.2cm} Final approval and acceptance of this thesis is contingent upon the candidate's
submission of the final copies of the thesis to HBNI.
\vskip -0.1cm
\hspace{0.2cm} I/We hereby certify that I/we have read this thesis prepared under my/our
direction and recommend that it may be accepted as fulfilling the thesis requirement.

\vskip 0.0cm 
{\bf Date:} 
\vskip -0.2cm 
{\bf {Place:} \hspace{6cm} Guide:{\underline{ \hspace{5.0cm}}}
\vskip -0.8cm 
\hspace{8.65cm}}

\newpage
\centerline{{\bf {\large STATEMENT BY AUTHOR}}}
\vskip 1.00cm
%
%
This dissertation has been submitted in partial fulfillment of
requirements for an advanced degree at Homi Bhabha National Institute
(HBNI) and is deposited in the Library to be made available to borrowers
under rules of the HBNI.
\vskip 0.6cm
Brief quotations from this dissertation are allowable without special
permission, provided that accurate acknowledgement of source is made.
Requests for permission for extended quotation from or reproduction of
this manuscript in whole or in part may be granted by the Competent
Authority of HBNI when in his or her judgment the proposed use of the
material is in the interests of scholarship. In all other instances,
however, permission must be obtained from the author.

\vskip 1.5cm


$~$\hspace{10.2cm}Aritra Bandyopadhyay
\newpage
%
\vskip 1.2cm
\centerline{{\bf{\large{DECLARATION}}}}
\vskip 1.2cm
I, hereby declare that the investigation presented in the thesis has been
carried out by me. The work is original and has not been submitted
earlier as a whole or in part for a degree / diploma at this or any
other Institution / University.
\vskip 2.0cm
%
%
\rightline{Aritra Bandyopadhyay \hspace{0.9cm}}
%
\newpage
\cleardoublepage
\chapter*{List of Publications Arising from the Thesis}
\phantomsection
\begin{flushleft}
 \textbf{Journal:}
\end{flushleft}

\begin{enumerate}

\item[ 1.] ``Three-loop HTLpt thermodynamics at finite temperature and chemical potential"
     \\Najmul Haque, Aritra Bandyopadhyay, Jens O. Andersen, Munshi G. Mustafa, Michael Strickland and Nan Su
     \\ JHEP 1405 (2014) 027,
     \href{https://arxiv.org/abs/1402.6907}{[arXiv:1402.6907 [hep-ph]].}

\item[ 2.] ``Dilepton rate and quark number susceptibility with the Gribov action"
     \\ Aritra Bandyopadhyay, Najmul Haque, Munshi G. Mustafa and Michael Strickland
     \\Phys. Rev. D 93 (2016) 065004,
     \href{https://arxiv.org/abs/1508.06249}{[arXiv:1508.06249 [hep-ph]].}

\item[ 3.] ``Electromagnetic spectral properties and Debye screening of a strongly magnetized hot medium"
  \\Aritra Bandyopadhyay, Chowdhury Aminul Islam and Munshi G. Mustafa
  \\Phys. Rev. D 94 (2016) 114034,
  \href{https://arxiv.org/abs/1602.06769}{[arXiv:1602.06769 [hep-ph]].}

\item[ 4.] ``Power corrections to the electromagnetic spectral function and the dilepton rate in QCD plasma within operator product expansion in D=4"
     \\Aritra Bandyopadhyay and Munshi G. Mustafa
     \\ JHEP 1611 (2016) 183, 
     \href{https://arxiv.org/abs/1609.06969}{[arXiv:1609.06969 [hep-ph]].}

\item[ 5.] ``Effect of magnetic field on dilepton production in a hot plasma "
  \\Aritra Bandyopadhyay and S. Mallik
  \\Phys. Rev. D 95 (2017) 074019,
  \href{https://arxiv.org/abs/1704.01364}{[arXiv:1704.01364 [hep-ph]].}

  \item[ 6.] ``The pressure of a weakly magnetized deconfined QCD matter within one-loop Hard-Thermal-Loop perturbation theory"
     \\ Aritra Bandyopadhyay, Najmul Haque and Munshi G. Mustafa
     \\ Communicated
     \href{https://arxiv.org/abs/1702.02875}{[arXiv:1702.02875 [hep-ph]].}
\end{enumerate}
\vspace{0.001cm}

\begin{flushleft}
 \textbf{Chapters in Books and Lecture Notes}: None
\end{flushleft}

\vspace{0.001cm}
\begin{flushleft}
 \textbf{Conferences}:
\end{flushleft}

\begin{enumerate} 
\item[ 1.]``Three loop HTL perturbation theory at finite temperature and chemical potential"
       \\Michael Strickland, Jens O. Andersen, Aritra Bandyopadhyay, Najmul Haque, Munshi G. Mustafa, Nan Su
      \\ XXIV Quark Matter, 2014
        \\Nucl.Phys. A931 (2014) 841-845
        \href{https://arxiv.org/abs/1407.3671}{[arXiv:1407.3671 [hep-ph]].}

\item[ 2.]``Three-Loop HTLpt Thermodynamics at Finite Temperature and Chemical Potential"
      \\Aritra Bandyopadhyay, N. Haque, M. G. Mustafa, M. Strickland, N. Su
       \\ XXI DAE-BRNS HEP Symposium, IIT Guwahati, December 2014
    \\Springer\ Proc.\ Phys. 174 (2016) 17-21,
     \href{https://arxiv.org/abs/1508.04291}{[arXiv:1508.04291 [hep-ph]].}
\end{enumerate}

\begin{flushleft}
 \textbf{Others}: 
\end{flushleft}
\begin{enumerate}
\item[ 1.]``Rho meson decay in presence of magnetic field "
      \\Aritra Bandyopadhyay and S. Mallik
    \\Communicated,  \href{https://arxiv.org/abs/1610.07887}{[arXiv:1610.07887 [hep-ph]].}
    
\item[ 2.]`` General structure of fermion two-point function and its spectral representation in a hot magnetised medium "
      \\Aritra Das, Aritra Bandyopadhyay, Pradip K Roy and Munshi G Mustafa
    \\Communicated,  \href{https://arxiv.org/abs/1709.08365}{[arXiv:1709.08365 [hep-ph]].}
\end{enumerate}

\vspace{1cm}

\begin{flushright}
Aritra Bandyopadhyay
\end{flushright}
\newpage


\centerline{{\bf{\large DEDICATION}}}
\vspace{4cm}
\begin{center}
{{\large To my parents \\ without whom \\ none of my achievements were possible}}
\end{center}
\cleardoublepage


\input{./text/ackn}


\normalfont
\phantomsection
\clearpage
\begin{KeepFromToc}
  \tableofcontents
\end{KeepFromToc}
\clearpage
\pagestyle{plain}
\pagenumbering{roman}

%
%
  \input{./text/Abbrv}
  \addcontentsline{toc}{chapter}{Abbreviation}
  \cleardoublepage


  \input{./text/synopsis}
  \addcontentsline{toc}{chapter}{Synopsis}


\clearpage
\listoffigures
\clearpage
\cleardoublepage
\pagenumbering{arabic}
\pagestyle{plain}

\setcounter{page}{1}


\input{./text/intro}


 \input{./text/DPR}


 \input{./text/ope}


 \input{./text/gribov}


 \input{./text/mag_dpr}


 \input{./text/ds}


 \input{./text/qns}


\input{./text/conclusion}

\appendix

 \input{./text/app1}


\bibliography{thesis}{}
\bibliographystyle{JHEP}

\end{document}

%% file: text/ackn.tex
\centerline{{\bf{\large ACKNOWLEDGEMENTS}}}
\vspace{1cm}

My journey in physics started ten years ago with a panicky heart and an inquisitive mind, when I joined St. Xavier's College, Kolkata as a Physics Hons student. Today after ten years, here I am, writing my thesis acknowledgement. Within or beyond these ten years many people have been involved directly or indirectly to make this thesis a reality. I am deeply indebted to all of them for their support, be it continuous or transitory. I feel I should talk about some of them, who are close to my heart. 

First and foremost I would like to convey my earnest gratitude to my supervisor Munshi Golam Mustafa for his continuous support throughout my PhD tenure. I can proudly say that if there exists an ideal scholar-guide relationship, my relationship with my guide would perfectly fit into that. He would never overdo anything except caring for his students. I consider myself really lucky to have a cool and caring supervisor like him who never intervened into my freedom and helped me whenever I reached out to him. Besides my supervisor I am also grateful to my Doctoral committee members Pradip Kumar Roy, Debades Bandyopadhyay, Sourav Sarkar and Asit Kumar De for their constant push regarding giving my thesis a good shape.   

After Munshi da, the first name that comes to my mind is of Najmul Haque, my academic senior who has been not a bit less than a guide to me throughout these five years. Whenever I faced the slightest of problems while doing any kind of calculation, Najmulda would grace me with his superhuman calculation skills and resolve the problem in no time. Taking this opportunity I would also like to thank all my other collaborators, specially Michael Strickland and Samir Mallik for their valuable discussions and suggestions. 

Being a part of Munshi da's vibrant group has immensely helped me to develop my research mentality during these five years. I already talked about Najmulda. Apart from him, I feel blessed to have seniors like Raktim Abir, Sarbani Majumder and Chowdhury Aminul Islam. I have learned a great deal about life from them, specially from Aminul da, who has gradually become a friend from a perfectionist senior over the course of our PhD lives. We have played football, watched movies, visited so many places during national conferences and most importantly shared our opinion about different academic and non-academic issues. Our group also consists of humble juniors like Aritra Das and Bithika Karmakar, both exciting prospects of near future. 

I feel privileged to be a part of the Theory division of Saha Institute of Nuclear Physics. Direct or indirect contributions from all the faculties, staffs and students of this division are gratefully acknowledged. Senior faculties like Palash Baran Pal, Asit Kumar De and Avaroth Harindranath were always available for any kind of healthy discussion, be it academic or nonacademic. Divisional staffs Dola di, Pradyut da and Sangita di stood by my side whenever needed. I will definitely miss the spirited atmosphere of the Theory division's scholar room, first 363 and then 3319 which has been an inseparable part of my PhD life. I will also miss the lunch time's, tea time's and any other time's tittle-tattle in the institute canteen which provided me the all too necessary breaks from monotonous academic work. Mentioning this I would also like to thank the cooks of the institute canteen who have served me good food for the past five years. Among my seniors in the institute I am very much thankful to have Manindra da, Prasanta da, Goutam da, Arijit da, Avijit da, Sourav da, Dipankar da, Souvik da, Atanu da, Suvankar da, Avirup da, Anirban da, Mainak da, Ravindra da, Arindam da, Manas da, Somdeb da and Lab da because of all the lively discussions we had during our overlapping years. Similarly I am also lucky to have some amazing juniors like Rome, Sukannya, Shramana, Sudeshna, Rohit, Biswajit, Mugdha, Abhishek, Maireyee, Arghya, Bankim, Avik, Augnivo, Samrat, Aranya, Supriyo, Dipak, Anisa and Kajari.             

During my five years of PhD I stayed in the institute housing MSA-II where I have so many memorable experiences to cherish about. I want to thank the institute for providing me the accommodation and other necessary commodities which made my life easier in many ways. I am also grateful to our hostel cooks Shyamal da, Haran da and Biswajit for aptly satisfying my appetite. Contributions from helpful seniors like Niladri da, Rajani da and Sudip da are also appreciated. 

Friends have always occupied a large part of my life starting from the school days. Arup has been one friend who is singlehandedly carrying our friendship for the past 10 years starting from our graduation days. So, after the completion of my masters degree when I came to SINP, it is quite an anticipative fact that I became friends with my classmates within a very short period of time. Achyut has been a very dear friend of mine since then. Together we organized or tried to organize many tours during these five years which has always been great fun. Tapas, the most flamboyant person I know is one such friend with whom I can share the deepest of my secrets without having second thoughts in my mind. He has this ability to create a relaxed ambiance around him that has always pleased me. Naosad has been an indivisible part of my PhD life. Somehow I  became really attached with his quiet and carefree presence. We have spent so many silent moments together within the institute during which I found the required mental peace. Anshu and Kumar have also been very close to my heart with their amiable presence. I will miss all the trips we have organized, all the get togethers we have arranged. I will miss Gouranga's jokes, Pankaj's humor, Suvankar and Ashim's coordination, Kuntal Mondal's leadership, Satyajit and Kamalendu's simplicity, Sabuj's intellectuality, Debajyoti's immaturity, Sayanee's panic, Sukanta's informality, Sanjib's experiment, Arpan's sincerity and Mily's Sudip. I apologize to all of my other friends whose name has not been taken here. You will always be in my heart. 

Coming back to our hostel lives, I feel charmed enough to be a part of an elite group of friends. We would just talk through all those sleepless birthday nights together. This group consists of Amrita, Binita, Barnamala, Kuntal, Tirthankar, Chiranjib and Sanjukta. Tirthankar or Tirtha being my old friend from B.Sc, we were like comrades from the very first day of our PhD. Sometimes our wavelength matched so accurately, it left me almost speechless! One of us would seldom say no in case of a movie proposal from the other and hence we have watched uncountable movies together. Along with Tirtha, Chiranjib or Chiru, me and Sanjukta, we four formed another exclusive subgroup within the hostel. We would share all our happiness and sorrows with each other. Every time something noteworthy happened in any of our lives, we would sit together and announce it. Like Tirtha, Chiru is also in many ways similar to me. Apart from being close friends we were also joint secretaries of our Research Fellows' Association for one and a half years. It is not very common when you have friends like Chiru or Tirtha who would spontaneously agree for a costly trip to Goa, according to my wish. Last but not the least I would like to thank Sanjukta, my best friend or more than a friend, who would care for me like I am a part of her own family. We have had our differences in many occasions, but that did strengthen our friendship day by day, year by year. Her straightforwardness and inherent simplicity have taught me many subtle things and gradually made me a better person. She has even made valuable remarks about some parts of this manuscript.    

I would like to thank my parents, my teachers and my family for their valuable lessons throughout my life. Dadu and Nawdadu, I hope I can see life at least in part as you have seen it and inspired me to live it. Dinda dida has been an epitome of incessant affection and love since I was a day old and I hope I always get her by my side. I have never felt inhibited while traversing through my chosen path, and all the credit goes to my parents for sure. My father always understood and respected my opinion. Only once did he oppose it and as a result I joined IIT Kanpur for masters instead of IIT Bombay, my primary choice, which eventually turned out to be a great decision as I would have missed some of my lifetime friends Abhi, Satadru, Arif, Kamalika, Suchita, Rini and Santanu otherwise. My mother has always been this invisible source of energy for me which has helped me concentrate in my work alone. Without their support and goodwill I would never be pursuing my dreams! Subhro, my younger brother, will always be that toddler for me I poured all my love and care on. I would also like to mention about my cousins Dimpudada, Rahul and Om who are an integral part of my life. One of my favorite cousin brother Akash, who passed away few years ago only at the age of 21 has left a big void in my life since then. I hope, wherever he is, may his soul rest in peace.     

Finally I would like to thank the DST, INSPIRE for funding me through my graduation and master's days; Department of Atomic Energy, Govt of India for funding me throughout my PhD days and all others, whose names could not be taken here.

\vspace{2cm}

$~$\hspace{10.2cm}Aritra Bandyopadhyay

%% file: text/Abbrv.tex
\newpage
\chapter*{Abbreviations}

\begin{center}
\begin{tabular}{|c|c|}
\hline
{\bf AGS} & Alternating Gradient Synchrotron\\
{\bf BNL} & Brookhaven National Laboratory\\
{\bf BPY} & Braaten-Pisarski-Yuan \\
{\bf BW} & Breit-Wigner \\
{\bf CBM} & Compressed Baryonic Matter \\
{\bf CERN} & Conseil Europ{\'e}en pour la Recherche Nucl{\'e}aire\\
{\bf CF} & Correlation Function \\
{\bf COM} & Center Of Mass \\
{\bf DPR} & Dilepton Production Rate \\
{\bf EoS} & Equation of State \\
{\bf FAIR} & Facility for Antiproton and Ion Research \\
{\bf GSI} & Gesellschaft f{\"u}r Schwerionenforschung\\
{\bf GZ} & Gribov-Zwanziger \\
{\bf HIC} & Heavy Ion Collisions \\
{\bf HTLpt} & Hard Thermal Loop perturbation theory \\
{\bf IM} & Intermediate Mass \\
\hline
\end{tabular}
\end{center}

\begin{center}
\begin{tabular}{|c|c|}
\hline
{\bf IR} & InfraRed \\
{\bf (I/R)TF} & (Imaginary/Real) Time Formalism \\
{\bf JINR} & Joint Institute for Nuclear Research \\
{\bf LHC} & Large Hadron Collider \\
{\bf (L)LL} & (Lowest) Landau Level \\
{\bf (L/P)QCD} & (Lattice/Perturbative) QCD \\
{\bf LT} & Lowest Threshold \\
{\bf MEM} & Maximum Entropy Method \\
{\bf NICA} & Nuclotron-based Ion Collider fAcility \\
{\bf NJL} & Nambu-Jona-Lasinio \\
{\bf NS} & Neutron Star \\
{\bf OPE} & Operator Product Expansion \\
{\bf PL} & Polyakov Loop \\
{\bf PNJL}& Polykov-Nambu-Jona-Lasinio \\
{\bf (P)(N)LO} & (Perturbative) (Next-to) Leading Order \\
{\bf QED} & Quantum Electro Dynamics \\
{\bf QCD} & Quantum Chromo Dynamics \\
{\bf QGP} & Quark Gluon Plasma \\
{\bf QNS} & Quark Number Susceptibilities \\
{\bf RGE} & Renormalization Group Equations \\
{\bf RHIC} & Relativistic Heavy Ion Collider \\
{\bf SF} & Spectral Function \\
{\bf SPS} & Super Proton Synchrotron \\
{\bf ST} & Slavnov-Taylor \\
{\bf SVZ} & Shifman-Vainshtein-Zakharov \\
 \hline
\end{tabular}
\end{center}
\newpage

%% file: text/synopsis.tex
\chapter*{Synopsis}

It is a well known fact that there are four kinds of fundamental interactions in the universe: Electromagnetic, Weak, Strong and Gravity. Barring the attempts that are going on to unify these interactions using a Grand Unified Theory (first three) and String Theory (all four), individually each of these scintillating and rich interaction has been studied vividly over the ages using fundamental theories. Quantum Chromo Dynamics is one such theory which governs the dynamics of the strong interaction. Plenty of remarkable properties of strong interaction have been already explored using QCD. One such property is the asymptotic freedom, i.e. the decrease in the effective coupling constant of the strong interaction at large energies, making the quarks and gluons asymptotically free. Discovery of asymptotic freedom subsequently led to the search for a new phase, which resides in the domain of high temperature and/or density and in which quarks and gluons, normally confined within hadrons, are expected to form a deconfined state of matter. Physicists call this state as Quark Gluon Plasma~\cite{Muller:1983ed}, a plasma by its virtue, made by weakly interacting quarks and gluons.  

Direct motivations for studying this phase comes from Cosmology and Astrophysics. In standard theory of Big Bang, it is theoretically established that the younger the universe was, the hotter it was. After the period of inflation, when the age of universe was less than fractions of a microsecond, then the temperature was higher than 200 MeV. This very fact frames high temperature QGP as an essential phase in the evolution of the universe. In modern universe also, after undergoing various supernova explosion regular stars eventually form heavy and superdense objects, known as Neutron Stars. The density in the core of a NS, which is expected to be several orders higher than the normal nuclear matter density, drives the hadrons to be so closely packed that quarks and gluons no more remain confined within the hadrons, leading to a high density QGP phase. 

Theoretical modeling and experimental observations complement each other fascinatingly well while studying the diverse properties of QGP. To assess the theoretical modeling based on QGP phenomenology and QCD, one needs to have experimental observations via relativistic heavy ion collisions, which are taking place in different parts of the world now a days. In relativistic HIC, highly accelerated spherical nuclei get Lorentz contracted and its constituent hadrons lose their identity after the collisions. Due to asymptotic freedom these effectively become quark-quark interactions and most of the energetic quarks pass through each other creating secondary partons in the middle in the form of a fireball. The fireball then thermalizes and cools as it expands. The fireball in its initial condition is expected to attain a high temperature QGP phase, once it is thermalized. Excluding technicality, this mimics the early universe. So investigating this system helps us to know more about early universe. RHIC~\cite{Arsene:2004fa,Adams:2005dq,Adcox:2004mh} at BNL and LHC~\cite{Carminati:2004fp,Alessandro:2006yt} at CERN are exploring this type of QGP phase within their limits. On the other hand to mimic the core of the NS, slightly different kinds of experiments, i.e. fixed target experiments are being built in FAIR at GSI and NICA at JINR, Dubna~\cite{Senger:2004jw,Friman:2011zz,Toneev:2007yu,Sissakian:2009zza}. In these type of experiments, colliding nuclei tend to stay with each other after collision making the baryon densities higher in the center. There, the high density and relatively low temperature part of the QCD phase diagram will be explored.

Returning back to the theoretical modeling, there are broadly two ways to probe the characteristics of QGP, i.e. perturbative and nonperturbative. Perturbative QCD are analytical methods which works well in the domain of relatively higher temperature (where the running coupling $g$, in strong interaction is small) by taking the higher order radiative corrections into account. Naively, the
asymptotic freedom of QCD leads us to expect that perturbation theory should be a reliable guide at high temperature and/or high density. In fact, it has been recognized early on that this is not so. Technically, infrared divergences plague the calculation of observables at finite temperature, preventing the determination of higher order corrections. In order to cope with this difficulty, whose origin is the presence of massless particles, it has been suggested to reorganize perturbation theory, by performing the expansion around of a system of massive quasiparticles, because thermal fluctuations can generate a mass. The Hard Thermal Loop perturbation theory is one of these techniques~\cite{Andersen:1999fw,Haque:2014rua}. It amounts to a resummation of a class of loop diagrams, where the loop momenta are of the order of the temperature. Such diagrams are those which contribute to give the excitations a thermal mass. In QCD, there is an infinity of such diagrams, whose sum can be elegantly represented by a non-local effective action. The main idea of HTLpt is to use this effective action as the zeroth order of a systematic expansion. It seems to work well at a temperature of approximately 2 $T_c$ ($T_c$ is the pseudocritical transition temperature $\approx 160$ MeV) and above to calculate various physical quantities associated with QGP. But the time-averaged temperature generated at RHIC and LHC energies is quite close to $T_c$, where the running coupling $g$ is large and QGP could be completely non-perturbative in nature in the vicinity of phase transition. Thus it is necessary to consider the nonperturbative physics associated with QCD in this regime. There are various nonperturbative methods which are distinctly useful in this situation. Among them Lattice QCD, a numerical technique based on first principle QCD leads the way~\cite{Wilson:1974sk,Kogut:1982ds}. There also exist many effective QCD models eg, Colorsinglet model~\cite{Islam:2012kv}, Nambu-Jona-Lasinio model~\cite{Nambu:1961tp}, its Polyakov loop extended version~\cite{Fukushima:2003fw} and so on which try to imitate the actual situation. By and large, in this dissertation an effort has been made to exploit nonperturbative methods and use them to study different observables which characterizes QGP.  However, some of the works presented in this thesis bridges between perturbative and nonperturbative methods.

Different signals of the transient stages of QGP are needed to characterize this locally equilibrated state of the plasma. These are dilepton and photon emission from QGP, $J/\psi$ suppression because of Debye screening, strangeness enhancement, jet quenching and anisotropic flow. Most of the signals can be obtained via the spectral properties of the finally detected particles or the $n$-point correlation function, which eventually is related to various physical quantities associated with the deconfined state of matter. In this thesis we mainly focus on one such signal, the dilepton production rate in the nonperturbative domain of the QGP. The spectral function or the spectral discontinuity of the electromagnetic correlator is directly related to the production rate of dileptons (virtual photon) and real photons~\cite{Braaten:1990wp}. When a quark interacts with its antiparticle to form a virtual photon which subsequently decays into a lepton-antilepton pair, it is termed as dilepton. The dileptons created in the QGP phase, being colorless, do not suffer from any final state interaction and carry least contaminated information of the locally equilibrated QGP. This very fact makes DPR a desirable candidate for studying QGP. We also scrutinize some other important quantities, e.g the quark number susceptibilities, which is directly associated with the conserved number fluctuation of QGP and can be computed easily from the temporal part of the correlation function.

The dilepton spectra is a space-time integrated observable which has contributions coming from  various stages of the collisions and it is quite difficult to disentangle the contribution from different stages. Dileptons with higher (invariant) mass are less informative about QGP because of the dominance of Drell-Yan processes~\cite{Drell:1969km} and charmonium decays~\cite{Dominguez:2009mk} in that regime. But the low and intermediate mass dilepton production has optimized contribution from the QGP phase, in which the nonperturbative contributions could be important and substantial. 

As mentioned earlier, while talking about nonperturbative methods, one first turns to LQCD. Unfortunately the lattice techniques, which are solely applicable in the Euclidean spacetime, face some complications while computing the spectral function, an inherently Minkowskian object. Recently LQCD studies have provided critically needed information about the thermal dilepton rate using a probabilistic method known as Maximum Entropy method~\cite{Asakawa:2000tr,Nakahara:1999vy}. But nevertheless, given the uncertainty associated with lattice computation of dynamical quantities, e.g. spectral functions,  dilepton rate, and transport coefficients,  it is always desirable to have alternative approaches to include nonperturbative effects in DPR. 

A few such approaches are available in the literature. One approach is to use the well known fact, that the non-perturbative fluctuations of the QCD vacuum can be traced via phenomenological quantities, known as vacuum condensates. In standard perturbation theory values of such condensates vanish by definition. But in reality they are non-vanishing~\cite{Lavelle:1988eg} and thus the idea of the nonperturbative dynamics of QCD is signaled by the emergence of power corrections in physical observables through the inclusion of nonvanishing quark and gluon condensates in combination with the Green functions in momentum space.  In this thesis IM DPR using this nonperturbative power corrections will be presented~\cite{Bandyopadhyay:2016inp}.

There can also be other sources of nonperturbative contributions in the aforementioned observable, like the magnetic scale $g^2T$. Unlike electric scale magnetic scale is still a challenge for the theoreticians because the physics associated with the magnetic scale remains completely non-perturbative, being related to the confining properties of the theory. In another recent approach quark propagation in a deconfined medium has also been studied by taking into account this non-perturbative magnetic screening scale using the Gribov-Zwanziger action~\cite{Su:2014rma}. The gluon propagator with the GZ action is IR regulated and mimics confinement.  Interestingly, the resulting HTL-GZ quark collective modes consist of a new massless spacelike excitation along with the standard HTL dispersions. This new quark collective excitation may have important consequences for various physical quantities relevant for the study of QGP. In light of this, the DPR and the QNS using the non-perturbative GZ action~\cite{Bandyopadhyay:2015wua} will also be presented.  

A recently revealed captivating nature of non-central HIC is discussed in this thesis. In such collisions, a very strong anisotropic magnetic field is generated in the direction perpendicular to the reaction plane, due to the relative motion of the ions themselves~\cite{Shovkovy:2012zn, DElia:2012ems, Fukushima:2012vr}. The initial magnitude of this magnetic field can be very high ($eB\approx m_\pi^2$ at RHIC and $eB\approx 10m_\pi^2$ at LHC; $e$ is the electric charge of fermion, $B$ is the magnitude of the magnetic field and $m_\pi$ is the pion mass) at the time of the collision and then it decreases very fast, being inversely proportional to the square of time~\cite{Bzdak:2012fr,McLerran:2013hla}. The presence of magnetic field introduces another scale in the system in addition to temperature ($T$). These two separate scales, i.e., $T$ and $B$ present us to pursue two types of situations. Initially when the value of the magnetic field is large ($T^2 \ll eB$), one works with the strong field approximation. In later stages, by the time the quarks and gluons thermalize in a QGP medium, the value of the magnetic field decreases rapidly and one can work in the regime of weak field approximation  ($T^2 \gg eB$).  

This thesis also includes the effect of the magnetic field in the QGP state produced in HIC. Using Schwinger formalism, we obtain the spectral representation of the electromagnetic correlation function vis-a-vis the DPR completely analytically in presence of both strong~\cite{Bandyopadhyay:2016fyd} and weak~\cite{Bandyopadhyay:2017raf} background magnetic fields at finite temperature. In addition, we also discuss another interesting topic, namely the Debye screening,  which reveals some of the intriguing properties of the medium in presence of strong and weak magnetic field~\cite{Bandyopadhyay:2016fyd,Bandyopadhyay:2017cle}. 

%% file: text/intro.tex
\newpage
\chapter{Introduction}
\label{th_intro}

\begin{center}
\begin{figure}[t]
\begin{center}
\includegraphics[scale=0.8]{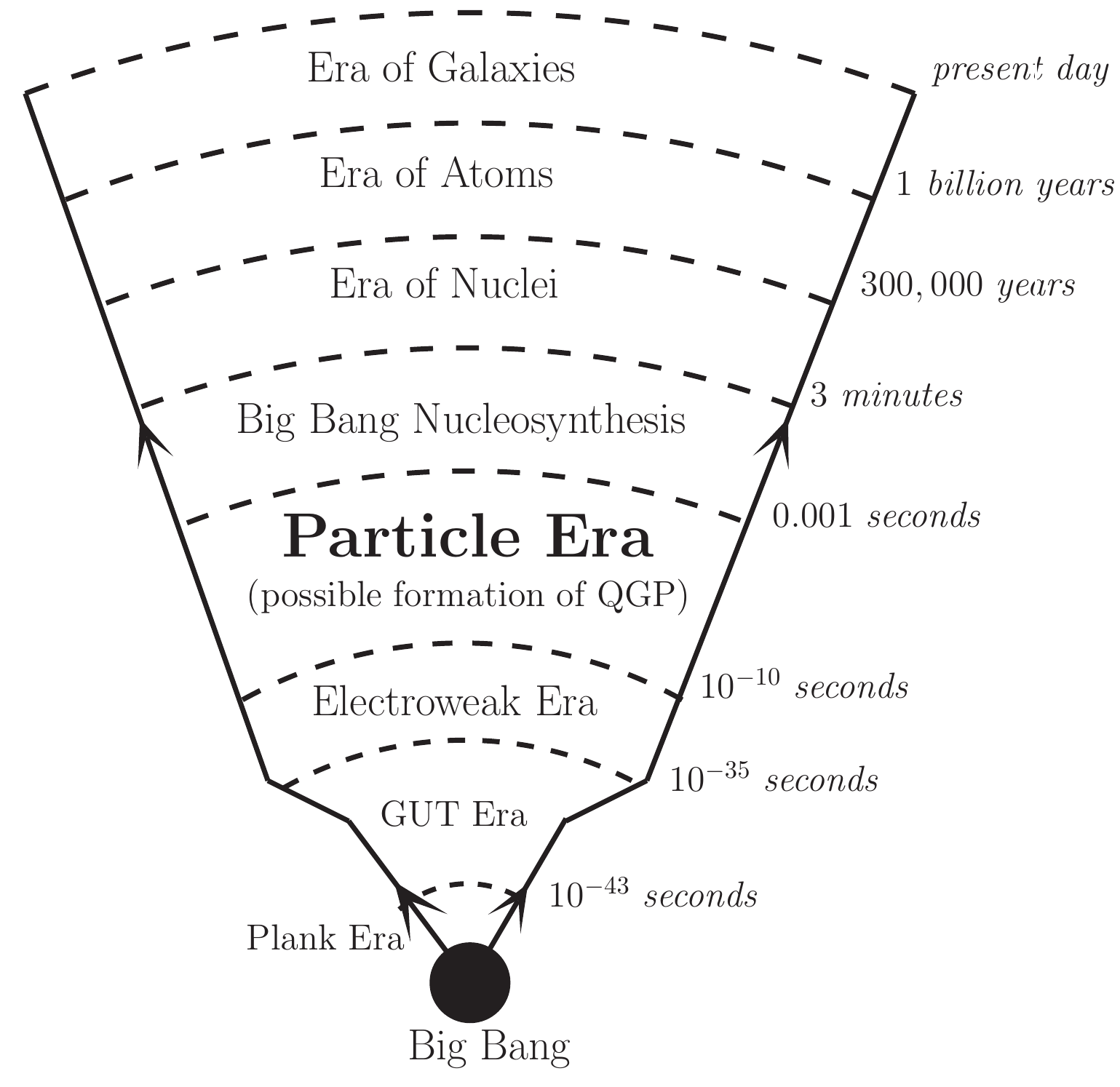}
\caption{Conjected timeline of the universe}
\label{bb1}
\end{center}
\end{figure}
\end{center}

Simply by looking at the chronicles of particle physics one cannot help but notice the reverse routes of knowledge and reality. This is as if, we are gradually traversing back in spacetime resolving all the enigmatic obstacles strewn across the path. The timeline of Avogadro's molecular theory (in 1815) to Rutherford-Bohr model (in 1913) and then the completion of the standard model through the discoveries of other elementary particles, e.g. leptons, hadrons and quarks is a pretty well known fact among the physicists. On the other hand in Fig.~\ref{bb1} one can look at the conjectured timeline of our expanding universe since the so called big bang to see that it is in correspondence with the aforementioned statement. 

In Fig.~\ref{bb1} one can see that the timeline of our universe is allocated into different eras which can be characterized by their different distinct features. But there are four fundamental interactions which spanned throughout this conjectured timeline as the four pillars of strength, governing whole of the universe since the very beginning. These are namely Gravitational, Weak, Strong and Electromagnetic interactions. In this dissertation our focus solely reside on the properties of {\it Strong interactions} in the particle era. More specifically, we will study a unique deconfined phase appearing in this condition, namely Quark Gluon Plasma.

\section{What is QGP?}
\label{intro_qgp}

QGP is a thermalized state of matter in which quasi-free quarks and gluons are deconfined from hadrons, so that the color degrees of freedom are manifested over a larger volume, as compared to the mere hadronic volume. To understand QGP, first one has to comprehend the asymptotic freedom. We know that QCD is the guiding theory for physicists to explore the strong interaction. The asymptotic freedom is a unique aspect of this non-Abelian gauge theory involving color charges.  Usually the participants of strong interactions, i.e. quarks and gluons are confined within the hadrons. But it was investigated that as the typical length scale associated with the system decreases, or the energy scale increases, the effective coupling strength of the strong interaction decreases, making the constituent quarks and gluons asymptotically free with respect to each other. This particular feature is known as asymptotic freedom, happening due to the anti-screening of color charges. It is just the opposite case of QED which deals with the screening of electric charges. Now, theoretically speaking, if one extrapolates this asymptotic freedom and keeps on increasing the energy of the system, at one point the quarks and gluons will no more remain confined within the hadrons and they will form the deconfined state of matter, QGP. 

\begin{center}
\begin{figure}[h]
\begin{center}
\includegraphics[scale=0.4]{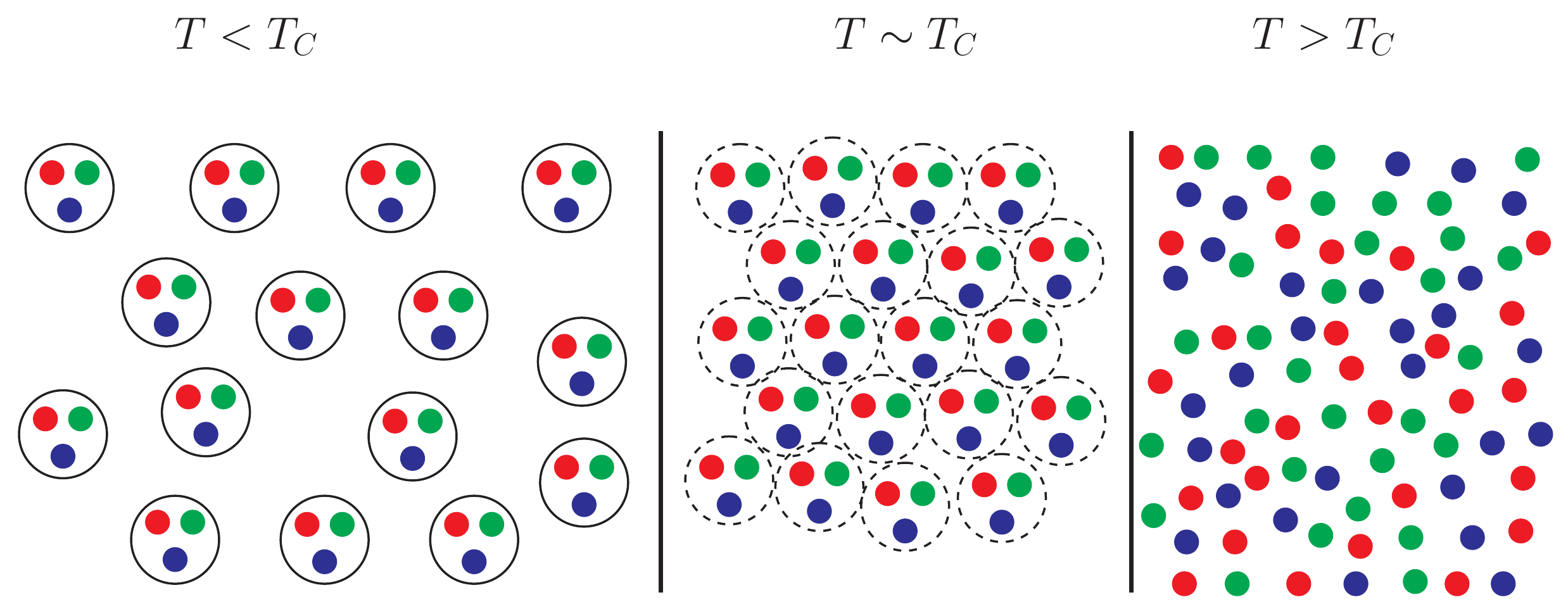}
\caption{Creation of high temperature QGP}
\label{htqgp}
\end{center}
\end{figure}
\end{center}
\begin{center}
\begin{figure}[h]
\begin{center}
\includegraphics[scale=0.4]{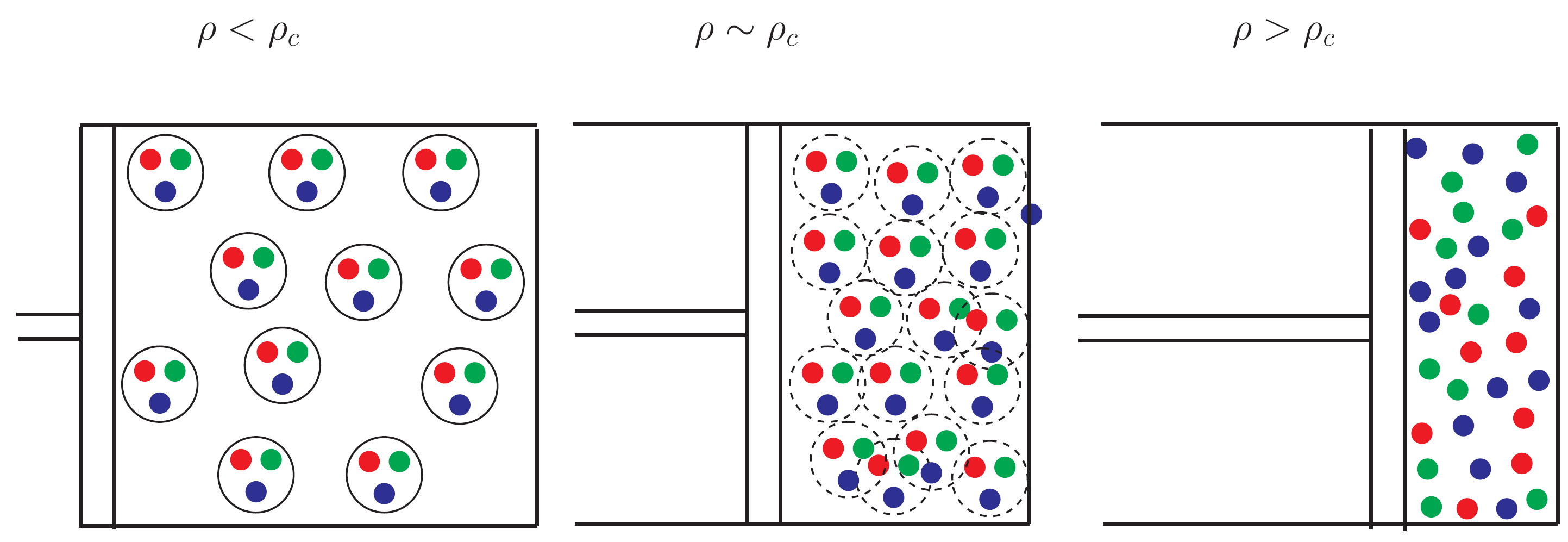}
\caption{Creation of high density QGP}
\label{hdqgp}
\end{center}
\end{figure}
\end{center}

Asymptotic freedom readily suggests two possibilities for the creation of QGP, i.e. at high temperature and/or at high density. If temperature of the QCD vacuum is increased gradually, after attaining a certain critical temperature hadrons of roughly similar size start to overlap with each other. Surpassing the critical temperature eventually causes the hadronic system to dissolve into QGP, a system of quarks and gluons (see Fig.~\ref{htqgp}). Similar deconfined state also develops when the nuclear matter density of the QCD vacuum is increased above a certain critical baryon density via compression (see Fig.~\ref{hdqgp}).

Reading up to this point, one can now expect to find QGP in the regimes of sufficiently high temperature and/or density. Proposed timeline of our universe suggests that in the early stages (a few microseconds) after the big bang, the temperature was well above the critical temperature. So, prior to hadronization there is a significant likelihood of finding QGP phase in the Particle Era (see Fig\ref{bb1}). Again the center of some compact stars like neutron star are known to be more dense than the critical baryon density invoking the scope of finding a high density QGP phase there. Apart from these two conjectural prospects one can also create QGP in the laboratory by colliding highly energetic heavy ions. In the next section we shall discuss more about this heavy ion collisions, it being the main experimental motivation behind the study of QGP.

\section{Mimicking early universe; what happens in HIC?}
\label{whinRHIC}

 In the previous section we observed that there are mainly two kinds of situation which can provide us the QGP state, i.e. very high temperature or very high density. Both of these situations can be achieved in heavy ion collisions. The main aim of the HIC experiments is to accelerate heavy and stable ions with as much energy as required for making them travel at a relativistic speed and then collide them. The basic intention of these experiments is not to focus on energy but to focus on energy density created and on the collective physics instead of the precision physics. 
 \begin{center}
\begin{figure}[h]
\begin{center}
\includegraphics[scale=0.45]{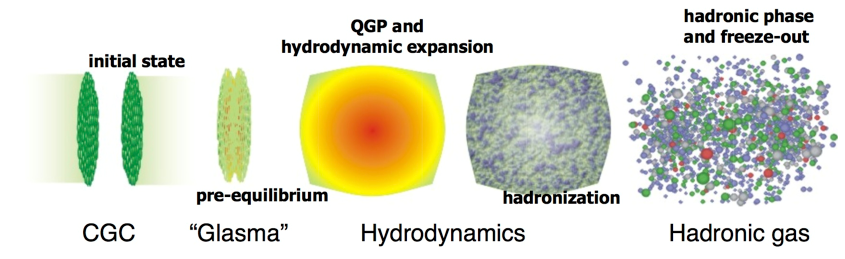}
\caption{Different stages of HIC}
\label{hic2}
\end{center}
\end{figure}
\end{center}
 Different stages of a typical HIC experiment is shown in Fig.~\ref{hic2}, which will now be discussed in some details. Generally in high energy heavy ion collisions highly energetic spherical nuclei get Lorentz contracted. They lose their identity and the interaction between nuclei becomes effectively quark-quark interactions producing numerous secondary particles in the process. When the energy of the colliding ions are upto $\mathcal{O}$($10$) GeV per nucleon, the incoming nucleons lose all their kinetic energies in the collisions and the colliding ions effectively stop each other. Also the quark-quark interaction is stronger at lower energies. So, in this case higher baryon densities can be achieved at the center, motivating us to look for the existence of high density QGP phase. The following experiments were/are/will be done in these kind of relatively lower energies. 
 \begin{itemize}
  \item AGS (1986-1996) @ BNL, USA : Colliding ions Si, Au ; $\sqrt{s} = 2.5-5.5$ GeV/A ; First generation experiment in which the energy density barely reached that required for QGP formation. 
  \item SPS @ CERN : Colliding ions O, Si, Pb ; $\sqrt{s} = 158-200$ GeV ; Also a first generation experiment which provided the indication of a possible deconfined state of QCD matter. 
  \item FAIR @ GSI, Germany : This future CBM experiment is designed to explore at lower bombarding energy of $(10-45)$ GeV/A, leading to $\sqrt{s} = 2.7-8.3$ GeV/A. It is expected to be operational soon and designed to create QCD matter at lower temperature and sufficiently high baryon density~\cite{Senger:2004jw,Friman:2011zz}. 
  \item NICA @ JINR, Dubna, Russia : Also a future experiment to study the properties of dense baryonic matter~\cite{Toneev:2007yu,Sissakian:2009zza}. 
 \end{itemize}
The produced matter in these future experiments is expected to be similar to that of the core of a neutron star, which could help us in understanding the low temperature and high baryon density domain of QCD. Precise theoretical knowledge of various quantities like equation of state, in-medium mass modification and decay of light vector mesons, transport coefficients, collective flow of charmonium and multistrange hyperons at this domain has important significance for the analysis of these kind of HIC experiments. 

 At much larger energies, in the so-called ultra relativistic heavy ion collisions, due to asymptotic freedom the strong interaction becomes feeble. As a result most of the quarks and hence the Lorentz contracted nuclei pass through each other and the central region becomes less dense and very hot, forming a fireball. Eventually the system thermalizes as secondary partons are created in the fireball. The density of those secondary partons grows due to multiple scatterings and the central thermalized system gradually cools as it expands. If the initial temperature of the thermalized fireball is above the critical temperature $T_c$ (according to LQCD $\sim 160$ MeV), one can expect that it should be in an equilibrated high temperature QGP state. Subsequently it will form hadronic matter via phase transition. This second system mimics the early universe (apart from technical differences). Thus investigating this lab-made system allows us to probe a part of the particle era in the timeline of our universe (Fig.~\ref{bb1}). These type of ultra relativistic heavy ion collisions are being carried out in the following experiments: 
 \begin{itemize}
  \item RHIC (2000-ongoing) @ BNL, USA : Colliding ions/nuclei d, 3He, Cu, Au, U;  $\sqrt{s} = 7.7-200$ GeV/A ; Some evidences of deconfinement and partonic degrees of freedom were found~\cite{Arsene:2004fa,Adams:2005dq,Adcox:2004mh}. 
  \item LHC (2010-ongoing) @ CERN : Pb-Pb and p-Pb Collisions ;  $\sqrt{s} = 2.76-5.5$ TeV/A; There are some clear evidences of the formation of the QGP phase~\cite{Carminati:2004fp,Alessandro:2006yt}. 
 \end{itemize}

 \begin{center}
\begin{figure}[t]
\begin{center}
\includegraphics[scale=0.8]{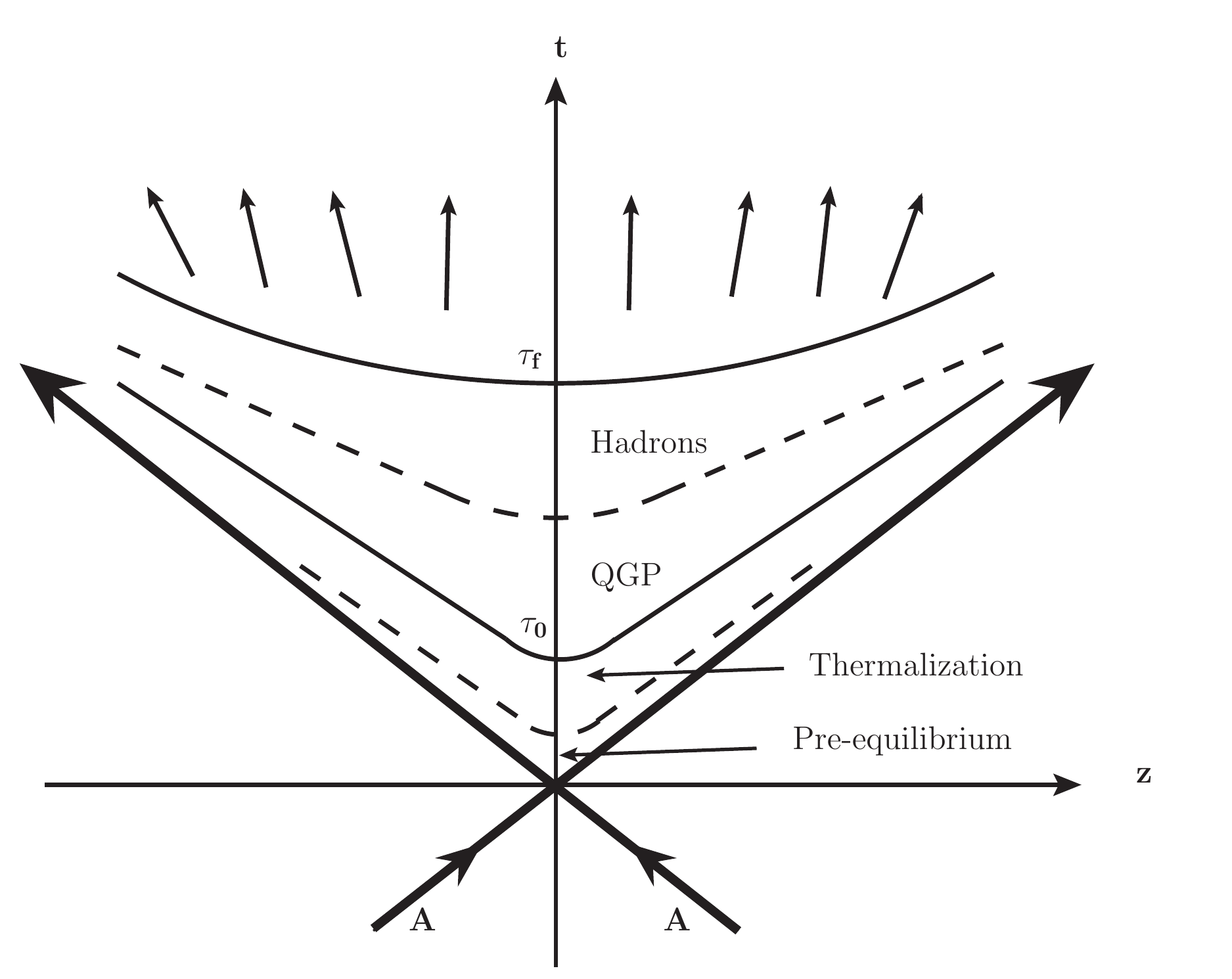}
\caption{Light cone diagram of a typical HIC}
\label{hic1}
\end{center}
\end{figure}
\end{center}

In Fig.~\ref{hic1} the timeline of an ultra relativistic heavy ion (AA) collision is shown through a light cone diagram. Just like the conjectured big bang timeline, here also we can split the timeline up into three separate stages depending on two characteristic timescale.
\begin{enumerate}
 \item $0<\tau<\tau_0$ is called the pre-equilibrium stage with $\tau_0$ representing the characteristic proper time of local thermalization in the system. Several models like Color Glass Condensate tries to explore this stage of the HIC.
 \item $\tau_0<\tau<\tau_f$ is the stage in which locally equilibrated QGP is formed. Then subsequently hydrodynamical evolution takes place leading to hadronization and eventually the system prepares for the freeze-out with $\tau_f$ being the freezeout time. All the theoretical methods to probe this particular stage of HIC will be discussed later in section \ref{methods_qgp}.
 \item Lastly $\tau>\tau_f$ stage constitutes the freezeout and post equilibrium. The freezeout happens in two phases. First the chemical freezeout where the number of each species is frozen and then the kinetic freezeout after which the kinetic equilibrium is no longer maintained. Post-equilibrium the hadronic interaction can be described by the Boltzmann equation. 
\end{enumerate}
In the next section we will discuss briefly about non-central HIC, which has the fascinating prospect of having a large anisotropic magnetic field in the colliding system.

\section{Non-central HIC; Generation of magnetic fields}
\label{nc_hic}

\begin{center}
\begin{figure}[h]
\begin{center}
\includegraphics[scale=0.6]{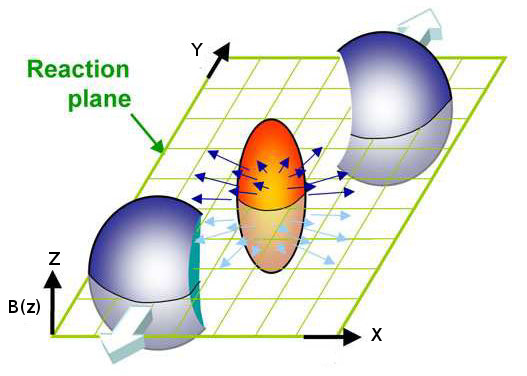}
\caption{Non-central HIC - schematic diagram}
\label{nchic1}
\end{center}
\end{figure}
\end{center}

\begin{center}
\begin{figure}[h]
\begin{center}
\includegraphics[scale=1.]{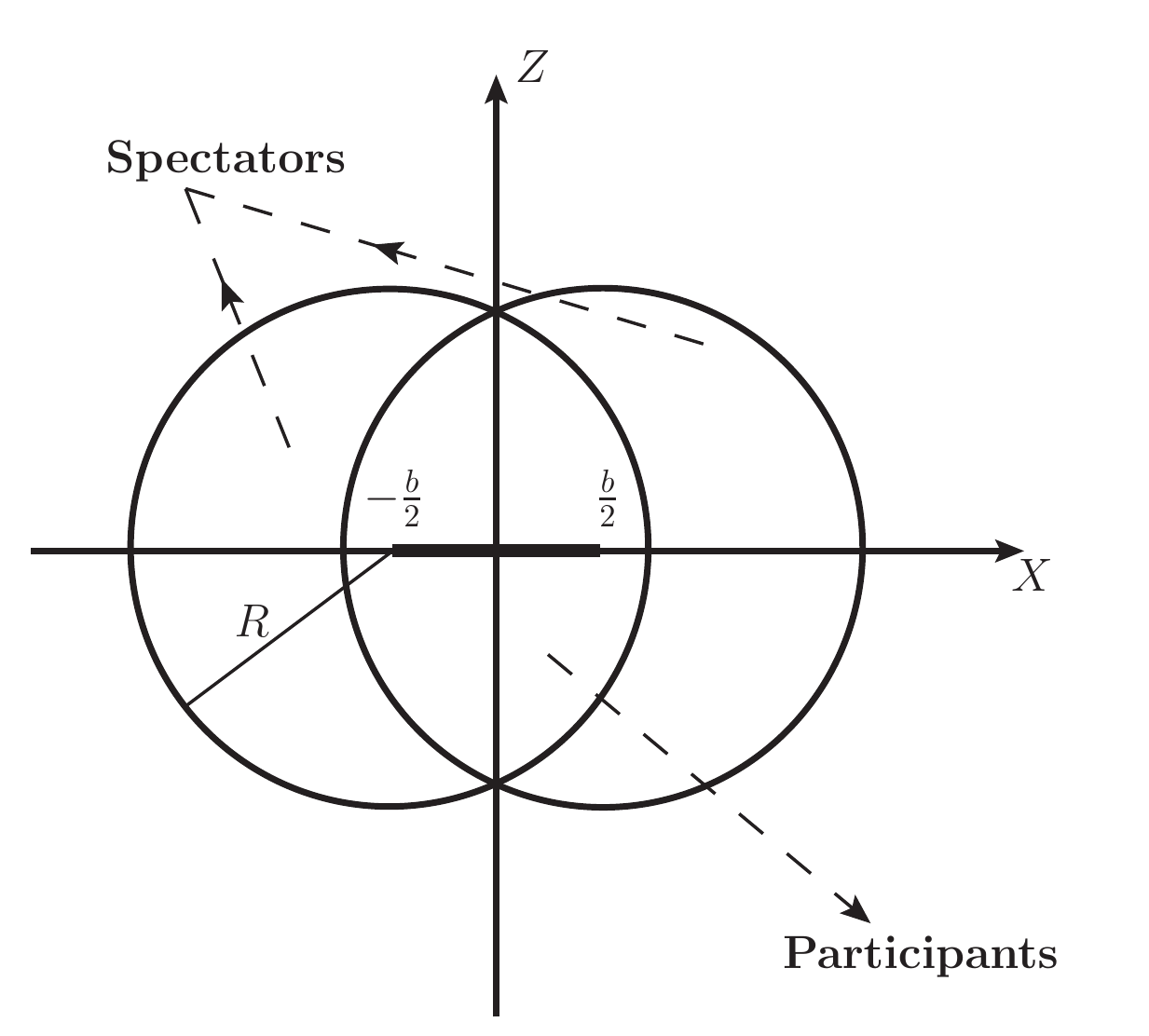}
\caption{Crosssectional view of a non-central HIC along the beam axis.}
\label{nchic}
\end{center}
\end{figure}
\end{center}

As mentioned in the last section, in HIC, the positively charged heavy ions move at relativistic speed. In this section we will discuss about the non-central HIC. A schematic diagram of a typical non-central HIC is shown in Fig.~\ref{nchic1}. The distance between the centers of the two colliding nuclei is called the impact parameter (distance between $-b/2$ and $b/2$ in Fig.~\ref{nchic}).  Larger the impact parameter is, greater are the chances that the collision becomes more non-central. In Fig.~\ref{nchic}, a cross-sectional view of a non-central HIC along the beam axis ($Y$ axis in Fig.~\ref{nchic}) is shown. The $Z=0$ plane is the reaction plane. The region where the two nuclei overlap contains the participant particles which form the fireball. Rest of the particles are called spectators, which do not scatter at all and travel almost with the same rapidity as the beam rapidity. Because of this relative motion between the colliding participants and the collision unaffected spectators, in non-central HIC a huge anisotropic magnetic field can be generated in the direction perpendicular to the reaction plane.  

The initial magnitude of the produced anisotropic magnetic field is estimated to be of the order of $eB_z \approx 10^{-1} m_\pi^2$ ($m_\pi^2 =10^{18}$ Gauss) at the SPS energies, $eB_z\approx m_\pi^2$ at RHIC and $eB_z\approx 15m_\pi^2$ at LHC. Such scales are relevant to QCD and that is why non-central HIC is gaining more and more attention from the heavy ion community. Recently various studies of the magnetized medium revealed several novel effects like the chiral magnetic effect, magnetic catalysis and inverse magnetic catalysis, thereby nourishing the demand of theoretical embodiment of non-central HIC. In chapter \ref{th_mag} and chapter \ref{th_ds} of this thesis we will study some of the required modifications of the present theoretical tools to appropriately investigate QGP in presence of an external magnetic field in the medium. 

Presently in the next section we will briefly review the present theoretical tools that are used in the literature for studying various properties of QGP.

\section{Methods of studying QGP}
\label{methods_qgp}

The discovery and characterization of the properties of QGP remains one of the best orchestrated international efforts in modern nuclear physics. This subject is presently actively studied at particle accelerators, where one collides heavy nuclei, moving at nearly the speed of light, in order to produce the hot and dense deconfined state of matter in the laboratory, as discussed in the previous sections. The RHIC and LHC studying the collisions of heavy nuclei at relativistic energies continue to generate a wealth of data which is being analyzed to provide valuable information about the nature of the ephemeral matter thus created. This calls for a better theoretical understanding of particle properties of hot and dense deconfined matter, which reflect both the static and the dynamical properties of QGP. The hot and dense matter produced in high energy heavy-ion collisions is a many particle system whose study seeks theoretical tools from an interface of particle physics and high energy nuclear physics. This requires the systematic use of QCD (both perturbative and nonperturbative methods) with a strong overlap from (i) Finite temperature and density field theory, (ii) Relativistic fluid dynamics, (iii) Kinetic or transport theory, (iv) Quantum collision theory, (v) String theory and (vi) Statistical mechanics and thermodynamics. 

In the following part of this section we will discuss two methods, i.e. perturbative and nonperturbative methods of QCD, mainly used to study the characteristics of QGP.

\subsection{Perturbative methods}
\label{methods_pert_qgp}

PQCD are analytical methods which take into account the radiative corrections as higher order perturbations. Because higher order radiative corrections depend on higher and higher powers of the running coupling $g$, in strong interaction, PQCD works well in the domain of relatively higher temperature (where the running coupling $g$ is small). Naively, because of the asymptotic freedom in QCD one naturally expects that bare perturbation theory should be a reliable guide at high temperature and/or high density. But eventually it has been realized that this is not the case. Infact, bare PQCD is accompanied by the infrared divergences which plague the calculation of observables at finite temperature, thereby preventing the determination of higher order corrections. It was recognized early on that the origin of this difficulty is the presence of massless particles in the system.  Another major problem that hindered the progress of bare PQCD was the apparent gauge dependence of the gluon damping rate. Braaten and Pisarski~\cite{Braaten:1989kk} resolved this problem of gluon damping rate, hence concluding that one-loop bare perturbative calculations are incomplete and in order to get the correct result one must resum an infinite subset of diagrams. To cope with this, different reorganized perturbation techniques have been proposed. 

Before going into the reorganized perturbative techniques, let us discuss the separation of scales when the temperature is higher than any intrinsic mass scale and the running coupling $g$ is lower than $1$. There are mainly three kind of scales as listed below :
\begin{itemize}
 \item {\bf Hard Scale} (length scale $\lambda_{\rm hard} = 1/2\pi T$) :  Due to the thermal fluctuations the typical momenta in the hard scale is of the order of $T$. In this regime, purely perturbative contribution to the QCD thermodynamics ($g^{2n}$, $n$ is the loop order) comes out to be
$n_B(E)g^2 \approx g^2T/E \sim g^2$. Here $n_B(E)$ is the Bose Einstein distribution function.
 \item {\bf Soft or electric scale} ($\lambda_{\rm soft} \sim 1/gT$) : Due to the chromoelectric fluctuations the typical momenta in the soft scale is of the order of $gT$. This is called the electric scale because of the electric (Debye) screening mass, which is also of the same order. In this regime resummation of an infinite subset of diagrams is possible for a given order in $g$. In this case, the perturbative contribution to the QCD thermodynamics will be $n_B(E)g^2 \approx g^2T/E \sim g$, resulting odd powers of $g$ and $\ln g$ to creep into the theory. 
 \item {\bf Ultra-soft or magnetic scale} ($\lambda_{\rm mag} \sim 1/g^2T$) : Due to the chromomagnetic fluctuations, the typical ultra-soft momenta is of the order of $g^2T$. Because the magnetic screening mass is of the similar order, this is called the magnetic scale. Within this scale $n_B(E)g^2 \approx g^2T/E \sim g^0$, which shows that the magnetic scale is non-perturbative in nature. The physics associated with this scale is largely unexplored and related to the confining properties of the theory. 
\end{itemize}
HTLpt is one of the reorganized perturbative techniques~\cite{Andersen:1999fw,Haque:2014rua} which separates the hard ($\sim T$) and soft ($\sim gT, g<1$) scales, and amounts to a resummation of a class of loop diagrams. Such diagrams are those which contribute to give the excitations a thermal mass, thus remedying the infrared problem due to massless particles. The sum of such infinite diagrams can be elegantly represented by a non-local effective action in QCD. Main idea of HTLpt is to use this effective action as the zeroth order of a systematic expansion. Various HTLpt calculations and their agreement with other available sources suggest that it works well at a temperature of approximately 2 $T_c$ and above to calculate various physical quantities associated with QGP. 
 
There also exist efficient alternatives to HTLpt which are based on dimensional reduction~\cite{Mogliacci:2013mca,Andersen:2012wr,Ipp:2006ij, Vuorinen:2002ue,Vuorinen:2003fs} technique within effective field theory approaches. In dimensional reduction three of the aforementioned scales are well separated. The main idea of this technique is to integrate out the non-static massive modes, thus eventually becoming an effective theory of static electric modes. Dimensional reduction have been successfully used to evaluate the thermodynamic quantities. In this thesis, among effective/resummed perturbative methods, only HTLpt will be used. 

\subsection{Nonperturbative methods}
\label{methods_npert_qgp}

As mentioned in the previous subsection, perturbative methods work well in the regime of relatively higher temperature. But the time-averaged temperatures generated at RHIC and LHC energies are quite close to $T_c$, where the running coupling $g$ is quite large. So the lab-made QGP is expected to be completely non-perturbative in nature in the vicinity of phase transition. Thus nonperturbative methods play a really strong role in this regime. 

There are several nonperturbative methods which are distinctly useful in this situation. Among them LQCD, a numerical technique based on first principle QCD leads the way~\cite{Wilson:1974sk,Kogut:1982ds}. LQCD has been used to probe the behavior of QCD in the vicinity of $T_c$, where matter undergoes a phase transition from the hadronic phase to the deconfined QGP phase. At this point, 
the QCD thermodynamic functions and some other relevant quantities associated with the fluctuations of conserved charges at finite temperature and zero chemical potential have been very reliably computed using LQCD (see  e.g.~\cite{Borsanyi:2011sw,Borsanyi:2012cr,Bazavov:2013dta,Bazavov:2013uja,Bernard:2004je,Bazavov:2009zn,Bazavov:2012jq,Datta:2012pj}).  In 
addition, quenched LQCD has also been used to study the structure of vector meson correlation functions.  Such studies have provided critically needed information about the thermal dilepton rate and various transport coefficients at zero momentum~\cite{Ding:2010ga,Kaczmarek:2011ht,Aarts:2002cc,Karsch:2001uw} and finite momentum~\cite{Aarts:2005hg}.  But LQCD has its own difficulties while dealing with the finite chemical potential due to the infamous sign problem and uncertainties while evaluating dynamical quantities like spectral functions. This does not change the fact that it is always desirable to have alternative approaches to include non-perturbative effects in the system. 

Among the other non-perturbative methods, there are non-perturbative effective QCD models which are constructed with inputs from QCD symmetry and the first principle LQCD data. There exist many such models e.g., Colorsinglet model~\cite{Islam:2012kv}, NJL model~\cite{Nambu:1961tp, Nambu:1961fr}, its Polyakov loop extended version PNJL~\cite{Fukushima:2003fw,Ratti:2005jh,Islam:2014sea}, functional methods like Dyson-Schwinger Equation, chiral perturbation theory~\cite{Leutwyler:1993iq}, quasi-particle model~\cite{Peshier:1995ty} and so on. There are other analytical nonperturbative methods also, like Operator Product Expansion which include non-perturbative effects that can be handled in a similar way as in perturbation theory.  Some of these methods will gradually unfold with the progression of this dissertation. 

In the next section we will briefly discuss how some of the observables produced in high energy HIC, can be used to characterize QGP.

\section{Different signatures of QGP}
\label{sign_qgp}

In section \ref{whinRHIC} the evolution of the matter produced in the high or ultra high energetic HIC is discussed. The energy scale of the system suggests that during this evolution the produced matter first leave the hadronic phase to make a short trip to the deconfined QGP phase and then return back to the hadronic phase via subsequent expansion and cooling. This expansion dynamics is generally studied by ideal/viscous hydrodynamics with suitable initial conditions. Particles interacting during this short time when the matter is in the QGP phase should provide information about this locally equilibrated state of the plasma. 

Because of the collective behavior of the plasma, there is no unique signal which can identify QGP. Rather, accumulation of various signals of the transient stages of QGP are needed to characterize this deconfined state of matter. Broadly there are two classes of quantities which can characterize this QGP phase. One is the static quantities or the snap shot properties of QGP, generally evaluated for a given value of temperature and chemical potential. Another is the dynamic quantities obtained through the spacetime evolution of hot and dense fireball which are heavily linked with various experimental observations. However, phenomenology of HIC and their microscopic description is too large a subject to be discussed here. We list below some of the snap shot properties and the corresponding experimentally observed dynamic quantities: 
\begin{center}
\begin{tabular}{|c|c|}
\hline
Snap shot properties (statics) & Experimental observation (dynamics) \\
\hline 
 Thermodynamic quantities (Pressure, EoS)   &  Inputs to Hydrodynamics \\
 Transport coefficients & Inputs to non-ideal Hydrodynamics \\
 Various susceptibilities & Fluctuations, Critical point\\
 Screening of Plasma & Quarkonia ($J/\psi$, $\Upsilon$) suppression \\
 Collisional and radiative energy loss & Jet quenching, Anisotropic flow \\
 Dilepton/ Photon production rate & Dilepton/ Photon spectra \\
 \hline
\end{tabular}
\end{center}

The observables combining these static inputs and their corresponding dynamic quantities are known as the signatures for the QGP. Different signatures for the QGP phase, extensively used by the heavy-ion-community are briefly discussed below: 
\begin{itemize}
\item {\bf Quarkonia suppression due to Debye screening} : 
In QGP, the color charge becomes screened because of the presence of quarks and gluons in the plasma. This phenomenon is known as the Debye screening. Quarkonia is a bound state of a heavy quark and antiqaurk in the plasma. At high temperature Debye screening weakens the interaction between them and also because of the deconfined nature of the plasma, quarkonia gradually dissociates leading to the suppression of its production. Though the quarkonia suppression is out of the scope of this thesis, Debye screening is discussed as a by product of our calculation of correlation function, which subsequently reveals some of the interesting properties of the magnetized medium. 
\item {\bf Strangeness enhancement} :
In normal nuclear matter, the valance quarks are made up of mainly up and down quarks. So, the amount of the strangeness is relatively small in hadronic phase. But during the HIC experiments $s\bar{s}$ pairs are formed which subsequently becomes quasi-free and interacts with the neighboring quarks and antiquarks to form other strange particles. This way, in the transient deconfined phase, the value of the strangeness is enhanced. This is known as the strangeness enhancement which acts as a probe to identify the QGP phase.  
\item {\bf Jet quenching} :
In HIC experiments, the collision between particles with relativistic speed can produce jets of elementary particles which emerge from these collisions. The jets consist of partonic degrees of freedom which gradually form the hadronic state of matter through hadronization.  The high energetic partons produced initially in the form of jets undergo multiple interactions inside the fireball during which the energy of the partons is reduced through collisional and radiative energy loss. Jets are also detected by experimentally observed peaks in angular correlation with trigger jets. As a result of that, the $P_T$ spectra of hadronic $AA$ collision is degraded with respect to that in a $p-p$ collision and hence the jet is quenched. 
\item {\bf Anisotropic flow} :
In HIC, anisotropic flow is a measure of the non-uniformity of the energy, momentum or the number density with respect to the beam direction which affects the different secondary particles differently. It generally occurs in non-central collisions because of the nonzero value of impact parameter and the fireball gets deformed into an almond like shape. Thus its expansion in the transverse plane leads to the elliptic flow.
\item {\bf Dilepton and Photon spectra from QGP} : 
 A quark interacts with its antiparticle forming a real or virtual photon in the process, out of which the latter subsequently decays into a lepton-antilepton pair. This pair is termed as dilepton. The dileptons created in the QGP phase are colorless and interacts only through the electromagnetic interaction. Consequently, their mean free path is expected to be large enough such that they are very less likely to suffer from any final state interaction. Carrying least contaminated information of the locally equilibrated QGP makes dilepton/photon spectra a desirable candidate for studying QGP.  
\end{itemize}

This thesis will mainly focus on the dilepton production rate from the QGP phase which acts as an input to the dilepton spectra. The production rate of dileptons~\cite{Braaten:1990wp} is directly related to the spectral function or the spectral discontinuity of the electromagnetic correlator. Chapter \ref{th_dpr} will contain a more detailed discussion about the spectral functions and DPR.

In this dissertation, we shall also throw light on another important quantity to characterize QGP. The quark number susceptibilities are directly associated with the conserved number fluctuation and hence the charge fluctuation, a quantity which can distinctly differentiate between QGP and hadronic phase. Also QNS can be computed easily from the temporal part of the correlation function. 

We will finish the introduction by mentioning the scope of the present thesis in the next section.

\section{Scope of the thesis}

In chapter \ref{th_dpr} we discuss about the basic ingredients required for the thesis i.e. basics of QCD, Imaginary and Real Time Formalism, HTLpt, GZ action, the Correlation Function along with the Spectral Function and OPE. We will also discuss the scope of DPR both in absence and in presence of an external anisotropic magnetic field, it being the primary area of interest throughout this dissertation. We conclude the chapter by discussing some generalities about QNS. 

In chapter \ref{th_ope} we use the well known fact, that the non-perturbative fluctuations of the QCD vacuum can be traced via phenomenological quantities, known as vacuum condensates to evaluate the intermediate mass DPR in a QCD plasma~\cite{Bandyopadhyay:2016inp}. OPE is used to incorporate the nonperturbative dynamics of QCD in physical observables through the inclusion of non-vanishing quark and gluon condensates in combination with the Green functions in momentum space. 

In chapter \ref{th_gribov} we discuss another important source of non-perturbative contribution in DPR, namely the magnetic scale $g^2T$ in the HTL perturbation theory, which is related to the confining properties of the theory. This non-perturbative magnetic screening scale can be taken into account using the GZ action~\cite{Su:2014rma}. Interestingly a new spacelike mode was obtained while studying the resulting HTL-GZ quark collective modes. In view of probable important consequences of this new exciting mode, we evaluate the DPR using the non-perturbative GZ action~\cite{Bandyopadhyay:2015wua} in this chapter.  

Electromagnetic spectral properties and DPR in presence of a magnetized medium is studied in chapter \ref{th_mag}. The presence of magnetic field ($B$) introduces another scale in the system in addition to temperature. As the initial magnitude of the magnetic field produced in non-central HIC can be very high at the time of the collision and then decreases very fast, we pursue two types of situations in this chapter.  In initial stages, for $T^2 \ll eB$ one works with the strong field approximation. In later stages, by the time the quarks and gluons thermalize in a QGP medium, we work in the regime of weak field approximation ($T^2 \gg eB$). Using Schwinger formalism, we obtain the electromagnetic correlation function and hence the DPR completely analytically in presence of both strong~\cite{Bandyopadhyay:2016fyd} and weak \cite{Bandyopadhyay:2017raf} background magnetic fields at finite temperature. For the weak field case we use real time formalism unlike the rest of the thesis, which is based on imaginary time formalism.  

In chapter \ref{th_ds} we discuss the Debye screening in a hot and magnetized medium, which reveals some of the intriguing properties of the medium in presence of both strong and weak magnetic field~\cite{Bandyopadhyay:2016fyd,Bandyopadhyay:2017cle}. 

In chapter \ref{th_qns} we explore the GZ action within HTL once more to discuss another important observable which characterizes QGP, namely QNS, which is directly related with the number fluctuation and hence charge fluctuation. We compare our findings with that of our previous perturbative results and lattice data before concluding in chapter \ref{th_conclu} where we have also discussed some future directions.  

\newpage

%% file: text/DPR.tex
\chapter{Prerequisites}
\label{th_dpr}
 
 As mentioned in section \ref{sign_qgp}, this thesis will mainly focus on computation of different quantities, such as DPR, QNS and Debye Screening from the QGP phase. So, in this chapter we mainly discuss some preliminaries about the main ingredients used in this dissertation while computing the aforementioned observables. In section \ref{qcd_basics} we discuss basic formulation of QCD. Section \ref{itf_basics} and \ref{rtf} contains imaginary and real time formalism respectively as part of the formulation of finite temperature field theory. Basic formulations of HTLpt and GZ action are recalled in section \ref{htl} and \ref{gz_action}. As a bridging between perturbative and non-perturbative methods, the basic structure of OPE is discussed in section \ref{ope_intro}. Section \ref{mag_fermion} provides a prelude to the magnetized medium. As an inseparable part of every computation in this dissertation, the formulation of correlation function and spectral function is studied in section \ref{cf_sf}. Following sections \ref{dpr_intro}, \ref{dpr_formu} and \ref{dil_mag} contain a detailed discussion about the importance and the formulation of DPR both in presence and in absence of an external magnetic field. This discussion will act as the foundation for the next three chapters (\ref{th_ope}, \ref{th_gribov} and \ref{th_mag}). We have also discussed different techniques of evaluation of DPR and their scope in section \ref{dpr_eval}. We conclude the chapter by briefly mentioning some generalities about QNS in section \ref{qns_smgen}, which will act as the starting point of chapter \ref{th_qns}.

 
 \section{Quantum Chromo Dynamics}
 \label{qcd_basics}
 
 As already mentioned in section \ref{intro_qgp}, QCD is the governing theory of the strong interaction. QCD is a non-Abelian gauge theory which ideally belongs to the group $SU(3)_c$ when we consider that quark of a particular flavor can have three kinds of color associated with it ($N_c=3$) and as mediator of strong interaction there are eight ($d_A=N_c^2-1=8$) types of non-Abelian guage fields or gluons.
 
 The QCD Lagrangian is given by
 \bea
 \mathcal{L}_{\rm QCD} = \sum\limits_f\bar{\psi}_f\left(i\slashed{D}-m_f\right)\psi_f - \frac{1}{4}F_c^{\mu\nu}F^c_{\mu\nu}
 \label{qcd_lag}
 \eea
 where $F_c^{\mu\nu}$ is the field strength tensor for the non-Abelian case
 \bea
 F^c_{\mu\nu} = \partial_\mu A_\nu^c -\partial_\nu A_\mu^c + g~f_{abc}A_\mu^a A_\nu^b,
 \label{fst_nabelian}
 \eea
 $A_\mu^c$ is the non-Abelian gauge field with color index $c$ and $D_\mu = \partial_\mu - ig~T_cA_{\mu c}$ is the covariant derivative. $T_c$'s are the generators of the non-Abelian $SU(3)_c$ group which satisfies the Lie algebra
 \bea
 \left[T_a,T_b\right] = if_{abc}T_c,
 \eea
 with $f_{abc}$ as the antisymmetric structure constants of the group. Also, $f$ is the flavor index for quarks and in the 3-dimensional fundamental representation of $SU(3)_c$, $\psi_f$ can be represented as 
 \bea
 \psi_f = \begin{pmatrix}
 \psi^{red}\\
 \psi^{blue}\\
 \psi^{green}
 \end{pmatrix}_f
 \eea
where red, blue and green are the three typical colors associated with a quark of mass $m_f$. 

Appearance of the third term in Eq.~(\ref{fst_nabelian}) suggests that $F_c^{\mu\nu}F^c_{\mu\nu}$ term in the Lagrangian has terms like $g~\partial_\nu A_\mu^a f^{abc}A^{\mu b}A^{\nu c}$ and $g^2~f^{abc}f^{ajk}A^b_\mu A^c_\nu A^{\mu j}A^{\nu k}$ which correspond to three and four point gluonic vertices in Feynman diagrams, respectively. This shows that QCD is a self-interacting theory in contrast with QED, where photons have no self interaction. To study different aspects of QCD in medium one needs to use finite temperature field theory. In the next two sections, we will briefly discuss two such ways to incorporate the medium effects within the regime of strong interaction.


\section{Imaginary time formalism}
\label{itf_basics}

The imaginary time formalism was first proposed by Matsubara~\cite{Matsubara:1955ws} which is achieved by the substitution $t=i\tau$ with $t$ being the Minkowski or real time and $\tau$ being the Euclidean or imaginary one. Physically it is realized by the Wick rotation in the complex time plane as shown in Fig. \ref{wick_rot}. Below we list some of the important points of ITF: 

\begin{figure}[t]
\begin{center}
\includegraphics[height=7cm]{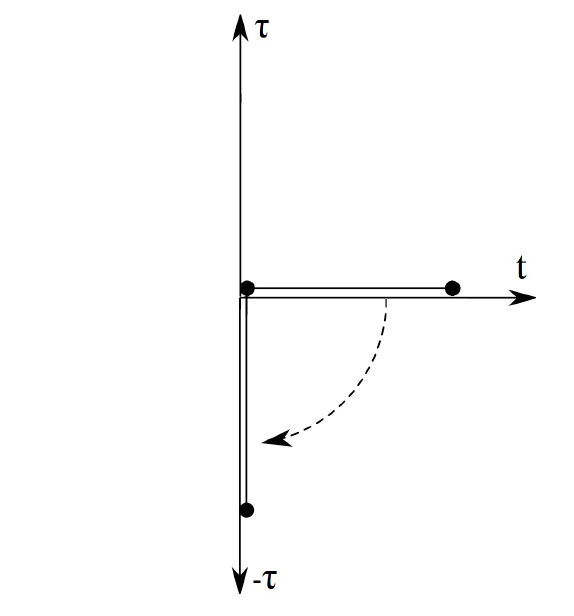}
\end{center}
\caption{Wick rotation of real time to imaginary time in complex time plane.}
\label{wick_rot}
\end{figure}

\begin{enumerate}
\item[$\bullet$]
The inverse temperature $\beta=1/T$ is related with $\tau$ because the evolution operator $e^{\beta {\cal H}}$ has the form of a time 
evolution operator ($e^{-i{\cal H}t}$) which implies $\beta=-it = \tau$ through analytic continuation. 
\item[$\bullet$] 
Because of the compactness of $\beta$, $\tau$ becomes finite : $0\le \tau\le \beta$. It amounts to the decoupling of space and time and the theory no longer remains Lorentz invariant.
\item[$\bullet$]
The thermal Green's function for $\tau > \tau'$ satisfies the following condition
\bea
G_\beta(\vec{x},\vec{x}';\tau,\tau'))&=& \pm \ G_\beta(\vec{x},\vec{x}';\tau ,\tau'+\beta),
\label{kms_greens}
\eea
which shows that within ITF the Dirac fields are anti-periodic in nature whereas the bosonic fields are periodic.
\end{enumerate}

The Feynman rules in finite temperature field theory are exactly the same as in zero-temperature except that the imaginary time $\tau$ is now compact with an extent $1/T$. To go from $\tau$ to frequency space, one has to perform a Fourier series decomposition rather than a Fourier transform. That accounts for the only difference with zero-temperature Feynman rules as loop frequency integrals are now replaced by loop frequency sums
\begin {equation}
  \int \frac{d^4P}{(2\pi)^4} \; \rightarrow \; iT \sum_{p_0} \int\frac{d^3p}{(2\pi)^3} \equiv T \sum_{\omega_n} \int\frac{d^3p}{(2\pi)^3},
\label{intro_fs}
\end {equation}
over the discrete imaginary-time frequencies known as Matsubara frequencies
\bea
p_0 = i\omega_n \!\!& = &\!\! 2 ni \pi T \hspace{2cm}\mbox{ for bosons} \;, \\
p_0= i\omega_n \!\!& = &\!\! (2n+1)i \pi T + \mu\hspace{1cm}\mbox{for fermions} \;,
\eea
which in turn implement the periodic and anti-periodic boundary conditions respectively.

Next we define the dimensionally regularized bosonic and fermionic sum-integrals, which will be extensively used later in chapter \ref{th_gribov}, as
\bea
  \sumintb_{P}& \;\equiv\; &
  \left(\frac{e^{\gamma_E}\Lambda^2}{4\pi}\right)^\epsilon\;
  iT\sum_{p_0=2ni\pi T}\:\int \frac{d^{3-2\epsilon}p}{(2 \pi)^{3-2\epsilon}}\;,\\ 
  \sumintf_{\{P\}}& \;\equiv\; &
  \left(\frac{e^{\gamma_E}\Lambda^2}{4\pi}\right)^\epsilon\;
  iT\sum_{p_0=(2n+1)i\pi T+\mu}\:\int \frac{d^{3-2\epsilon}p}{(2 \pi)^{3-2\epsilon}}\;,
\label{sumint-def}
\eea
where $3-2\epsilon$ is the dimension of space, $\gamma_E\approx0.577216$ the Euler-Mascheroni
constant, $\Lambda$ is an arbitrary momentum scale, $P=(p_0,p)$
is the bosonic loop momentum, and $\{P\}$ is the fermionic loop momentum. Because of the factor
$(e^{\gamma_E}/4\pi)^\epsilon$, after minimal subtraction 
of the poles in $\epsilon$ due to ultraviolet divergences, $\Lambda$ coincides 
with the $\overline{\rm MS}$ renormalization scheme.

Now the frequency sum in Eq.~(\ref{intro_fs}) for bosonic case is given by 
\be
T\sum\limits_{p_0}f(p_0=i\omega_n=2n\pi i T)=\frac{T}{2\pi i}\oint\limits_C
dp_0\frac{\beta}{2}f(p_0)\coth\frac{\beta p_0}{2}\ ,
\label{freq_boson}
\ee
where the contour $C$ can be deformed as $C\equiv C_1\cup C_2$, as shown in Fig.~(\ref{contour1}a):
\begin{figure}[h]
\begin{center}
\includegraphics[height=7cm]{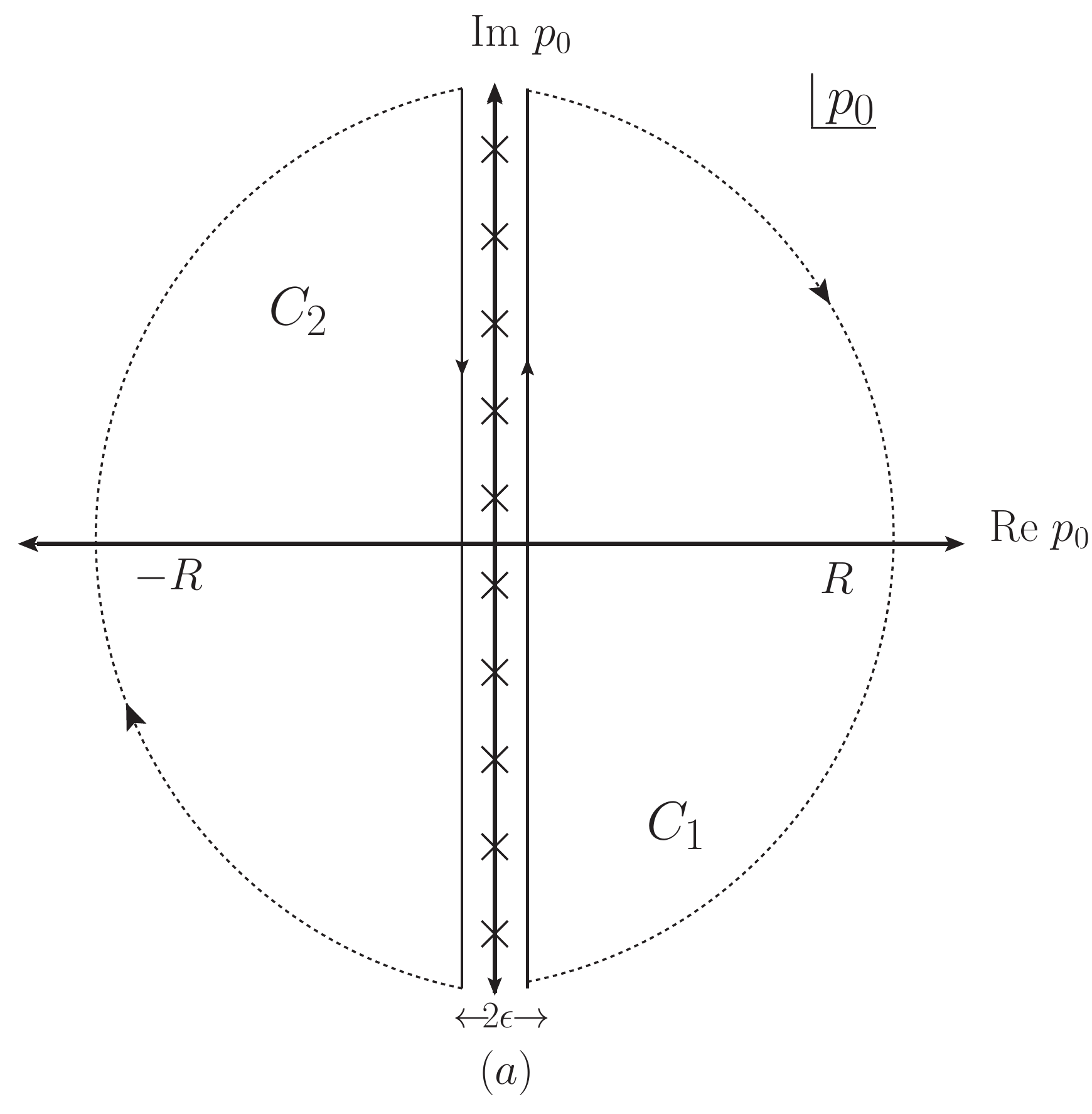}\hspace{1cm}\includegraphics[height=7cm]{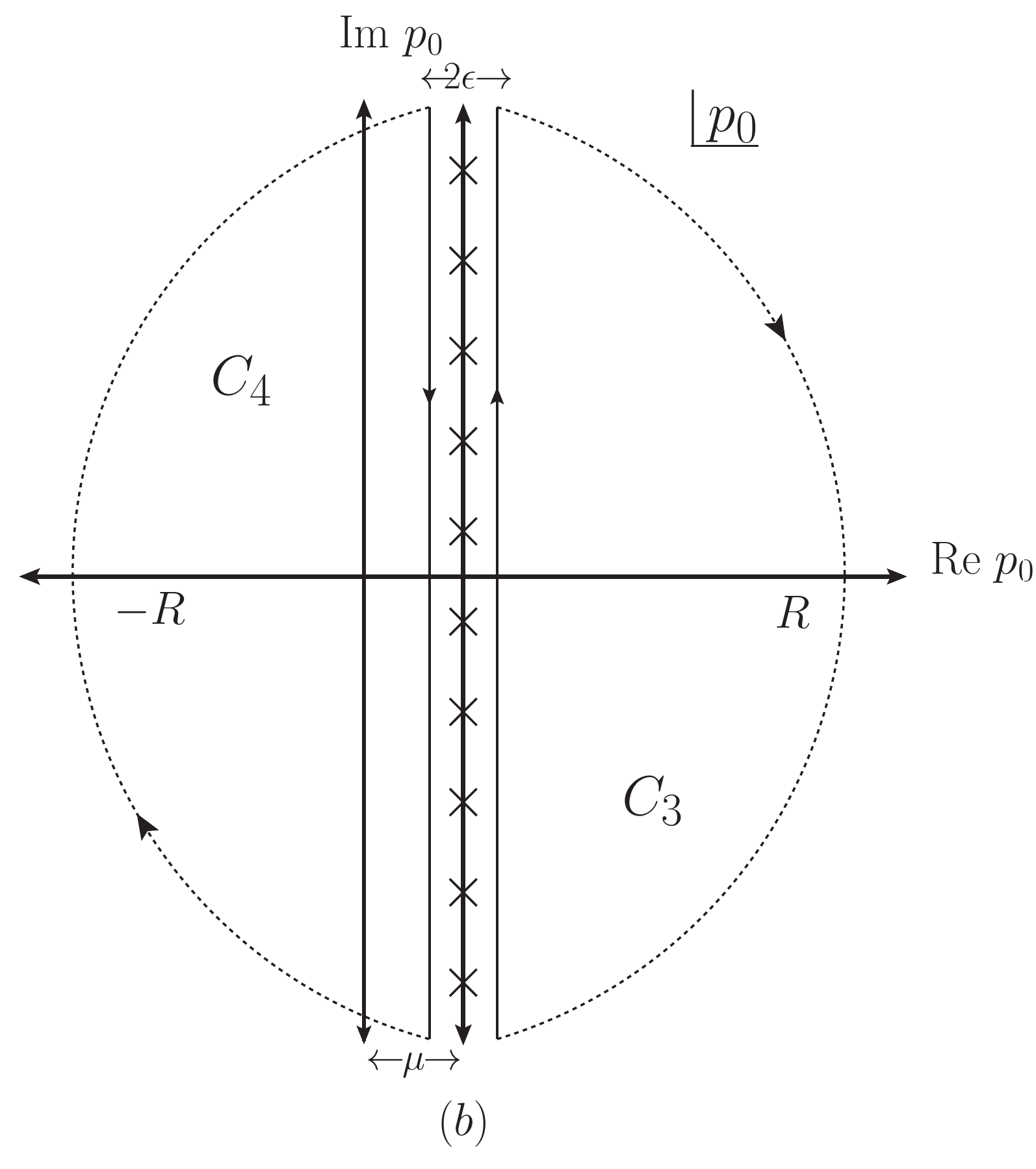}
\end{center}
\caption[Contours for bosonic and fermionic frequency sums]{Contours for (a) bosonic and (b) fermionic frequency sums.}
\label{contour1}
\end{figure}
The function $\frac{\beta}{2} \coth\frac{\beta p_0}{2}$ has poles at $p_0 = 2\pi n i T $ as shown by the crosses and is everywhere
else bounded and analytic. So with the deformed contour, the frequency sum in Eq.~(\ref{freq_boson}) can be rewritten as
\bea
&&T\sum\limits_{p_0}f(p_0=i\omega_n=2n\pi i T)\nn\\
&=&\frac{1}{2\pi i}\oint\limits_{C_1}
dp_0f(p_0)\frac{1}{2}\coth\frac{\beta p_0}{2}+\frac{1}{2\pi i}\oint\limits_{C_2}
dp_0f(p_0)\frac{1}{2}\coth\frac{\beta p_0}{2}
\nn\\
\!\!\!\!&=&\!\!\!\!\frac{1}{2\pi i}\int\limits_{-i\infty+\epsilon}^{i\infty+\epsilon}\!\!\!\!
dp_0f(p_0)\left[\frac{1}{2}+n_B(p_0)\right]
-\frac{1}{2\pi i}\int\limits_{-i\infty-\epsilon}^{i\infty-\epsilon}\!\!\!\!
dp_0f(p_0)\left[\frac{1}{2}+n_B(p_0)\right]
\eea
where $n_B(p)=1/\left(\exp(\beta p)-1\right)$ is Bose-Einstein distribution function, consisting of the medium effects.

One can also evaluate the frequency sum (\ref{intro_fs}) for fermionic case in the similar manner as
\bea
T\sum\limits_{p_0}f(p_0=i\omega_n=(2n+1)\pi i T + \mu)=\frac{T}{2\pi i}\oint\limits_{C'}
dp_0\frac{\beta}{2}f(p_0)\tanh\frac{\beta(p_0-\mu)}{2}
\label{freq_fermion}
\eea
where the contour $C'$ can be deformed in the similar manner as $C'\equiv C_3\cup C_4$, as shown in the Fig.~(\ref{contour1}b).

The function $\frac{\beta}{2} \tanh\frac{\beta (p_0-\mu)}{2}$ has poles at $p_0 = (2n+1)\pi i T +\mu$, as shown by the crosses and is everywhere
else bounded and analytic. With the deformed contour, the frequency sum in Eq.~(\ref{freq_fermion}) can be rewritten as
\bea
&&T\sum\limits_{p_0}f(p_0=i\omega_n=(2n+1)\pi i T+\mu)\nn\\
&=&\frac{1}{2\pi i}\oint\limits_{C_3}
dp_0f(p_0)\frac{1}{2}\tanh\frac{\beta(p_0-\mu)}{2}
+\frac{1}{2\pi i}\oint\limits_{C_4}
dp_0f(p_0)\frac{1}{2}\tanh\frac{\beta(p_0-\mu)}{2}\nn\\
\!\!\!\!&=&\!\!\!\!\frac{1}{2\pi i}\!\!\!\!\!\!\int\limits_{-i\infty+\mu+\epsilon}^{i\infty+\mu+\epsilon}\!\!\!\!\!\!
dp_0f(p_0)\!\!\left[\frac{1}{2}\!-\!n_F(p_0\!-\!\mu)\right]\!\!
+\!\frac{1}{2\pi i}\!\!\!\!\!\!\int\limits_{-i\infty+\mu-\epsilon}^{i\infty+\mu-\epsilon}\!\!\!\!\!\!
dp_0f(p_0)\!\!\left[\frac{1}{2}\!-\!n_F(\mu\!-\!p_0)\right]
\eea
 where $n_F(p)=1/\left(\exp(\beta p)+1\right)$ is Fermi-Dirac distribution function that takes into account the finite temperature effects.
 
 
 \section{Real time formalism}
 \label{rtf}
 
 Real time formalism was introduced after ITF, first by Schwinger and Keldysh~\cite{Schwinger:1960qe,Keldysh:1964ud} and then reformulated by Umezawa~\cite{Semenoff:1982ev,Takahashi:1996zn}. Similarly as in ITF, in RTF also, the Boltzmann weight is compared with the time evolution operator. One can write $e^{-\beta H} = e^{-iH(\tau-i\beta-\tau)}$, which can be thought to help the system evolve from time $\tau$ to $\tau-i\beta$, where $\tau$ may take complex values.
 \begin{center}
\begin{figure}[tbh]
\begin{center}
\includegraphics[scale=0.75]{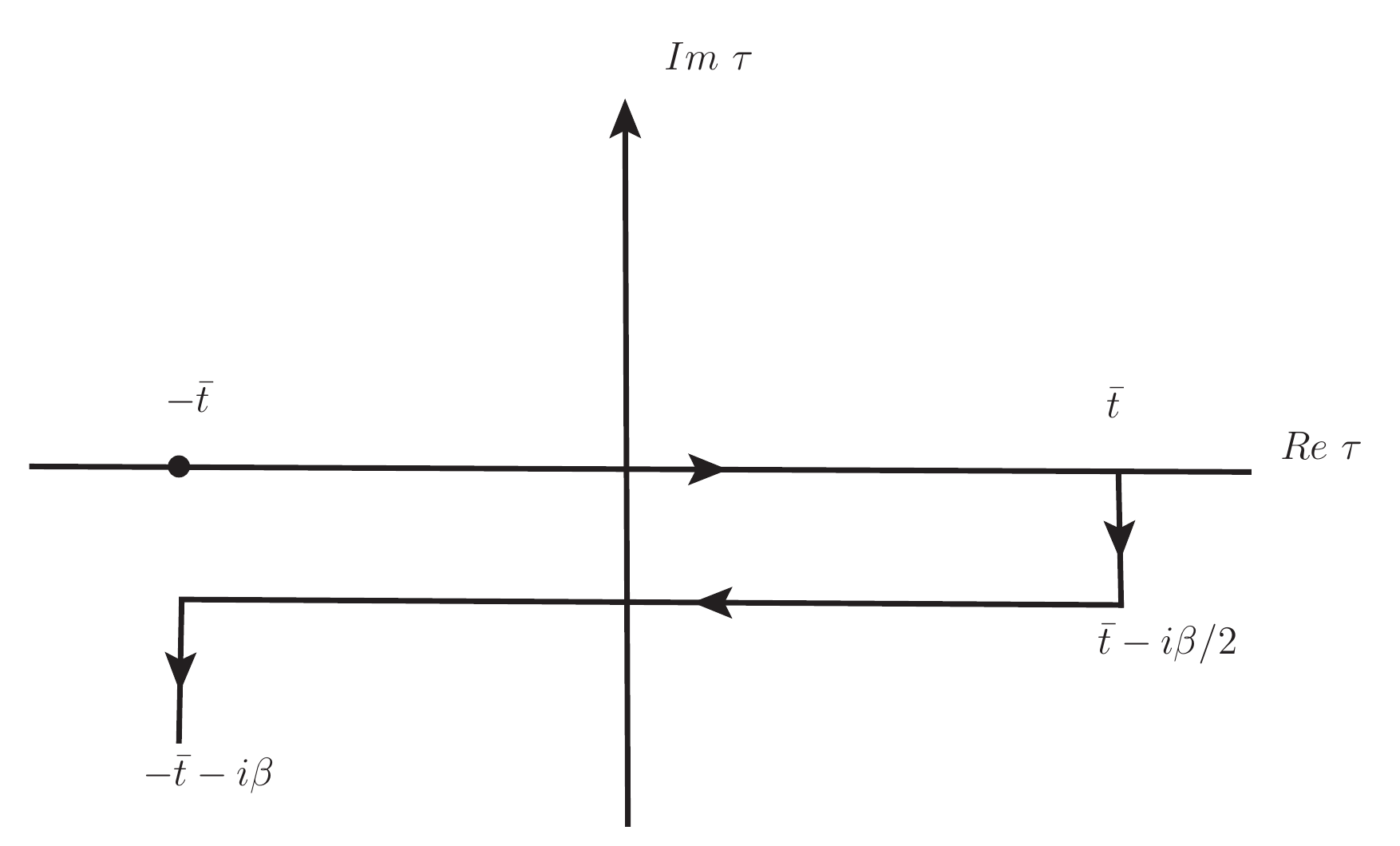} 
\caption{Path of the time contour in the Real time formalism}
\label{rtfc}
\end{center}
\end{figure}
\end{center} 
The properties of thermal propagator imposes certain conditions which become restricted by the domain 
 \bea
 -\beta \le {\rm Im}(\tau-\tau')\le 0.
 \eea
 Out of the various contour choices which satisfy this condition, ITF originates when time runs from $0$ to $-i\beta$. However, RTF would require one segment of the contour to run over the whole of real axis, with the two infinite intervals of time giving rise to a $2\times 2$ matrix as the propagator. In Fig. \ref{rtfc} a suitable time path is shown as it traverses the real axis from $-\bar{t}$ to $+\bar{t}$, then continues parallel to the imaginary axis up to $\bar{t}-i\beta/2$, going parallel to the real axis again up to $-\bar{t}-i\beta/2$ and finally again vertically to $-\bar{t}-i\beta$. This choice, shown in Fig. \ref{rtfc}, gives rise to a symmetric propagator for the case of scalar fields as :
 \bea
 \begin{pmatrix}
 D(\vec{x},\vec{x}';t,t') & D(\vec{x},\vec{x}';t,t'-i\beta/2) \\
 D(\vec{x},\vec{x}';t-i\beta/2,t') & D(\vec{x},\vec{x}';t',t) 
 \end{pmatrix}
 \!\!=\!\! \int\!\!\frac{d^4P}{(2\pi)^4} e^{-iP\cdot (X-X')} D(\vec{p},p_0)
 \eea
 where the momentum space scalar propagator is given by
 \bea
 D(\vec{p},p_0) \!\!=\!\!  \begin{pmatrix} \Delta_F + 2\pi i n_B ~\delta(P^2-m^2) & 2\pi i \sqrt{n_B(1+n_B)}~\delta(P^2-m^2) \\ 2\pi i \sqrt{n_B(1+n_B)}~\delta(P^2-m^2) & -\Delta_F^* + 2\pi i n_B ~\delta(P^2-m^2)\end{pmatrix}
 \eea
 with $n_B=n_B(\omega)$ is the Bose-Einstein distribution function and $\Delta_F$ is the free scalar propagator. 
 The fermionic propagator in RTF, will be used later in chapter \ref{th_mag}. In the next section we will briefly discuss about one of the leading perturbative methods to study QCD at finite temperature and finite chemical potential, namely HTLpt.


 \section{HTL perturbation theory}
 \label{htl}
The basic idea and the importance of HTLpt is already discussed in subsection \ref{methods_pert_qgp}. In this section we will briefly discuss the mathematical formulation of HTLpt as a foundation for chapters \ref{th_gribov} and \ref{th_qns}. After the idea of Hard Thermal Loop given by Braaten and Pisarski~\cite{Braaten:1989mz}, HTLpt was gradually developed by Andersen, Braaten and Strickland~\cite{Andersen:1999sf,Andersen:1999fw}. The HTLpt Lagrangian density can be written as a rearrangement of the in-medium perturbation theory for QCD. It reads as
\be
 {\cal L}=\left.({\cal L}_{\rm QCD}+{\cal L}_{\rm HTL})\right|_{g\rightarrow\sqrt{\delta}g}+\Delta{\cal L}_{\rm HTL} \, , 
\label{total_lag} 
\ee
where $\Delta{\cal L}_{\rm HTL}$ is the HTL counterterm, ${\cal L}_{\rm QCD}$ is given in Eq.~(\ref{qcd_lag}) and the added HTL term is~\cite{Andersen:2003zk}
\be
 {\cal L}_{\rm HTL}=(1-\delta)i m_q^2\bar\psi\gamma^\mu\left\langle\frac{x_\mu}{x\cdot\! D}\right\rangle_x\psi-\frac{1}{2}(1-\delta)
 m_D^2 {\rm \Tr}\left( G_{\mu\alpha}\left\langle\frac{x^\alpha x_\beta}{(x\cdot\! D)^2}\right\rangle_x G^{\mu\beta}\right) \, ,
\label{htl_lag}
\ee
where $x^\mu = (1, {\bf\hat{x}})$ is a light-like four-vector. The angular bracket indicates an average over the direction of the three dimensional unit vector ${\bf\hat{x}}$. $m_D$ and $m_q$ can be recognized as the Debye screening mass and the thermal quark mass, respectively. They account for the screening effects, which we shall discuss in chapter \ref{th_ds} also. The one-loop running strong coupling, $g^2=4\pi\alpha_s$, is 
\bea
g^2(T)=\frac{48\pi^2}{(33-2N_f)\ln\left (\frac{Q^2_0}{\Lambda_0^2} \right )} , \label{alpha_s}
\eea
where $N_f$ is the number of quark flavors and $Q_0$ is the renormalization scale, which is usually chosen to be $2\pi T$ unless specified.  We fix the scale $\Lambda_0$ by requiring that $\alpha_s$(1.5 GeV) = 0.326, as obtained from lattice measurements~\cite{Bazavov:2012ka}. For one-loop running, this procedure gives $\Lambda_0 = 176$ MeV.

A HTLpt is formulated by treating $\delta$ as a formal expansion parameter. As we can see from Eq.~(\ref{htl_lag}) and Eq.~(\ref{total_lag}), the HTLpt Lagrangian reduces to the QCD Lagrangian if we set $\delta=1$.  In HTLpt physical observables are calculated first by expanding in powers of $\delta$, then truncating at some specified order, and finally setting $\delta = 1$. For example, to obtain the one-loop or the leading order (LO) results, one has to expand the corresponding observable up to order $\delta^0$.  Here we also note that HTLpt is gauge invariant order-by-order in the $\delta$ expansion. So, eventually, the results obtained are independent of the gauge-fixing parameter. 

\begin{figure}[h]
\begin{center}
\includegraphics[height=14cm, width=15cm]{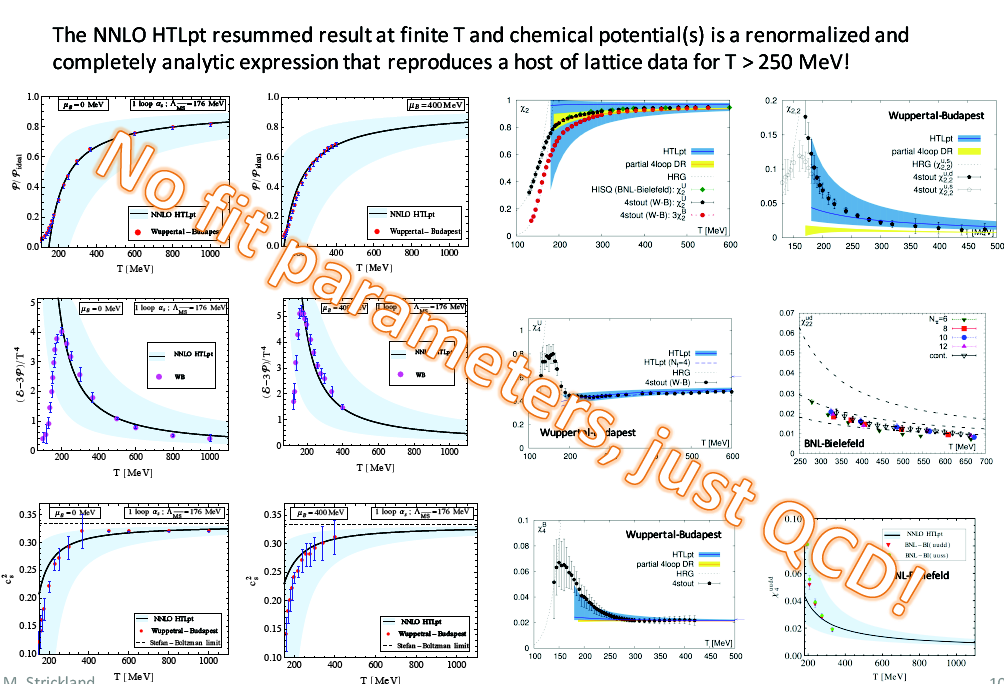} 
\caption[3-loop HTLpt Thermodynamic results in one frame]{All thermodynamic results using 3-loop HTLpt compared with the lattice data. Results are quoted from~\cite{Haque:2014rua} and most of them are used in the thesis of Najmul Haque~\cite{Haque:2014cda}.}
\label{3loop_td}
\end{center}
\end{figure}
In recent years, several novel results have been obtained using HTLpt, the most widespread result being the one-loop, two-loop and three-loop QCD equation of state and corresponding thermodynamic quantities~\cite{Haque:2012my,Haque:2013qta,Haque:2013sja, Haque:2014rua,Strickland:2014zka, Andersen:2015eoa,Mogliacci:2013mca,Andersen:2011sf, Andersen:2010wu,Andersen:2010ct,Andersen:2009tc, Andersen:2009tw,Andersen:2003zk, Andersen:2002ey, Andersen:2000yj,Andersen:1999va,Andersen:1999sf,Andersen:1999fw,Blaizot:2001ev,Blaizot:2000fc,Blaizot:1999ap,Blaizot:1999ip}. In addition, HTLPt has also been used to calculate several physical quantities which are relevant to the deconfined state of matter.  These include the DPR~\cite{Braaten:1990wp,Greiner:2010zg}, QNS~\cite{Chakraborty:2001kx, Chakraborty:2003uw}, photon production rate~\cite{Baier:1993zb}, single quark and quark anti-quark potentials~\cite{Mustafa:2004hf,Chakraborty:2006md,Chakraborty:2007ug,Laine:2006ns, Dumitru:2007hy,Dumitru:2009fy,Thakur:2012eb}, fermion~\cite{Pisarski:1993rf}, photon~\cite{Abada:2011cc} and gluon~\cite{Braaten:1989kk, Braaten:1990it} damping rate, jet energy loss~\cite{Braaten:1991we,Thoma:1990fm,Romatschke:2003vc,Romatschke:2004au,Mustafa:2004dr}, plasma instabilities~\cite{Rebhan:2008uj,Attems:2012js}, and lepton asymmetry during leptogenesis~\cite{Kiessig:2011fw,Kiessig:2011ga}. In Fig. \ref{3loop_td} from one of our recent publication we demonstrate a set of thermodynamic quantities obtained in three-loop HTLpt, which agree remarkably well with the available LQCD data. In view of this we will use HTLpt along with the non-perturbative Gribov-Zwanziger action to evaluate the DPR in chapter \ref{th_gribov} and the QNS in chapter \ref{th_qns}. We will briefly introduce the GZ action and its consequences in the next section. 


\section{Gribov-Zwanziger action and its consequences}
\label{gz_action}

Gribov showed in 1978~\cite{Gribov:1977wm} that in a non-Abelian gauge theory, fixing the divergence of the potential does not commute with the gauge fixing. Unfortunately, the solutions of the differential equations, which specify the gauge fixing with vanishing divergence, can have several copies (Gribov copies) or none at all. This is known as Gribov ambiguity. To resolve this ambiguity, the domain of the functional integral has to be restricted within a fundamental modular region, bounded by Gribov horizon. Following this, in 1989 Zwanziger~\cite{Zwanziger:1989mf} derived a local, renormalizable action for non-Abelian gauge theories which fulfills the idea of restriction. He also showed that by introducing this GZ action the divergences may be absorbed by suitable field and coupling constant renormalization. The GZ action is given by~\cite{Vandersickel:2011zc}
\bea
S_{GZ} &=& S_0 + S_{\gamma_G} ; \\
S_0 &=& S_{YM} + S_{gf} + \int d^Dx\left(\bar{\phi}_\mu^{ac}\partial_\nu D_\nu^{ab} \phi_\mu^{bc}-\bar{\omega}_\mu^{ac}\partial_\nu D_\nu^{ab} \omega_\mu^{bc}\right) + \Delta S_0;\\
S_{\gamma_G} &=& \gamma_G^2\int d^Dx~g~f^{abc}A_\mu^a\left(\phi_\mu^{bc}+\bar{\phi}_\mu^{bc}\right) + \Delta S_\gamma ;
\eea
where $(\phi_\mu^{bc},\bar{\phi}_\mu^{bc})$ and $(\omega_\mu^{bc},\bar{\omega}_\mu^{bc})$ are a pair of complex conjugate bosonic and Grassmann fields respectively, introduced due to localization of the GZ action. $S_{YM}$ and $S_{gf}$ are the normal Yang-Mills and the gauge fixing terms of the action and $D$ is the dimension of the theory. $\Delta S_0$ and $\Delta S_{\gamma_G}$ are the corresponding counterterms of the $\gamma_G$ independent and dependent parts of the GZ action. $\gamma_G$ is called the Gribov parameter. In practice, $\gamma_G$ can be self-consistently determined using a one-loop gap equation and at asymptotically high temperatures it takes the following form\footnote{Equation (\ref{Gribov_para}) is a one-loop result.  In the 
vacuum, the two-loop result has been determined \cite{Gracey:2005cx} and the 
Gribov propagator form (\ref{modified_gluon_prop}) is unmodified.  Only 
$\gamma_G$ itself is modified to take into account the two-loop correction.  To 
the best of our knowledge, this would hold also at finite temperature.}  ~\cite{Su:2014rma,Zwanziger:2006sc,Fukushima:2013xsa} 
\bea
\gamma_G = \frac{D-1}{D}\frac{N_c}{4\sqrt{2}\pi}g^2T.  \label{Gribov_para}
\eea

We know that gluons play an important role in confinement. Using the GZ action~\cite{Gribov:1977wm,Zwanziger:1989mf} the issue of confinement is usually tackled kinematically with the gluon propagator in covariant gauge taking the form~\cite{Gribov:1977wm,Zwanziger:1989mf}
\bea
D^{\mu\nu}(P)=\left[\delta^{\mu\nu}-(1-\xi)\frac{P^\mu P^\nu}{P^2}\right]\frac{P^2}{P^4+\gamma_G^4}\, ,
\label{modified_gluon_prop}
\eea
where $\xi$ is the gauge parameter. Inclusion of the term involving $\gamma_G$ in the denominator moves the poles of the gluon propagator off the energy axis so that there are no asymptotic gluon modes.  Naturally, to maintain the consistency of the theory, these unphysical poles should not be considered in the exact correlation functions of gauge-invariant quantities.  This suggests that the gluons are not physical excitations.  In practice, this means that the inclusion of the Gribov parameter results in the effective confinement of gluons.  

Though the Gribov ambiguity renders perturbative QCD calculations ambiguous, but the dimensionful Gribov parameter appearing above can acquire a well-defined meaning if the topological structure of the $SU(3)$ gauge group is made to be consistent with the theory.  Very recently, this has been argued and demonstrated by Kharzeev and Levin~\cite{Kharzeev:2015xsa} by taking into account the periodicity of the $\theta$-vacuum~\cite{Jackiw:1977ng} of the theory due to the compactness of the $SU(3)$ gauge group.  The recent work of Kharzeev and Levin indicates that the Gribov term can be physically interpreted as the topological susceptibility of pure Yang-Mills theory and that confinement is built into the gluon propagator in Eq.~(\ref{modified_gluon_prop}), indicating non-propagation and screening of color charges at long distances in the running coupling.  This also reconciles with the original view Zwanziger had regarding $\gamma_G$ being a statistical parameter \cite{Zwanziger:1989mf}.


 \section{Operator Product Expansion}
\label{ope_intro}

In 1969, while dealing with the short distance behavior within strong interactions, Wilson proposed~\cite{Wilson:1969zs} that an ordinary product of two local fields $A(X)$ and $B(Y)$ can be expanded in the following form when the four vector $Y$ is near $X$, as
\bea
A(X)B(Y)= \sum\limits_nW_n(X-Y)~O_n(X),
\label{ope_wilson}
\eea 
where $O_n(X)$ are a set of local fields at $X$. The coefficient functions $W_n(X-Y)$, later termed as Wilson coefficients, involve powers of $(X-Y)$. Though the complete expansion of Eq. (\ref{ope_wilson}) would require an infinite number of local fields, but upon restricting upto a finite order in $(X-Y)$, the number of fields will be finite. This multipurpose proposal later became famous by the name of Operator Product Expansion. 

Shifman-Vainshtein-Zakharov first argued~\cite{Shifman:1978bx, Shifman:1978by}  that OPE is  valid in presence of the non-perturbative effects~\cite{Novikov:1980uj}. By using OPE judiciously one can exploit both perturbative and non-perturbative domains separately~\cite{Hubschmid:1982pa, Mallik:1983nn, Bagan:1992tg}. Unlike QED a favorable situation occurs particularly in QCD that allows us to do the power counting~\cite{Novikov:1984rf,Novikov:1984ac}. OPE basically assumes a separation of large and short distance effects via condensates and Wilson coefficients. Also according to SVZ, the less effective nature of ordinary perturbation theory at relatively low invariant mass is a manifestation of the fact that nonperturbative vacuum condensates are appearing as power corrections in the OPE of a Green's function. So in view of OPE, in the large-momentum (short-distance) limit,  a two point current-current correlation function~\cite{Shifman:1978bx, Shifman:1978by, Narison:2002pw} can be expanded in terms  of local composite operators  and $c$-numbered Wilson coefficients as
\bea
C_{\mu\nu}(P) &\abcom{=}{z \rightarrow 0}& i\int e^{iP\cdot Z}d^4Z \big \langle {\cal T} 
\left\{J_\mu(Z) J_\nu(0)\right\}\big \rangle \nn\\
&=& \left(\frac{P_\mu P_\nu}{P^2}-g_{\mu\nu}\right)
W(P^2,\nu^2)\langle O\rangle_D, \label{corr_func}
\eea
provided $P^2 \gg \Lambda^2$, where $\Lambda$ is the QCD scale and ${\cal T}$ is the time ordered product. $D$ is the dimension of the composite operators (condensates) $\langle O\rangle$ and those may have non-zero expectation values which are absent to all orders in perturbation theory.  $\nu$ is a scale that separates long and short distance dynamics. The power corrections appear through the  Wilson coefficients $W$ that contain all information about large momentum (short distance) physics above the scale $\nu$, implying that those are free from any infrared and nonperturbative long distance effects. We note that for computing a correlator in vacuum(medium) one should first calculate it in a background of quark and gluonic fields and then average it with respect to these fields in the vacuum(medium) to incorporate the power corrections through relevant condensates. This is precisely the procedure we will follow in chapter \ref{th_ope}. 


\section{Fermions in presence of constant magnetic fields}
\label{mag_fermion}

The Dirac Lagrangian density for fermions in a constant magnetic field $B$, anisotropic along the $z$ direction, has the following form, 
\bea
{\cal L}_{\rm Dirac} = \sum\limits_f \bar{\psi}_f\left(i\slashed{\tilde{D}}-m_f\right)\psi_f,
\label{Dlag_mag}
\eea
where\footnote{$\tilde{D}_\mu$ is not to be confused with the $D_\mu$ used in Eq. (\ref{qcd_lag}).}
\bea
\tilde{D}_\mu = \partial_\mu + i q_f A_\mu^{\rm ext},
\eea
where $q_f$ is the absolute charge of the fermion.
The external gauge field $A$ is given by 
\bea
A_\mu^{\rm ext} &=& \frac{B}{2}\left(0,-y,x,0\right)\hspace{2cm}\mbox{Symmetric gauge},\nn\\
&=& B\left(0,-y,0,0\right)\hspace{2cm}\mbox{Landau gauge}.
\label{Aext}
\eea

The equation of motion for $\psi$ can be deduced from Eq.~(\ref{Dlag_mag}), using
\bea
\frac{\partial {\cal L}}{\partial \bar{\psi}}-\partial_\mu\left(\frac{\partial {\cal L}}{\partial\left(\partial_\mu\bar{\psi}\right)}\right) = 0, 
\eea
as 
\bea
\left(i\slashed{\partial} - q_f \slashed{A}^{\rm ext}-m_f\right)\psi_f = 0.
\eea

\begin{center}
\begin{figure}[tbh]
\begin{center}
\includegraphics[scale=0.6]{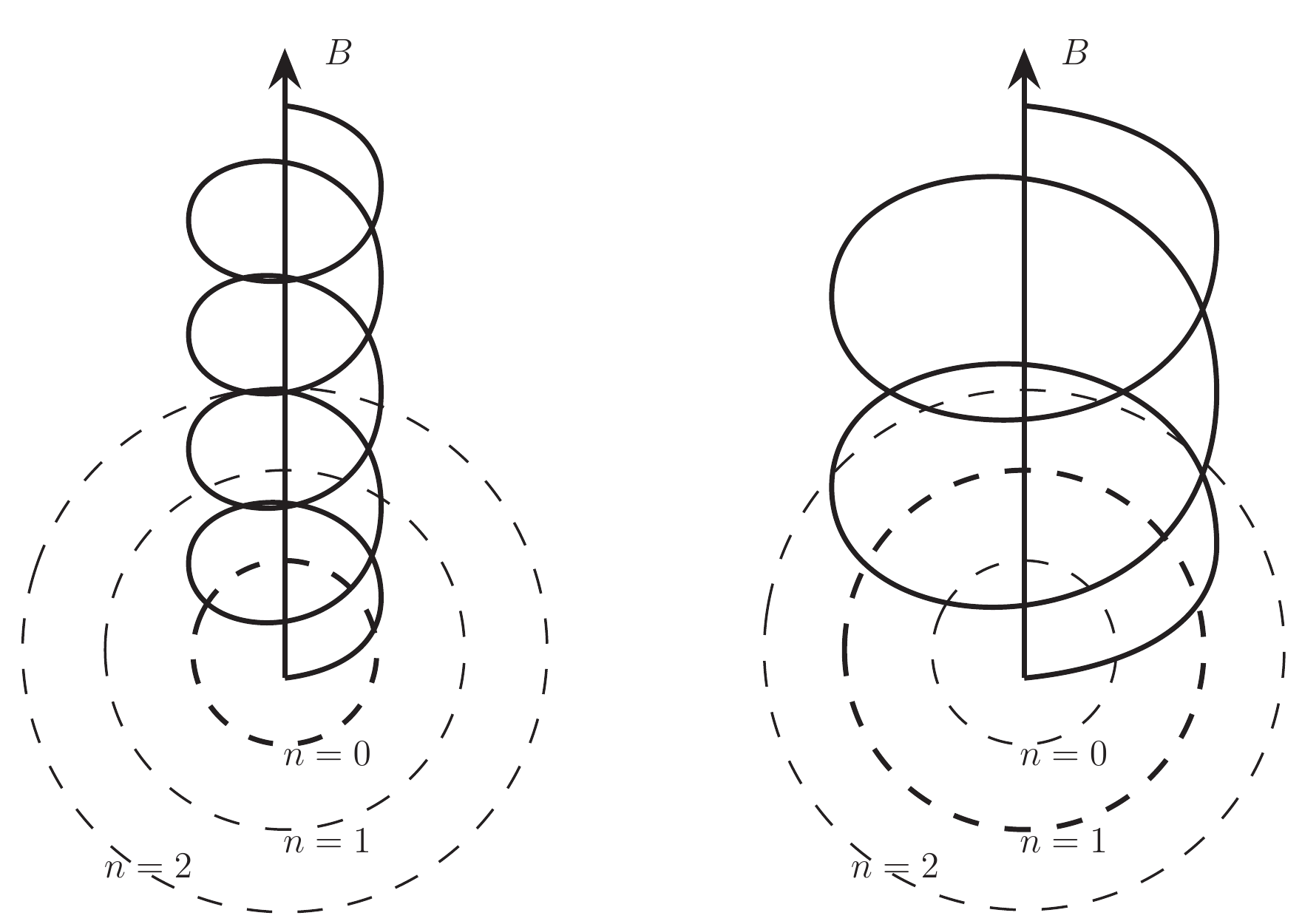} 
\caption{Orientation of Landau levels in presence of a magnetic field.}
\label{ll_orient}
\end{center}
\end{figure}
\end{center}

Now, choosing stationary state solution for the wave function $\psi_f$ and putting the value of $A_\mu^{\rm ext}$ from Eq. (\ref{Aext}), one can get the energy spectrum as 
\bea
E_n(p_z) = \pm \sqrt{p_z^2+m_f^2+2nq_fB}
\label{espec}
\eea
where $n=0,1,2, ....$ are known as the degenerate Landau levels. As seen from Eq.~(\ref{espec}), the transverse momenta $p_x$ and $p_y$ become quantized, in presence of an anisotropic magnetic field along the $z$ direction. The situation is pictorially depicted in Fig. \ref{ll_orient}. So, a fermion in the lowest Landau level will be quantized along the transverse direction, represented by the circle with the lowest radius among the concentric circles depicting the other Landau levels. These Landau levels can affect the quantum fluctuations of the charged fermions in the Dirac sea at $T=0$ and thermal fluctuations at $T\ne 0$. We shall discuss more about Landau levels in chapter \ref{th_mag}.


\section{Correlation function and Spectral function}
\label{cf_sf}

Correlation generally means a mutual relationship between multiple things. For our purpose, correlation functions are such kind of Green's functions which describe how microscopic variables co-vary with response to one another. A CF in coordinate space ($X\equiv(t,\vec{x})$) can be represented as
\bea
C_{AB}(X) = \langle \mathcal{T} \hat{A}(X)\hat{B}(0)\rangle = \int \frac{d^4Q}{(2\pi)^4}e^{iQ\cdot X} C_{AB}(Q),
\eea
where $\mathcal{T}$ is the time ordering between the operators $\hat{A}$ and $\hat{B}$. 

The vacuum properties of any particle changes while propagating through medium. Be it hot and dense medium or magnetized medium or both, its dispersion properties get modified via new attributes due to the presence of the medium. CF are extensively used in several branches of physics because all the modifications arising due to the presence of a medium are reflected in the CF of that particle.
There are various kinds of CF in the literature. But we will extensively use the current-current CF which is directly related to most of the observables that will be computed in this thesis to characterize QGP. 
As we know, a many particle system can be characterized by studying both its static and dynamic properties. Static properties like the thermodynamic functions can be computed from the QCD equation of state itself. But to study the dynamical properties of such systems one typically provides an external
perturbation to disturb the system slightly from equilibrium and correspondingly studies both the quantum fluctuations at $T=0$ and thermal fluctuations at $T\ne 0$. Both of these fluctuations are related to various current-current CF. At finite temperature the current-current CF is given by 
\bea
C(t,\vec{x})= \langle \mathcal{T} J(t,\vec{x})J^\dagger(0,\vec{0})\rangle_\beta = iT\sum_{q_0}\int \frac{d^3q}{(2\pi)^3}e^{iq_0t-i\vec{q}\cdot \vec{x}} C(q_0,\vec{q}),
\eea
where $J(t,\vec{x}) = \bar{\psi}(t,\vec{x})\Gamma\psi(t,\vec{x})$ with $\Gamma=1,\gamma_5,\gamma_\mu$ and $\gamma_\mu\gamma_5$ for scalar, pseudoscalar, vector and pseudovector channels respectively. $\langle~\rangle_\beta$ represents the thermal average.
In Euclidean space putting $t=i\tau$ and $q_0=i\omega_n$ we obtain
\bea
C(\tau,\vec{x})=  T\sum_{\omega_n}\int \frac{d^3q}{(2\pi)^3}e^{-i(\omega_n\tau+\vec{q}\cdot \vec{x})} C(i\omega_n,\vec{q}),
\label{cf_est}
\eea
Fourier transform of which yields
\bea
C(i\omega_n,\vec{q}) = \int\limits_0^\beta d\tau \int d^3\vec{x}~ e^{i(\omega_n\tau+\vec{q}\cdot \vec{x})}\langle \mathcal{T} J(\tau,\vec{x})J^\dagger(0,\vec{0})\rangle_\beta.
\label{cf_est_msp}
\eea
Using the Kramers-Kronig relation, one can relate the real and imaginary parts of a complex function. Thus, following Eq.~(\ref{cf_est_msp}) we can write down the momentum space current-current CF as
\bea
{\rm Re~} C(i\omega_n,\vec{q}) = \frac{1}{\pi} P\left(\int\limits_{-\infty}^{\infty} d\omega \frac{{\rm Im~} C(i\omega_n = \omega + i\epsilon,\vec{q})}{\omega-i\omega_n}\right),
\label{kram_kron}
\eea
where $P$ represents the Cauchy principle value. Now, with the help of Eq.~(\ref{kram_kron}) and writing $C(i\omega_n,\vec{q})$ instead of ${\rm Re~} C(i\omega_n,\vec{q})$, we get the spectral representation of the momentum space current-current CF as
\bea
C(i\omega_n,\vec{q}) = -\int\limits_{-\infty}^{\infty} d\omega \frac{\rho(\omega,\vec{q})}{i\omega_n - \omega}.
\eea
Here $\rho(\omega,\vec{q})$ is called the spectral function, which is extracted from the momentum space correlator by analytic continuation and can be defined as the following
\bea
\rho(\omega,\vec{q}) = \frac{1}{\pi} {\rm Im~} C(i\omega_n = \omega + i\epsilon,\vec{q}),
\label{spec_def}
\eea
identifying spectral function as the discontinuity or the imaginary part of a CF. As we know that in zero temperature imaginary parts play a crucial role in evaluating the decay rates. Likewise in medium by evaluating the discontinuities of a particular current-current CF, a SF generally carries all the information about the hadronic spectrum, coupled to the aforementioned current. This is the reason why SF is related with various physical quantities in hot and dense medium, like DPR, photon production, various susceptibilities, transport coefficients and damping rates. Importance of SF will be more and more lucid with the advancement of this dissertation.  
 
 \begin{center}
\begin{figure}[tbh]
\begin{center}
\includegraphics[scale=0.75]{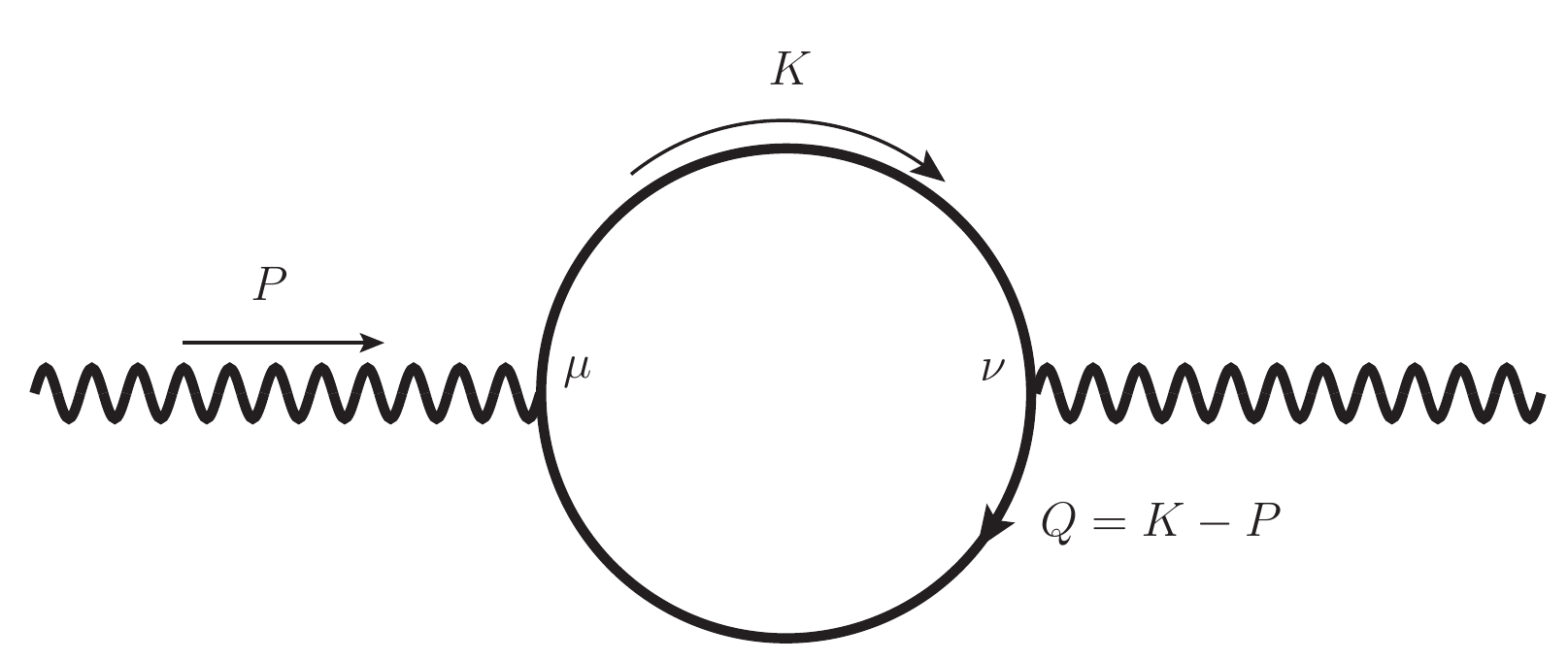} 
\caption[One loop photon self energy diagram]{One loop photon self energy diagram. The wavy line corresponds to photon.}
\label{photon_se}
\end{center}
\end{figure}
\end{center} 
 
The computation of SF is easily handled in the perturbative QCD using the Feynman diagrams. As a demonstration, in Fig \ref{photon_se} the electromagnetic polarization tensor or the one loop self energy diagram for photon is shown.  In one loop level in QED it can be expressed as
\bea
\Pi_{\mu\nu}(P) = -i\sum_{f} q_f^2\int \frac{d^4K}{(2\pi)^4} 
\textrm{Tr}_{c}\left[\gamma_\mu S(K) \gamma_\nu S(Q)\right] , \label{pola}
\eea
where $P$ is the external momentum, $K$ and $Q=K-P$  are the loop momenta. $\textrm{Tr}_{c}$ represents both  color and Dirac traces whereas the $\sum_{f}$ is over flavor.
\begin{center}
\begin{figure}[tbh]
\begin{center}
\includegraphics[scale=0.55]{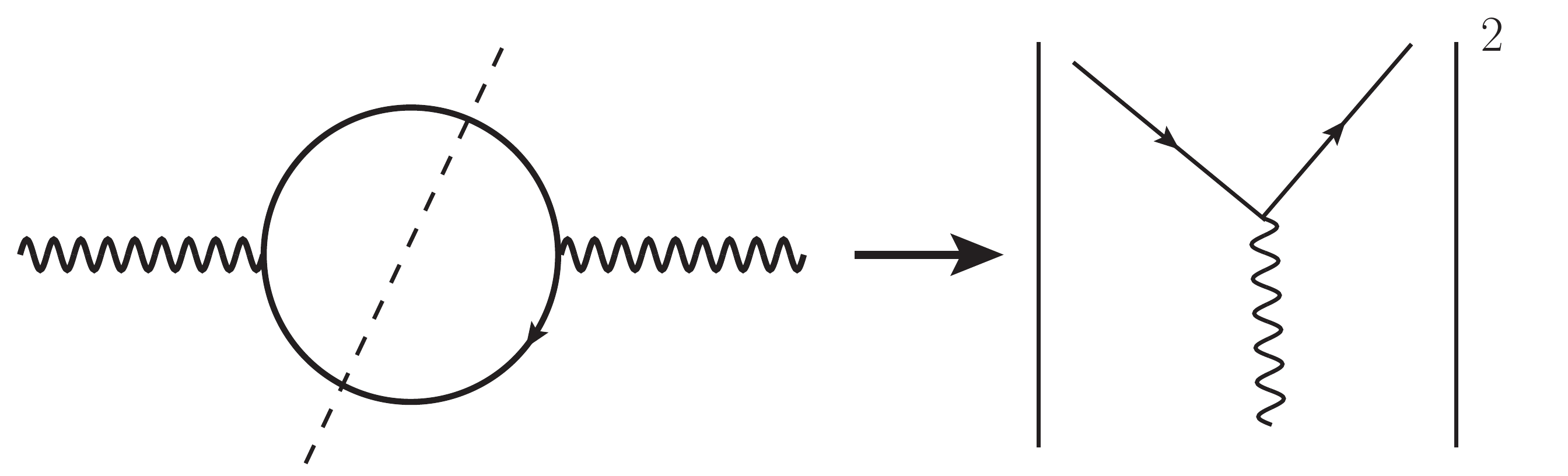} 
\caption[Discontinuity of one loop photon self energy diagram]{Discontinuity of one loop photon self energy diagram, from which the information about hadronic spectrum can be extracted.}
\label{photon_se_cut}
\end{center}
\end{figure}
\end{center} 
The two point current-current CF $C_{\mu\nu}(P)$ is related to photon self-energy as 
\bea
q_f^2C_{\mu\nu} (P) &=&  \Pi_{\mu \nu} (P), \label{corr_func1}
\eea
with $q_f$ is the electric charge of a given quark flavor $f$.
The electromagnetic spectral function for a given flavor $f$, $\rho_f\left(P\right)$, is extracted from the timelike discontinuity of the two point correlation function $C_\mu^\mu(P)$ as
\bea  
\rho_f\left(P\right) = \frac{1}{\pi}\, \mathrm{Im} \left (C^{\mu}_{\mu}(P)\right )_f =\frac{1}{\pi} 
\mathrm{Im}~\left(\Pi^\mu_\mu(P)\right)_f/q_f^2,
\label{spec_func}
\eea   
which is shown in Fig \ref{photon_se_cut}. It is also very clear from Fig. \ref{photon_se_cut}, how the spectral function can be used as an information carrier of the hadronic sector of any process.

However in LQCD the situation is quite different. Unfortunately the lattice techniques are solely applicable in Euclidean spacetime, while the spectral function is an inherently Minkowskian object. Though it can be obtained from the Euclidean correlator in principle, but the process of analytic continuation in the regime of lattice is ill-defined due to limited set of data. Because of this complication, in LQCD the spectral function is not defined by Eq.~(\ref{spec_def}). Instead they proceed by the evaluation of Euclidean CF in the following way : 
\bea
C(i\omega_n,\vec{q}) = \int\limits_{-\infty}^{\infty} d\omega \frac{\rho(\omega,\vec{q})}{\omega-i\omega_n} = \int\limits_{-\infty}^{\infty} d\omega~\rho(\omega,\vec{q})~\left(1+n_B(\omega)\right) \frac{e^{-\beta\omega}-1}{i\omega_n-\omega}.
\eea
For bosons $\omega_n = 2n\pi T$ and one can further write
\bea
C(i\omega_n,\vec{q}) &=& \int\limits_{-\infty}^{\infty} d\omega~\rho(\omega,\vec{q})~\left(1+n_B(\omega)\right) \frac{e^{\beta(i\omega_n-\omega)}-1}{i\omega_n-\omega}\nn\\
&=& \int\limits_0^\beta d\tau' e^{i\omega_n\tau'} \int\limits_{-\infty}^{\infty} d\omega~\rho(\omega,\vec{q})~\left(1+n_B(\omega)\right) e^{-\omega\tau'}.
\eea
Now, Fourier transform of the temporal part yields
\bea
C(\tau,\vec{q}) &=& T\sum\limits_{n=-\infty}^{\infty} C(i\omega_n,\vec{q}) e^{-i\omega_n\tau}\nn\\
&=& \int\limits_{-\infty}^{\infty} d\omega~\rho(\omega,\vec{q})~\left(1+n_B(\omega)\right)\int\limits_0^\beta d\tau' \delta(\tau'-\tau)e^{-\omega\tau'}\nn\\
&=& \int\limits_{-\infty}^{\infty} d\omega~\rho(\omega,\vec{q})~\left(1+n_B(\omega)\right)e^{-\omega\tau}.
\label{lqcd_sf_1}
\eea
After breaking the limits in Eq.~(\ref{lqcd_sf_1}) and some straightforward mathematical manipulation one can get
\bea
C(\tau,\vec{q}) &=& \int\limits_{-\infty}^{\infty} d\omega~\rho(\omega,\vec{q})~\frac{\cosh\left(\omega(\tau-\beta/2)\right)}{\sinh\left(\omega\beta/2\right)}.
\label{lqcd_sf}
\eea
Eq.~(\ref{lqcd_sf}) is then inverted to extract the SF vis-a-vis the spectral properties using a probabilistic Maximum Entropy method~\cite{Asakawa:2000tr,Nakahara:1999vy}, which is also to some extent error prone~\cite{Cuniberti:2001hm}. MEM requires an ansatz for the spectral function and hence lower energy part of the spectral function becomes Ansatz dependent. More discussion about the LQCD calculations of spectral properties will be discussed in section \ref{dpr_eval}. 


\section{Importance of DPR}
\label{dpr_intro}

As we know from the HIC timeline shown in Fig. \ref{hic1}, that the locally equilibrated plasma is short-lived in the collision. However, there are always some initial or final state interactions that may contaminate an observable one is interested in. In this respect the electromagnetic emissivity of the plasma in the form of real or virtual photon is particularly important.  After a quark interacts with its antiparticle, real or virtual photos are created in the plasma. Real photon escapes unperturbed and virtual photon decays into a dilepton in the process. The momentum distribution of the dilepton depends on the momentum distribution of the quark and the antiquark which is eventually governed by the thermodynamic condition of the plasma. So the dilepton carries information about the transient plasma state at the moment of their production. Again, since the produced dilepton are colorless and interact only through the electromagnetic interaction their mean free path become quite large. So, it is very less probable that they would suffer further collisions after they are produced. The very fact that they do not suffer from final state interactions and carry least contaminated information of the local equilibrium makes real or virtual photon production a desirable candidate for studying QGP. This is why, the DPR from QGP phase has been studied vividly in the last three decades~\cite{Weldon:1983jn,Weldon:1990iw, McLerran:1984ay, Hwa:1985xg, Kajantie:1986dh, Kajantie:1986cu, Cleymans:1986na,Cleymans:1992gb, Gale:1987ey, Gale:1987ki,Braaten:1990wp, Karsch:2000gi, Bandyopadhyay:2015wua, Greiner:2010zg, Aurenche:1998nw, Mustafa:1999dt, Karsch:2001uw,  Ding:2010ga, Ding:2016hua, Islam:2014sea, Gale:2014dfa, Islam:2015koa, Bandyopadhyay:2016fyd, Sadooghi:2016jyf, Tuchin:2013bda, Srivastava:2002ic, Kvasnikova:2001zm,Chatterjee:2007xk, Laine:2013vma, Ghisoiu:2014mha, Ghiglieri:2014kma}. 

Even though the lepton pairs behave as free particles after production, they are produced in every stage of the collisions. So, disentangling the particular set of dileptons which originate from the QGP phase is quite tricky. To avoid this hindrance one usually talks about the total DPR all through the collision time. Thus ideally the observable to characterize QGP is the Dilepton spectrum which hydrodynamically takes care of the spacetime evolution of the DPR through the whole range of collision. But discussion of the dilepton spectrum is out of the scope of this dissertation and hence we will focus solely on the DPR in QGP phase in this thesis. 


\section{Formulation of DPR}
\label{dpr_formu}

Previously in section \ref{sign_qgp} we made a statement that the DPR~\cite{Braaten:1990wp} is directly related to the spectral function of the electromagnetic CF. In this section we will frame that relation. 

\begin{center}
\begin{figure}[tbh]
\begin{center}
\includegraphics[scale=0.95]{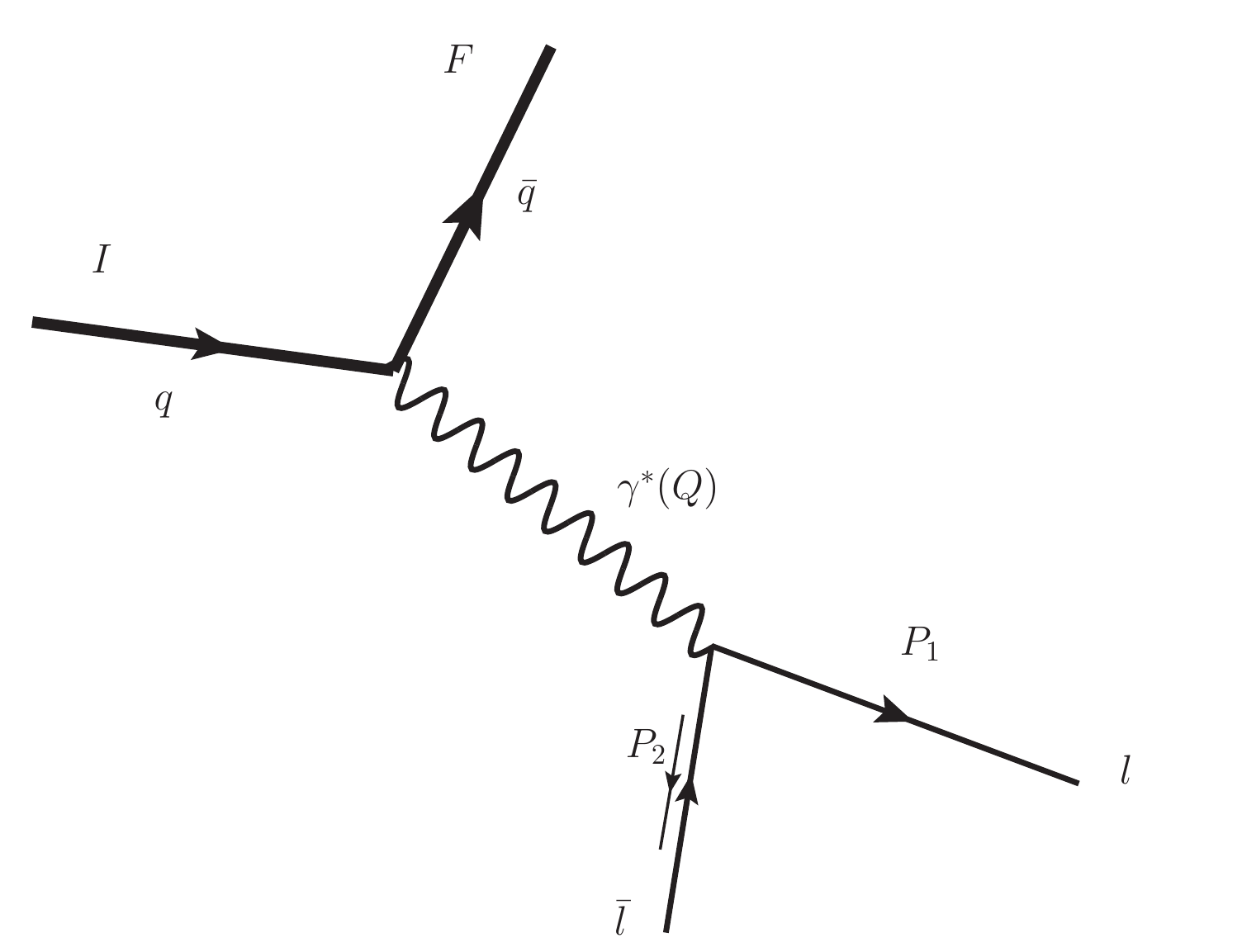} 
\caption[Schematic diagram of dilepton production]{Dilepton production amplitude in QGP phase. The states I and F consist of 
quarks, while $l\bar{l}$ is a dilepton. The wavy line corresponds to photon.}
\label{dil_pro}
\end{center}
\end{figure}
\end{center}

In Fig. \ref{dil_pro} a dilepton production process is shown. During the collision of two heavy nuclei, a quark ($q$) and an antiquark ($\bar{q}$) interact to produce a virtual photon $\gamma^*$, which in turn produces a $l\bar{l}$ pair as dilepton. The transition amplitude (Fig. \ref{dil_pro}) from an initial state $I$, composed of quarks and gluons to a final state $F$ of similar composition, along with the emission of a dilepton $l(P_1)$ and $\bar{l}(P_2)$ is given as
\bea
\langle F, l(P_1), \bar{l}(P_2)| S | I\rangle,
\label{trans_ampl}
\eea
where the scattering matrix operator $S$ is given by the interaction Lagrangian
\bea
\mathcal{L}_{\mathrm{int}} = \left(J_l^\mu(X)+J_q^\mu(X)\right)A_\mu(X),
\eea
of lepton and quark currents coupled to the electromagnetic field $A_\mu(X)$. This eventually yields 
\bea
\langle F, l(P_1), \bar{l}(P_2)| S | I\rangle = \frac{e_0 \bar{u}(P_2) \gamma_\mu v(P_1)}{V\sqrt{4E_1E_2}} \int d^4X e^{i Q\cdot X} \langle F| A^\mu(X) | I\rangle,
\label{dil_transamp}
\eea
where $e_0$ is the unrenormalized charge and $\bar{u}(P_2) (v(P_1))$ represents the incoming (outgoing) antilepton (leptons) in Fig. \ref{dil_pro}. 

In ultra relativistic HIC the process of dilepton production rapidly thermalizes within a very short period of time. Thermalization discards the possibility of a single specific initial state and proposes an ensemble average over all possible initial states, each weighted by a Boltzmann factor. For high energetic dileptons produced by a single virtual photon with energy $q_0=E_1+E_2$ and spatial momentum $q=p_1+p_2$, the thermally averaged dilepton multiplicity in the local rest frame of the plasma is given by 
\bea
N = \sum\limits_I\sum\limits_F |\langle F, l(P_1), \bar{l}(P_2)| S | I\rangle |^2~ \frac{e^{-\beta q_0}}{Z}~ \frac{Vd^3p_1}{(2\pi)^3} \frac{Vd^3p_2}{(2\pi)^3},
\label{dil_mult}
\eea
where, $Z={\rm Tr}\left[e^{-\beta H}\right]$ is the canonical partition function.
Now, after some straightforward mathematical steps following Eq.~(\ref{dil_transamp}) and Eq.~(\ref{dil_mult}), the dilepton multiplicity per unit space-time volume can be obtained as~\cite{Weldon:1990iw}
\bea
\frac{dN}{d^4X}&=& 2\pi e^2 e^{-\beta q_0}L_{\mu\nu}\rho^{\mu\nu}\frac{d^3p_1}{(2\pi)^3E_1}\frac{d^3p_2}{(2\pi)^3E_2}.
\label{dlp_mult}
\eea
As expected from Fig. \ref{dil_pro}, the multiplicity consists of a hadronic part $\rho^{\mu\nu}$, a leptonic part $L_{\mu\nu}$ and a phase space part. The hadronic part is given by,
\bea
\rho^{\mu\nu}(q_0,{\bf q}) &=& -\frac{1}{\pi}\frac{e^{\beta q_0}}{e^{\beta q_0}-1}\mathrm{Im}\left[D_R^{\mu\nu}(q_0,{\bf q})\right]\nn\\
&\equiv& -\frac{1}{\pi}\frac{e^{\beta q_0}}{e^{\beta q_0}-1}~\frac{e_e^2}{Q^4}\mathrm{Im}\left[C^{\mu\nu}(q_0,{\bf q})\right],
\eea
where $C^{\mu\nu}$ is the photonic two point current-current CF, whereas $D_R^{\mu\nu}$ represents the photon propagator and they are related by the well known Dyson-Schwinger equation.~\cite{Weldon:1990iw,McLerran:1984ay}. Also the notation $e_e^2$ represents the effective charge of the system. 

Again, the leptonic part is given by the spin sum over the Dirac spinors as
\bea
L_{\mu\nu} &=& \frac{1}{4}\mathrm{Tr}\left[\bar{u}(P_2)\gamma_\mu v(P_1)\bar{v}(P_1)\gamma_\nu u(P_2)\right]\nn\\
&=& P_{1\mu}P_{2\nu}+P_{1\nu}P_{2\mu}-(P_1\cdot P_2+m_l^2)g_{\mu\nu}.
\label{lmunu_unmagnetized}
\eea

So, putting together the hadronic and the leptonic part, the expression for the dilepton multiplicity comes out to be
\bea
\frac{dN}{d^4X}&=& 2\pi e^2 e^{-\beta q_0}L_{\mu\nu}\rho^{\mu\nu}\frac{d^3p_1}{(2\pi)^3E_1}\frac{d^3p_2}{(2\pi)^3E_2}\nn\\
&=& 2\pi e^2 e^{-\beta q_0}\int d^4Q\delta^4(P_1+P_2-Q)L_{\mu\nu}\rho^{\mu\nu}\frac{d^3p_1}{(2\pi)^3E_1}\frac{d^3p_2}{(2\pi)^3E_2}.
\eea

After getting the expression of dilepton multiplicity, now the differential DPR can be obtained by differentiating the multiplicity by the four momentum $Q$ of the virtual photon $\gamma^*$ and is expressed as 
\bea
\frac{dN}{d^4Xd^4Q} &=& 2\pi e^2 e^{-\beta q_0}\rho^{\mu\nu}\int\frac{d^3p_1}{(2\pi)^3E_1}\frac{d^3p_2}{(2\pi)^3E_2}\delta^4(P_1+P_2-Q)L_{\mu\nu},\nn\\
&=&\frac{\alpha_{\textrm{em}}}{8\pi^4} e^{-\beta q_0}\rho^{\mu\nu}I_{\mu\nu},
\label{dpr_d4q_1}
\eea
where,
\bea
I_{\mu\nu}(Q) &=& \int\frac{d^3p_1}{E1}\int\frac{d^3p_2}{E2}~\delta^4(P_1+P_2-Q)L_{\mu\nu},
\label{imunu_defn}
\eea
and $\alpha_{\textrm em}$ is the electromagnetic fine structure constant. 

The evaluation of the integral $I_{\mu\nu}$ for nonzero leptonic mass is done in the Appendix \ref{imunu}. As throughout this dissertation we will work with massless leptons, here we present the expression of $I_{\mu\nu}$ for $m_l=0$ as
\bea
I_{\mu\nu}(Q)\Big\vert_{m_l=0} &=& \frac{2\pi}{3}\left(Q_\mu Q_\nu-Q^2g_{\mu\nu}\right)
\label{unmagnetized_imn_massive}
\eea

Now, putting the expression of \ref{dpr_d4q_1}, we obtain the finite temperature differential DPR as
\bea
\frac{dN}{d^4Xd^4Q}
&=&\frac{\alpha_{\textrm{em}}}{8\pi^4} e^{-\beta q_0}\rho^{\mu\nu}\frac{2\pi}{3}\left(Q_\mu Q_\nu-Q^2g_{\mu\nu}\right)\nn\\
&=&\frac{\alpha_{\textrm{em}}}{12\pi^3} e^{-\beta q_0}\rho^{\mu\nu}\left(Q_\mu Q_\nu-Q^2g_{\mu\nu}\right)\nn\\
&=&\frac{\alpha_{\textrm{em}}}{12\pi^3} \frac{e_e^2}{e^{\beta q_0}-1}\frac{\left(Q^2g_{\mu\nu}-Q_\mu Q_\nu\right)}{Q^4}\left(\frac{1}{\pi}\mathrm{Im}\left[C^{\mu\nu}(q_0,{\bf q})\right]\right)\nn\\
&=&\frac{\alpha_{\textrm{em}}e_e^2}{12\pi^3}\frac{n_B(q_0)}{Q^2}\left(\frac{1}{\pi}\mathrm{Im}\left[C^{\mu}_{\mu}(q_0,{\bf q})\right]\right)
\label{diff_dpr}
\eea
 The invariant mass of the lepton pair is $M^2 =Q^2= q_0^2 - q^2$. The dilepton rate in Eq.~(\ref{diff_dpr}) is valid only 
at leading order in $\alpha_{\textrm{em}}$ but to all orders in strong coupling constant $\alpha_s$. The lepton masses are neglected in 
Eq.~(\ref{diff_dpr}).

Now if we consider a two-flavor case, $N_f=2$,
\bea
e_e^2 &=& \frac{1}{N_f}\sum\limits_f e_f^2 = \frac{5}{18}e^2 = \frac{5\times 4\pi\alpha_{\textrm{em}}}{18}.\nn\\
\therefore\frac{dN}{d^4Xd^4Q} &=& \frac{5\alpha_{\textrm{em}}^2}{54\pi^2} \frac{n_B(q_0)}{M^2}\left(\frac{1}{\pi}\mathrm{Im}\left[C^{\mu}_{\mu}(q_0,{\bf q})\right]\right).
\label{dpr_unmagnetized_final}
\eea


\section{Dilepton rate in presence of strong external constant magnetic field}
\label{dil_mag}

As mentioned in chapter \ref{th_intro}, in non-central HIC an anisotropic magnetic field is expected to be generated in the direction perpendicular to the reaction plane, due to the relative motion of the heavy-ions themselves (participants and spectators). In chapter \ref{th_mag} of this thesis the electromagnetic spectral properties of a hot magnetized medium will be explicitly discussed. As a stepping stone to that computation, the modification in DPR in the presence of an external magnetic field will be discussed in this section. The dilepton production from a magnetized hot and dense matter can generally be dealt with three different scenarios~\cite{Tuchin:2013bda,Sadooghi:2016jyf}:
\begin{itemize}
\item Only the quarks move in a magnetized medium but not the final lepton pairs, 

\item Both quarks and leptons move in a magnetized medium, 

\item Only the final lepton pairs move in the magnetic field. 
\end{itemize}
Based on our recent effort Ref~\cite{Bandyopadhyay:2016fyd}, we discuss below all three cases one by one.

\subsection{Quarks move in a magnetized medium but not the final lepton pairs}

\begin{center}
\begin{figure}[h]
\begin{center}
\includegraphics[scale=0.75]{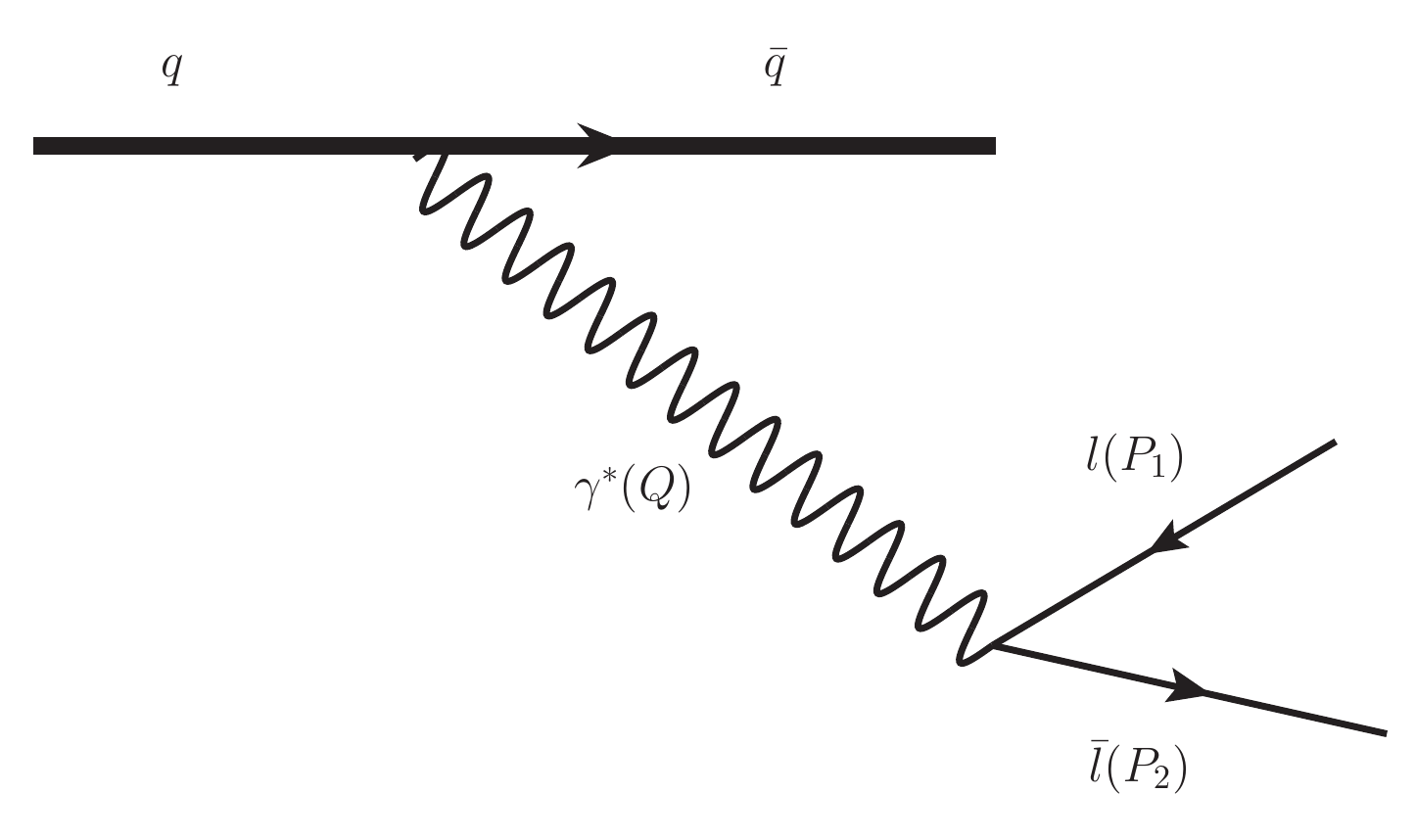} 
\caption[Dilepton production when quarks move in a magnetized medium but lepton pairs remain unmagnetized]{Dilepton production when quarks move in a magnetized medium but lepton pairs remain unmagnetized. Bold fermionic line represents magnetized quarks.}
\label{dil_pro_mag_case1}
\end{center}
\end{figure}
\end{center} 

The first case is shown in Fig.~\ref{dil_pro_mag_case1}. We hereby emphasize that this scenario is interesting and extremely relevant to noncentral heavy-ion collisions, especially for the scenario of fast decaying magnetic field~\cite{Bzdak:2012fr,McLerran:2013hla} and also for lepton pairs produced late or at the edges of hot and dense magnetized medium so that they are unaffected by the magnetic field. 
In this scenario only the hadronic part $\rho^{\mu\nu}$ in Eq.~(\ref{dlp_mult}) will be modified by the background constant magnetic field whereas the leptonic tensor $L_{\mu\nu}$ and the phase space factors will remain unaffected. So, the dilepton rate for massless ($m_l=0$) leptons can  then be written from Eq.~(\ref{diff_dpr}) as~\cite{Bandyopadhyay:2016fyd}
\bea
\frac{dN^m}{d^4Xd^4Q}
&=&\frac{\alpha_{\textrm{em}}e_e^2}{12\pi^3}\frac{n_B(q_0)}{Q^2}\left(\frac{1}{\pi}\mathrm{Im}\left[\tilde{C}^{\mu}_{\mu}(Q_\shortparallel,Q_\perp)\right]\right),\label{dlpm_case1}
\eea
where $\tilde{C}(Q_\shortparallel,Q_\perp)$ represent the modified CF in presence of external anisotropic magnetic field. Detailed computations of the modification in CF for the case of both strongly and weakly magnetized hot medium are discussed in chapter \ref{th_mag}.

\subsection{Both quark and lepton move in magnetized medium}

\begin{center}
\begin{figure}[h]
\begin{center}
\includegraphics[scale=0.75]{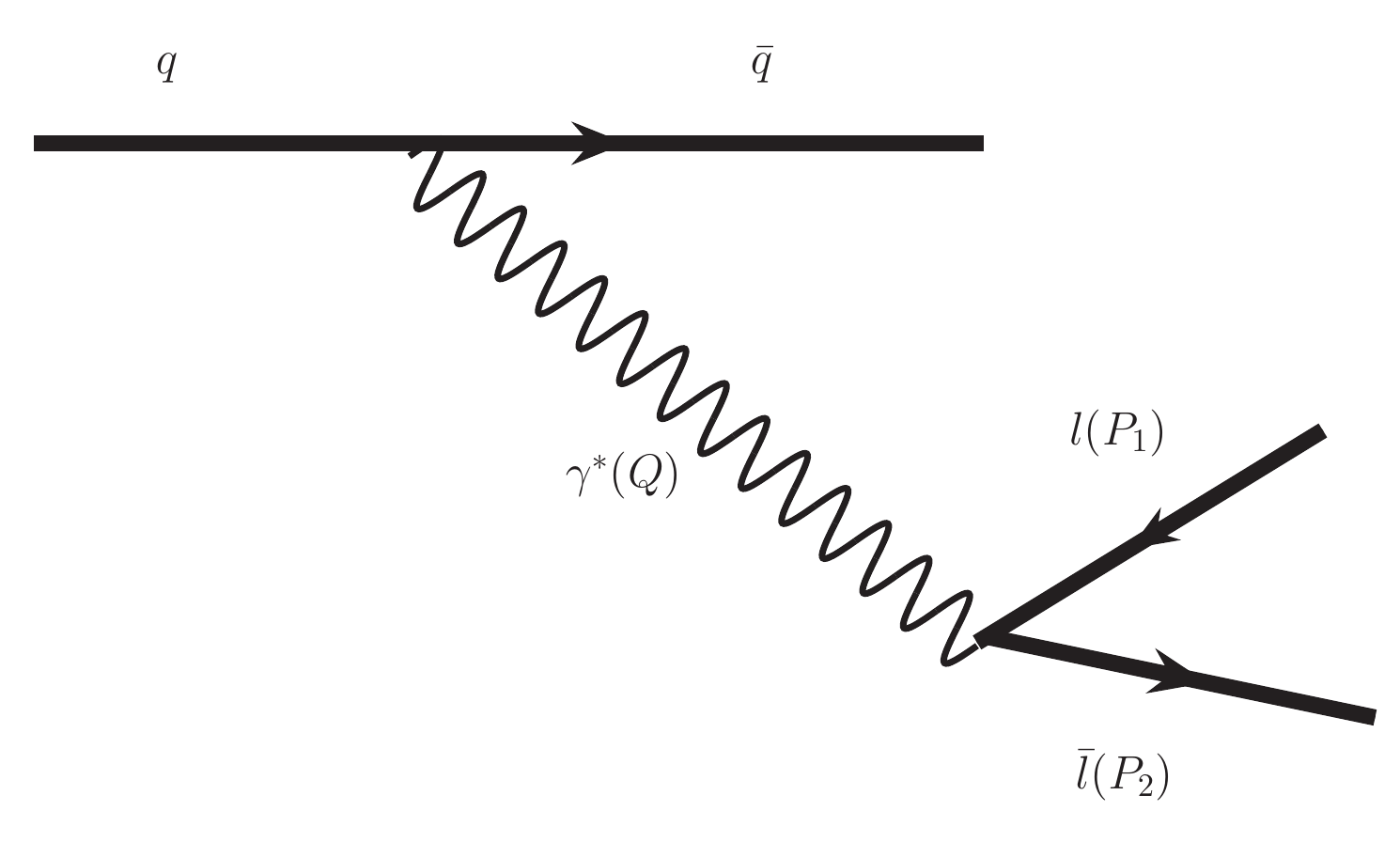} 
\caption[Dilepton production when both quarks and the final lepton pair move in a magnetized medium]{Dilepton production when both quarks and the final lepton pair move in a magnetized medium. Bold fermionic line represents magnetized quarks and lepton pair.}
\label{dil_pro_mag_case2}
\end{center}
\end{figure}
\end{center} 

This scenario, shown in Fig.~\ref{dil_pro_mag_case2}, is expected to be the most general one. To consider such a scenario the usual DPR given in Eq.~(\ref{diff_dpr}) has to be supplemented with the appropriate modification of the hadronic and leptonic tensor along with the phase space factors in a magnetized medium.
Later in chapter \ref{th_mag} this scenario will be dealt with sufficiently strong magnetic field, which focuses only on the lowest landau level. So, in this subsection though we will briefly outline the modifications in the DPR for any landau level, but for simplicity the explicit calculation will be done only for LLL. We list below all the required modifications

\begin{enumerate}

\item[$\bullet$]
The phase space factor in presence of magnetized medium gets modified as
\bea
\frac{d^3p}{(2\pi)^3E} & \rightarrow& \frac 
{|eB|}{(2\pi)^2}\sum\limits _{n=0}^{\infty} \frac{dp_z}{E}.
\eea
where $d^2P_\perp=2\pi|eB|$, $e$ is the electric charge of the lepton and $\sum\limits_{n=0}^{\infty}$ is over all the landau levels. For strong magnetic field one is confined within LLL and $n=0$ only. The factor $|eB|/(2\pi)^2$ is the density of states in the transverse direction and true for LLL~\cite{Gusynin:1995nb}.

\item[$\bullet$] The hadronic part gets modified as before. The unmagnetized electromagnetic CF $C$ gets modified to magnetized $\tilde{C}$. The modification of electromagnetic CF and SF for LLL will be discussed in detail in chapter \ref{th_mag}.

\item[$\bullet$] In presence of constant magnetic field  the spin of 
fermions is aligned along the field direction and the usual 
Dirac spinors $u(P)$ and $v(P)$ in Eq.~(\ref{lmunu_unmagnetized}) get modified~\cite{Schwinger:1951nm,Gusynin:1995nb} 
by $\mathcal{P}_n u(\tilde{P})$ and  $\mathcal{P}_n v(\tilde{P})$  with  
$\tilde{P}^\mu=\left(p^0,0,0,p^3\right)$. $\mathcal{P}_n$ is the projection operator at the 
$n$th LL. For LLL  it takes a simple form
\bea
\mathcal{P}_0 = \frac{1-i\gamma_1\gamma_2}{2}. \label{d11}
\eea
Now, the modification in the leptonic part in presence of a strong magnetic field 
can be carried out as
\bea
L^m_{\mu\nu} &=& \frac{1}{4} 
\sum\limits_{\mathrm{spins}}\mathrm{Tr}\left[\bar{u}(\tilde{P}_2)\mathcal{P}_0\gamma_\mu 
\mathcal{P}_0v(\tilde{P}_1)\bar{v}(\tilde{P}_1)\mathcal{P}_0\gamma_\nu \mathcal{P}_0 u(\tilde{P}_2)\right]\nn\\
&=& \frac{1}{4} 
\mathrm{Tr}\Bigl[(\slashed{\tilde{P}}_1+m_l)\left(\frac{1-i\gamma_1\gamma_2}{2}
\right)\gamma_\mu 
\left(\frac{1-i\gamma_1\gamma_2}{2}\right)\times\nn\\
&&~~~~~~~~~~~~~~~~~~~~~~(\slashed{\tilde{P}}_2-m_l)\left(\frac{
1-i\gamma_1\gamma_2}{2}\right) 
\gamma_\nu\left(\frac{1-i\gamma_1\gamma_2}{2}\right)\Bigr]\nn\\
&=&\!\!\!\!\!\frac{1}{2} 
\!\!\left[P^\shortparallel_{1\mu}P^\shortparallel_{2\nu}\!+\!P^\shortparallel_{1\nu}
P^\shortparallel_{2\mu}\!\!-\!\!((P_1\cdot P_2)_\shortparallel + 
m_l^2)(g_{\mu\nu}^\shortparallel-g_{\mu\nu}^\perp-g_{1\mu}g_{1\nu}-g_{2\mu}g_{2\nu})\right
].\nn
\eea
\item[$\bullet$] Similarly as in the previous section here also we define an integral as
\bea
I^m_{\mu\nu}(Q) &=& \int\frac{d^3p_1}{2E1}\int\frac{d^3p_2}{E2}~\delta^4(P_1+P_2-Q)\times\nn\\
\!\!\!\!\!\!\!\!\!\!\!\!\!\!\!\!\!\!\!\!\!\!\!\!\!\!\!\!\!\!\!\!&&\!\!\!\!\!\!\!\!\!\!\!\!\!\!\!\!\!\!\!\!\!\!\!\!\!\!\!\!\!\!\!\left[P^\shortparallel_{1\mu}P^\shortparallel_{2\nu}+P^\shortparallel_{1\nu}P^\shortparallel_{2\mu}-((P_1\cdot P_2)_\shortparallel + m_l^2)(g_{\mu\nu}^\shortparallel-g_{\mu\nu}^\perp-g_{1\mu}g_{1\nu}-g_{2\mu}g_{2\nu})\right].
\label{immunu_defn}
\eea
The evaluation of this integral is done in Appendix \ref{immunu} yielding
\bea
I^m_{\mu\nu}(Q) &=& \frac{4\pi}{(Q_\shortparallel^2)^2} F_2(m_l,Q_\shortparallel^2)\left(Q^\shortparallel_\mu Q^\shortparallel_\nu-Q_\shortparallel^2g^\shortparallel_{\mu\nu}\right),
\eea
where from Appendix \ref{immunu} we know that
\bea
F_2(m_l,Q_\shortparallel^2) = |eB|m_l^2\left(1-\frac{4m_l^2}{Q_\shortparallel^2}\right)^{-\frac{1}{2}}.
\label{F2_mag}
\eea
\end{enumerate}

Putting all these together, we finally obtain the dilepton production rate from Eq.~(\ref{diff_dpr}) for LLL as~\cite{Bandyopadhyay:2016fyd}
\bea
\frac{dN^m}{d^4Xd^4Q}
&=&\frac{\alpha_{\textrm{em}}e_e^2}{2\pi^3}\frac{n_B(q_0)}{Q_\shortparallel^2Q^4}F_2(m_l,
Q_\shortparallel^2)\left(\frac{1}{\pi} \mathrm{Im} \left[\tilde{C}^{\mu}_{\mu}(Q_\shortparallel,
Q_\perp)\right]\right).\label{dlpm_case2}
\eea

\subsection{Only the final lepton pairs move in the magnetic field}

\begin{center}
\begin{figure}[tbh]
\begin{center}
\includegraphics[scale=0.75]{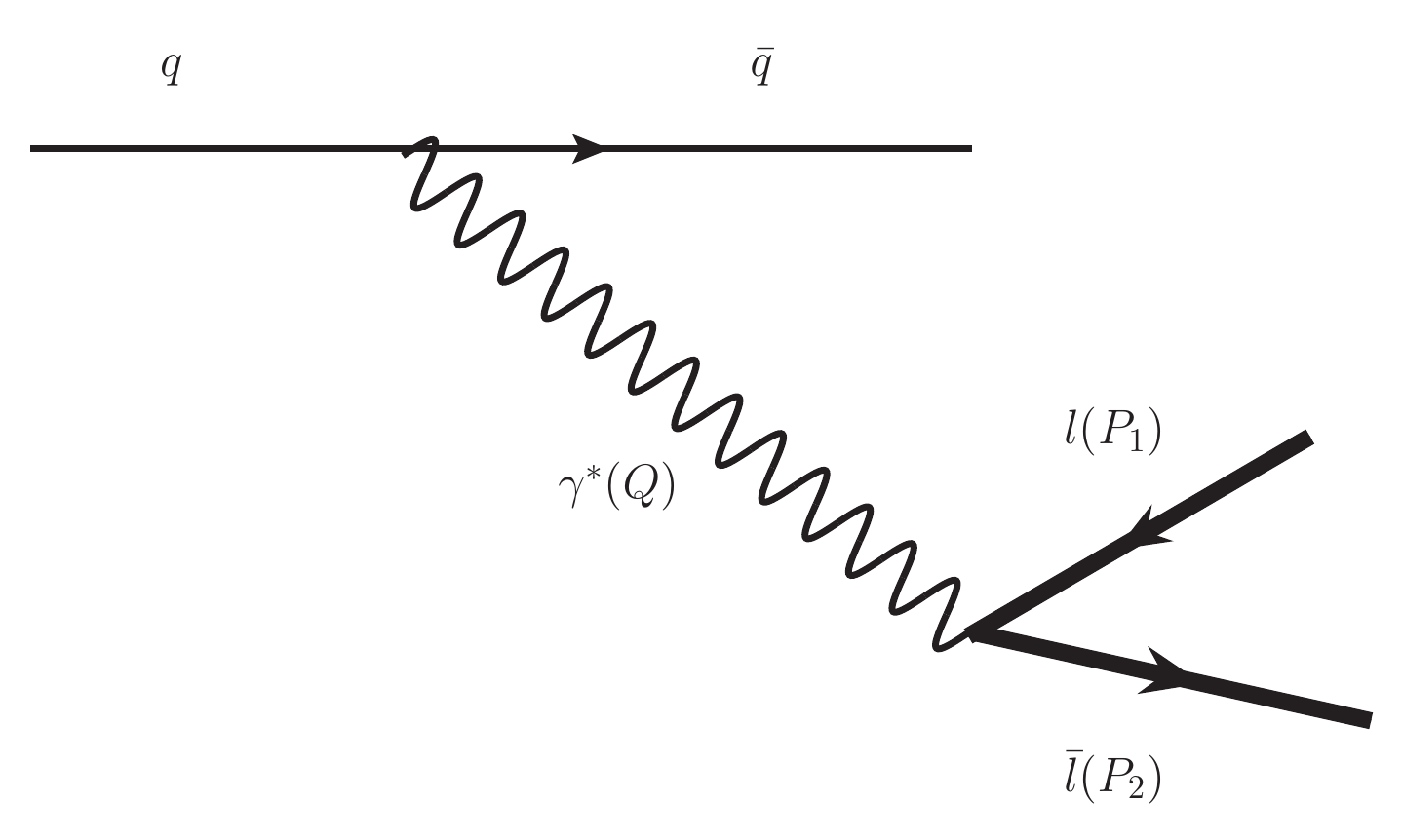} 
\caption[Dilepton production when quarks remain unmagnetized but the final lepton pair move in a magnetized medium]{Dilepton production when quarks remain unmagnetized but the final lepton pair move in a magnetized medium. Bold fermionic line represents magnetized lepton pair.}
\label{dil_pro_mag_case3}
\end{center}
\end{figure}
\end{center} 

Fig.~\ref{dil_pro_mag_case3} represents the final scenario. The production rate for this relatively rare case requires modification of the leptonic tensor in a magnetized medium but the hadronic part vis-a-vis the CF remains unmagnetized. Since we have already discussed the most general case, now it can easily be obtained as 
\bea
\frac{dN^m}{d^4Xd^4Q}
&=&\frac{\alpha_{\textrm{em}}e_e^2}{2\pi^3}\frac{n_B(q_0)}{Q_\shortparallel^2Q^4}F_2(m_l,
Q_\shortparallel^2)\left(\frac{1}{\pi} \mathrm{Im} \left[C^{\mu}_{\mu}(Q)\right]\right).
\eea

\section{Evaluation of DPR}
\label{dpr_eval}

After providing a basic introduction to DPR in the three previous sections (\ref{dpr_intro}, \ref{dpr_formu} and \ref{dil_mag}), in this section, different ways of evaluating DPR will be discussed. Here we would also 
like to mention that the next three chapters (chapter \ref{th_ope}, \ref{th_gribov} and \ref{th_mag}) of this thesis will be devoted to detailed 
computation of DPR via three different methods under three different circumstances.

As explicitly stated in section \ref{dpr_intro}, dileptons are produced in every stage 
of HIC. One of the ways to differentiate between them from the final yield is to classify 
them with respect to their invariant mass $M$. The high mass dileptons are mostly produced 
due to collision between hard partons and they are not particularly very informative about QGP. 
This is because the Drell-Yan processes~\cite{Drell:1969km} and charmonium decays~\cite{Dominguez:2009mk, 
Dominguez:2010mx} are the major processes in that regime. On the other hand the low mass 
dilepton production is enhanced~\cite{Adare:2009qk} compared to  all known sources of 
electromagnetic decay of the hadronic particles and the contribution of a radiating
QGP. So, the low mass dileptons ($\le 1$ GeV) possibly indicate, some nonhadronic sources 
and the intricacies are discussed in the literature in a more phenomenological way
~\cite{Greiner:2010zg,Islam:2014sea,Gale:2014dfa}. There also exists dilepton production 
~\cite{Shuryak:1978ij} in the intermediate range of invariant mass (say $1-3$ GeV) with 
optimized contribution from the QGP phase, which is not dominated by hadronic processes, 
but still can be treated via perturbative methods. We emphasize that the higher order 
perturbative calculations~\cite{Aurenche:1998nw, Laine:2013vma, Ghisoiu:2014mha, 
Ghiglieri:2014kma} of the dilepton rate do not converge in a small strong coupling ($g$) 
limit. This is because the temperatures attained in recent heavy-ion collisions are not so 
high to make perturbative calculations applicable. However, the leading order perturbative 
quark-antiquark annihilation is the only dilepton rate from the QGP phase that has been used 
extensively in the literature. Nevertheless, this contribution is very appropriate at large 
invariant mass but not in low and intermediate invariant mass. In this mass regime one expects 
that the nonperturbative contributions could be important and substantial. 

The non-perturbative effects of QCD are taken care of by the LQCD computations. As mentioned 
in chapter \ref{th_intro}, LQCD has very reliably computed the nonperturbative effects 
associated with the bulk properties (thermodynamics and conserved density fluctuations) 
of the deconfined phase, around and above the pseudocritical transition temperature. 
Efforts have also been made in lattice within the quenched  approximation of QCD~\cite{Ding:2010ga, 
Ding:2016hua, Karsch:2001uw, Aarts:2002cc, Kaczmarek:2011ht,Aarts:2005hg} and in full 
QCD~\cite{Amato:2013naa, Aarts:2014nba} for studying the structure of vector 
correlation functions and their spectral representations as discussed in section \ref{cf_sf}.  Nevertheless, such studies 
have provided \textit{only} critically needed information about various transport 
coefficients both at zero~\cite{Aarts:2002cc, Kaczmarek:2011ht} and finite~\cite{Aarts:2005hg} momentum, 
and the thermal dilepton rate~\cite{Karsch:2001uw, Ding:2010ga, Ding:2016hua}. Employing a free-field spectral function as an 
ansatz, the spectral function in the quenched approximation of QCD was obtained earlier 
and found to approach zero in the low-energy limit~\cite{Karsch:2001uw}.  In the same work, the 
authors found that the lattice dilepton rate approached zero at low invariant 
masses~\cite{Karsch:2001uw}.  In a more recent LQCD calculation with larger lattice 
size, the authors used a Breit-Wigner form for low-energies plus a 
free-field form for high-energies as their ansatz for the spectral 
function~\cite{Ding:2010ga}.  The low-energy BW form of their ansatz gave a finite 
low-energy spectral function and low-mass dilepton rate. Nevertheless, the above 
discussion indicates that because of its limitations the computation of low and intermediate 
mass dilepton rate in LQCD is indeed a difficult task and it is also not very clear if there 
are structures in the low-mass dilepton rate similar to those found in the HTLpt calculation~\cite{Braaten:1990wp}.

Given the uncertainty associated with lattice computation of dynamical 
quantities, e.g. spectral functions, dilepton rate, and transport coefficients, 
it is desirable to have some alternative approaches to evaluate the DPR as well as 
include the non-perturbative effects that can be handled analytically in a similar 
way as in resummed perturbation theory. A few such approaches are available in the 
literature:  one approach is a semi-empirical way to incorporate non-perturbative 
aspects by introducing a gluon condensate in combination with the Green's functions 
in momentum space, which has been proposed in e.g. Refs.~\cite{Schaefer:1998wd,Mustafa:1999jz,
Peshier:1999dt,Chakraborty:2011uw,Chakraborty:2012kx,Chakraborty:2013dda}. An important aspect of the phase structure of QCD is to 
understand the effects of different condensates, which serve 
as order parameters of the broken symmetry phase.  These condensates are non-perturbative 
in nature and their connection with bulk properties of QCD matter is provided by 
LQCD.  The gluon condensate has a potentially substantial impact on the bulk 
properties, e.g., on the equation of state of QCD matter, compared to the quark 
condensate. In this approach, the effective $n$-point functions are related by 
Slavnov-Taylor identities which contain gluon condensates in the deconfined 
phase as hinted from lattice measurements in pure-glue QCD~\cite{Boyd:1996bx}.  The 
dispersion relations with dimension-four gluon condensates in medium exhibit 
two massive modes~\cite{Schaefer:1998wd} (a normal quark mode and a plasmino mode) 
similar to HTL quark dispersion relations.  This feature leads to sharp 
structures (van Hove singularities, energy gap, etc.) in the dilepton production 
rates~\cite{Mustafa:1999dt,Peshier:1999dt} at zero momentum, qualitatively similar to the 
HTLpt rate~\cite{Braaten:1990wp}. In chapter \ref{th_ope} a similar approach will be applied 
to incorporate the non-perturbative effects via dimension-four gluon and quark condensates 
within OPE to compute the intermediate mass DPR. 

Using quenched LQCD, Refs.~\cite{Kitazawa:2009uw,Kaczmarek:2012mb} calculated the 
Landau-gauge quark propagator and its corresponding spectral function by 
employing a two pole ansatz corresponding to a normal quark and a plasmino mode
following the HTL dispersion relations~\cite{Braaten:1990wp}. In a very recent 
approach~\cite{Kim:2015poa}, a Schwinger-Dyson equation  
has been constructed with the aforementioned Landau-gauge propagator obtained 
using quenched LQCD~\cite{Kitazawa:2009uw,Kaczmarek:2012mb} and a vertex function related 
through ST identity.  Using this setup the authors computed the dilepton rate 
from the deconfined phase and found that it has the characteristic van-Hove 
singularities but does not have an energy gap. In chapter \ref{th_gribov} DPR 
will be evaluated following another such recent observation using the Gribov-Zwanzigor 
action to include the confining and nonperturbative effects in the gluon propagator.


\section{QNS - some generalities}
\label{qns_smgen}

Apart from DPR or photon production, screening and fluctuations of conserved quantities have been considered as an important probe for the transient QGP Phase. In contrast with the hadronic phase, where charges are in integer units, in the deconfined QGP phase charges are associated with individual quarks in fractional units. This leads to the difference in charge fluctuations between the two phases, which are related to the corresponding susceptibilities. The QNS is of direct experimental relevance because it is related with the charge fluctuation of a system through the number fluctuation. It can be interpreted as the response of the conserved quark number density $n$, with infinitesimal variation in the quark chemical potentials $\mu+\delta\mu$. In QCD thermodynamics it is defined as the second order derivative of pressure ${\cal P}$ with respect to quark chemical potential $\mu$. But again, using the fluctuation-dissipation (FD) theorem, the QNS for a given quark flavor can also be defined from the time-time component of the current-current correlator in the vector channel~\cite{Haque:2011iz,Chakraborty:2001kx, Blaizot:2001vr,Blaizot:2002xz,Chakraborty:2003uw,Kubo:1957mj,Kunihiro:1991qu}. 

The preceding discussion suggests the formulation of QNS in the following way. QNS can be generally expressed as
\bea
\chi_q(T)&=& \! \! \left. \frac{\partial n}{\partial\mu}\right|_{\mu\rightarrow 0}
\!\!\! = \!\!\!\left. \frac{\partial^2 {\cal P}}{\partial^2\mu}\right|_{\mu\rightarrow 0}
\!\!\! =\int d^4X\langle J_0(0,{\vec x})J_0(0,{\vec 0})\rangle \nn \\
&=&
\beta\int\limits_{-\infty}^{\infty}\frac{d\omega}{2\pi}\frac{-2}{1-e^{-\beta\omega}}
~\mathrm{Im}~\Pi_{00}(\omega, {\vec 0}), \label{defn_qns}
\eea
where $J_0$ is the temporal component of the vector current and $\Pi_{00}$ is the time-time component of the vector correlator or self-energy with external four-momenta $Q \equiv (\omega, \vec q)$. The above relation in Eq.~(\ref{defn_qns}) is known as the  thermodynamic sum rule~\cite{Kubo:1957mj,Kunihiro:1991qu} where the thermodynamic derivative with respect to the external source $\mu$, is related to the time-time 
component of static correlation function in the vector channel. We will discuss more about QNS in chapter \ref{th_qns} where we will calculate it within HTLpt using the non-perturbative GZ action.

%% file: text/ope.tex
\chapter{Dilepton Production Rate from QGP within Operator Product Expansion in $D=4$}
\label{th_ope}

The main aim of the present chapter is to obtain the in-medium electromagnetic SF incorporating the power corrections within OPE in $D=4$ dimension and analyze its effect on the thermal DPR from QGP. To obtain the in-medium electromagnetic SF one needs to calculate the two-point CF  via OPE corresponding to the $D=4$ gluonic and quark operators (condensates) in hot QCD medium. The power corrections appear in the spectral function through the nonanalytic behavior of the correlation function in powers of $P^{-D/2}$ (P is the external four momentum) or logarithms in the Wilson coefficients within OPE in $D$ dimension. This chapter is based on \textit{Power corrections to the electromagnetic spectral function and the dilepton rate in QCD plasma within operator product expansion in D=4} by Aritra Bandyopadhyay and Munshi G. Mustafa, {\bf JHEP 1611 (2016) 183}.

\section{Introduction}
\label{pc_intro}
Over the last couple of decades, with the international efforts from the relativistic HIC experiments in SPS to LHC, we already have some profound signatures of the high temperature QGP. As we already know from chapter \ref{th_intro} and \ref{th_dpr}, DPR is one of them. The importance of having an alternative approach to include nonperturbative effects in DPR is explained earlier in section \ref{dpr_eval}. It is well known that the QCD vacuum has a nontrivial structure consisting of non-perturbative fluctuations of the quark and gluonic fields. These fluctuations can be traced by a few phenomenological quantities, known as vacuum condensates~\cite{GellMann:1968rz}. In standard perturbation theory for simplicity one works with an apparent vacuum and the theory becomes less effective with relatively lower invariant mass.  The vacuum expectation values of such condensates vanishes in the perturbation theory by definition. But in reality they are non-vanishing~\cite{Vainshtein:1978wd,Lavelle:1988eg} and thus the idea of the nonperturbative dynamics of QCD is signaled by the emergence of power corrections in physical observables through the inclusion of nonvanishing vacuum expectation values of local quark and gluonic operators such as the quark and gluon condensates.  In present calculation we intend to compute the IM dilepton production using non-perturbative power corrections within the regime of OPE, the basics of which has already been discussed in section \ref{ope_intro}.

Before going into our calculation we would like to note the following points in OPE: The general and important issue in OPE is the separation of various scales. At finite temperature the heatbath introduces a scale $T$, and then OPE  has three scales: $\Lambda$, $T$ and $P$ beside the factorization scale\footnote{ One chooses the factorization scale $\nu$ as $\Lambda\sim T\lesssim \nu \ll P$ above which the state dependent fluctuations reside in the expectation values of the operators~\cite{CaronHuot:2009ns}.} $\nu$. Based on this one can have\footnote{We also note that there can be another one: $\Lambda$ soft, $P$ hard and $T$ super-hard ($\Lambda < P < < T$). In this case there is a double scale separation and one does not need it for OPE.} either (i) $\Lambda$ and $T$ soft but $P$ hard ($\Lambda\sim T < P$) or (ii)  $\Lambda$ soft  but $P$ and $T$ hard ($\Lambda < P\sim T$). 
\begin{enumerate}
 \item The general belief~\cite{Shuryak:1988ck,Bochkarev:1985ex} is that the Wilson coefficients $W(P^2,\nu^2)$ are c-numbered and remain same 
irrespective of the states considered. This means if one takes vacuum average $\langle \cdots \rangle_0$ or thermal average $\langle \cdots \rangle_\beta$ of Eq.(\ref{corr_func}),  the Wilson coefficient functions, $W(P^2,\nu^2)$, remain temperature independent whereas the temperature dependence resides only in $\langle O \rangle_\beta$. In other words,  OPE is an expansion in $1/P$ where $P$ is the typical momentum scale. But thermal effects are essentially down by thermal factors $\exp(-P/T)$. For $\Lambda\sim T < P$, this does not contribute to any order in $1/P$ in OPE. It is like $\exp(-1/x)$ for which all coefficients in the Taylor expansion in $x(=T/P)$ vanish. 

\item  There are also efforts to extend the OPE to a system with finite temperature~\cite{Furnstahl:1989ji,Hansson:1990tt,Bochkarev:1985ex,Bochkarev:1984zx,Dosch:1988vt}. At nonzero $T$ the heatbath introduces perturbative contributions to the matrix element in OPE, $\langle O \rangle_\beta$ in addition to the nonperturbative contributions of finite dimension composite operator. One needs to determine a temperature $T\sim P$ above which the perturbative calculation of thermal 
corrections is reliable. Usually these perturbative thermal corrections are incorporated by making the Wilson coefficients temperature 
dependent through systematically resummed infinite order in the expansion that comes out to be $\sim T$ (but not $\sim gT$). This means if one
takes the thermal average of Eq.(\ref{corr_func}), then one requires contributions to infinite order in the expansion to get $W$'s temperature dependent, and it becomes  an expansion of $\Lambda/P$ whereas the expansion in $1/(P/T)$ is already resummed~\cite{Furnstahl:1989ji}. Nevertheless, this resummation is appropriate when $T\gtrsim P$,  but QCD sum rule approach may break down and lose its predictive power~\cite{Hansson:1990tt}.  

However,  for low temperature ($\Lambda \sim T< P$) such resummation does not make much sense. Thus for low  $T$  the temperature acts as an infrared effect and cannot change the Wilson coefficients. Since our calculation  is intended for low temperature ($\pi T< P<\omega$), the temperature dependence is only considered in the condensates based on the above point 1 vis-a-vis scale separation as in case (i).
\end{enumerate}

The plan of the rest of this chapter is as follows. In section \ref{ope_setup} we outline some generalities needed for the purpose. In sections \ref{quark_op} and  \ref{gluon_op} we discuss how in-medium quark and gluonic composite operators in $D=4$, respectively, can be included in electromagnetic polarization diagram. We then obtain the two point correlators in terms of Wilson coefficients and those composite operators for the case of light quarks. We also demonstrate how the mass singularity appearing in the correlator is absorbed by using minimal subtraction via operator mixing. In section \ref{spectral}  we discuss about the thermal spectral function and its modification due to incorporation of leading order power correction, particularly in the range of intermediate invariant mass. As a spectral property the dilepton production is discussed  in section \ref{dilepton_rate} and we compare our results with some other known perturbative and nonperturbative results. Then we conclude in section \ref{conclusion}.

\section{Some generalities about composite operators}
\label{ope_setup}

\begin{center}
\begin{figure}[tbh]
 \begin{center}
 \includegraphics[scale=0.4]{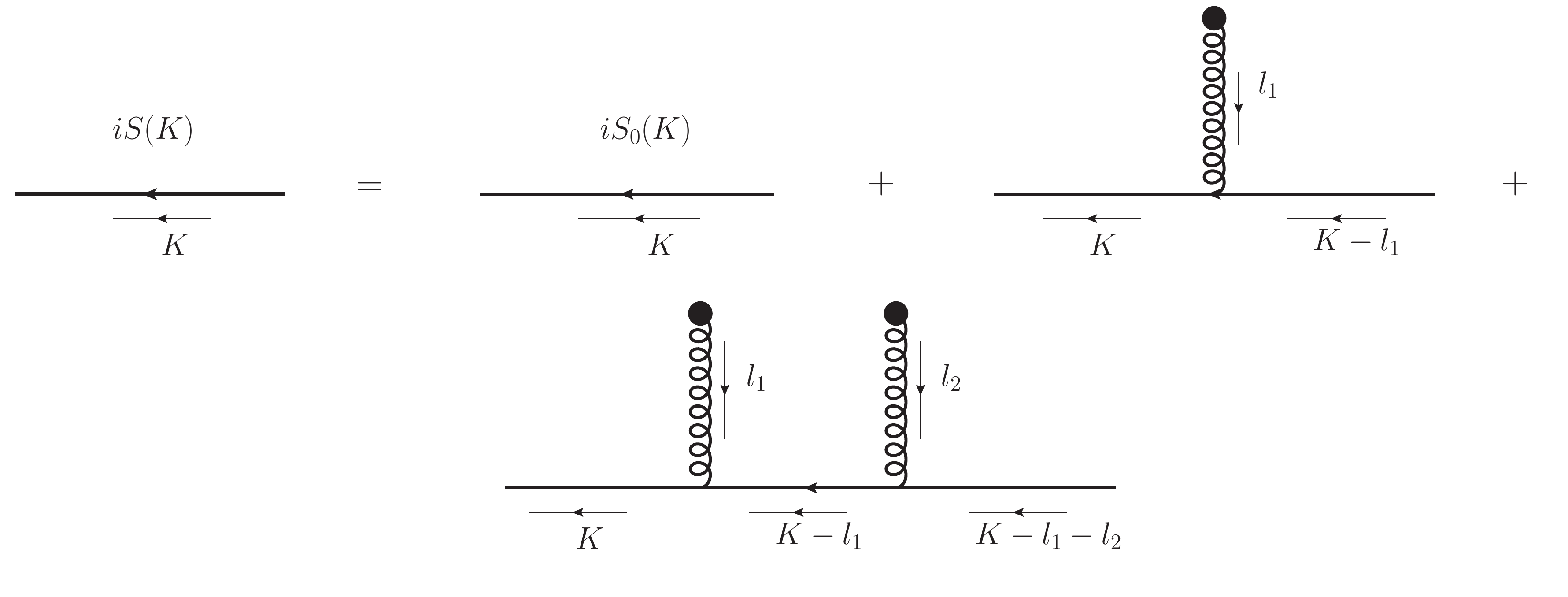} 
 \caption{Effective quark propagator in the background gluon field.}
  \label{quark_propag}
 \end{center}
\end{figure}
\end{center}

In this section some essential ingredients of the following calculation is briefly outlined. In usual notation the nonabelian field tensor in $SU(3)$ is defined as
\bea
G_{\mu\nu}^a &=& \partial_\mu A_\nu^a - \partial_\nu A_\mu^a + 
gf^{abc}A_\mu^bA_\nu^c,\nn\\
 G_{\mu\nu}&=&G_{\mu\nu}^at^a, \nn\\
D_\alpha &=& \partial_\alpha-igt^aA_\mu^a,\
\label{vacuum_fields}
\eea
where $a,b,c$ are color indices, $t^a$ are the generators and 
$g_{\mu\nu}=\text{diag}(1,-1,-1,-1)$. Consequently in vacuum it satisfies the projection 
relation for composite operator
\bea
\Big\langle G_{\mu\nu}^a(0) G_{\alpha\beta}^b(0)\Big \rangle &=& \frac{1}{96}\delta^{ab}
(g_{\mu\alpha}g_{\nu\beta}-g_{\mu\beta}g_{\nu\alpha})\Big \langle G_{\rho\sigma}^c 
G^{c~\rho\sigma} \Big \rangle.
\label{vacuum_projection}
\eea
Choosing the Fock-Schwinger aka the fixed point gauge ($X^\mu A_\mu^a(X)=0$) for 
convenience, the gauge field $A_\mu^a(X)$ can be expressed easily in terms of 
gauge covariant quantities~\cite{Smilga:1982wx, 
Reinders:1984sr,Novikov:1984rf,Novikov:1984ac} as 
\bea
A_\mu(X)=\int\limits_0^1 \sigma ~ d\sigma ~G_{\nu\mu}(\sigma X)X^\nu =\frac{1}{2} 
X^\nu G_{\nu\mu}(0)+\frac{1}{3}X^\alpha X^\nu D_\alpha G_{\nu\mu}(0)+\cdots \, , \nn
\eea
where first the gauge field $G(y)$ has been Taylor expanded in the small $\sigma$ limit 
and then the integration over $\sigma$ has been performed. Now in momentum space it reads
\bea
\!\!\!\!\!\!\!\!&&A_\mu(K)= \int A_\mu(X) e^{iK\cdot X} d^4X\nn\\
\!\!\!\!\!\!\!\!&=&\!\!\!\!\frac{-i(2\pi)^4}{2}G_{\nu\mu}(0)\frac{\partial}{\partial K_\nu}\delta^4(K)+
\frac{(-i)^2(2\pi)^4}{2}(D_\alpha G_{\nu\mu}(0))\frac{\partial^2}{\partial K_\nu \partial 
K_\alpha}\delta^4(K)+\cdots,
\label{gauge_field}
\eea
where each background gluon line will be associated with a momentum integration as we 
will see below.
Using this one can now evaluate the effective quark propagator in presence of 
background gluon lines~\cite{Novikov:1984rf,Novikov:1984ac} by 
expanding the number of gluon legs attached to the bare quark line as in 
Fig.~\ref{quark_propag}. So, it can be written as
\bea
S_{\textrm{eff}}&=& S_0 + S_1 + S_2 + \cdots , \label{quark_propagator_eff} 
\eea
where the bare propagator for a quark with mass $m$ reads as
\bea
S_0&=&\frac{i}{\slashed{K}-m}. \label{quark_propagator1}
\eea
With one gluon leg attached to the bare quark (Fig.~\ref{quark_propag}) the expression 
reads as
\bea
S_1&=& \frac{-i}{\slashed{K}-m}\int 
\frac{d^4l_1}{(2\pi)^4}\frac{\slashed{A}(l_1)}{\slashed{K}-\slashed{l_1}-m}\nn\\
&=&-\frac{i}{4}gt^aG_{\mu\nu}^a(0)\frac{1}{(K^2-m^2)^2}\{\sigma^{\mu\nu}(\slashed{K}
+m)+(\slashed{K}+m)\sigma^{\mu\nu}\} \label{quark_propagator2},
\eea
where 
\bea
\sigma^{\mu\nu} &=& \frac{i}{2}[\gamma^\mu,\gamma^\nu], \nonumber
\eea
and the background gauge field $\slashed{A}(l_1)$ is replaced by the first term of 
the gauge field as given in Eq.~(\ref{gauge_field}).

Similarly for the diagram where two gluon legs are attached to the bare quark, we get
\bea
S_2 &=& \frac{i}{\slashed{K}-m}\int 
\frac{d^4l_1}{(2\pi)^4}\frac{\slashed{A}(l_1)}{\slashed{K}-\slashed{l_1}-m}\int 
\frac{d^4l_2}{(2\pi)^4}\frac{\slashed{A}(l_2)}{\slashed{K}-\slashed{l_1}-\slashed{l_2}-m}
\nn\\
&=&-\frac{i}{4}g^2t^at^bG_{\alpha\beta}^a(0)G_{\mu\nu}^b(0)\frac{(\slashed{K}+m)}{
(K^2-m^2)^5}(f^{\alpha\beta\mu\nu}+f^{\alpha\mu\beta\nu}+f^{\alpha\mu\nu\beta}), 
\label{quark_propagator3}
\eea
where
\bea
f^{\alpha\beta\mu\nu} &=& \gamma^\alpha(\slashed{K}+m)\gamma^\beta(\slashed{K}+m) 
\gamma^\mu(\slashed{K}+m)\gamma^\nu(\slashed{K}+m).\nonumber
\eea
In presence of a medium, however, a four-vector $u_\mu = (1,0,0,0)$ is usually introduced 
to restore Lorentz invariance in the rest frame of 
the heat bath. So, at finite 
temperature additional scalar operators can be constructed so that the vacuum operators 
are generalized to in-medium ones. The projection relation of composite operator in 
(\ref{vacuum_projection}) gets modified in finite temperature~\cite{Mallik:1997pq, 
Mallik:1983nn, Antonov:2004mt, 
Basar:2014swa} as 
\bea
\big \langle G_{\mu\nu}^a(0) G_{\alpha\beta}^b(0) \big \rangle_T &=& 
\big [g_{\mu\alpha}g_{\nu\beta}-g_{\mu\beta}g_{\nu\alpha}\big ]A  
 - \big [(u_\mu u_\alpha g_{\nu\beta}-u_\mu u_\beta g_{\nu\alpha}-u_\nu u_\alpha 
g_{\mu\beta}  \nn \\
&&  +
u_\nu u_\beta 
g_{\mu\alpha})-\frac{1}{2}(g_{\mu\alpha}g_{\nu\beta}-g_{\mu\beta}g_{\nu\alpha})\big ]B 
 + i\epsilon_{\mu\nu\alpha\beta}C ,
\label{medium_projection}
\eea
where $A,B \, \textrm{and} \, C$ are, respectively,  given as
\bea
A = \frac{\delta^{ab}}{96}\big \langle G^2 \big \rangle_T,~ B = 
\frac{\delta^{ab}}{12}\big \langle u\Theta^gu \big \rangle_T,~ C = 
\frac{\delta^{ab}}{96}\big \langle {\cal E} \cdot {\cal B} \big \rangle_T \nn
\eea
with ${\cal E}$ and ${\cal B}$ are, respectively, the electric and magnetic fileds.  
The traceless gluonic  stress tensor, $\Theta^g_{\mu\nu}$,  is given by 
\bea
\Theta^g_{\mu\nu} = 
-G_{\mu\rho}^aG_\nu^{a\rho}+\frac{1}{4}g_{\mu\nu}G^a_{\rho\sigma}G^{\rho\sigma~a}.
\eea
QCD vacuum consists of both quark and gluonic fields. We note that the composite 
operators involving quark fields will be defined below whenever necessary. 

\section{Composite Quark Operators}
\label{quark_op}

\begin{center}
\begin{figure}[h]
 \begin{center}
 \includegraphics[scale=0.3]{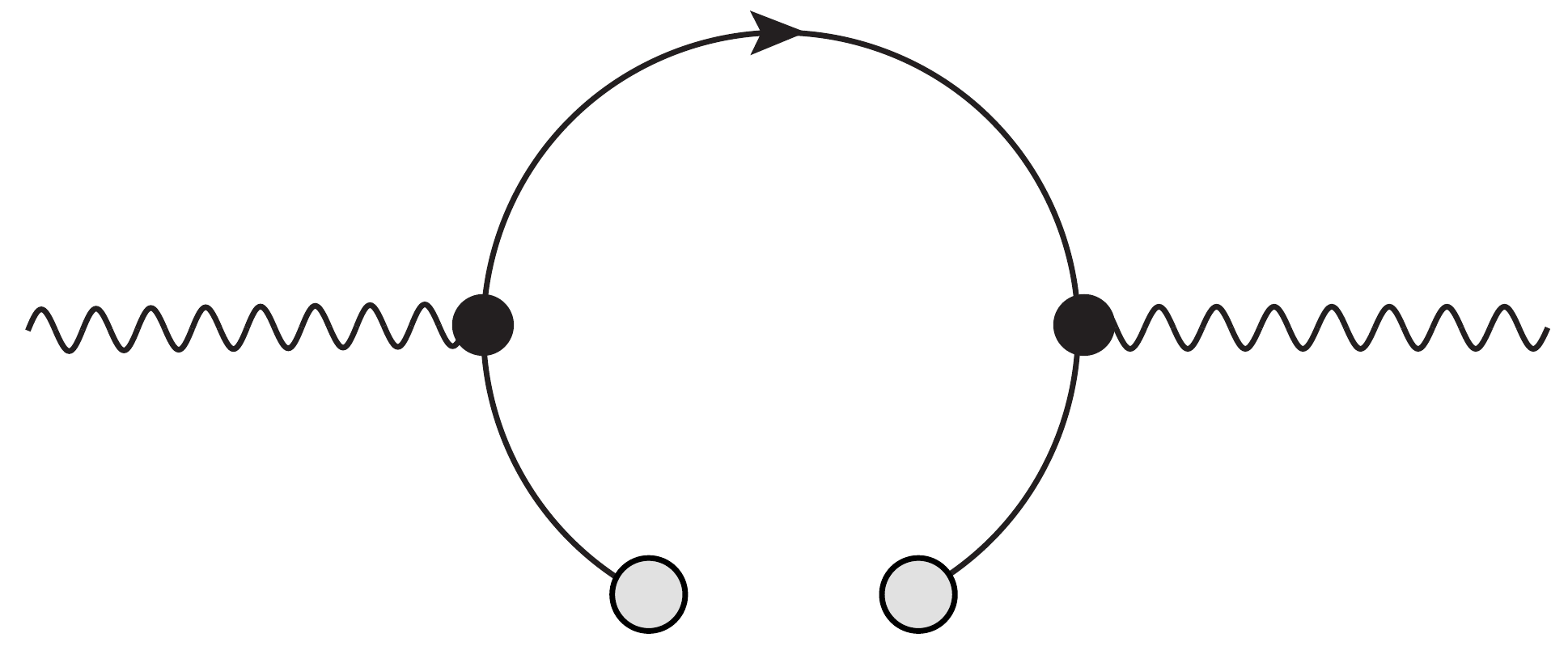}\hspace*{0.5cm} 
 \includegraphics[scale=0.3]{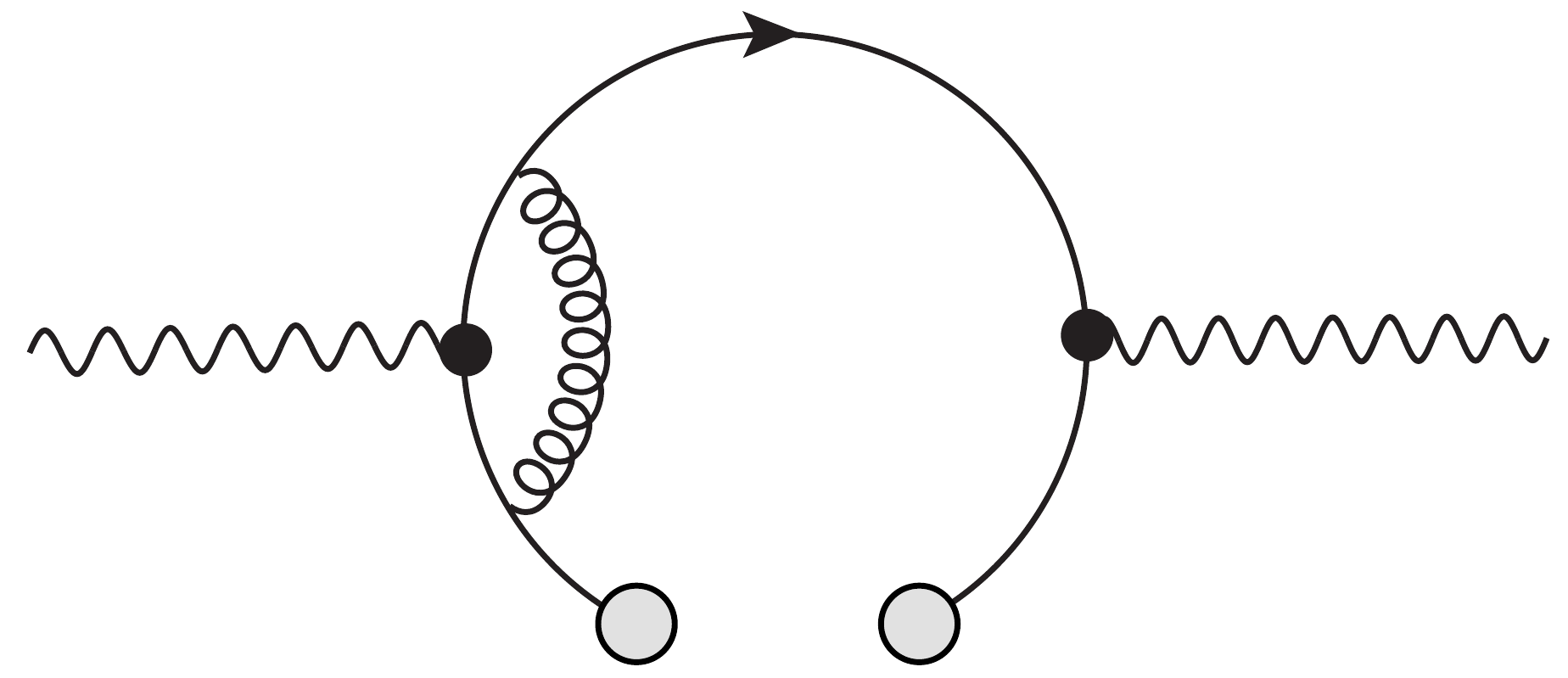} \\
 \vspace*{0.6cm}
 \includegraphics[scale=0.3]{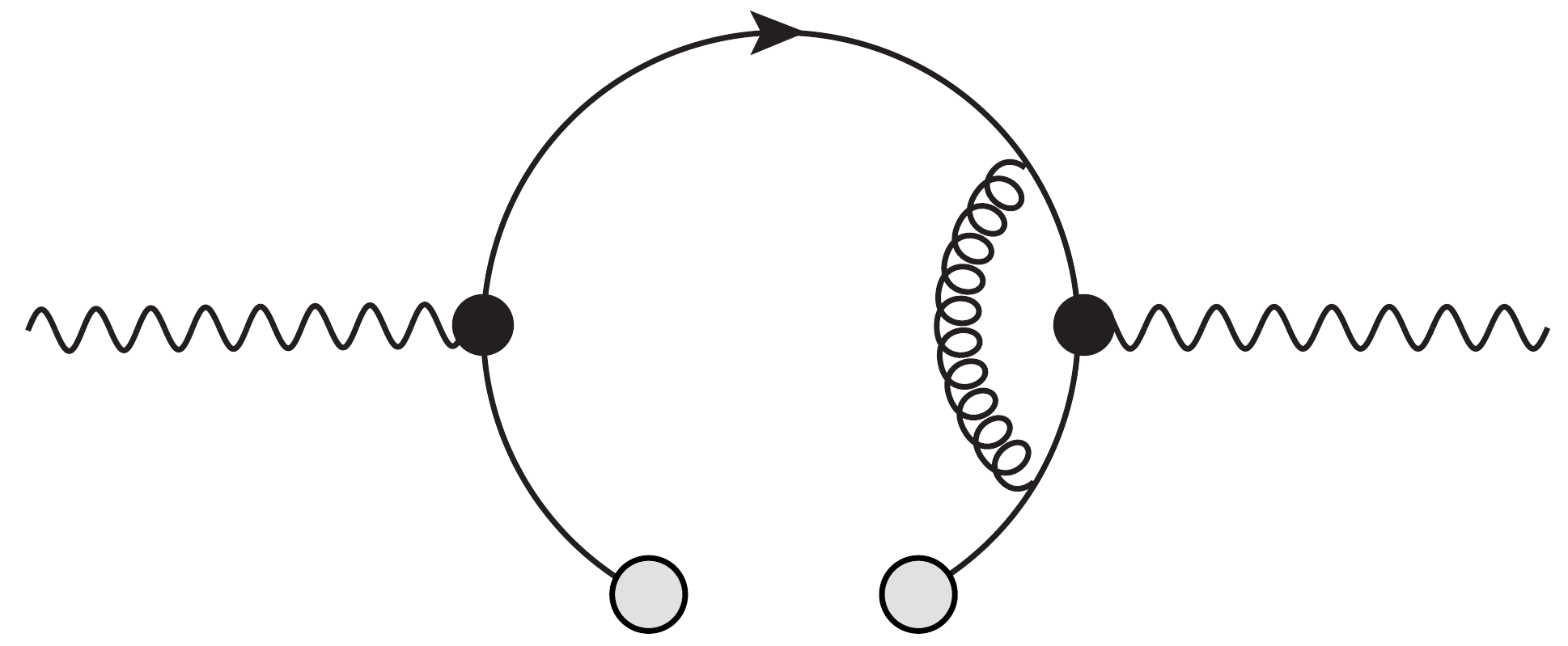}\hspace*{0.5cm} 
 \includegraphics[scale=0.3]{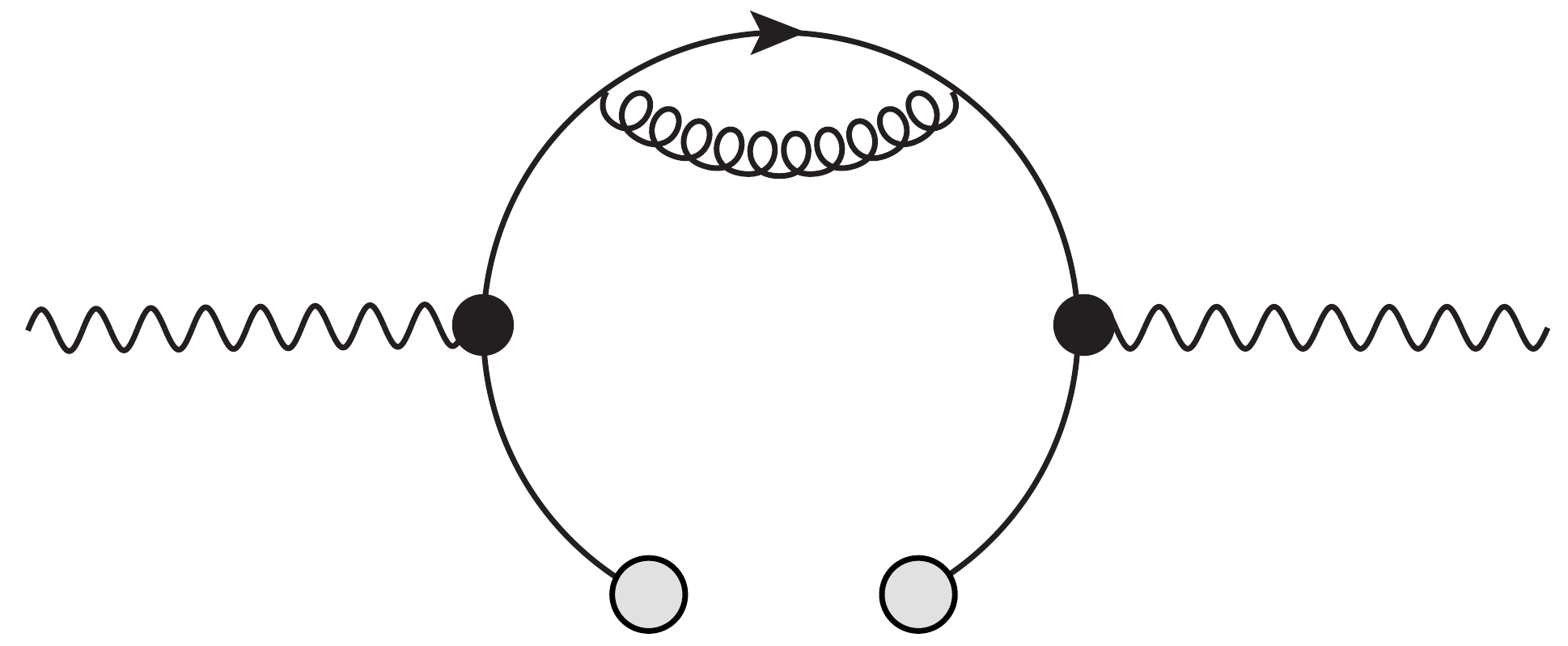} 
 \caption[Topologies corresponding to $D=4$ composite quark operators]{Topologies corresponding to  $D=4$ composite quark operator that can 
contribute to the power correction in the one-loop electromagnetic polarization. The composite quark operator  
is denoted by two gray blobs separated by a gap in the internal quark line. Top left one 
contributes to the leading order (LO:$e^2$) whereas the remaining three are due to  
gluonic corrections contribute to non-leading order (NLO: $e^2g^2$).}
  \label{quark_condensate}
 \end{center}
\end{figure}
\end{center}

In this section we would like to discuss  the power 
corrections using  composite quark operators (condensates) in OPE. In presence of quark 
condensate the leading order contribution comes from the top left 
panel in Fig.~\ref{quark_condensate} where one of the internal quark line  is soft in 
the polarization diagram  represented by two gray blobs with a gap. The 
corresponding contributions can be obtained as
\bea
\Big[C^{\textrm{LO}}_{\mu\nu}(P)\Big ]_q &=& i\int e^{iP\cdot Z}d^4Z \big \langle 
{\cal T}\{j_\mu(Z)j_\mu(0)\}\big \rangle\nn\\
= -N_cN_f\int\!\!\! e^{iP\cdot Z}\!\!\!\!\!\!\!\!\!&&\!\!\!\!\!d^4Z ~\textrm{Tr}\big[\bar{\psi}(Z)\gamma_\mu S(Z,0)  
\gamma_\nu \psi(0) + \bar{\psi}(0)\gamma_\nu S(0,Z) \gamma_\mu \psi(Z)\big] \label{lo_Q}
\eea
where  $N_f$ is the number of quark flavor and $N_c$ is the number of color for a 
given flavor. We also note that the soft quark lines are represented by Heisenberg 
operators $\psi(Z)$ and $\psi(0)$. In the large $P$ limit, $\psi(Z)$ can be expanded as
\bea
\psi(Z)=\psi(0)+ Z^\mu D_\mu \psi(0).\label{exp_heisen}
\eea
Now considering only the first term in the expansion of (\ref{exp_heisen}) 
in (\ref{lo_Q}), one gets
\bea
\Big[C^{\textrm{LO}}_{\mu\nu}(P)\Big]^1_q &=& -N_cN_f\frac{1}{12}\big \langle 
\bar{\psi}\psi\big \rangle\textrm{Tr}\big[\gamma_\mu S(P) \gamma_\nu + \gamma_\nu S(P) 
\gamma_\mu\big ], \label{lo_Q1}
\eea
which, as expected, vanishes in the chiral limit. This is because of the 
appearance of the chiral condensate, $\langle \bar{\psi}\psi\rangle$  which 
is proportional to  the quark mass $m$. On the other hand, choosing the second term in 
the expansion of (\ref{exp_heisen}) we get 
\bea
\Big[C^{\textrm{LO}}_{\mu\nu}(P)\Big]^2_q\!\! &=&\!\! -N_cN_f \!\! \int e^{iP\cdot Z} \, 
\, d^4Z \, \, \times\nn\\ 
&&\!\!\!\!\!\!\!\!\!\!\!\!\!\textrm{Tr}\left[\bar{\psi}(0)Z^\rho \overleftarrow{D}_\rho\gamma_\mu S(Z,0) \gamma_\nu 
\psi(0) + \bar{\psi}(0)\gamma_\nu S(0,Z) \gamma_\mu Z^\rho 
\overrightarrow{D}_\rho\psi(0)\right].\label{lo_Q2}
\eea
Now the most general decomposition of $\langle \bar{\psi}iD_\rho \psi \rangle$ 
for the massless case at finite temperature~\cite{Mallik:1997pq} is  given  as
\bea
\langle \bar{\psi}iD_\rho \psi \rangle_T &=& \left(-\frac{1}{12} 
\gamma_\rho+\frac{1}{3}u_\rho \slashed{u}\right) \langle  u\Theta^fu \rangle_T, 
\label{quark_ope} 
\eea
where $\Theta^f_{\mu\nu}$ is traceless fermionic stress tensor and in the 
massless limit it is given by
\bea
\Theta^f_{\mu\nu} = \bar{\psi}\gamma_\mu i D_\nu \psi.
\label{quark_condensate_relation1}
\eea
Using (\ref{quark_ope}) in (\ref{lo_Q2}), and performing $Z$-integration  one gets
\bea
\!\!\!\!\Big[C^{\textrm{LO}}_{\mu\nu}(P)\Big]^2_{q,T} \!\!\!\!\!&=&\!\!\!\!\frac{-N_cN_f}{12}\frac{\partial}{\partial 
P^\rho}\textrm{Tr}\Big[\!\big(\gamma_\mu S(P) \gamma_\nu+\gamma_\nu S(P) 
\gamma_\mu\big)\big(\!\!\!-\gamma_\rho+4u_\rho \slashed{u} \big)\!\Big]
\!\big \langle u\Theta^fu \big \rangle_T.\nn
\eea
Now treating the Wilson coefficients to be temperature independent, the LO 
contribution is obtained as
\bea
\Big[C^{\mu \, \textrm{LO}}_\mu(P)\Big]_{q,T} &=& \Big[C^{\mu \, \textrm{LO}}_\mu(P)\Big]^2_{q,T} 
= 
\frac{8N_cN_f}{3P^2}\left(1-4\frac{\omega^2}{P^2}\right)\big\langle u\Theta^fu 
\big\rangle_T. \label{lo_q2_final} 
\eea
We note that the contributions from NLO order gluonic corrections (Fig. 
\ref{quark_condensate}) to  quark vacuum condensates are already evaluated 
in~\cite{Pascual:1981jr}. Following the same prescription as LO   
the total NLO in-medium contributions from remaining three diagrams in 
Fig.~\ref{quark_condensate} in the massless limit is obtained as 
\bea
\Big[C^{\mu \, \textrm{NLO}}_\mu(P)\Big]_{q,T} &=& \frac{8N_cN_f}{3P^2}\big\langle u\Theta^fu 
\big\rangle_T\left(1-4\frac{\omega^2}{P^2}\right)
\frac{2g^2}{9\pi^2}\left(1-\ln\left(\frac{-P^2}{\Lambda^2}\right)\right).
\label{nlo_Q}
\eea
Combining (\ref{lo_q2_final}) and (\ref{nlo_Q}) one obtains  power 
correction upto NLO due to quark operator in the electromagnetic correlation function at 
finite temperature as 
\bea
\Big [C_\mu^\mu(P) \Big ]_{q,T} \!\!\!\!&=&\!\!\!\! \frac{8N_cN_f}{3P^2}\big\langle u\Theta^fu 
\big\rangle_T\left(1-4\frac{\omega^2}{P^2}\right)
\left[1+\frac{2g^2}{9\pi^2}\left(1-\ln\left(\frac{-P^2}{\Lambda^2}\right)\right)\right].
\label{quark_operator_massless}
\eea


\section{Composite Gluonic Operators}
\label{gluon_op}
\begin{center}
\begin{figure}[h]
 \begin{center}
 \includegraphics[scale=0.5]{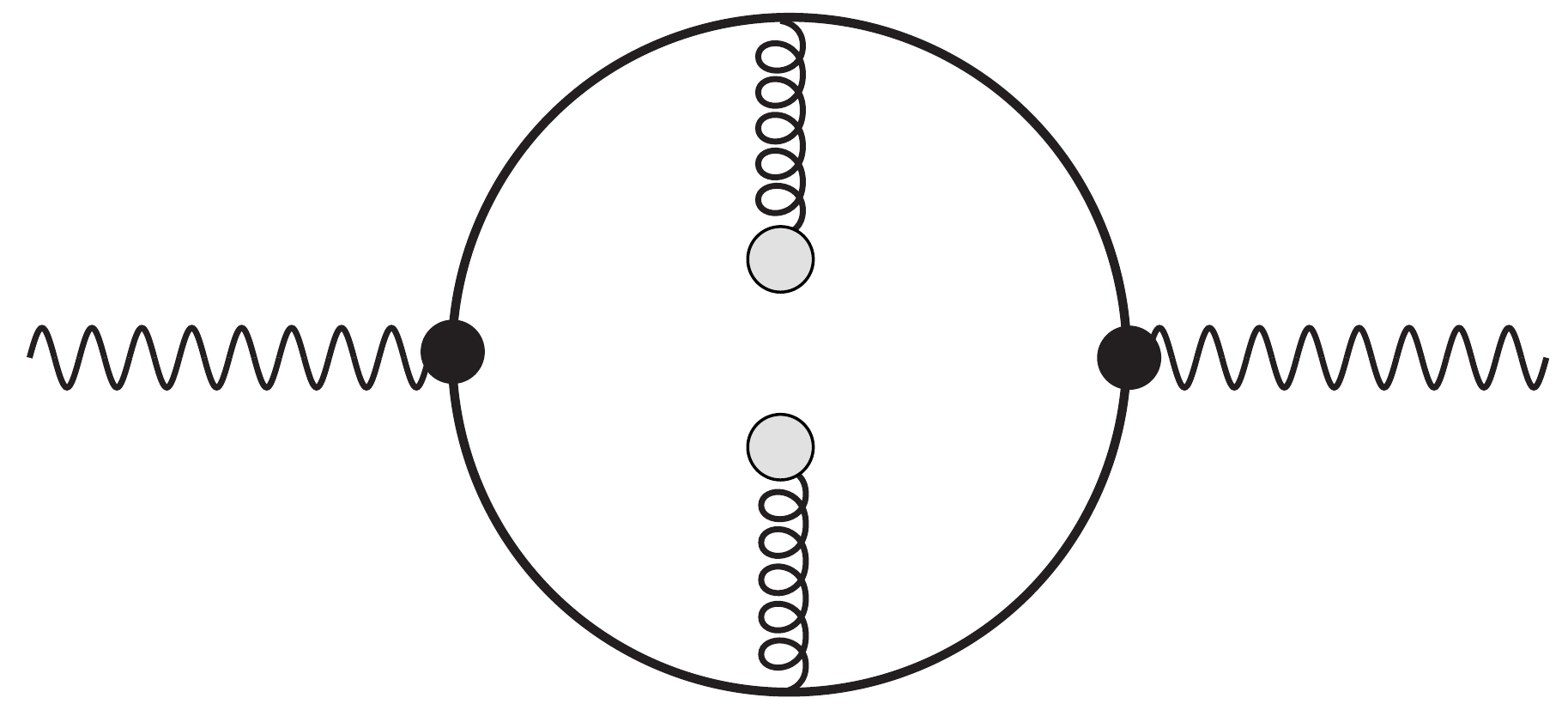} 
 \caption[$D=4$ gluonic operators, Topology-I]{(Topology-I) Vertex correction  where one soft gluon line is attached to 
each internal quark line  in the electromagnetic polarization diagram.}
  \label{topology_1}
 \end{center}
\end{figure}
\end{center}

\begin{center}
\begin{figure}[h]
 \begin{center}
\includegraphics[scale=0.5]{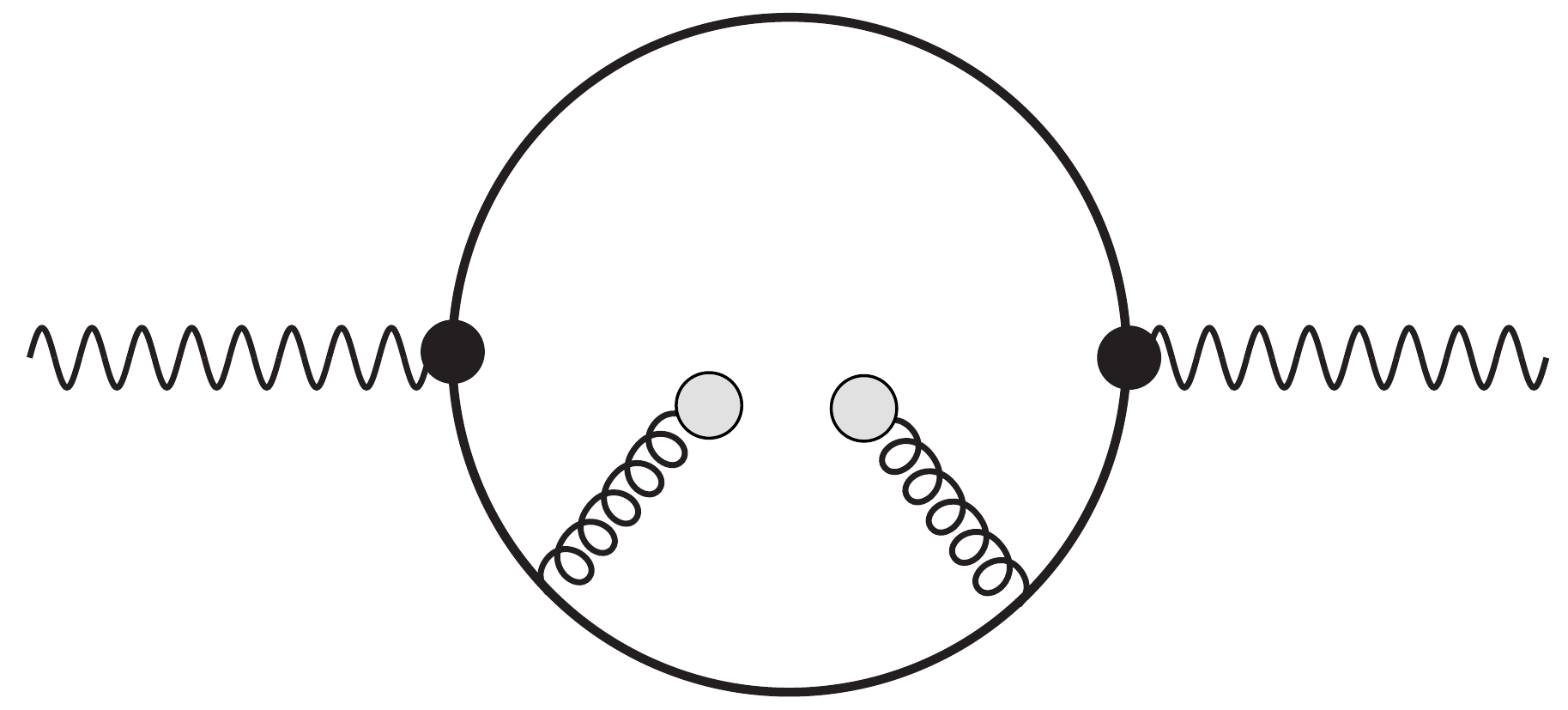}
\caption[$D=4$ gluonic operators, Topology-II]{(Topology-II) Self-energy correction where two soft gluon lines are attached to 
one internal quark line in the electromagnetic polarization diagram. A similar topology 
will also arise when two soft gluon lines are attached to the other quark line.}
\label{topology_2}
\end{center}
\end{figure}
\end{center}

In this section we compute the power correction to the electromagnetic correlation 
function  from the $D=4$ composite gluonic operator by considering the soft  
gluon lines attached to the internal quark lines in the electromagnetic polarization diagram. 
There are two such topologies, as shown in Figs. \ref{topology_1} and \ref{topology_2}, 
depending upon how the soft gluon line is attached to the internal quark lines in 
electromagnetic polarization diagram.

Using Eqs.~(\ref{quark_propagator1})-(\ref{quark_propagator3}),
the contribution of the vertex correction diagram (Fig.~\ref{topology_1}) in vacuum can be written 
as 
\bea
\Big[C_{\mu}^{\mu}(P)\Big]_{g}^{\textrm{I}} &=& iN_cN_f\int 
\frac{d^4K}{(2\pi)^4}  
\textrm{Tr} \Big [ \gamma_\mu S_1(K)\gamma^\mu S_1(Q)\Big ] \nn\\
\!\!\!\!&=&\!\!\!\! -\frac{iN_cN_f}{16}g^2t^at^b\Big\langle G_{\rho\sigma}^a(0) G_{\alpha\beta}^b(0) 
\Big\rangle \nn\\
\times \int \frac{d^4K}{(2\pi)^4} \!\!\!\!\!\!\!&&\!\!\!\!\!\!\!\frac{\textrm{Tr} \Big[\gamma_\mu 
\left(\sigma^{\rho\sigma} 
(\slashed{K}+m)+(\slashed{K}+m)\sigma^{\rho\sigma}\right)\gamma^\mu 
\left(\sigma^{\alpha\beta}(\slashed{Q}+m)+(\slashed{Q}+m)\sigma^{\alpha\beta}\right)\Big 
]} { (K^2-m^2)^2(Q^2-m^2)^2}\nn\\
\!\!\!\!&=&\!\!\!\! 
-iN_cN_f\Big\langle g^2G^2 \Big\rangle \int \frac{d^4K}{(2\pi)^4} \frac{K\cdot Q} 
{(K^2-m^2)^2(Q^2-m^2)^2}, \label{topologyI_vacuum}
\eea
where $Q=K-P$  and  we have also used Eq.~(\ref{vacuum_projection}) in the last step after 
evaluating the trace.

Similarly, the contribution from the topology-II (including 
the one where two gluon lines are attached to the other quark propagator in 
Fig.~\ref{topology_2}) in vacuum can be written as
\bea
\Big[C_{\mu}^{\mu}(P)\Big]_{g}^{\textrm{II}} &=&  2 iN_cN_f\int \frac{d^4K}{(2\pi)^4} 
\textrm{Tr} \Big[\gamma_\mu S_2(K)\gamma^\mu S_0(Q)\Big]\nn\\
&=& \frac{iN_cN_f}{4}g^2t^at^b\Big\langle G_{\rho\sigma}^a(0) G_{\alpha\beta}^b(0) 
\Big\rangle \nn\\
&&\times \int \frac{d^4K}{(2\pi)^4} \frac{\textrm{Tr} \Big[ \gamma_\mu 
(\slashed{K}+m)(f^{\alpha\beta\mu\nu}
+f^{\alpha\mu\beta\nu}+f^{\alpha\mu\nu\beta})\gamma^\mu 
\left(\slashed{Q}+m\right)\Big]}{(K^2-m^2)^5(Q^2-m^2)}\nn\\
 &=& 
-iN_cN_f\Big\langle g^2G^2 \Big\rangle \int \frac{d^4K}{(2\pi)^4} 
\frac{4m^2 (K\cdot Q-2K^2)}{(K^2-m^2)^4(Q^2-m^2)}.\label{topologyII_vacuum}
\eea

Now, we would like to compute both topologies in presence of $D=4$ composite gluonic 
operators at finite $T$. For the purpose,  unlike vacuum case one requires to use 
in-medium gluon condensates as given in Eq.~(\ref{medium_projection}). The
vacuum contributions in Eqs.~(\ref{topologyI_vacuum}) and (\ref{topologyII_vacuum}) are, 
respectively, modified at finite $T$ as
\bea
\Big[C_{\mu}^{\mu}(P)\Big]_{g,T}^{\textrm{I}}
\!\!\!\!&=&\!\!\!\! -\frac{iN_cN_f}{16}g^2t^at^b\Big\langle G_{\rho\sigma}^a(0) G_{\alpha\beta}^b(0) 
\Big\rangle_T \nn\\
\times\int \frac{d^4K}{(2\pi)^4}\!\!\!\!\!\!\!&&\!\!\!\!\!\!\! \frac{\textrm{Tr} \Big[\gamma_\mu 
\big(\sigma^{\rho\sigma}(\slashed{K}+m)+(\slashed{K}+m)\sigma^{\rho\sigma}
\big)\gamma^\mu 
\big(\sigma^{\alpha\beta}(\slashed{Q}+m)+(\slashed{Q}+m)\sigma^{\alpha\beta}\big)\Big]}
{ (K^2-m^2)^2(Q^2-m^2)^2}\nn\\
\!\!\!\!&=&\!\!\!\! -iN_cN_f\Big\langle g^2G^2 \Big\rangle_T 
\int \frac{d^4K}{(2\pi)^4} \frac{K\cdot Q}{(K^2-m^2)^2(Q^2-m^2)^2}\nn\\
\!\!\!\!&&\!\!\!\! +\frac{4iN_cN_f}{3}\Big\langle g^2 u\Theta^gu \Big\rangle_T 
\int \frac{d^4K}{(2\pi)^4} \frac{(K\cdot 
Q-4k_0q_0)}{(K^2-m^2)^2(Q^2-m^2)^2}, \label{topologyI_medium}
\eea
and 
\bea
\Big[C_{\mu}^{\mu}(P)\Big]_{g,T}^{\textrm{II}}
&=& \frac{iN_cN_f}{4}g^2t^at^b\Big\langle G_{\rho\sigma}^a(0) G_{\alpha\beta}^b(0) 
\Big\rangle_T \nn\\
&&\times\int \frac{d^4K}{(2\pi)^4} \frac{\textrm{Tr} \Big[\gamma_\mu 
(\slashed{K}+m)(f^{\alpha\beta\mu\nu}+f^{\alpha\mu\beta\nu}+f^{\alpha\mu\nu\beta}
)\gamma^\mu \left(\slashed{Q}+m\right)\Big]}{(K^2-m^2)^5(Q^2-m^2)}\nn
\eea
\bea
\!\!\!\!&=&\!\!\!\! -4iN_cN_fm^2\Big\langle g^2G^2 \Big\rangle_T 
 \int \frac{d^4K}{(2\pi)^4} \frac{(K\cdot 
Q-2K^2)}{(K^2-m^2)^4(Q^2-m^2)}+\frac{32iN_cN_f}{3}\Big\langle g^2 u\Theta^gu 
\Big\rangle_T\nn\\
&&\times
\int \frac{d^4K}{(2\pi)^4} \frac{K\cdot 
Q(2k_0^2-\frac{1}{2}m^2)+m^2(K^2-4k_0^2)-2k_0q_0(K^2-m^2)}{(K^2-m^2)^4(Q^2-m^2)}.
\label{topologyII_medium}
\eea
Now, in the short-distance or large-momentum limit of nonperturbative power correction, 
one can work with massless quarks without loss of generality. In the massless limit  
Eq.~(\ref{topologyI_medium}) reduces to
\bea
\Big[C_{\mu}^{\mu}(P)\Big]_{g,T}^{\textrm{I}} &\abcom{=}{m\rightarrow 0}& 
-iN_cN_f\Big\langle g^2G^2 \Big\rangle_T 
\int \frac{d^4K}{(2\pi)^4} \frac{K\cdot Q}{(K^2)^2(Q^2)^2}\nn\\ 
&&+\frac{4iN_cN_f}{3}\Big\langle g^2 u\Theta^gu \Big\rangle_T 
\int \frac{d^4K}{(2\pi)^4} \frac{(K\cdot Q-4k_0q_0)}{(K^2)^2(Q^2)^2}, 
\label{topologyI_medium_1}
\eea
whereas for Eq.~(\ref{topologyII_medium}) the coefficient for $\langle g^2G^2 \rangle_T$ 
vanishes and it takes a simple form 
\bea
\Big[C_{\mu}^{\mu}(P)\Big]_{g,T}^{\textrm{II}} &\abcom{=}{m\rightarrow 0}& 
\frac{32iN_cN_f}{3}\Big\langle g^2 u\Theta^gu \Big\rangle_T 
\int \frac{d^4K}{(2\pi)^4} \frac{2(K\cdot 
Q~k_0^2-k_0q_0k^2)}{(K^2)^4(Q^2)}.\label{topologyII_medium_1}
\eea
Here we emphasize the fact that the vacuum correlation function corresponding to 
topology-II (self-energy correction) in Eq.~(\ref{topologyII_vacuum}) vanishes in the 
massless limit. But in medium, one obtains a finite contribution in the massless limit  
as found in Eq.~(\ref{topologyII_medium_1}) due to the in-medium 
condensates. Now, the integrals in the above expressions can 
be expressed in terms of standard Feynman integrals $\mathcal{I}_{mn}$, 
$\mathcal{I}_{mn}^\mu$ and $\mathcal{I}_{mn}^{\mu\nu}$ which  have been evaluated in 
Appendix \ref{massless_feynman}. Using those results in Appendix \ref{massless_feynman}, 
Eqs.~(\ref{topologyI_medium_1}) and (\ref{topologyII_medium_1}) in the 
massless limit ($p_0=\omega$), respectively, become
\bea
\Big [C_{\mu}^{\mu}(P)\Big ]_{g,T}^{\textrm{I}} &\abcom{=}{m\rightarrow 0}& -iN_cN_f\langle 
g^2G^2 \rangle_T ~
\frac{1}{2}(2\mathcal{I}_{12}-P^2\mathcal{I}_{22})\nn\\
&& +\frac{4iN_cN_f}{3}\langle g^2 u\Theta^gu \rangle_T~
\frac{1}{2}(2\mathcal{I}_{12}-P^2\mathcal{I}_{22}+8p_0\mathcal{I}_{22}^0-8\mathcal{I}_{22}
^{00})\nn\\
&=&  \langle G^2 \rangle_T~ \frac{g^2N_cN_f}{16\pi^2 P^2} -\langle u\Theta^gu 
\rangle_T~\frac{g^2N_cN_f}{3\pi^2 P^2}
\left[\frac{\omega^2}{P^2}-\frac{1}{4}\right], \label{topologyI_massless} \\
\textrm{\ \ }\ \ \ \ \ \ \ \ \ \ \ \ \ \ && \ \ \ \ \ \ \ \ \ \ \ \ \ \ \ \ \ \ \ \ \ \ \ 
\ \ \ \ \ \ \ \ \ \ \ \ \ \ \  \ \ \ \ \ \nn  \\ 
\textrm{and}\ \ \ \ \ \ \ \ \ \ \ \ \ \ \ \ \ \ \ \ \ \ \  && \ \ \ \ \ \ \ \ \ \ \ \ \ \ 
\ \ \ \ \ \ \ \ \ 
\ \ \ \ \ \ \ \ \ \ \ \ \ \ \  \ \ \ \ \ \nn  \\ 
\textrm{\ \ }\ \ \ \ \ \ \ \ \ \ \ \ \ \ && \ \ \ \ \ \ \ \ \ \ \ \ \ \ \ \ \ \ \ \ \ \ \ 
\ \ \ \ \ \ \ \ \ \ \ \ \ \ \  \ \ \ \ \ \nn  \\ 
\Big[C_{\mu}^{\mu}(p)\Big]_{g,T}^{\textrm{II}} &\abcom{=}{m\rightarrow 0}& 
 \frac{32iN_cN_f}{3}\langle g^2 u\Theta^gu \rangle_T~ 
(2p_0\mathcal{I}_{31}^0-\mathcal{I}_{31}^{00}-P^2\mathcal{I}_{41}^{00})\nn\\
&=& -\frac{g^2N_cN_f}{9\pi^2P^2}\langle u\Theta^gu \rangle_T~
\left[\frac{1}{\tilde{\epsilon}}\left(1-\frac{4\omega^2}{P^2}\right)+2-6\frac{\omega^2}{
P^2}\right].\label{topologyII_massless}
\eea
We note that Eq.~(\ref{topologyII_massless}) has a mass singularity as 
${1}/{\tilde{\epsilon}} = 
{1}/{\epsilon}-\ln\left({-P^2}/{\Lambda^2}\right)$, the reason for which could be 
understood in the following way:  while computing the self-energy correction 
corresponding to topology-II in Fig.~\ref{topology_2}, one actually overcounts a 
contribution from quark condensate. This is because the quark line 
in-between two soft gluon lines in Fig.~\ref{topology_2} becomes soft, leading to 
quark condensate. So the actual contribution from the gluonic operators can only be 
obtained after minimally subtracting the quark condensate 
contribution~\cite{Generalis:1983hb,Broadhurst:1984rr}
which should cancel the mass singularity arising in the massless 
limit\cite{Nikolaev:1982rq, Nikolaev:1982ra, Nikolaev:1981ff}.

To demonstrate this we begin by considering finite quark mass in which a correlator 
containing  quark condensates ($\mathcal{Q}_k$) can be expressed via gluon condensates 
($\mathcal{G}_n$)~\cite{Grozin:1994hd} in mixed representation as
\bea
\mathcal{Q}_k = \sum_n c_{kn}(m)\mathcal{G}_n, \label{mixed_rep}
\eea
where, $c_{kn}(m)$ is an expansion in $1/m$. Then one can also represent a correlator 
with gluon condensates ($\mathcal{G}_n$) as 
\bea
\Big[C_{\mu}^{\mu}(P)\Big]_g &=& \sum_n a_n(P^2,m)\mathcal{G}_n, \label{gc_basis}
\eea
whereas for quark condensates ($\mathcal{Q}_k$) it can be written as
\bea
\Big[C_{\mu}^{\mu}(P)\Big]_q &=& \sum_k b_k(P^2,m)\mathcal{Q}_k, \label{qc_basis}
\eea
with $a_n$ and $b_k$ are the corresponding coefficients for the gluon and quark 
condensates, respectively. Now, in general a correlator with  minimal subtraction
using Eqs.~(\ref{mixed_rep}), (\ref{gc_basis}) and (\ref{qc_basis}) can now be written as 
\bea
\!\!\Big[C_{\mu}^{\mu}(P)\Big]_{g}^{\textrm{a}} \!\!\!\!\!&=&\!\!\!\!\! 
\Big[C_{\mu}^{\mu}(P)\Big]_{g} \!\!\!-\!\!\! \Big[C_{\mu}^{\mu}(P)\Big]_{q} 
\!\!=\!\! \sum_n a_n(P^2,m)\mathcal{G}_n \!\!-\!\! \sum_{n,k} b_k(P^2,m)c_{kn}(m)\mathcal{G}_n.
\label{grozin_massive}
\eea
We note here that after this minimal subtraction with massive correlators
and then taking  the massless limit renders the resulting correlator finite.
Since we are working in a massless limit, one requires an appropriate 
modification~\cite{Broadhurst:1984rr, Broadhurst:1985js} of 
Eq.~(\ref{mixed_rep}). The difference between a renormalized quark condensate 
and a bare one can be written as~\cite{Grozin:1994hd},
\bea
\mathcal{Q}_k-\mathcal{Q}_k^{b} = -\frac{1}{\epsilon} \sum_{d_n\le 
d_k}m^{d_k-d_n}\gamma_{kn}\mathcal{G}_n, \label{difference_mixed_rep}
\eea
where $d_n$ and $d_k$ represents the dimensions of $\mathcal{G}_n$ and $\mathcal{Q}_k$  
and $\gamma_{kn}$ are the mixing coefficients of $\mathcal{Q}_k$ with $\mathcal{G}_n$.
Now if one wants to go to $m\rightarrow 0$ limit, $\mathcal{Q}_k^{b}$ vanishes 
because there is no scale involved in it. Also only $d_n=d_k$ term survives producing
\bea
\mathcal{Q}_k = -\frac{1}{\epsilon} \sum_{d_n= 
d_k}\gamma_{kn}\mathcal{G}_n. \label{modified_mixed_rep}
\eea
Using Eq.~(\ref{modified_mixed_rep}) in the first line of  Eq.~(\ref{grozin_massive}) one 
can write
\bea
\Big[C_{\mu}^{\mu}(P)\Big]_{g}^{\textrm{a}} = 
\Big[C_{\mu}^{\mu}(P)\Big]_{g} + \frac{1}{\epsilon}\sum\limits_{d_n=d_k} 
b_k(P)\gamma_{kn}\mathcal{G}_n, \label{modified_min_sub}
\eea
where, $b_k(P)$ is the coefficient of the quark condensate $\mathcal{Q}_k$ in the 
massless limit, which is of similar dimension as $\mathcal{G}_n$.

So, for minimal subtraction of  the quark condensate contribution 
overestimated in Eq.~(\ref{topologyII_massless}), the  in-medium quark condensate 
(appearing in Eq.~(\ref{quark_operator_massless})) has to be expressed in terms of the 
in-medium gluon condensates of the same dimension as~\cite{Zschocke:2011aa}
\bea
\big \langle \bar{\psi}\gamma_\mu i D_\nu \psi \big \rangle_T &=& \big \langle : 
\bar{\psi}\gamma_\mu i D_\nu \psi : \big \rangle_T 
+\frac{3}{16\pi^2}m^4 g_{\mu\nu}\left(\ln\frac{\mu^2}{m^2}+1\right)
-\frac{g_{\mu\nu}}{48}\Big \langle \frac{g^2}{4\pi^2}G^2 \Big \rangle_T\nn\\
&&-\frac{1}{18}(g_{\mu\nu}-4u_\mu u_\nu)\left(\ln\frac{\mu^2}{m^2}-
\frac{1}{3}\right)\Big \langle \frac{g^2}{4\pi^2}u\Theta^gu \Big \rangle_T,
\label{hilger}
\eea
where the first term in the right hand side represents the normal ordered condensate.
After contracting Eq.~(\ref{hilger}) by $u^\mu u^\nu$ and applying 
Eq.~(\ref{quark_condensate_relation1}) we obtain,
\bea
\langle u \Theta^f u \rangle &=& \textrm{Other nonrelevant terms~} + 
\frac{1}{6}\left(\ln\frac{\mu^2}{m^2}-
\frac{1}{3}\right)\Big \langle \frac{g^2}{4\pi^2}u\Theta^gu \Big \rangle . 
\label{qcon_via_gcon}
\eea
Now comparing Eqs.~(\ref{qcon_via_gcon}) and (\ref{modified_mixed_rep}) we find 
\bea
\gamma_{kn} = \frac{1}{6},~~b_k(P) 
= \frac{8N_cN_f}{3P^2}\left(1-\frac{4\omega^2}{P^2}\right).\label{coeff_min_sub}
\eea
Therefore, the electromagnetic correlator with gluon condensates for self-energy 
correction (topology-II) in the massless limit can now be written as
\bea
\Big[C_{\mu}^{\mu}(P)\Big]_{g,T}^{\textrm{II,a}}&=\atop{m\rightarrow 0}&\Big[C_{\mu}^{\mu}
(P)\Big ] _ { g , T } ^ {
\textrm{II}}+ \frac{1}{\epsilon}\sum\limits_{d_n=d_k} 
b_k(P)\gamma_{kn}\mathcal{G}_n,\nn\\
 \!\!\!\!&=&\!\!\!\!\! -\frac{g^2N_cN_f}{9\pi^2P^2}\langle u\Theta^gu \rangle_T
\left[\frac{1}{\tilde{\epsilon}}\left(1-\frac{4\omega^2}{P^2}\right)+2-6\frac{\omega^2}{
P^2}\right]\nn\\
&&+
\frac{1}{\epsilon}\frac{g^2N_cN_f}{9\pi^2P^2}\left(1-4\frac{\omega^2}{P^2}\right)\langle 
u\Theta^gu \rangle_T\nn\\
\!\!\!\!&=&\!\!\!\!\!\!\! -\frac{g^2N_cN_f}{9\pi^2P^2}\langle u\Theta^gu \rangle_T 
\left[-\ln\left(\frac{-P^2}{\Lambda^2}\right)\left(1-\frac{4\omega^2}{P^2}\right)+2-6\frac
{\omega^2}{P^2}\right].
\label{minimal_subtracted_topology_II}
\eea
So, the minimal subtraction eventually cancels the divergence from the expression of 
gluonic operators. Now combining Eq.~(\ref{topologyI_massless}) and 
Eq.~(\ref{minimal_subtracted_topology_II}), the final expression for the gluonic 
contribution in the self-energy power correction is given by,
\bea
\!\!\!\!&&\!\!\!\!\Big[C_{\mu}^{\mu}(P)\Big]_{g,T} =
\frac{g^2N_cN_f}{\pi^2P^2} \times \nn\\
&&\left[\frac{1}{9}\big\langle u\Theta^gu \big\rangle_T
\left(\ln\left(\frac{-P^2}{\Lambda^2}\right)\!\left(1-\frac{4\omega^2}{P^2}\right)+9\frac{
\omega^2}{P^2}-\frac{11}{4}\right)-\frac{1}{16}\big\langle G^2 \big\rangle_T\right].
\label{gluonic_final}
\eea

\section{Electromagnetic spectral function}
\label{spectral}

The correlation function with power corrections from both quark and gluonic composite 
operators can now be written from Eq.~(\ref{quark_operator_massless}) and 
Eq.~(\ref{gluonic_final}) as 

\bea
\Big[C_{\mu}^{\mu}(P)\Big]_{T} \!\!\!\!&=&\!\!\!\!\! \Big[C_{\mu}^{\mu}(P)\Big]_{g,T} 
+\Big[C_{\mu}^{\mu}(P)\Big]_{q,T} \nn\\ 
\!\!\!\!&=&\!\!\!\!\!\frac{g^2N_cN_f}{\pi^2P^2}\left[\frac{1}{9}\big\langle u\Theta^gu \big\rangle_T
\left(\ln\left(\frac{-P^2}{\Lambda^2}\right)\!\left(1-\frac{4\omega^2}{P^2}\right)+9\frac{
\omega^2}{P^2}-\frac{11}{4}\right)-\frac{1}{16}\big\langle G^2 \big\rangle_T\right]\nn\\
\!\!\!\!&+&\!\!\!\!\! \frac{8N_cN_f}{3P^2}\big\langle u\Theta^fu 
\big\rangle_T\left(1-4\frac{\omega^2}{P^2}\right)
\left[1+\frac{2g^2}{9\pi^2}\left(1-\ln\left(\frac{-P^2}{\Lambda^2}\right)\right)\right].
\label{correlator_final}
\eea
The contribution to spectral 
function comes from  nonanlytic behavior of $\ln\left(\frac{-P^2}{\Lambda^2}\right)$ 
having a discontinuity of $2\pi$. Following Eq.~(\ref{spec_func}) the 
electromagnetic spectral function with leading  non-perturbative power 
corrections in the OPE limit $\pi T<  P< \omega$  can be written as
\bea
\rho^{\textrm{pc}}_f(P)=-\frac{16N_c \alpha_s}{9 
P^2\pi}\left(1-4\frac{\omega^2}{P^2}\right)\left[\frac{8}{3}\Big\langle \Theta^f_{00} 
\Big\rangle_T-\frac{1}{2}\Big\langle \Theta^g_{00} \Big\rangle_T\right],
\label{spec_func_pc}
\eea
where the power corrections ($P^{D/2}$) from 
the QCD vacuum, resides in the denominator of the  Wilson coefficient as we have 
considered the $D=4$ dimensional composite operators. Now, $\Theta_g^{00}$ and 
$\Theta_f^{00}$ are respectively the gluonic and fermionic part 
of the energy density $\mathcal{E}$, and in the Stefan-Boltzmann limit given by
\bea
\Big\langle \Theta^g_{00} \Big\rangle^{\textrm{SB}}_T &=& \frac{\pi^2T^4}{15}d_A,\nn\\
\Big\langle \Theta^f_{00} \Big\rangle^{\textrm{SB}}_T &=& 
\frac{7\pi^2T^4}{60}d_F,\label{opera_sb}
\eea
where, $d_A=N_c^2-1$ and $d_F=N_cN_f$. So the leading correction to the electromagnetic spectral function is ${\cal O}(g^2T^4)$ and  is in conformity with those obtained in Ref.\cite{CaronHuot:2009ns} using Renormalization Group equations. Now the perturbative leading order result, or the free spectral function, is given by
\bea
\rho^{\textrm{PLO}}(p)=\frac{N_c N_f TP^2}{4 \pi^2 p}\ln\left[\frac{\cosh\left(\frac{\omega + p}{4T}\right)}{\cosh\left(\frac{\omega - 
p}{4T}\right)}\right].
\label{spec_func_lop}
\eea

Before analyzing the spectral function we would like to note that the evaluation 
of the composite quark and gluon operators (condensates) in Eq.~(\ref{spec_func_pc}),  
$\langle \Theta_g^{00}\rangle_T$ and $\langle \Theta_f^{00}\rangle_T$, respectively, at 
finite $T$  should proceed via nonperturbative methods of QCD. Unfortunately, the present
knowledge of these in-medium operators are very meagre in the existing literature and 
some preliminary estimation should come from LQCD techniques. Nevertheless, we rely on 
the Stefan-Boltzmann limits as given in Eq.~(\ref{opera_sb}) for these composite operators 
to have limiting information from the power corrections.

\begin{center}
\begin{figure}[h]
 \begin{center}
 \includegraphics[scale=1.0]{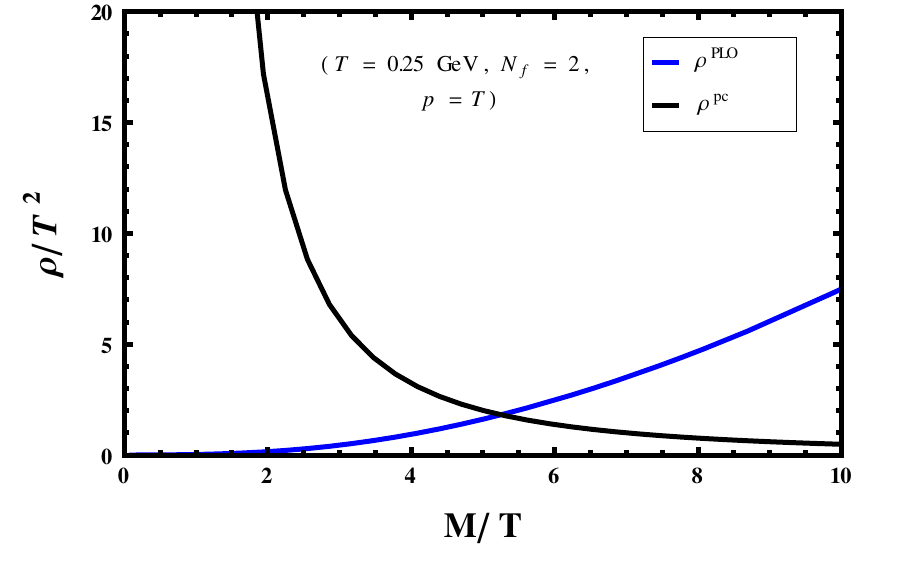}
 \caption{Comparison of the  electromagnetic spectral function betwen  
the perturbative leading order (PLO)  and power corrections from $D=4$.}
  \label{spec_plot}
 \end{center}
\end{figure}
\end{center}

We now demonstrate the importance of the power corrections  in 
the thermal spectral function. Figure \ref{spec_plot} displays a comparison between 
the perturbative leading order contribution in Eq.~(\ref{spec_func_lop}) and the power 
corrections contribution in Eq.~(\ref{spec_func_pc}). As is seen from the figure, the nature of 
the two spectral functions are drastically different to each other as a function
of $M/T$, the scaled invariant mass with respect to temperature. While the perturbative 
leading order result increases with the increase of $M/T$, the leading order 
power correction starts with a very high value but falls off very rapidly. We emphasize here, that the low invariant  mass region is excluded in the OPE  limit, 
$\pi T < P <\omega$. On the other hand the vanishing contribution of the power 
corrections 
at large invariant mass ($M\approx 10T$) is expected because of the appearance of 
$P^{-2}$ due to dimensional argument as discussed after  Eq.~(\ref{spec_func_pc}). So, at 
large invariant mass the perturbative calculation becomes more effective as can be seen 
in Fig.~\ref{spec_plot}. 

\begin{center}
\begin{figure}[h]
 \begin{center}
 \includegraphics[scale=0.8]{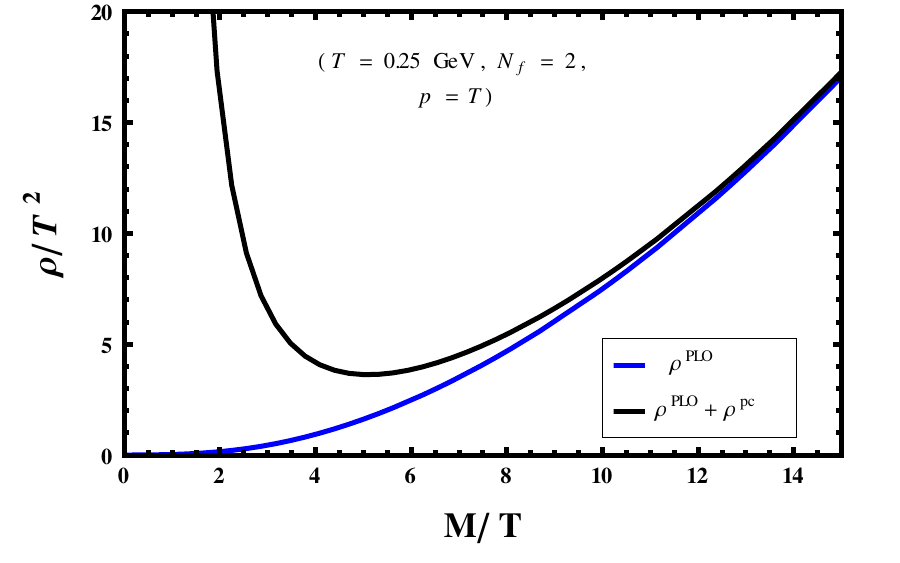}
\includegraphics[scale=0.8]{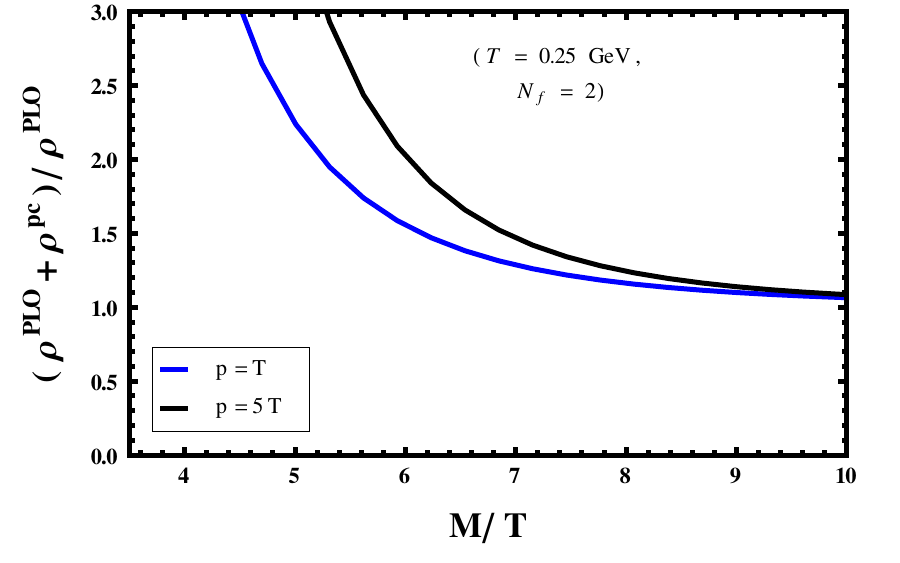}
 \caption{ Comparison between  $\rho^{\textrm{PLO}}$ in Eq.~(\ref{spec_func_lop}) and   
$\rho^{\textrm{PLO}} + \rho^{\textrm{pc}}$ in the \textit{left panel}  and  their ratio 
in the \textit{right panel}. }
  \label{spec_plot_2}
 \end{center}
\end{figure}
\end{center}

In figure \ref{spec_plot_2}, a comparison (left panel) between the $\rho^{\textrm{PLO}}$ 
in Eq.~(\ref{spec_func_lop}) and   
$\rho^{\textrm{PLO}} + \rho^{\textrm{pc}}$ 
[Eq.~(\ref{spec_func_lop})+Eq.~(\ref{spec_func_pc})] and their ratio are displayed, 
respectively. From the 
left panel one finds that in the intermediate mass regime,  $M\approx 4T$ to  $10T$, 
\textrm{i.e.}, (1 to 2.5) GeV, there is a clear indication of enhancement in the  
electromagnetic spectral function due to the leading order power corrections in $D=4$ 
dimension.  This is also reflected in the ratio plot in the right panel. Both plots 
assure that the power corrections becomes important in the intermediate mass range of 
the electromagnetic spectral function. 

\begin{center}
\begin{figure}[h]
 \begin{center}
\includegraphics[scale=0.9]{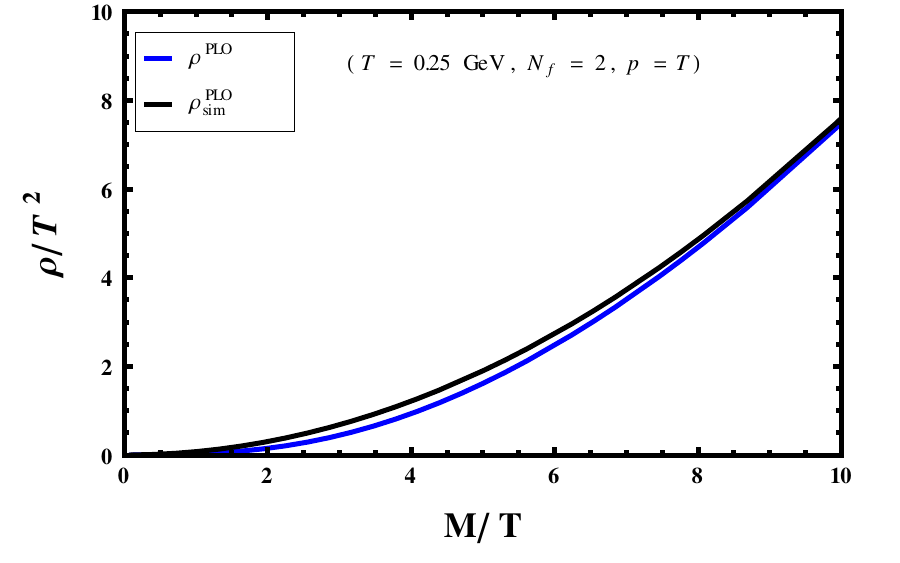}
 \caption{ Comparison between  $\rho^{\textrm{PLO}}$ in Eq.~(\ref{spec_func_lop}) and  
simplified one  $\rho^{\textrm{PLO}}_{\textrm{sim}}$ in Eq.~(\ref{plo_ope}).}
\label{plo_ope_limit}
 \end{center}
\end{figure}
\end{center}

For convenience the PLO spectral function in Eq.~(\ref{spec_func_lop}) can be simplified 
in the OPE limit as
\bea
\rho^{\textrm{PLO}}_{\textrm{sim}}(P)&=&\frac{N_cN_fP^2}{4 \pi^2}. \label{plo_ope}
\eea
and is also justified through Fig.~\ref{plo_ope_limit}.

The total spectral function with  the power 
correction in the OPE limit can now be written as
\bea
\rho(P)\vert^{OPE}&=& \rho^{\textrm{PLO}}_{\textrm{sim}}(P)+\rho^{\textrm{pc}}(P) \nn \\
&=& \frac{N_cN_fP^2}{4 \pi^2}-\frac{16N_cN_f \alpha_s}{9\pi 
P^2}\left(1-4\frac{\omega^2}{P^2}\right)\left[\frac{8}{3}\Big 
\langle \Theta_f^{00}\Big \rangle_T -\frac{1}{2} \Big \langle\Theta_g^ { 
00}\Big \rangle_T\right
] .
\label{spec_func_ope}
\eea
Now we note that the virtual photon will decay into two leptons and the features 
observed in the electromagnetic spectral function will also be reflected in the dilepton 
production rate, which will be discussed in the next 
section. 

\section{Dilepton Rate}\label{dilepton_rate}

The modified PLO  differential dilepton production rate in presence of leading power 
correction in OPE with $D=4$  is now obtained in a compact form by combining 
Eqs.~(\ref{dpr_unmagnetized_final}) and (\ref{spec_func_ope}) for $N_f=2$ as
\bea
\label{dilepton_rate_final}
{\frac{dN}{d^4Xd^4P}}\Bigg\vert^{\textrm{OPE}} \!\!\!\!&=&\!\!\!\!  \frac{5\alpha^2_{\textrm 
em}}{27 \pi^2 M^2}
n_B\left(\omega\right) \nn \\ 
\!\!\!\!&&\!\!\!\!\times \left[ \frac{N_cP^2}{4 \pi^2} -\frac{16N_c \alpha_s}{9\pi 
P^2}\left(1-4\frac{\omega^2}{P^2}\right)\left(\frac{8}{3}\Big 
\langle \Theta_f^{00}\Big \rangle -\frac{1}{2} \Big \langle\Theta_g^ { 
00}\Big \rangle\right ) \right ],
\eea
where we have used $\sum_f q_f^2 =5/9$, for massless $u$ and $d$ quarks.
The leading power corrections within  OPE in $D=4$ dimension is of ${\cal 
O}(\alpha^2_{em} \alpha_s)$ to the PLO of ${\cal O}(\alpha^2_{em})$.

\begin{center}
\begin{figure}[h]
 \begin{center}
 \includegraphics[scale=1.]{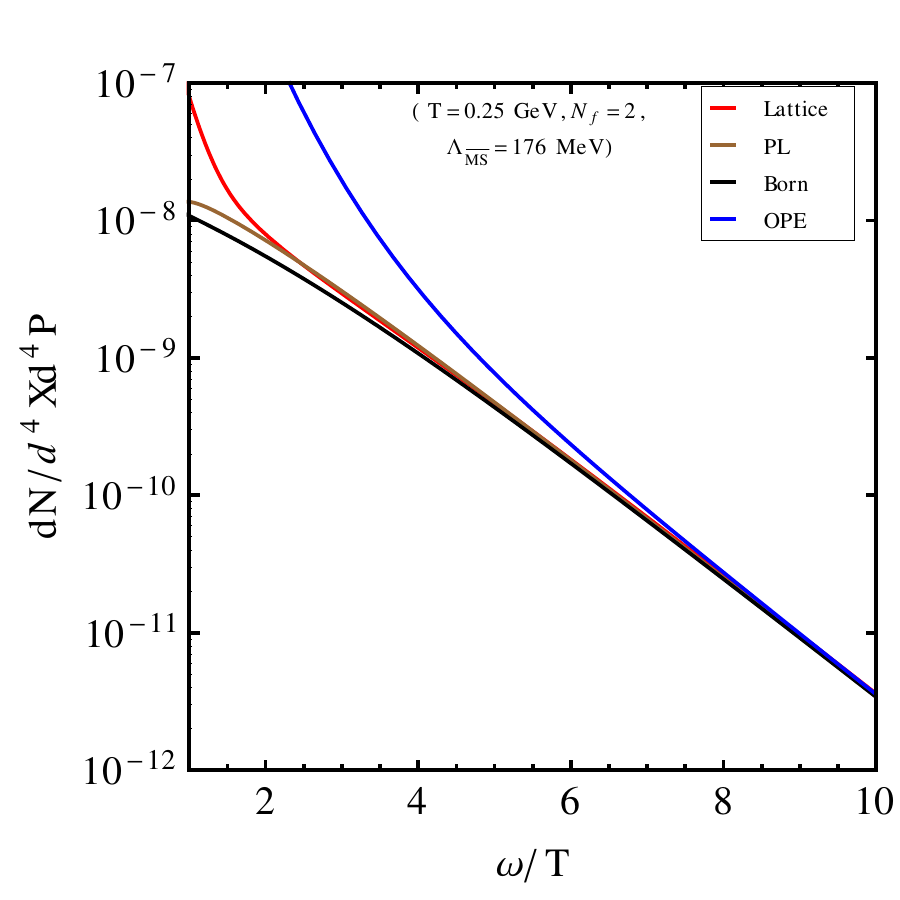} 
 \caption[Comparison between different dilepton rates as a function of  $\omega/T$]{Comparison between different dilepton rates as a function of  $\omega/T$ with $T=250$ MeV, obtained from Lattice simulations~\cite{Ding:2010ga,Ding:2016hua}, PL based  
model 
calculations~\cite{Islam:2014sea,Gale:2014dfa}, Born rate and the nonperturbative power 
corrections}
  \label{dilepton_final}
 \end{center}
\end{figure}
\end{center}

In  Fig. \ref{dilepton_final}, a comparison is displayed among various thermal dilepton 
rates as a function of $\omega/T$ with $T=250$ MeV and zero external three momentum. 
The various dilepton rates considered here are  Born (PLO)~\cite{Greiner:2010zg, 
Cleymans:1992gb}, PLO plus power 
corrections within OPE in Eq.~(\ref{dilepton_rate_final}), 
LQCD~\cite{Ding:2010ga,Ding:2016hua} and 
Polyakov Loop based models in an effective QCD 
approach~\cite{Islam:2014sea,Gale:2014dfa}. The dilepton rate from  PL based models
and LQCD  for 
$\omega/T>4$ becomes simply perturbative in nature whereas  it is so for 
$\omega/T \ge 10$ 
in case of the PLO with power corrections in OPE. For $\omega/T>4$, in PL based  models 
the confinement effect due to Polyakov loop becomes very weak whereas in LQCD the 
spectral 
function is replaced by the PLO one. On the other hand the enhancement of the dilepton 
rate  at low energy ($\omega/T < 4$) for both PL based models and LQCD is due to the
presence of some nonperturbative effects ({\textrm{e.g.,}} residual confinement effect 
etc) whereas in that region the power corrections within OPE is not 
applicable\footnote{In principle one can approximate the dilepton rate in the low mass, 
$\omega/T \le 4$, region by the results from PNLO
~\cite{Laine:2013vma, Ghisoiu:2014mha, Ghiglieri:2014kma} and 1-loop HTL resummation~\cite{Braaten:1990wp}, which agree to each other in order of 
magnitudes. We also note that in the low mass regime (soft-scale)
the perturbative calculations break down as the loop expansion has its generic 
convergence problem in the limit of small coupling ($g \le 1$). On the other hand  PLO, 
PNLO,  HTL resummation and OPE agree in the hard scale, \textit{i.e.}, in the very large 
mass $\omega/T\ge 10$.}. However, in the intermediate domain ($4 <\omega/T < 10 $) the 
dilepton rate is enhanced compared to PL based models and LQCD due to the
presence of the nonperturbative composite quark and gluon operators that incorporates 
power corrections within OPE in $D=4$.  We note that 
the power corrections in OPE considered here may play an important role for intermediate
mass dilepton spectra from high energy heavy-ion collisions in RHIC and LHC. 

\section{Conclusion}
\label{conclusion}

QCD vacuum has a nontrivial structure due to the fluctuations of the quark and gluonic 
fields which generate some local composite operators of quark and gluon fields, 
phenomenologically known as condensates. In perturbative approach by definition such 
condensates do not appear in the observables. However, the nonperturbative dynamics of 
QCD is evident through the power corrections in physical observables by considering the 
nonvanishing vacuum expectation values of such local quark and gluonic composite 
operators.
In this chapter we, first, have made an attempt to compute the nonperturbative 
electromagnetic spectral function in QCD plasma by taking into account the power 
corrections and the nonperturbative condensates within the framework of the OPE 
in $D=4$ dimension. The power corrections appear in the in-medium 
electromagnetic spectral function through the nonanalytic behavior 
of the current-current correlation function in powers of $P^{-D/2}$ or logarithms in the 
Wilson coefficients within OPE in $D=4$ dimension. The $c$-numbered Wilson coefficients 
are computed through Feynman diagrams by incorporating the various condensates. 
In the massless limit of quarks, the self-energy diagram involving local gluonic operator 
(topology-II in Fig.~\ref{topology_2}) encounters mass singularity. By exploiting the 
minimal subtraction through operator mixing this mass singularity cancels out, which 
renders the Wilson coefficients free from any infrared singularity and hence finite.
This result is in conformity with the RGE analysis.

The lepton pairs are produced through the electromagnetic interaction 
in every stage of the hot and dense medium created in high energy 
heavy-ion collisions. They are considered to be an important probe of QGP formation 
because  they leave, immediately after their production,  the hot and dense medium 
almost without any interaction.
As a spectral property of the electromagnetic spectral function, we then
evaluated  the differential dilepton production rate from QCD plasma in the 
intermediate mass range to analyze the effects of power corrections and nonperturbative 
condensates. The power correction contribution is found to be  ${\cal 
O}(\alpha^2_{\textrm{em}}\alpha_s)$ to the PLO, ${\cal 
O}(\alpha^2_{\textrm{em}})$. Further, we note that the intermediate mass range is 
considered because the low mass regime ($ M \le 4T \sim 1$ GeV; $T=0.25$ GeV) is 
prohibited by OPE whereas high mass regime ($ M \ge 10T \sim 2.5$ GeV) is well described 
by the perturbative approach. The intermediate mass range ($4T\le M \le 10T$) dilepton 
in presence of power corrections is found to be enhanced compared to other nonpertubative 
approaches, \textit{i.e.}, LQCD and 
effective QCD models. However,  we note that the 
power corrections in differential dilepton rate through OPE considered here could be 
important to describe the intermediate mass dilepton spectra from heavy-ion collisions. 

Finally,  we would like to note that there is no estimate available in the 
present literature for the composite quark and gluon operators (condensates),  
$\langle \Theta_g^{00}\rangle_T$ and $\langle \Theta_f^{00}\rangle_T$, respectively, at 
finite $T$. Since the present knowledge of these in-medium operators is very insufficient in 
the literature, we have exploited  the Stefan-Boltzmann limits for these composite 
operators to have some limiting information of the nonperturbative effects in the 
electromagnetic spectral function and its spectral properties. We expect that in near 
future the computation of such phenomenological quantities should be possible by 
nonperturbative methods of QCD in lattice and some definite estimation of 
the power corrections within OPE can only be made for spectral function and its spectral 
properties.

%% file: text/gribov.tex
\chapter{Dilepton Production Rate with the Gribov-Zwanziger action}
\label{th_gribov}

In this chapter we use a recently obtained resummed quark propagator at finite temperature which takes into account both the chromoelectric scale $gT$ and the chromomagnetic scale $g^2T$ through the GZ action to compute the non-perturbative DPR at vanishing three-momentum and at one-
loop order. The resulting rate has a rich structure at low energies due to the inclusion of the non-perturbative magnetic scale. This chapter is  based on the part of the paper \textit{Dilepton rate and quark number susceptibility with the Gribov action} by Aritra Bandyopadhyay, Najmul Haque, Munshi G. Mustafa and Michael Strickland, {\bf Phys.Rev. D93 (2016) no.6, 065004}.

\section{Introduction}
\label{intro}

The quark-gluon plasma may be strongly coupled at low temperatures, however, at high temperature there is evidence that resummed perturbation theory can be used to understand the properties of the QGP.  To perturbatively study the QGP one needs to have an in-depth understanding of the various collective modes associated with different thermal scales. The majority of studies in the literature have focused on the hard and electric scales since the magnetic scale is related to the difficult non-perturbative physics of confinement. 

The time-averaged temperature of the QGP generated at RHIC and LHC energies is quite close to the pseudocritical transition temperature as mentioned in section \ref{methods_qgp}. In order to make some progress at these temperatures, it is necessary to consider the non-perturbative physics associated with the QCD magnetic scale in order to assess its role. The fact that the $\mathcal{O}(g^2T)$ correction to the Debye mass receives non-perturbative contributions indicates that the background physics associated with the magnetic scale is fundamentally nonperturbative ~\cite{Nadkarni:1986as,Arnold:1995bh}.   

In a very recent approach~\cite{Su:2014rma} quark propagation in a deconfined medium including both electric- and magnetic-mass effects has also been studied by taking into account the non-perturbative magnetic screening scale by using the GZ action~\cite{Gribov:1977wm,Zwanziger:1989mf}, which regulates the magnetic IR behavior of QCD. Since the gluon propagator with the GZ action is IR regulated, this mimics confinement, making the calculations more compatible with results of LQCD and functional methods~\cite{Maas:2011se}.  Interestingly, the resulting HTL-GZ quark collective modes consist of two massive modes (a normal quark mode and a plasmino mode) similar to the standard HTL dispersions along with a \textit{new} massless spacelike excitation which is directly related to the incorporation of the magnetic scale through the GZ action. This new quark collective excitation results in a long range correlation in the system, which may have important consequences for various physical quantities relevant for the study of deconfined QCD matter. In light of this, we would like to compute the DPR from the deconfined QGP using the non-perturbative GZ action.  

This chapter is organized as follows. In section~\ref{gribov_setup} we briefly outline the setup for quark propagation in a deconfined medium using GZ action. In section~\ref{gribov_dilepton} we calculate the non-perturbative dilepton rate and discuss the results. In section~\ref{gribov_conclu} we summarize and conclude.

\section{Quark spectrum in GZ action}
\label{gribov_setup}

To study the properties of a hot QGP using (semi-)perturbative methods, the 
effective quark propagator is an essential ingredient.  After resummation, the 
quark propagator can be expressed as 
\bea
i S^{-1}(P) &=& \slashed{P}-\Sigma(P), \label{resum_prop} 
\eea
where $\Sigma(P)$ is the quark self energy. One can calculate $\Sigma$ using the 
GZ action modified gluon propagator 
(\ref{modified_gluon_prop}) in the high-temperature limit to 
obtain~\cite{Su:2014rma}
\bea
\Sigma(P) &=& (ig)^2C_F \sumint_{\;\;\;\;\;\;\;\;\;\;\;\;\{K\}} \!\! \gamma_\mu S_f(K) \gamma_\nu D^{\mu\nu}(P-K)
\approx -(ig)^2C_F\sum_\pm\int\limits_0^\infty \frac{dk}{2\pi^2}k^2\int \frac{d\Omega}{4\pi} \nn \\
&& \times \frac{\tilde{n}_\pm(k,\gamma_G)}{4E_\pm^0}\left[\frac{i\gamma_0+\hat{\bf{k}}\cdot\bf{\gamma}}{iP_0+k-E_\pm^0+
\frac{\bf{p}\cdot\bf{k}}{E_\pm^0}}
 +\frac{i\gamma_0-\hat{\bf{k}}\cdot\bf{\gamma}}{iP_0-k+E_\pm^0-\frac{\bf{p}\cdot\bf{k}}{E_\pm^0}}\right],
\label{modified_quark_self}
\eea
where $\Sigma_{\{K\}}\!\!\!\!\!\!\!\!\!\!\!\!\!\int\;\;\;\;\;\;$ is a fermionic sum-integral, $S_f(K)$ is the bare quark propagator, and 
\bea
\tilde{n}_\pm(k,\gamma_G)&\equiv& n_B\!\left(\sqrt{k^2 \pm i\gamma_G^2}\right)+n_F(k) \nn ,\\
E_\pm^0 &=& \sqrt{k^2 \pm i\gamma_G^2}\ , \label{freq}
\eea
where $n_B$ and $n_F$ are Bose-Einstein and Fermi-Dirac distribution functions, respectively.
The  modified  thermal quark  mass in presence  of the  Gribov term can also be obtained  as
\bea
m_q^2(\gamma_G) = \frac{g^2C_F}{4\pi^2}\sum_\pm\int\limits_0^\infty dk \, \frac{k^2}{E_\pm^0} \, \tilde{n}_\pm(k,\gamma_G).
\label{tmass}
\eea
Using the modified quark self energy given in Eq.~(\ref{modified_quark_self}), it is now easy to write down 
the modified effective quark propagator in presence of the Gribov term as 
\bea
i S^{-1}(P) &=& A_0\gamma_0 - A_s \gamma\cdot \hat{\bf{p}}, \label{propa0as}
\eea
where, keeping the structure typically used within the HTL approximation, $A_0$ and $A_s$ are defined as~\cite{Su:2014rma}
\bea
A_0(\omega,p) &=& \omega-\frac{2g^2C_F}{(2\pi)^2}\sum_\pm\int dk \, k \, \tilde{n}_\pm(k,\gamma_G) 
\left[Q_0(\tilde{\omega}_1^\pm,p)+Q_0(\tilde{\omega}_2^\pm,p)\right]\nn , \\
A_s(\omega,p) &=& p+\frac{2g^2C_F}{(2\pi)^2}\sum_\pm\int dk \, k \, \tilde{n}_\pm(k,\gamma_G) 
\left[Q_1(\tilde{\omega}_1^\pm,p)+Q_1(\tilde{\omega}_2^\pm,p)\right] . \label{a0as}
\eea
Here the modified frequencies are defined as $\tilde{\omega}_1^\pm \equiv E_\pm^0(\omega+k-E_\pm^0)/k$ and 
$\tilde{\omega}_2^\pm \equiv E_\pm^0(\omega-k+E_\pm^0)/k$. The Legendre functions of 
the second kind, $Q_0$ and $Q_1$, are
\bea
Q_0(\omega,p) &\equiv& \frac{1}{2p}\ln \frac{\omega+p}{\omega-p} \\
Q_1(\omega,p) &\equiv& \frac{1}{p}(1-\omega Q_0(\omega,p)). \label{legd}
\eea
Using the {helicity representation, the modified effective fermion propagator can also be written as
\bea
i S(P) &=& \frac{1}{2} \frac{(\gamma_0 - \gamma\cdot \hat{\bf{p}})}{D_+}
+ \frac{1}{2} \frac{(\gamma_0 + \gamma\cdot \hat{\bf{p}})}{D_-} , \label{hprop}
\eea
where $D_\pm$ are obtained as
\bea
D_+(\omega,p,\gamma_G) &=& A_0(\omega,p) - A_s(\omega,p)\nn 
= \omega - p - \frac{2g^2C_F}{(2\pi)^2}\sum_\pm\int dk k \tilde{n}_\pm(k,\gamma_G) \nonumber \\
&& \hspace{2cm} \times 
\left[Q_0(\tilde{\omega}_1^\pm,p)+ Q_1(\tilde{\omega}_1^\pm,p)
+Q_0(\tilde{\omega}_2^\pm,p)+ Q_1(\tilde{\omega}_2^\pm,p)\right] ,\nn\\
D_-(\omega,p,\gamma_G) &=& A_0(\omega,p) + A_s(\omega,p)\nn 
= \omega + p - \frac{2g^2C_F}{(2\pi)^2}\sum_\pm\int dk k \tilde{n}_\pm(k,\gamma_G) \nonumber \\
&& \hspace{2cm} \times 
\left[Q_0(\tilde{\omega}_1^\pm,p)- Q_1(\tilde{\omega}_1^\pm,p)
+Q_0(\tilde{\omega}_2^\pm,p)- Q_1(\tilde{\omega}_2^\pm,p)\right]. \label{dpm}
\eea

\begin{figure}[t]
\begin{center}
\includegraphics[width=0.32\linewidth]{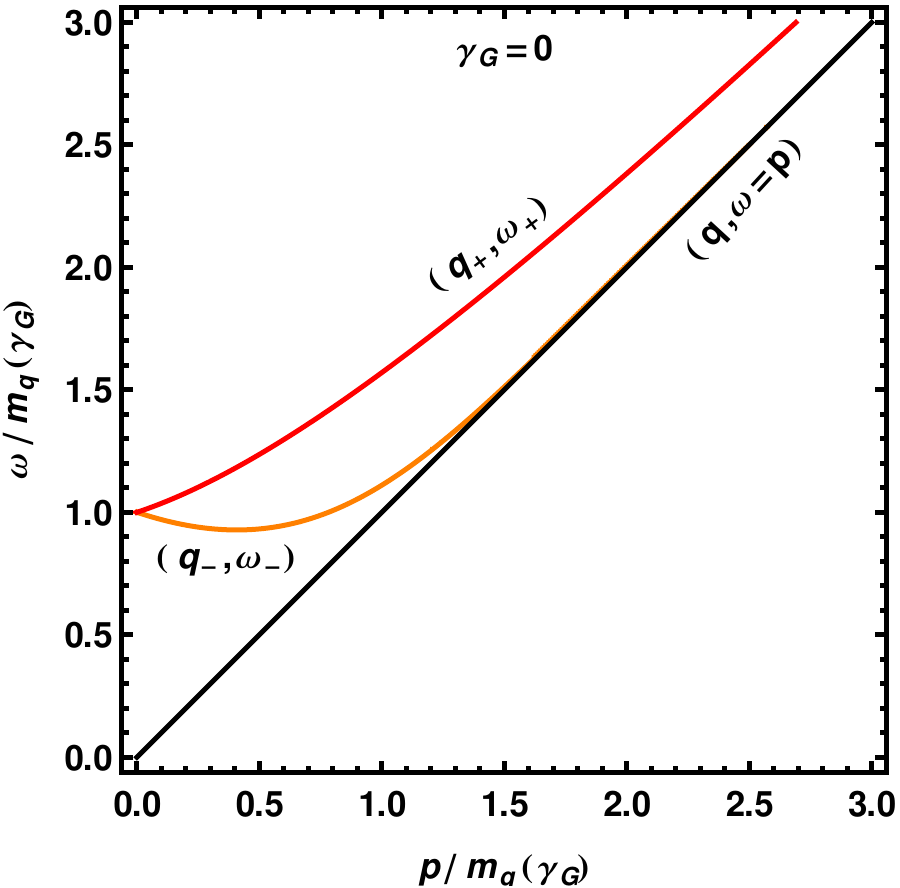}
\includegraphics[width=0.32\linewidth]{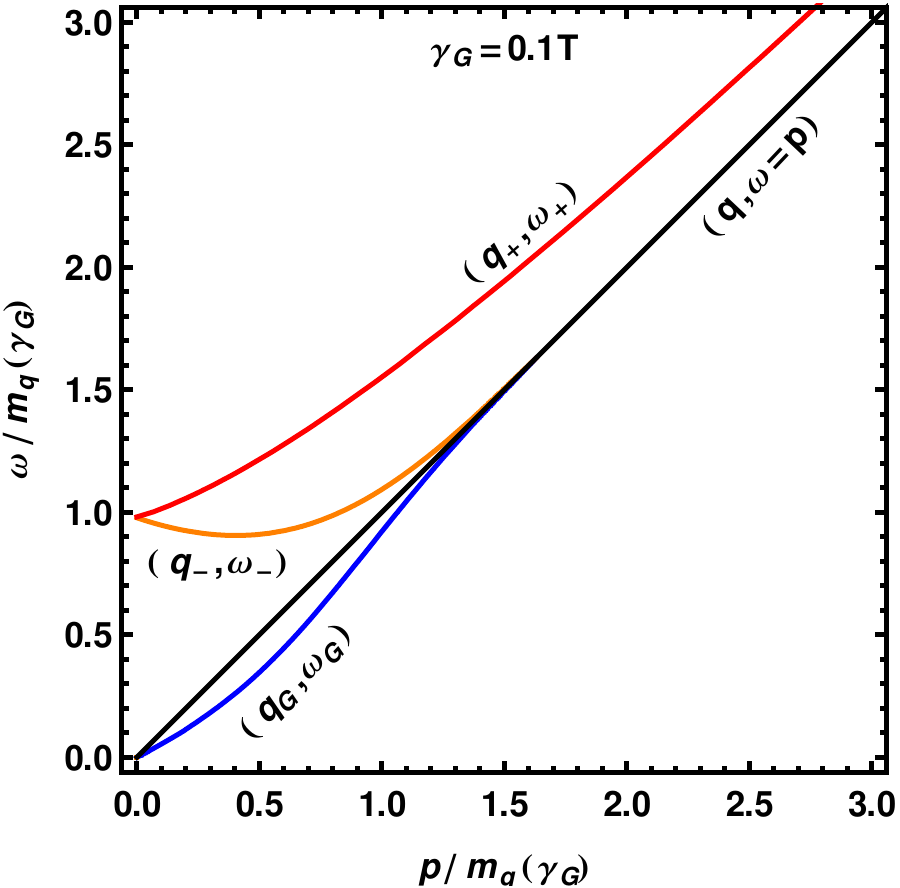}
\includegraphics[width=0.32\linewidth]{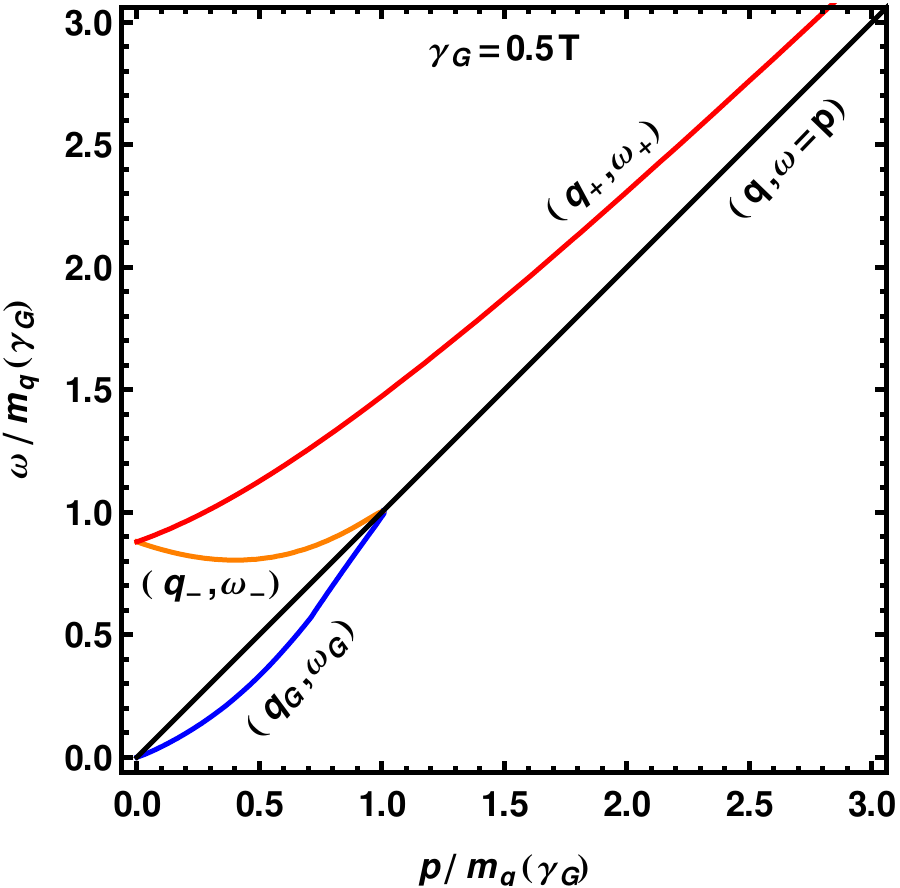} 
\end{center}
\caption[Plot of the dispersion relations for different values of $\gamma_G$]{Plot of the dispersion relations for different values of $\gamma_G$. In the parenthesis, the first one represents a collective excitation mode whereas 
the second one is the corresponding energy of that mode.}
\label{disp_rel}
\end{figure}

Solving for the zeros of $D_{\pm}^{-1}(P,\gamma_G)$ gives the dispersion 
relations for the collective excitations in the medium.  In Fig.~\ref{disp_rel} 
we show the resulting dispersion relations for three different values of the 
Gribov parameter $\gamma_G$.  In absence of the Gribov term (\textit{i.e,} 
$\gamma_G=0$), there are two massive modes corresponding to a normal quark mode 
$q_+$ with energy $\omega_+$ and a long wavelength plasmino mode $q_-$ with 
energy $\omega_-$ that quickly approaches free massless propagation in the 
high-momentum limit. These two modes are similar to those found in the HTL 
approximation~\cite{Braaten:1990wp}.  With the inclusion of the Gribov term, there is a 
massless mode $q_G$ with energy $\omega_G$, in addition to the two massive 
modes, $q_+$ and $q_-$~\cite{Su:2014rma}. The extra mode $q_G$ is due to the 
presence of the magnetic screening scale.  
This new massless mode~\footnote{We note that the slope of the 
dispersion relation for this 
massless extra spacelike mode $q_G$ exceeds unity in some domain of momentum. 
Thus, the group velocity, $d\omega_G/dp$, is superluminal for the spacelike 
mode $q_G$, which approaches the light cone ($d\omega/dp=1$)  from above (see Fig.~\ref{group_vel}).  
Since the mode is spacelike,  there is no causality problem but could be termed 
as anomalous dispersion as in the presence of  GZ action  the Landau damping is 
converted into amplification of spacelike dispersive mode} becomes 
lightlike at large momentum.  In this context, we note that in Ref.~\cite{Chakraborty:2013dda}, 
such an extra massive mode with significant spectral width was observed near 
$T_c$ in presence of dimension-four gluon condensates~\cite{Chakraborty:2013dda} in addition to 
the usual propagating quark and plasmino modes.  The existence of this extra 
mode could affect lattice extractions of the dilepton rate since even the most 
recent LQCD results~\cite{Kitazawa:2009uw,Kaczmarek:2012mb} assumed that there were only two 
poles (a quark mode and a plasmino mode) inspired by the HTL approximation.

In  HTL approximation ($\gamma_G=0$) the propagator contains a discontinuity in 
complex plane stemming from the logarithmic terms in (\ref{dpm}) due to 
spacelike momentum $\omega^2<p^2$.  Apart from two collective excitations 
originating from the in-medium dispersion as discussed above, there is also a 
Landau cut contribution in the spectral representation of the propagator due to 
the discontinuity in spacelike momentum.  On the other hand, for $\gamma_G \neq 
0$ the individual terms in (\ref{dpm}) possess  discontinuities at spacelike 
momentum but canceled out when all terms are summed owing to the fact that the 
poles come in complex-conjugate pairs.  As a result, there is no discontinuity 
in the complex plane.\footnote{Starting from the Euclidean expression 
(\ref{modified_quark_self}), we have numerically checked for discontinuities and 
found none.  We found some cusp-like structures at complex momenta, but $\Sigma$ 
was found to be $C^0$-continuous everywhere in the complex plane.}   This 
results in disappearance of the Landau cut contribution in the spectral 
representation of the propagator in spacelike domain. It appears as if the 
Landau cut contribution in spacelike domain  for $\gamma_G=0$ is replaced by 
massless spacelike dispersive mode in presence of  magnetic scale ($\gamma_G\ne 
0$). So the spectral function corresponding to the propagator $D_{\pm}^{-1}$  
for $\gamma_G \neq 0$ has only pole contributions.  As a result, one has
\bea
\rho_\pm^G(\omega,p)= \frac{\omega^2-p^2}{2m_q^2(\gamma_G)}\left[\delta(\omega\mp\omega_+)+\delta(\omega\pm\omega_-)
+\delta(\omega\pm\omega_G)\right], \label{gspect}
\eea
where  $D_+$ has poles at $\omega_+$, $-\omega_-$, and $-\omega_G$ 
and $D_-$ has poles at $\omega_-$, $-\omega_+$, and $\omega_G$ with a prefactor,  $(\omega^2-p^2)/2m_q^2(\gamma_G)$,  as the residue.  

\begin{figure}[t]
\begin{center}
\includegraphics[width=0.48\linewidth]{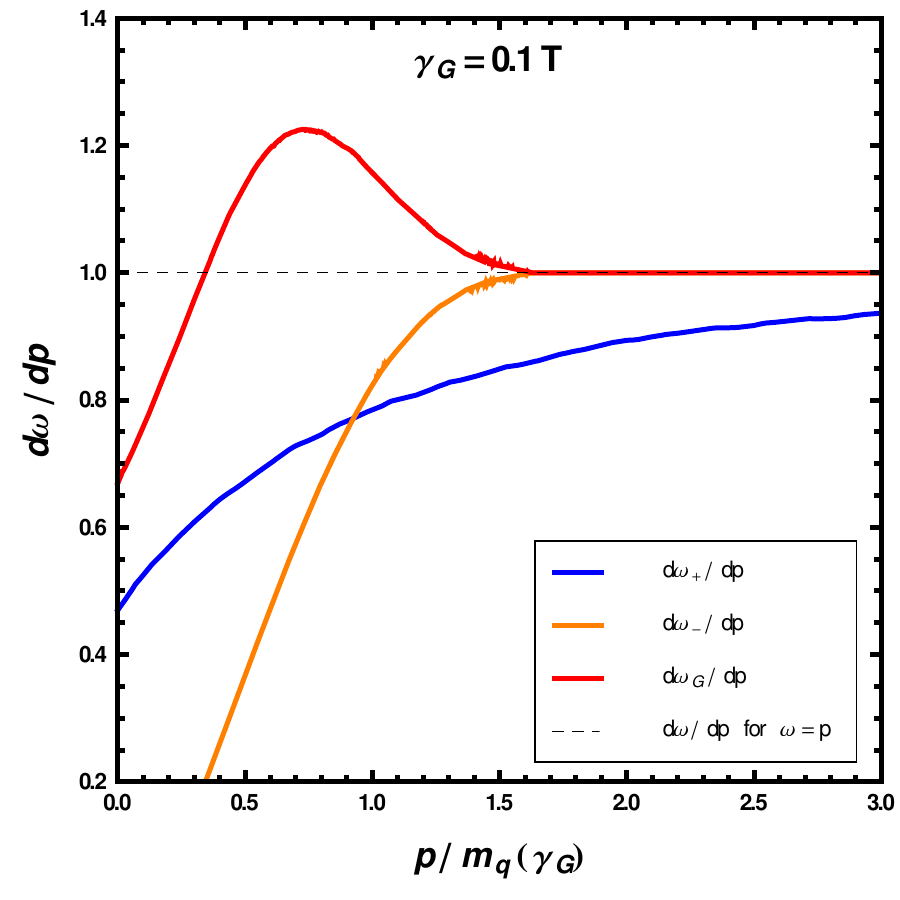}
\includegraphics[width=0.48\linewidth]{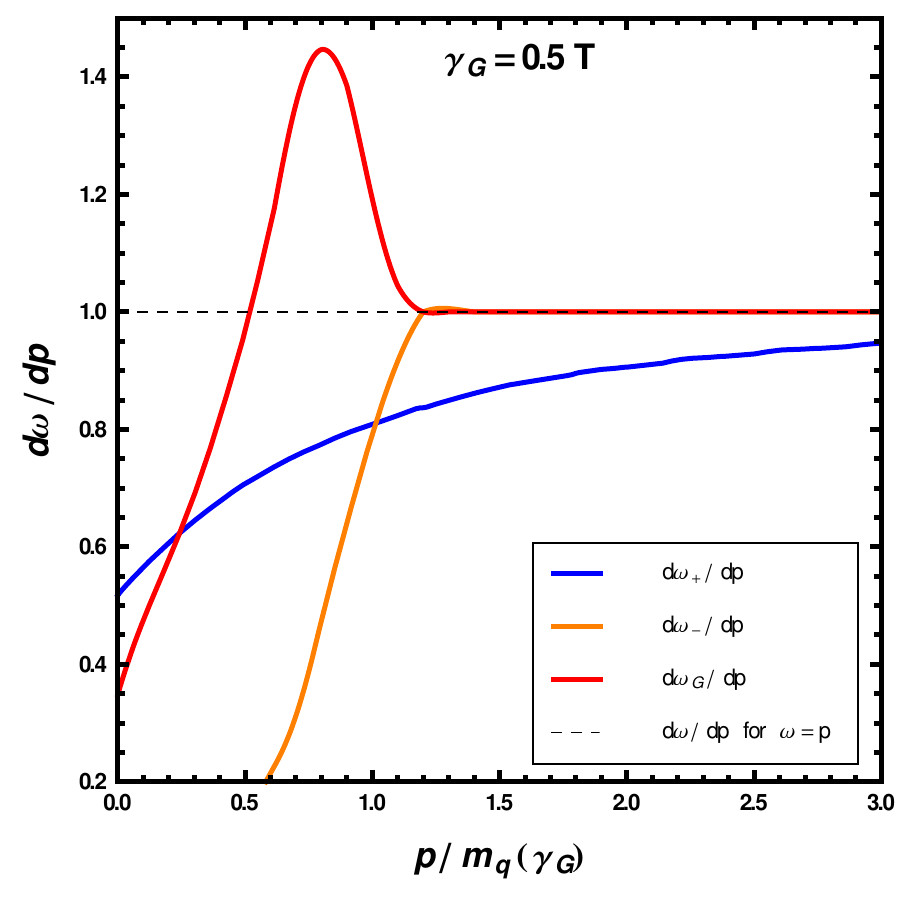} 
\end{center}
\caption[Plot of the group velocity for different values of $\gamma_G$]{Plot of the group velocity for different values of $\gamma_G$. The group velocity for the spacelike gribov mode $d\omega_G/dp$ becomes superluminal, as can be seen from both the plots.}
\label{group_vel}
\end{figure}

At this point we would like to mention that the non-perturbative quark spectral 
function obtained using the quark propagator analyzed in the quenched LQCD 
calculations of Refs.~\cite{Kitazawa:2009uw,Kaczmarek:2012mb,Kim:2015poa} and utilizing gluon 
condensates in Refs.~\cite{Schaefer:1998wd,Mustafa:1999jz,
Peshier:1999dt,Chakraborty:2013dda} also 
forbids a Landau cut contribution since the effective quark propagators in these 
calculations do not contain any discontinuities. This stems from the fact that 
the quark self-energies in 
Refs.~\cite{Schaefer:1998wd,Mustafa:1999jz,
Peshier:1999dt,Chakraborty:2013dda}  do not have any 
imaginary parts whereas in Refs.~\cite{Kitazawa:2009uw,Kaczmarek:2012mb,Kim:2015poa} an ansatz of two 
quasiparticles was employed for spectral function based on the LQCD quark 
propagator analyzed in quenched approximation. The spectral function obtained 
with the Gribov action (\ref{gspect}) also possesses only pole contributions but 
no Landau cut. These nonperturbative approaches here as well as 
elsewhere~\cite{Kitazawa:2009uw,Kaczmarek:2012mb,Kim:2015poa,Schaefer:1998wd,Mustafa:1999jz,
Peshier:1999dt}  completely remove the quasigluons responsible for 
Landau cut, which should have been present in the hot medium. The consequence 
of the absence of Landau cut part in nonperturbative approaches will be 
discussed later in details. 

Returning to the problem at hand, the spectral density in (\ref{gspect}) at 
vanishing three momentum ($p \equiv |\vec{p}| =0$) contains three delta function 
singularities corresponding to the two massive modes and one new massless Gribov 
mode.  To proceed, one needs the vertex functions in presence of the Gribov 
term.  These can be determined by explicitly computing the hard-loop limit of 
the vertex function using the Gribov propagator.  One can verify,
that the resulting effective quark-gluon vertex function satisfies the 
necessary ST identity
\bea
(P_1-P_2)_\mu\Gamma^\mu(P_1,P_2)= S^{-1}(P_1)-S^{-1}(P_2) \ .
\label{ward_id}
\eea
The temporal and spatial parts of the modified effective quark-gluon vertex can be written as
\bea
\Gamma^0 &=& a_G ~\gamma^0 +b_G~\bf{\gamma} \cdot \hat{\bf{p}},\nn\\
\Gamma^i &=& c_G ~\gamma^i +b_G~\hat{p}^i\gamma_0+d_G~\hat{p}^i\left(\bf{\gamma} \cdot \hat{\bf{p}}\right), \label{vertex}
\eea
where the coefficients are given by
\bea
a_G &=& 1-\frac{2g^2C_F}{(2\pi)^2}\sum_\pm\int dk \, k \, \tilde{n}_\pm(k,\gamma_G)\frac{1}{\omega_1-\omega_2}
\left[\delta Q_{01}^\pm+\delta Q_{02}^\pm\right],\nn\\
b_G &=& -\frac{2g^2C_F}{(2\pi)^2}\sum_\pm\int dk \, k \, \tilde{n}_\pm(k,\gamma_G)\frac{1}{\omega_1-\omega_2}
\left[\delta Q_{11}^\pm+\delta Q_{12}^\pm\right],\nn\\
c_G &=& 1+\frac{2g^2C_F}{(2\pi)^2}\sum_\pm\int dk \, k \, \tilde{n}_\pm(k,\gamma_G)\frac{1}{3(\omega_1-\omega_2)}
\left[\delta Q_{01}^\pm+\delta Q_{02}^\pm-\delta Q_{21}^\pm-\delta Q_{22}^\pm\right],\nn\\
d_G &=& \frac{2g^2C_F}{(2\pi)^2}\sum_\pm\int dk \, k \, \tilde{n}_\pm(k,\gamma_G)\frac{1}{\omega_1-\omega_2}
\left[\delta Q_{21}^\pm+\delta Q_{22}^\pm\right],
\label{gribov_vertex_coeff}
\eea
with
\bea
\delta Q_{n1}^\pm &=& Q_n(\tilde{\omega}_{11}^\pm,p)- Q_n(\tilde{\omega}_{21}^\pm,p){\rm{~for~}} n=0,1,2 \, \, ,\nn\\
\omega_{m1}^\pm &=& E_\pm^0(\omega_m+k-E_\pm^0)/k {\rm{~for~}} m=1,2\, \, , \nn\\
\omega_{m2}^\pm &=& E_\pm^0(\omega_m-k+E_\pm^0)/k {\rm{~for~}} m=1,2\,\, .\nn
\eea
Similarly, the four-point function can be obtained by computing the necessary 
diagrams in the hard-loop limit and it satisfies the following generalized ST 
identity 
\bea
P_\mu\Gamma^{\mu\nu}(-P_1,P_1;-P_2,P_2) = \Gamma^{\nu}(P_1-P_2,-P_1;P_2)-\Gamma^{\nu}(-P_1-P_2,P_1;P_2) \ . \label{wi_4pt}
\eea

\section{One-loop DPR with the GZ action}
\label{gribov_dilepton}

At one-loop order, the dilepton production rate is related to the two diagrams shown in Fig.~\ref{feyn_diag}, which can be written as

\bea
\Pi_\mu^\mu(Q)&=&\frac{5}{3}e^2\sum_{p_0}\int\frac{d^3p}{(2\pi)^3} \biggl\{\textrm{Tr}\biggl[S(P)
~\Gamma_\mu(K,Q,-P)~S(K)~\Gamma_\mu(-K,-Q,P)\biggr]\label{se_tr} \nn \\
&& \hspace{5cm} + \; \textrm{Tr}\biggl[S(P)~\Gamma_\mu^\mu(-P,P;-Q,Q)\biggr]\biggr\}, \label{dilep_tr}
\eea
where $K=P-Q$.  The second term in (\ref{dilep_tr}) is due to the tadpole 
diagram shown in Fig.~\ref{feyn_diag} which, in the end, does not contribute 
since $\Gamma_\mu^\mu=0$. However, the tadpole diagram is essential to satisfy 
the transversality condition, $Q_\mu\Pi^{\mu\nu}(Q)=0$ and thus gauge invariance
and charge conservation in the system.

\begin{figure}[h]
\begin{center}
\includegraphics[width=0.35\linewidth]{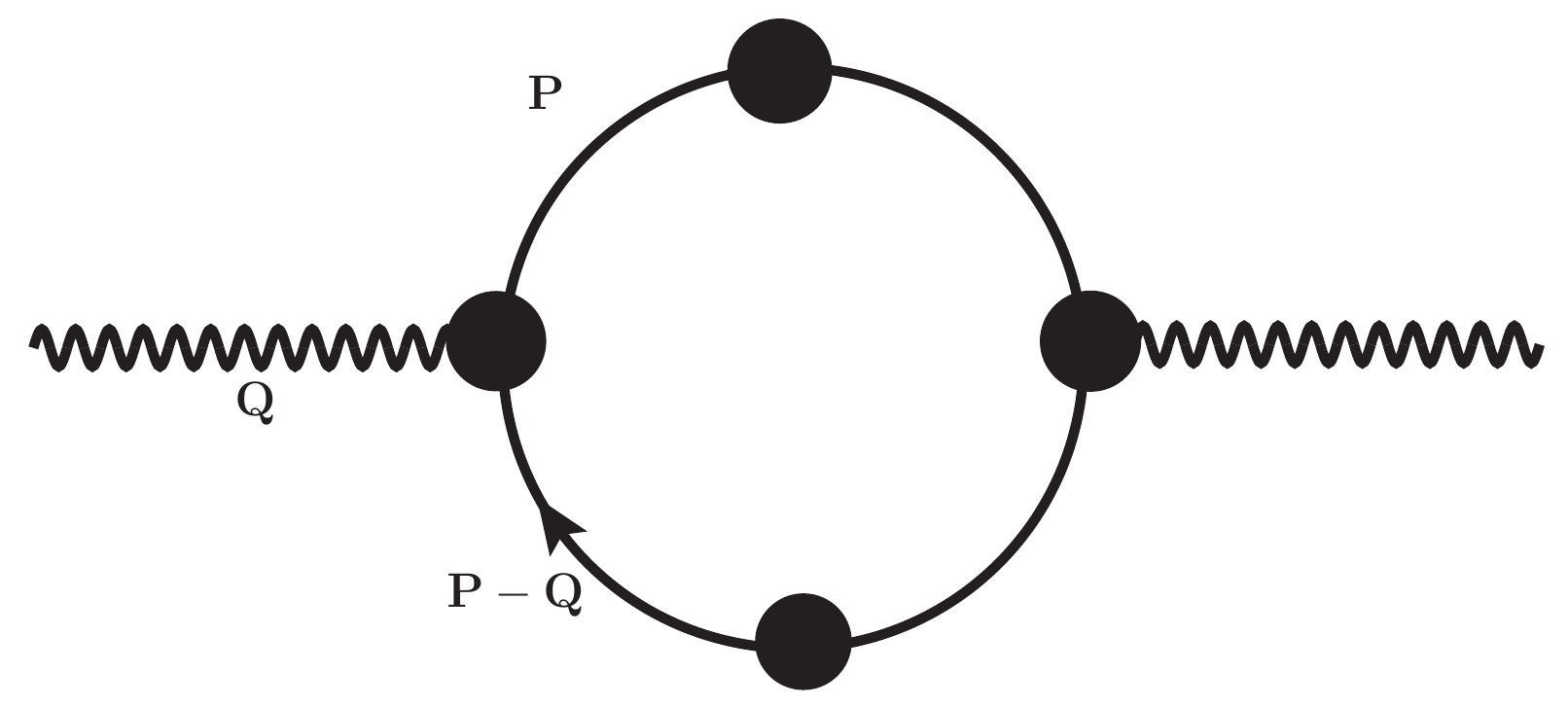}  
\hspace{5mm}
\includegraphics[width=0.3\linewidth]{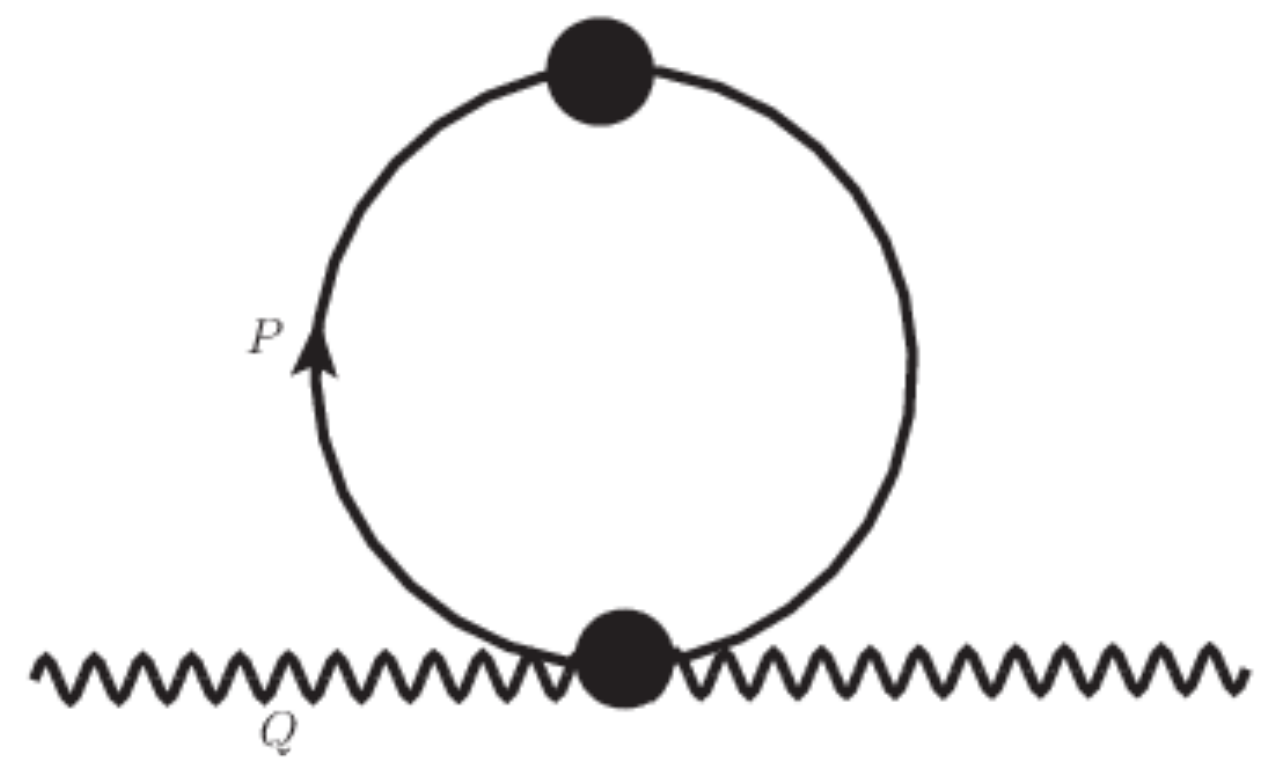}
\end{center}
\caption{The self-energy (left) and  tadpole (right) diagrams in one loop order.}
\label{feyn_diag}
\end{figure}

Using the $n$-point functions computed in sec.~\ref{gribov_setup} and performing traces, one obtains 
\bea
\Pi_\mu^\mu(\vec{q}=0)&=&\frac{10}{3}e^2T\sum_{p_0}\int\frac{d^3p}{(2\pi)^3}  \nn \\
\!\!\!\!&&\!\!\!\! \times \Biggl[\left\{\frac{(a_G+b_G)^2}{D_+(\omega_1,p,\gamma_G)D_-(\omega_2,p,\gamma_G)}+
\frac{(a_G-b_G)^2}{D_-(\omega_1,p,\gamma_G)D_+(\omega_2,p,\gamma_G)}\right\}\nn\\
\!\!\!\!&&\!\!\!\! - \left\{\frac{(c_G+b_G+d_G)^2}{D_+(\omega_1,p,\gamma_G)D_-(\omega_2,p,\gamma_G)}+
\frac{(c_G-b_G+d_G)^2}{D_-(\omega_1,p,\gamma_G)D_+(\omega_2,p,\gamma_G)}\right\}\nn\\
 -2c_G^2\!\!\!\!\!\!&&\!\!\!\!\!\!\left\{\frac{1}{D_+(\omega_1,p,\gamma_G)D_+(\omega_2,p,\gamma_G)}+\frac{1}{D_-(\omega_1,p,\gamma_G)D_-(\omega_2,p,\gamma_G)}\right\}\Biggr].
\label{dlep_trace_se}
\eea

The discontinuity can be obtained by Braaten-Pisarski-Yuan prescription~\cite{Braaten:1990wp}
\bea
\textmd{Disc~}T\sum_{p_0}f_1(p_0)f_2(q_0-p_0)&=& 2\pi i (1-e^{\beta \omega})\int d\omega_1 \int d\omega_2~n_F(\omega_1)
n_F(\omega_2)\nn \\
&& \times \ \delta(\omega-\omega_1-\omega_2) \ \rho_1(\omega_1)\rho_2(\omega_2), \label{bpy_pres}
\eea
which, after some calculation, allows one to determine the dilepton rate at zero three momentum
\bea
\frac{dN}{d^4Xd^4Q}\!\!\!\!&\equiv&\!\!\!\!\frac{dR}{d\omega d^3q}\Big\vert_{({\vec q}=0)}=\frac{10\alpha^2}{9\pi^4}\frac{1}{\omega^2}\int\limits_0^\infty\!\! p^2dp
\int\limits_{-\infty}^\infty \!\!d\omega_1 \int\limits_{-\infty}^\infty\!\! d\omega_2 n_F(\omega_1) n_F(\omega_2) 
\delta(\omega-\omega_1-\omega_2)\nn\\
\!\!\!\!\!\!&&\!\!\!\!\!\!\Bigg[4\left(1-\frac{\omega_1^2-\omega_2^2}{2p\,\omega}\right)^2 \rho_+^G(\omega_1,p) \rho_-^G(\omega_2,p)\nn\\
\!\!\!\!\!\!&&\!\!\!\!\!\!+\left(1+\frac{\omega_1^2+\omega_2^2-2p^2-2m_q^2(\gamma_G)}{2p\,\omega}\right)^2\rho_+^G(\omega_1,p) \rho_+^G(\omega_2,p)\nn\\
\!\!\!\!\!\!&&\!\!\!\!\!\!+\left(1-\frac{\omega_1^2+\omega_2^2-2p^2-2m_q^2(\gamma_G)}{2p\,\omega}\right)^2\rho_-^G(\omega_1,p) \rho_-^G(\omega_2,p)\Bigg]
.\label{dilep_spec}
\eea


Using (\ref{gspect}) and considering all physically allowed processes by the in-medium dispersion, 
the total contribution can be expressed as
\bea
\frac{dR}{d\omega d^3q}\Big\vert^{pp}({\vec q}=0)&=&\frac{10\alpha^2}{9\pi^4}\frac{1}{\omega^2}
\int\limits_0^\infty p^2\, dp \times \nn \\
&& \Biggl [\delta(\omega-2\omega_+)\ n_F^2(\omega_+)\left(\frac{\omega_+^2-p^2}{2m_q^2(\gamma_G)}\right)^2
\left\{1+\frac{\omega_+^2-p^2-m_q^2(\gamma_G)}{p~\omega}\right\}^2\nn\\
&&+~\delta(\omega-2\omega_-)\ n_F^2(\omega_-)\left(\frac{\omega_-^2-p^2}{2m_q^2(\gamma_G)}\right)^2
\left\{1-\frac{\omega_-^2-p^2-m_q^2(\gamma_G)}{p~\omega}\right\}^2\nn\\
&&+~\delta(\omega-2\omega_G)\ n_F^2(\omega_G)\left(\frac{\omega_G^2-p^2}{2m_q^2(\gamma_G)}\right)^2
\left\{1-\frac{\omega_G^2-p^2-m_q^2(\gamma_G)}{p~\omega}\right\}^2 \nn 
\eea

\bea
&& +4 \ \delta(\omega-\omega_+-\omega_-) \ n_F(\omega_+) \ n_F(\omega_-) 
\left(\frac{\omega_+^2-p^2}{2m_q^2(\gamma_G)}\right)
\left(\frac{\omega_-^2-p^2}{2m_q^2(\gamma_G)}\right) \nn \\
&& \times \left\{1-\frac{\omega_+^2-\omega_-^2}{2p\,\omega}\right\}^2 \nn \\
&& +\delta(\omega-\omega_++\omega_-) \ n_F(\omega_+)n_F(-\omega_-)
\left(\frac{\omega_+^2-p^2}{2m_q^2(\gamma_G)}\right)\left(\frac{\omega_-^2-p^2}{2m_q^2(\gamma_G)}\right) \nn \\
&& \times \left\{1+\frac{\omega_+^2+\omega_-^2-2p^2-2m_q^2(\gamma_G)}{2p\,\omega}\right\}^2 
~\Biggr]. \label{dilep_pp}
\eea
Inspecting the arguments of the various energy conserving $\delta$-functions in 
(\ref{dilep_pp}) one can understand the physical processes originating from the 
poles of the propagator.   The first three terms in (\ref{dilep_pp}) correspond 
to the annihilation processes of $q_+{\bar q}_+\rightarrow \gamma^*$, $q_-{\bar 
q}_-\rightarrow \gamma^*$, and $q_G{\bar q}_G\rightarrow \gamma^*$, 
respectively.  The fourth term corresponds to the annihilation of $q_+{\bar 
q}_-\rightarrow \gamma^*$.  On the other hand, the fifth term corresponds to a 
process, $q_+\rightarrow q_-\gamma^*$, where a $q_+$ mode makes a transition to a 
$q_-$ mode along with a virtual photon.  These processes involve soft quark 
modes ($q_+, \, q_-$, and $q_G$ and their antiparticles) which originate by 
cutting the self-energy diagram in Fig.~\ref{feyn_diag} through the internal 
lines without a ``blob''.  The virtual photon, $\gamma^*$, in all these five 
processes decays to lepton pair and can be visualized from the dispersion plot 
as displayed in the Fig.~\ref{dilepton_processes}.  The momentum integration in 
Eq.~(\ref{dilep_pp}) can be performed using the standard delta function 
identity 
\bea
\delta(f(x))&=& \sum_i \frac{\delta(x-x_i)}{\mid \! f'(x) \!\mid_{x=x_i}}, \label{delta-prop}
\eea
where $x_i$ are the solutions of $f(x_i)=0$.

\begin{figure}[t]
\begin{center}
\includegraphics[width=0.5\linewidth]{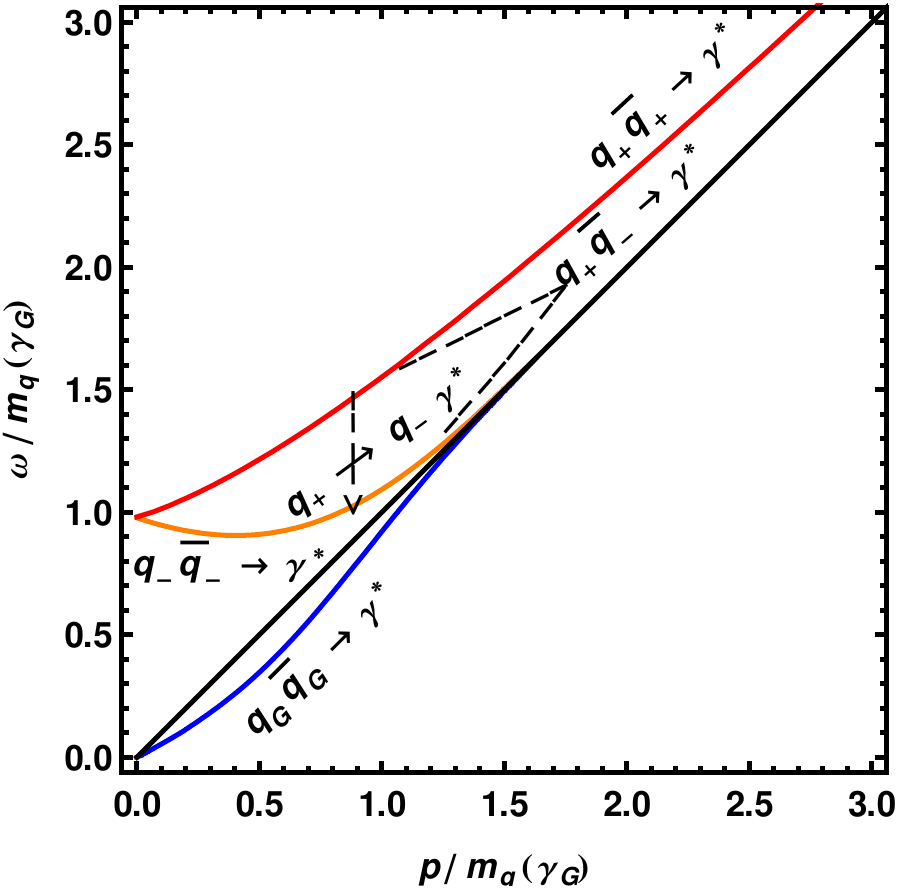}  
\end{center}
\caption{Various dilepton processes which originate from the in-medium dispersion with the Gribov term.}
\label{dilepton_processes}
\end{figure}

\begin{figure}[t]
\begin{center}
\includegraphics[width=0.75\linewidth]{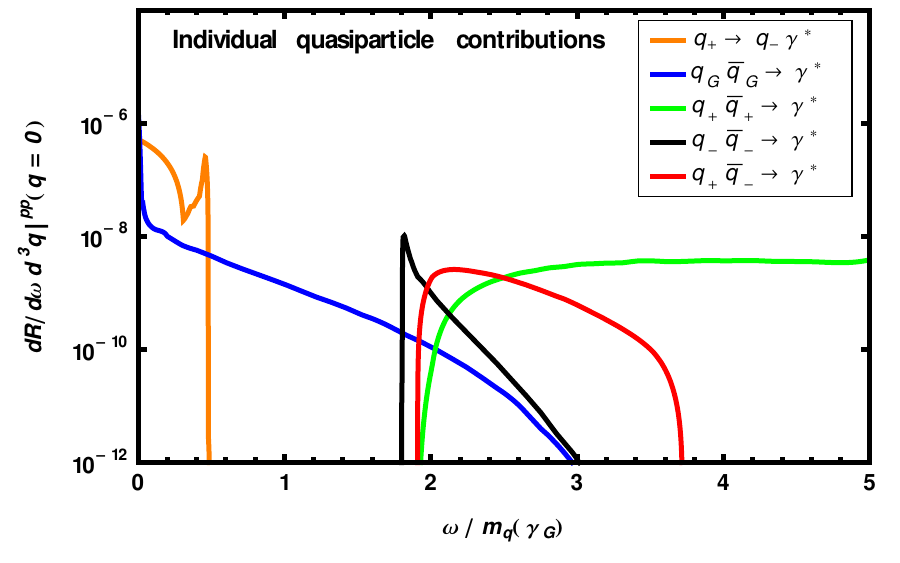} 
\end{center}
\caption{The dilepton production rates corresponding to quasiparticle processes in Fig.~\ref{dilepton_processes}.}
\label{dilepton_pp}
\end{figure}

\begin{figure}[t]
\begin{center}
\includegraphics[width=0.75\linewidth]{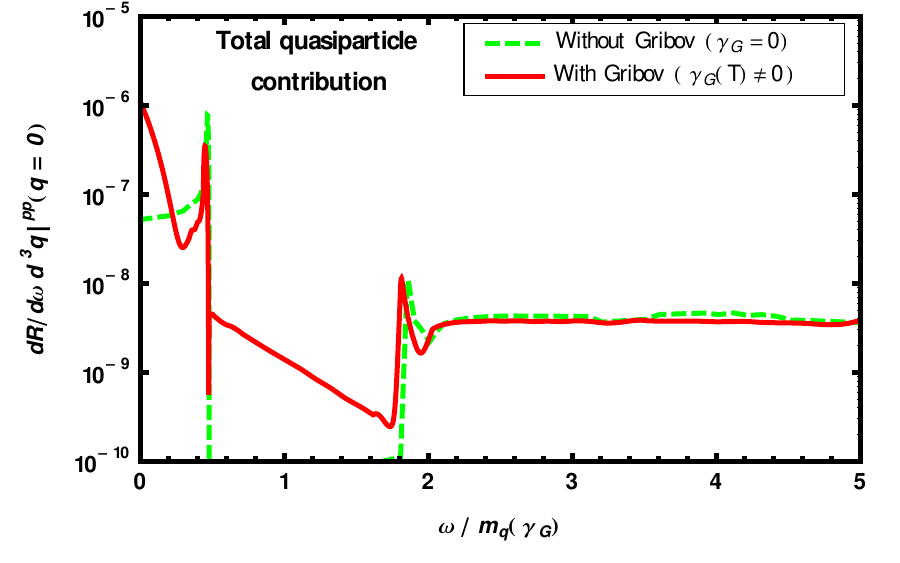} 
\end{center}
\caption{Comparison of dilepton production rates involving various quasiparticle modes with and without inclusion of $\gamma_G$.}
\label{dilepton_comparison}
\end{figure}

The contribution of various individual processes to  the dilepton production 
rate in presence of the Gribov term are displayed in the  
Fig.~\ref{dilepton_pp}.  Note that in this figure and in subsequent figures 
showing the dilepton rate, the vertical axis shows the dilepton 
rate $dR/d^4Q = dN/d^4Xd^4Q$ and the horizontal axis is scaled by the thermal 
quark mass so as to make it dimensionless. In Fig.~\ref{dilepton_pp} we see that 
the transition process, $q_+\rightarrow q_-\gamma*$, begins at the energy 
$\omega=0$ and ends up with a van-Hove peak\,\footnote{A van-Hove 
peak~\cite{Ashcroft:1976} appears where the density of states diverges as 
$f'(x)|_{x=x_0}=0$ since the density of states is inversely proportional to 
$f'(x)$.} where all of the transitions from $q_+$ branch are directed towards 
the minimum of the $q_-$ branch. The annihilation process involving the massless 
spacelike Gribov modes, $q_G{\bar q}_G\rightarrow \gamma^*$, also starts at 
$\omega=0$ and falls off very quickly.  The annihilation of the two plasmino 
modes, $q_-{\bar q}_-\rightarrow \gamma^*$, opens up with again a van-Hove peak 
at $\omega=2 \times $ the minimum energy of the plasmino mode. The contribution 
of this process decreases exponentially.  At $\omega=2m_q(\gamma_G)$, the 
annihilation processes involving usual quark modes, $q_+{\bar q}_+\rightarrow 
\gamma^*$, and that of a quark and a plasmino mode, $q_+{\bar q}_-\rightarrow 
\gamma^*$, begin.  However, the former one ($q_+{\bar q}_+\rightarrow \gamma^*$) 
grows with the energy and would converge to the usual Born rate (leading order 
perturbative rate)~\cite{Cleymans:1986na} at high mass whereas the latter one ($q_+{\bar 
q}_-\rightarrow \gamma^*$) initially grows at a very fast rate, but then 
decreases slowly and finally drops very quickly. The behavior of the latter 
process can easily be understood from the dispersion properties of quark and 
plasmino mode. Summing up, the total contribution of all theses five processes 
is displayed in Fig.~\ref{dilepton_comparison}. This is compared with 
similar dispersive contribution
when $\gamma_G=0$~\cite{Braaten:1990wp}, comprising processes $q_+\rightarrow q_-\gamma^*$, 
$q_+{\bar q}_+\rightarrow \gamma^*$, $q_-{\bar q}_-\rightarrow \gamma^*$ and  
$q_+{\bar q}_-\rightarrow \gamma^*$. We note that when $\gamma_G=0$, the 
dilepton rate contains both van-Hove peaks and an energy gap~\cite{Braaten:1990wp}. In 
presence of the Gribov term ($\gamma_G\ne 0$), 
the van-Hove peaks remain, but the energy gap disappears due to the annihilation 
of new massless Gribov modes, $q_G{\bar q}_G\rightarrow \gamma^*$.

\begin{figure}[t]
\begin{center}
\includegraphics[width=0.75\linewidth]{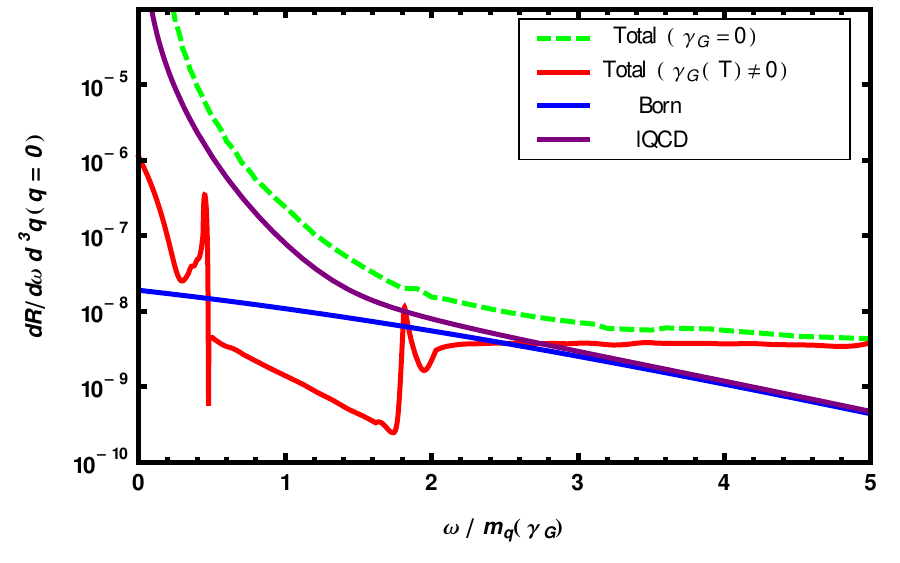} 
\end{center}
\caption{Comparison of various dilepton production  rates from  the deconfined matter.}
\label{dilepton_all}
\end{figure}

In Fig.~\ref{dilepton_all} we compare the rates obtained using various 
approximations:  leading-order perturbative (Born) rate~\cite{Cleymans:1986na}, quenched 
LQCD rate~\cite{Ding:2010ga,Karsch:2001uw}, and with and without the Gribov term.  The 
non-perturbative rate with the Gribov term shows important structures compared 
to the Born rate at low energies.  But when compared to the total HTLpt 
rate~\footnote{Since HTL spectral function (i.e, $\gamma_G=0$) has both pole and 
Landau cut contribution, so the HTLpt rate~\cite{Braaten:1990wp} contains an additional 
{\textit {higher order}} contribution due to the Landau cut stemming from 
spacelike momenta.} it is suppressed  in the low mass region due to the absence 
of Landau cut contribution for $\gamma_G \neq 0$.  It seems as if the higher 
order Landau cut contribution due to spacelike momenta for $\gamma_G=0$ is 
replaced by the soft process involving spacelike Gribov modes in the collective 
excitations for $\gamma_G\neq 0$. We also note that the dilepton rate~\cite{Kim:2015poa} 
using the spectral function constructed with two pole ansatz by analyzing LQCD 
propagator in quenched approximation~\cite{Kitazawa:2009uw,Kaczmarek:2012mb} shows similar 
structure  as found here for $\gamma_G\ne0$. On the other hand, such structure 
at low mass is also expected in the direct computation of dilepton rate from 
LQCD in quenched approximation~\cite{Ding:2010ga,Karsch:2001uw}. However, a smooth variation 
of the rate was found at low mass. The computation of dilepton rate in LQCD 
involves  
various intricacies and uncertainties.  This is because, as noted in 
sec.~\ref{intro}, the spectral function in continuous time is obtained from the 
correlator in finite set of discrete Euclidean time using a probabilistic MEM 
method~\cite{Asakawa:2000tr,Nakahara:1999vy} with a somewhat ad hoc continuous ansatz for the 
spectral function at low energy causing fundamental difficulties in performing 
the necessary analytic continuation in LQCD. Until LQCD overcomes the 
uncertainties and difficulties in the computation of the vector spectral 
function, one needs to depend, at this juncture, on the prediction of the 
effective approaches for dilepton rate at low mass in particular. We further 
note that at high-energies the rate for both  $\gamma_G=0$ and $\gamma_G\ne 0$ 
is higher than the lattice data and Born rate.  This is a consequence of using 
the HTL self-energy also at high-energies/momentum where the soft-scale 
approximation breaks down. Nevertheless, the low mass rate obtained here by 
employing the non-perturbative magnetic scale ($\gamma_G\ne 0$) in addition to 
the electric scale allows for a model-based inclusion of the effect of 
confinement and the result has a somewhat rich structure at low energy compared 
to that obtained using only the electric scale ($\gamma_G = 0$) as well in 
LQCD. 

We make some general comments of the dilepton rate below.
 If one looks at the dispersion plots in Fig.~\ref{disp_rel} for  $\gamma_G=0$ 
then one finds that $\omega_-$  falls off  exponentially  and approaches light 
cone whereas $\omega_+$ does not follow exponential fall off to light cone but 
follows a trend $[p+m^2_q(T)/p]$ for large $p$. On the other hand,  in 
presence of  $\gamma_G \ne 0$ both  $\omega_-$  and $\omega_G$ approach light 
cone very fast  but again $\omega_+$ has similar asymptotic behaviour as 
before.
This feature of $\omega_+$  makes the dilepton rate at large $\omega$ in 
Fig.~\ref{dilepton_all} leveled for both $\gamma_G=0$ and $\gamma_G \ne 0$ 
because the dominant contribution comes from the annihilation of two $\omega_+$
as discussed in Fig.~\ref{dilepton_pp}. In general the dilepton rate in 
Fig.~\ref{dilepton_all}, behaves as  $\sim \exp(-\omega/T)$ for  
$\gamma_G(T) = 0$  due to the Landau damping contribution coming from 
quasigluons in a hot and dense medium. As the Landau cut contribution is 
missing in $\gamma_G(T) \neq 0$ case, one finds a leveling off at low $\omega$ 
[In other words since the Landau damping contribution is absent for 
$\gamma_G(T) \ne 0$ the rate is again “leveled” with that of pole-pole 
contribution for  $\gamma_G=0$ as shown in Fig.~\ref{dilepton_comparison}  
except in the mass gap region]. We further note that LQCD 
rate~\cite{Ding:2010ga} matches  with Born rate at large $\omega$ just because it has 
assumed  a free spectral function for large $\omega$. On the other hand the 
LQCD spectral function~\cite{Ding:2010ga} at low $\omega$ is sensitive to the prior 
assumptions and  in such a case the spectral function extracted using  
MEM~\cite{Asakawa:2000tr,Nakahara:1999vy} analysis should carefully be interpreted with proper 
error analysis~\cite{Asakawa:2000tr}. Since the MEM analyses is sensitive to prior 
assumption but is not very sensitive to the structure of the spectral function 
at small $\omega$, so the error is expected to be significantly large at small 
$\omega$. The existence of fine structure such as van Hove singularities  at 
small $\omega$ cannot be concluded from the LQCD rate~\cite{Ding:2010ga}.

\section{Conclusions and outlook}
\label{gribov_conclu}

In this chapter we considered the effect of inclusion of magnetic screening in the 
context of the Gribov-Zwanziger picture of confinement.  In covariant gauge, 
this was accomplished by adding a masslike parameter, the Gribov parameter, to 
the bare gluon propagator resulting in the non-propagation of gluonic modes.  
Following Ref.~\cite{Su:2014rma} we obtained the resummed quark propagator taking 
into account the Gribov parameter.  A new key feature of the resulting resummed 
quark propagator is that it contains no discontinuities.  In the standard 
perturbative hard-thermal loop approach there are discontinuities at spacelike 
momentum associated with Landau damping which seem to be absent in the GZ-HTL 
approach.  Using the resulting quark propagator, we evaluated the spectral 
function, finding that it only contains poles for $\gamma_G \neq 0$.  We then 
used these results to compute the DPR at vanishing 
three-momentum.  For the dilepton 
production rate, we found that, due to the absence of Landau damping for 
$\gamma_G \neq 0$, the rate contains sharp structures, e.g. Van Hove 
singularities, which don't seem to be present in the lattice data.  That being 
said, since the lattice calculations used a perturbative ansatz for the spectral 
function when performing their MEM analysis\cite{Asakawa:2000tr} of the spectral 
function, it is unclear how changing the underlying prior assumptions about the 
spectral function would affect the final lattice results. 
Moreover, the error analsis for spectral function with MEM 
prescription~\cite{Nakahara:1999vy} has to be done carefully than it was done in LQCD 
calculation~\cite{Ding:2010ga}. Since the result is sensitive to the prior
assumptions, the error seems to become large and as a result no conclusion
can be drawn for fine structures at low mass dileptons from the LQCD result.   
The absence of quasigluons responsible for Landau cut  makes 
the results for the dilepton production drastically different from those in perturbative approaches. We conclude that 
the results with present GZ action is in conflict with those in perturbative 
approaches due to the absence of Landau cut contribution in the non-perturbative 
quark propagator.

%% file: text/mag_dpr.tex
\chapter{Dilepton Production Rate in a magnetized hot medium}
\label{th_mag}

In this chapter we explore the electromagnetic spectral properties and evaluate the DPR for the hot magnetized medium supposedly produced in non-central HIC. We will mainly discuss two limiting scenarios; the very strongly magnetized medium just after the collision takes place where non-perturbative strong field approximation has to be used and the weakly magnetized medium after the system thermalizes, where the modification due to the magnetic field comes as a perturbation. This chapter is based on two papers; some part of \textit{Electromagnetic spectral properties and Debye screening of a strongly magnetized hot medium} by Aritra Bandyopadhyay, Chowdhury Aminul Islam and Munshi G Mustafa, {\bf Phys.Rev. D94 (2016) no.11, 114034} and \textit{Effect of magnetic field on dilepton production in a hot plasma} by Aritra Bandyopadhyay and S. Mallik, {\bf Phys.Rev. D95 (2017) no.7, 074019}. 


\section{Introduction}
\label{mag_intro}

In section \ref{nc_hic} we briefly discussed about the non-central HICs and how magnetic field can be generated in these types of collisions. The presence of an external anisotropic field includes a new scale in the medium. So, the present theoretical tools require corresponding modification to be applied appropriately to investigate the hot and magnetized medium. An intense research activity is underway  to study the properties of strongly interacting matter in presence of an external magnetic field: resulting in the emergence of several novel phenomena, \textit{e.g.}, chiral magnetic effect~\cite{Kharzeev:2007jp,Fukushima:2008xe,Kharzeev:2009fn}, finite temperature magnetic catalysis~\cite{Alexandre:2000yf,Gusynin:1997kj,Lee:1997zj} and inverse magnetic catalysis~\cite{Bali:2011qj,Bornyakov:2013eya,Mueller:2015fka,Ayala:2014iba,Ayala:2014gwa,Ayala:2016sln,Ayala:2015bgv,Farias:2014eca}; chiral- and color-symmetry broken/restoration phase~\cite{Fayazbakhsh:2010bh,Fayazbakhsh:2010gc,Andersen:2013swa};  thermodynamic properties~\cite{Strickland:2012vu,Andersen:2014xxa}, refractive indices and decay constant~\cite{Fayazbakhsh:2012vr,Fayazbakhsh:2013cha} of mesons in hot magnetized medium; soft photon production from conformal anomaly~\cite{Basar:2012bp,Ayala:2016lvs} in HICs; modification of dispersion properties in a magnetized hot QED medium~\cite{Sadooghi:2015hha}; syncroton radiation~\cite{Tuchin:2013bda}, dilepton production from a hot magnetized QCD plasma~\cite{Tuchin:2012mf, Tuchin:2013bda, Tuchin:2013ie,Sadooghi:2016jyf} and in strongly coupled plasma in a strong magnetic field~\cite{Mamo:2013efa}. Also experimental evidences of photon anisotropy, provided by the PHENIX Collaboration~\cite{Adare:2011zr}, have posed a challenge for existing theoretical models. Subsequently some theoretical explanations are made by assuming the presence of a large anisotropic magnetic field in heavy ion collisions~\cite{Basar:2012bp}. This suggests that  there is clearly an increasing demand to study the effects of intense background magnetic fields on various aspects and observables of non-central heavy-ion collisions.

Processes like cyclotron emission which are usually abandoned in vacuum become active in presence of an external magnetic field~\cite{PerezRojas:1979jrk}. These processes affect the photon propagation and thus the spectral function. In vacuum, a full description of polarization tensor in presence 
of an external magnetic field have already been studied~\cite{Hattori:2012je,Hattori:2012ny,Chao:2014wla,Tsai:1974ap,Tsai:1974fa}. At this point we also note that the dilepton production rate under extreme magnetic fields have  been addressed  earlier by Tuchin~\cite{Tuchin:2012mf,Tuchin:2013bda,Tuchin:2013ie} in a more
phenomenological way. In order to estimate the dilepton production with logarithmic accuracy~\cite{Tuchin:2012mf,Tuchin:2013ie}, a semi-classical Weisz\"acker-Williams method~\cite{Jackson:1975,Dalitz:1957dd} was employed to obtain the dilepton production rate by a hard quark as a convolution of 
the real photon decay rate with the flux of equivalent photons emitted by a fast quark. In this calculation it was approximated that the virtuality of photon has negligible effect on photon emission and on dilepton production. Recently, Sadooghi and Taghinavaz~\cite{Sadooghi:2016jyf}
have  analyzed in details the dilepton production rate for magnetized hot and dense medium in a formal field theoretic approach using Ritus eigenfunction method~\cite{Ritus:1972ky}. In this chapter we use such formal field theoretic approach along with Schwinger method~\cite{Schwinger:1951nm} to obtain the electromagnetic spectral function and the dilepton rate in both strong and weak field approximation. 

The chapter is organized as follows: In section \ref{mag_setup} we will revise the Schwinger proper time formalism. In section \ref{sfa} we will explore the electromagnetic spectral properties of a strongly magnetized medium and hence compute the DPR. In section \ref{wfa} we will study the DPR in another limiting case, i.e. the weakly magnetized medium before concluding in section \ref{mag_conclu}. 


\section{Schwinger proper-time Formalism}
\label{mag_setup}

The effect of magnetic field in different processes arises through the altered propagation of particles in this field. A non-perturbative, gauge covariant expression for the Dirac propagator in an external electromagnetic field was derived long ago by Schwinger in an elegant way, using a proper-time parameter~\cite{Schwinger:1951nm}. It has since been re-derived and applied to many processes~\cite{Ritus:1972ky,Tsai:1974ap,Tsai:1974fa,PerezRojas:1979jrk,Gusynin:1995nb,Hattori:2012je, Hattori:2012ny, Ayala:2014gwa, Shovkovy:2012zn}.

The equation of motion for a fermion in presence of a magnetic field is shown in \ref{mag_fermion}. Following that, we know that the propagator 
\bea
S(X,X') = i\la 0 | T\psi(X)\bar{\psi}(X')| 0\ra
\eea
satisfies
\bea
\left[i\gm^\mu(\partial_\mu+iq_fA_\mu)-m_f\right]S(X,X') =-\delta^4(X-X').
\label{greens_function}
\eea
Here $ |0\ra$ is the vacuum state of the Dirac field (in presence of $A_\mu$). 
Defining states labeled by space-time coordinate (suppressing spinor
indices), we regard $S(X,X')$ as the matrix element of an operator $S$
\bea
S(X,X') = \la X | S |X'\ra.
\eea
Then Eq.~(\ref{greens_function}) can be written as
\bea
(\gm^\mu\Pi_\mu-m_f)S=-1,~~~  \Pi_\mu=P_\mu-q_fA_\mu,~~~P_\mu=i\partial_\mu
\eea
which has the formal solution \footnote{Another equivalent form follows by
writing $(\slashed{\Pi}+m)$ on the right in Eq.~(\ref{formalsoln})
\cite{Schwinger:1951nm}, but we shall not use it.}
\bea
S=\frac{1}{-\slashed{\Pi}+m_f}=(\slashed{\Pi}+m_f)\frac{1}{-\slashed{\Pi}^2+m_f^2}.
\label{formalsoln}
\eea
Schwinger relates these quantities to the dynamical properties 
of a particle with coordinate $X^\mu$ and canonical and
kinematical momenta $P^\mu$ and $\Pi^\mu$ respectively. Using their
commutation relations, we get $\slashed{\Pi}^2=\Pi^2-\frac{q_f}{2}\sg F$ .
Defining $H=-\Pi^2+m_f^2+\frac{q_f}{2}\sg F$ we can write Eq.~(\ref{formalsoln}) as 
\bea
S=(\slashed{\Pi}+m_f)i\int\limits_0^\infty ds U(s),~~~~ U=e^{-iHs}.
\eea
As the notation suggests, $U(s)$ may be regarded as the evolution operator of  
the particle with Hamiltonian $H$ in time $s$.

We now go to Heisenberg representation, where the operators $X_\mu$ and $\Pi_\mu$
as well as the base ket become time dependent,
\bea
X_\mu(s)=U^\dagger(s) X_\mu U(s),~~ \Pi_\mu(s)=U^\dagger(s) \Pi_\mu U(s),
~~~~|X';s\ra = U^\dagger(s)|X';0\ra
\label{heisenrep}
\eea
Then the construction of the propagator reduces to the evaluation of 
\bea
\la X''|U(s)| X'\ra = \la X'';s| X';0\ra,
\label{prop_construct}
\eea
which is the transformation function for a state, in which the operator $X_\mu(s=0)$ 
has the value of $X'_\mu$, to a state, in which $X_\mu(s)$ has the value $X''_\mu$. 

The equation of motion for $X_\mu(s)$ and $\Pi_\mu(s)$ following from 
Eq.~(\ref{heisenrep}) can be solved to get 
\bea
\Pi(s)=-\frac{1}{2}q_fFe^{-q_fFs}\sinh^{-1}(q_fFs)\left(X(s)-X(0)\right)
\label{pis}
\eea
which may also be put in the reverse order on using the antisymmetry of $F_{\mn}$. 
The matrix element $\la  X^{\prime\prime};s| \Pi^2(s)|  X^{\prime};0\ra$ can 
now be obtained by using the commutator $\left[X_\mu(s),X_\nu(0)\right]$ to reorder 
the operators $X_\mu(0)$ and $X_\nu(s)$. We then get
\bea
\la  X'';s| H(X(s),\Pi(s))|  X';0\ra = f(X'';X';s)\la X'';s| X';0\ra
\eea
where
\bea
f\!\!&=&\!\!(X''-X')K(X''-X')\!-\!\frac{i}{2}\textrm{Tr}[q_fF\coth(q_fFs)]\!-\!m_f^2\!
-\frac{q_f}{2}\sg F,\nn\\
K\!\!&=&\!\!\frac{(q_fF)^2}{4}\sinh^{-2}(q_fFs).
\label{fform}
\eea

We are now in a position to find the transformation function, which from 
Eq.~(\ref{prop_construct}) is found to satisfy
\bea
i\frac{d}{ds}\la X'';s| X';0\ra = \la X'';s| H |X';s\ra.
\eea
It can be solved as 
\bea
\la X'';s| X';0\ra &=& \phi(X'',X').\frac{i}{(4\pi)^2s^2}e^{-L(s)} \times \nn\\
&& \!\!\!\!\!\!\!\!\!\!\!\!\!\!\!\!\!\!\!\!\!\!\!\!\!\!\!\!\!\!\!\!\!\!\!\!\!\!\!\!\!\!\!\!\!\exp\left(-\frac{i}{4}(X''-X')q_fF\coth(q_fFs)(X''-X')\right)
\exp\left(-i(m_f^2+\frac{1}{2}q_f\sg F)\right)
\label{tfunc}
\eea
where
\bea
L(s)=\frac{1}{2}\textrm{Tr}\ln\left[(q_fFs)^{-1}\sinh(q_fFs)\right].
\eea
Here $\phi(X'',X')$ is a phase factor involving an integral over 
the potential $A_\mu$ on a straight line connecting $X'$ and $X''$. 
It will cancel out in our calculation. The spinor propagator is now given by
\bea
S(X'',X')&=& i\int\limits_0^\infty ds \la X''| (\slashed{\Pi}+m_f)U(s)| X'\ra\nn\\
&=& i\int\limits_0^\infty ds \left[\gm^\mu\la X'';s| \Pi_\mu(s)| X';0\ra
+m_f\la X'';s| X';0\ra\right]
\eea
with $\Pi_\mu(s)$ and $\la X'';s| X';0\ra$ given by 
Eqs. (\ref{pis}) and (\ref{tfunc}).

We now specialize the external electromagnetic field to magnetic field $B$ in the 
$z$ direction, $F^{12}=-F^{21}=B$. It is convenient to diagonalize the antisymmetric 
$2\times 2$ matrix $F^{ij}$ with eigenvalues $\pm iB$. Going over to spatial metric 
we get
\bea
S(X)=\frac{i}{(4\pi)^2}\int\!\!\frac{ds}{s}&&\!\!\!\!\!\!\!\!\!\!\frac{q_fB}{\sin(q_fBs)}\exp\left[
\frac{i}{4}X_\pp^2 q_fB\cot(q_fBs)-\frac{i}{4s^2}X_\pl^2-i(m_f^2+\frac{1}{2}q_f\sg F)s\right]\nn\\
&&\!\!\!\!\!\!\!\!\!\!\!\!\!\!\!\!\!\!\!\!\!\!\!\!\!\!\!\!\!\!\!\!\!\!\!\!\!\!\!\!\!\!\!\!\!\!\!\!\!\!\!\!\!\!\!\!\!\! \times\left[\left(\frac{1}{2s}(X\cdot \gamma)_\shortparallel + m_f\right)\left(\cos(q_fBs) - \gamma^1\gamma^2\sin(q_fBs)\right)-\frac{q_fB}{2\sin(q_fBs)}(X\cdot \gamma)_\perp
\right]
\eea
which can be Fourier transformed to 
\bea
S(P)&=&i\int\limits_0^\infty ds~e^{is(P^2-m_f^2+i\eps)}~e^{-isP_\pp^2\left(\frac{\tan(q_fBs)}{q_fBs}-1\right)}\times\nn\\
&& \left[(\slashed{P}_\pl +m_f)\left(1-\gm^1\gm^2\tan(q_fBs)\right)-\slashed{P}_\pp\left(1+\tan^2(q_fBs)\right)\right].
\label{schwinger_propertime}
\eea
Below we outline the notation we have used in~(\ref{schwinger_propertime}) and  are going to follow throughout as
\bea
&&a^\mu = a_\shortparallel^\mu + a_\perp^\mu;~~ a_\shortparallel^\mu = (a^0,0,0,a^3) ;~~  a_\perp^\mu = (0,a^1,a^2,0),\nn\\
&&g^{\mu\nu} = g_\shortparallel^{\mu\nu} + g_\perp^{\mu\nu};~~ g_\shortparallel^{\mu\nu}= \textrm{diag}(1,0,0,-1);~~ g_\perp^{\mu\nu} = \textrm{diag}(0,-1,-1,0),\nn\\
&&(a\cdot b) = (a\cdot b)_\shortparallel - (a\cdot b)_\perp;~~ (a\cdot b)_\shortparallel = a^0b^0-a^3b^3;~~ (a\cdot b)_\perp = (a^1b^1+a^2b^2),\nn
\eea
where $\shortparallel$ and $\perp$ are, respectively, the parallel and perpendicular components, which are now separated out in presence of the anisotropic external magnetic field. Note that the longitudinal and transverse directions are defined with respect to the direction 
of magnetic field, not the collision axis of ions.


\section{Strong field approximation}
\label{sfa}

At the time of the non-central HIC the initial magnitude of the produced magnetic field can be very high as measured in RHIC and LHC. We know that the energy levels of a moving charged particle in presence of a magnetic field get discretized, which are known as the Landau Levels (see \ref{mag_fermion}). One fascinating prospect of having a very strong background magnetic field is that only the LLL, whose energy is independent of the strength of the magnetic field, remains active in that situation. That is why, the LLL dynamics becomes solely important in the strong magnetic field approximation and the higher order contributions, \textit{i.e}, the radiative corrections play a significant role in this context, as it is the only way to get the $B$ dependence in the LLL energy. 

In this section we will mainly investigate the nature of the in-medium electromagnetic spectral function in presence of a very strong but constant magnetic field strength ($q_fB~\gg~T^2$), which could be relevant for initial stages of a non-central heavy-ion collision, as a high  intensity magnetic field is believed to be produced there. After that we will compute the DPR for this strongly magnetized medium. We will use the LLL approximation to implement the strong magnetic field in the system which will also analytically simplify our calculation.  Below we have considered a two-flavor system ($N_f=2$)  of equal current quark mass ($m_f=m_u=m_d=5$ MeV if not said otherwise).

\subsection{Fermion propagator within strong magnetic field}
\label{sfa_lll_prop}

After performing the proper time integration~\cite{Gusynin:1995nb}, the fermion propagator in (\ref{schwinger_propertime}) can be represented as sum over discrete energy spectrum of the fermion 
\bea
i S_m(K) = i e^{-\frac{K_\perp^2}{q_fB}} \sum_{n=0}^{\infty} \frac{(-1)^nD_n(q_fB, K)}{K_\shortparallel^2-m_f^2-2nq_fB},
\label{decomposed_propagator}
\eea
with Landau levels $n=0,\, 1,\, 2, \cdots$ and  
\bea
D_n(q_fB,K) &=& (\slashed{K}_\shortparallel+m_f)\Bigl((1-i\gamma^1\gamma^2)L_n\left(\frac{2K_\perp^2}{q_fB}\right)
-(1+i\gamma^1\gamma^2)L_{n-1}\left(\frac{2K_\perp^2}{q_fB}\right)\Bigr)\nn\\
&&- 4\slashed{K}_\perp L_{n-1}^1\left(\frac{2K_\perp^2}{q_fB}\right),
\label{d_n}
\eea
where $L_n^\alpha (x)$ is the generalized Laguerre polynomial written as
\bea
(1-z)^{-(\alpha+1)}\exp\left(\frac{xz}{z-1}\right) = \sum_{n=0}^{\infty} L_n^\alpha(x) z^n.
\eea
\begin{figure}
\begin{center}
\includegraphics[scale=1.2]{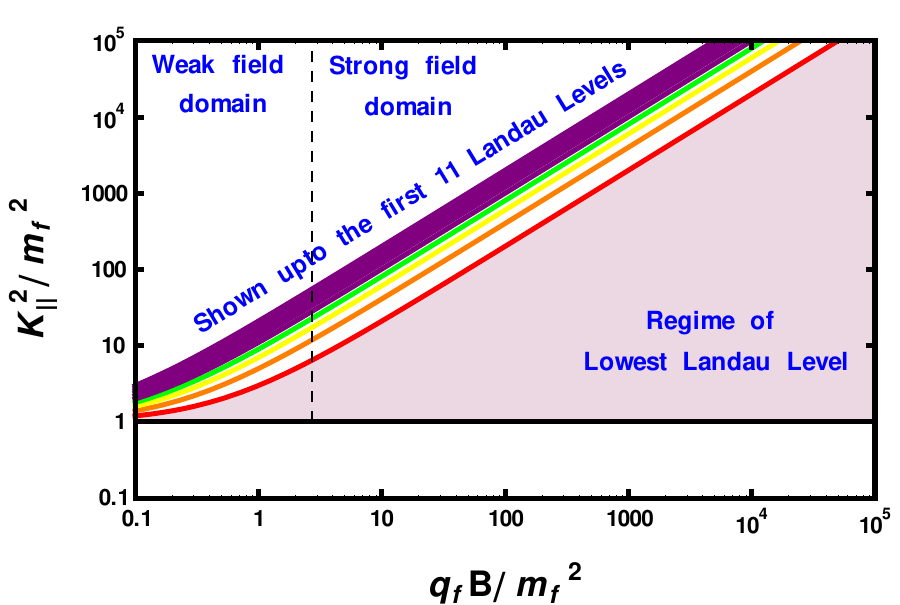}
\end{center}
\caption[Thresholds corresponding to a few Landau Levels ]{Thresholds corresponding to a few Landau Levels are displayed as a function of $q_fB/m_f^2$. This threshold plot is obtained by solving $\left(\omega^2-4m_f^2-8nq_fB\right)=0$ with 
zero photon momentum following energy conservation in a background magnetic field in 
general. Also the regime of the  LLL at strong magnetic field approximation is 
shown by the shaded area.}
\label{landau_levels}
\end{figure}

When the external magnetic field is very strong~\cite{Calucci:1993fi}, i.e. $q_fB\rightarrow \infty$, it pushes all the Landau levels ($n\ge 1$) to infinity compared to the LLL with $n=0$ (See Fig.~\ref{landau_levels}). For LLL approximation in the strong field limit the fermion propagator in (\ref{decomposed_propagator}) reduces to a simplified form as
\bea
iS_{ms}(K)=ie^{-{K_\perp^2}/{q_fB}}~~\frac{\slashed{K}_\shortparallel+m_f}{
K_\shortparallel^2-m_f^2}(1-i\gamma_1\gamma_2),
\label{prop_sfa}
\eea
where $K$ is four momentum and we have used the properties of generalized Laguerre polynomial, $L_n\equiv L_n^0$ and $L_{-1}^\alpha = 0$. The appearance of the projection operator $(1-i\gamma_1\gamma_2)$ in (\ref{prop_sfa}) indicates that the spin of the fermions in LLL are aligned along the field direction~\cite{Shovkovy:2012zn,Gusynin:1995nb}. As $K_\perp^2 << q_fB$,  one can see from (\ref{prop_sfa}) that an effective dimensional reduction from (3+1) to (1+1) takes place in the strong field limit within LLL approximation. 


\subsection{Photon polarization tensor in vacuum ($T=0$)}
\label{sfa_lll_pptv}

\begin{figure}
\begin{center}
\includegraphics[scale=0.6]{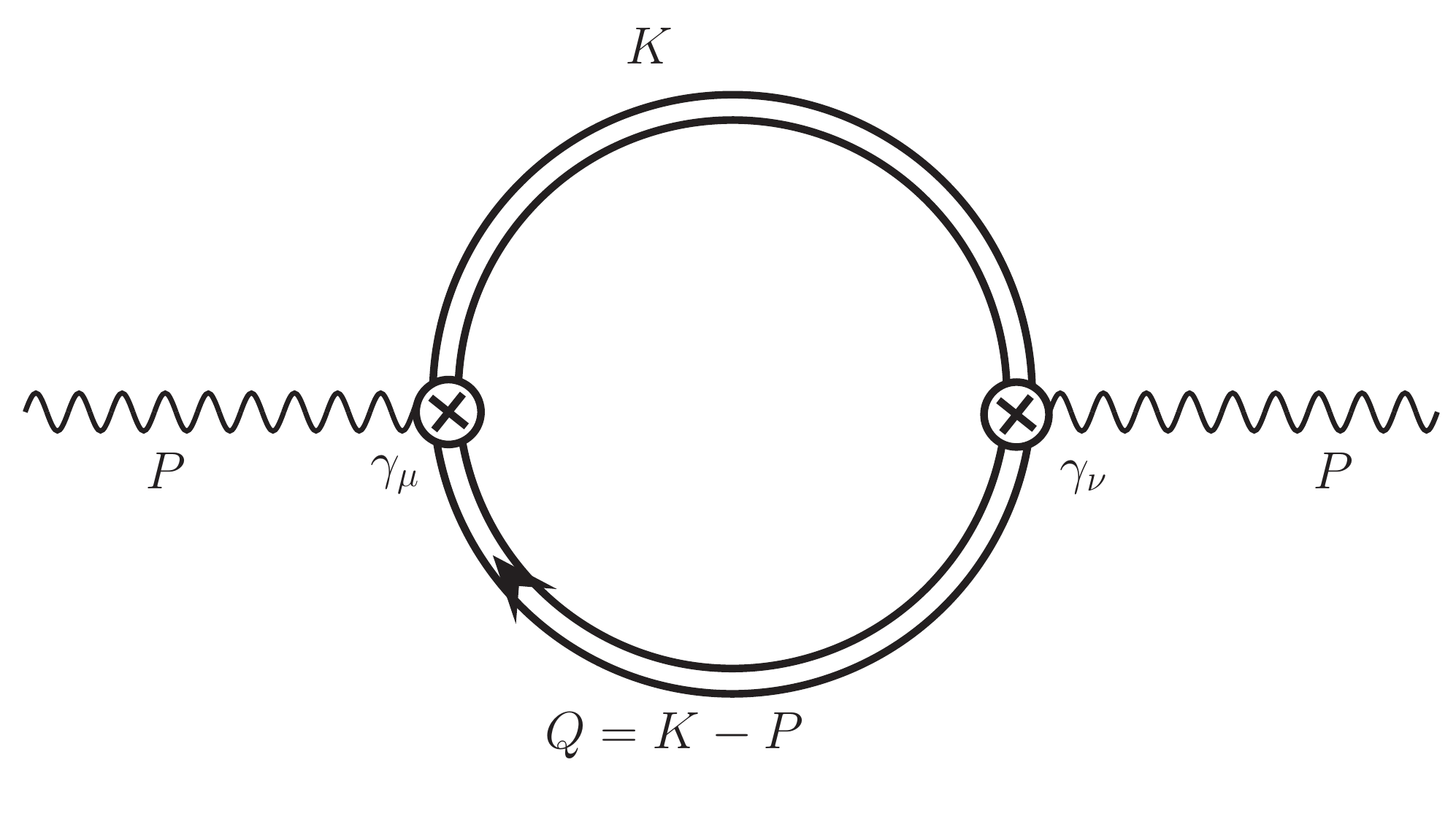}
\end{center}
\caption{Photon polarization tensor in the limit of strong field approximation.}
\label{sfa_self_energy}
\end{figure}

The effective dimensional reduction mentioned above in \ref{sfa_lll_prop} plays an important role in catalyzing the spontaneous chiral symmetry breaking~\cite{Shovkovy:2012zn,Gusynin:1995nb}. This is because of the fact that the fermion pairing taking place in LLL enhances the generation of fermionic mass through the chiral condensate in strong field limit at $T=0$. The pairing dynamics is essentially (1+1) dimensional where the fermion pairs fluctuate in the direction of magnetic field. It is also interesting to see how these fermionic pairs respond to the electromagnetic fields. The fluctuation of fermion pairs in LLL as shown in Fig.~\ref{sfa_self_energy} is a response to the polarization of the electromagnetic field and would reveal various properties of the system in presence of magnetic field. Also the response to the electromagnetic field at $T\ne 0$ due to the thermal fluctuation of charged fermion pairs in LLL would also be very relevant for the initial stages of the noncentral heavy-ion collisions where the intensity of the generated magnetic field is very high. 

Now in one-loop photon polarization in Fig.~\ref{sfa_self_energy} the effective fermionic propagator in strong field approximation is represented by a double line and the electromagnetic vertex remains 
unchanged~\footnote{This is not very apparent from the momentum space effective propagator in (\ref{prop_sfa}) because of the presence of the projection operator. In Ref.\cite{Ferrer:1998vw} the Ward-Takahashi identity in  LLL for fermion-antifermion-gauge boson in massless QED in presence of constant magnetic field was shown to be satisfied by considering the effective fermion propagator, bare vertex and free gauge boson propagator in ladder approximation through Dyson-Schwinger approach  in a representation where the fermion mass operator is diagonal in momentum space.} denoted  by a crossed circle. As mentioned earlier, the spin of the fermions in LLL are aligned in the direction of 
the magnetic field because of the projection operator in (\ref{prop_sfa}). In QED, vertex with two fermions from LLL make the photon spin equals to zero in the field direction~\cite{Gusynin:1995nb} and there is no polarization in the transverse direction.  Thus the longitudinal components ({\it i.e}, (0,3) components) of QED vertex would  only be relevant. 

Now in the strong field limit the self-energy in (\ref{pola}) can be computed as
\bea
\Pi_{\mu\nu}(P)\Big\vert_{sfa} &=& -i\sum_{f}q_f^2\int\frac{d^4K}{(2\pi)^4}\textrm{Tr}_c\left[\gamma_\mu S_{ms}(K)\gamma_\nu S_{ms}(Q)\right]\nn\\
&=& -iN_c\sum_{f}q_f^2 \int\frac{d^2K_\perp}{(2\pi)^2} \exp\left(\frac{-K_\perp^2-Q_\perp^2}{q_fB}\right)\nn\\
&&\!\!\!\!\!\!\!\!\times \int\frac{d^2K_\shortparallel}{(2\pi)^2} \textrm{Tr} \left[\gamma_\mu \frac{\slashed{K}_\shortparallel+m_f}
{K_\shortparallel^2-m_f^2}(1-i\gamma_1\gamma_2)\gamma_\nu \frac{\slashed{Q}_\shortparallel+m_f}{Q_\shortparallel^2-m_f^2}(1-i\gamma_1\gamma_2)\right]
\eea
where `\textit{sfa}' indicates the strong field approximation and $\textrm{Tr}$ represents only the Dirac trace. Now one can notice that the longitudinal and transverse parts are completely separated and the gaussian integration over the transverse momenta can be done trivially, which leads to
\bea
\Pi_{\mu\nu}(P)\Big\vert_{sfa} &=& -iN_c\sum_{f}~e^{{-P_\perp^2}/{2q_fB}}~~\frac{q_f^3 B}{\pi}\int\frac{d^2K_\shortparallel}{(2\pi)^2} 
\frac{S_{\mu\nu}}{(K_\shortparallel^2-m_f^2)(Q_\shortparallel^2-m_f^2)}, 
\label{pol_vacuum}
\eea
with the tensor structure $S_{\mu\nu}$ that originates from the Dirac trace  is 
\bea
S_{\mu\nu} = K_\mu^\shortparallel Q_\nu^\shortparallel + Q_\mu^\shortparallel K_\nu^\shortparallel 
- g_{\mu\nu}^\shortparallel \left((K\cdot Q)_\shortparallel -m_f^2\right),
\eea
where the Lorentz indices  $\mu$ and $\nu$ are restricted to longitudinal values. In vacuum, (\ref{pol_vacuum}) can be simplified using the Feynman parametrization technique \cite{Calucci:1993fi}, after which the structure of the photon polarization tensor can be written in compact form as
\bea
\Pi_{\mu\nu}(P) = \left(\frac{P^\shortparallel_\mu P^\shortparallel_\nu}{P_\shortparallel^2}-g^\shortparallel_{\mu\nu}\right)\Pi (P^2),
\label{lll_genstruc}
\eea
which directly implies that due to the current conservation, the two point function is transverse. 
The scalar function $\Pi (P^2)$ is given by, 
\bea
\Pi (P^2) \!\!\!\!&=&\!\!\!\!\sum_{f}\frac{N_c~q_f^3B}{8\pi^2 m_f^2}\, e^{-\frac{P_\perp^2}{2q_fB}}\!\!\!\left[4m_f^2+\frac{8m_f^4}{P_\shortparallel^2}
\left(1-\frac{4m_f^2}{P_\shortparallel^2}\right)^{\!\!-\frac{1}{2}}
\ln\frac{\left(1-\frac{4m_f^2}{P_\shortparallel^2}\right)^{\frac{1}{2}}\!\!+\!\!1}{\left(1-\frac{4m_f^2}{P_\shortparallel^2}\right)^{\frac{1}{2}}\!\!-\!\!1}\right].
\label{vacuum_check}
\eea
We note that the lowest threshold for a photon to decay into fermion and antifermion is provided by the energy conservation when photon momenta $P_\shortparallel^2=(m_f+m_f)^2= 4m_f^2$.  Interestingly $\Pi(P^2)$ is singular in presence of magnetic field at this threshold.  This is because of the appearance of the pre-factor $\sqrt{1-4m_f^2 /P_\shortparallel^2}$ in the denominator of the second term in (\ref{vacuum_check}) due to the dimensional reduction from (3+1) to (1+1) in presence of the strong magnetic field. This behavior is in contrast to that in absence of the magnetic field where a similar prefactor appears in the numerator~\cite{Peskin:1995}. Now, we explore $\Pi(P^2)$ physically in the following two domains around the LT, $P_\shortparallel^2=4m_f^2$ :
\begin{enumerate}
\item 
{\sf {Region-I}}~~$P_\shortparallel^2 < 4m_f^2$~:~~ In this case with 
$a=\sqrt{4m_f^2 /P_\shortparallel^2-1}$, 
let us write the logarithmic term in the second term of  (\ref{vacuum_check})  as
\bea
\ln\left(\frac{ai+1}{ai-1}\right) = \ln\left(\frac{re^{i\theta_1}}{re^{i\theta_2}}\right) = i(\theta_1-\theta_2),
\eea
where $r=\sqrt{(1+a^2)}$, $\theta_1 = \arctan(a)$ and $\theta_2 = \arctan(-a)$. Thus in (\ref{vacuum_check}) the logarithmic term is purely 
imaginary but overall $\Pi(P^2)$  is real because of the prefactor $\left(1-{4m_f^2}/{P_\shortparallel^2}\right)^{-{1}/{2}}$ being imaginary. 
Even if we choose the limit $P_\shortparallel^2<0$, then too the whole term is real again, since the denominator of the logarithmic term, 
$\sqrt{1-4m_f^2 /P_\shortparallel^2}$, is always greater than unity. So in the region $P_\shortparallel^2 < 4m_f^2$, $\Pi(P^2)$ is purely real. 
 
\item {\sf{Region-II}}~~$P_\shortparallel^2 > 4m_f^2$~:~~
Though in this limit the prefactor is real definite, but the denominator in the logarithmic term becomes negative and a complex number arises 
from it as $\ln(-x)=\ln\vert x\vert+i~\pi$. Thus we get both real and imaginary contributions, {\it i.e}, ${\mathcal{R}}e~\Pi(P^2)$ and $\textrm{Im}~\Pi(P^2)$, in this limit. The imaginary contribution is  relevant for studying the spectral function and its spectral properties.
\end{enumerate}

\begin{figure}
\begin{center}
\includegraphics[scale=0.8]{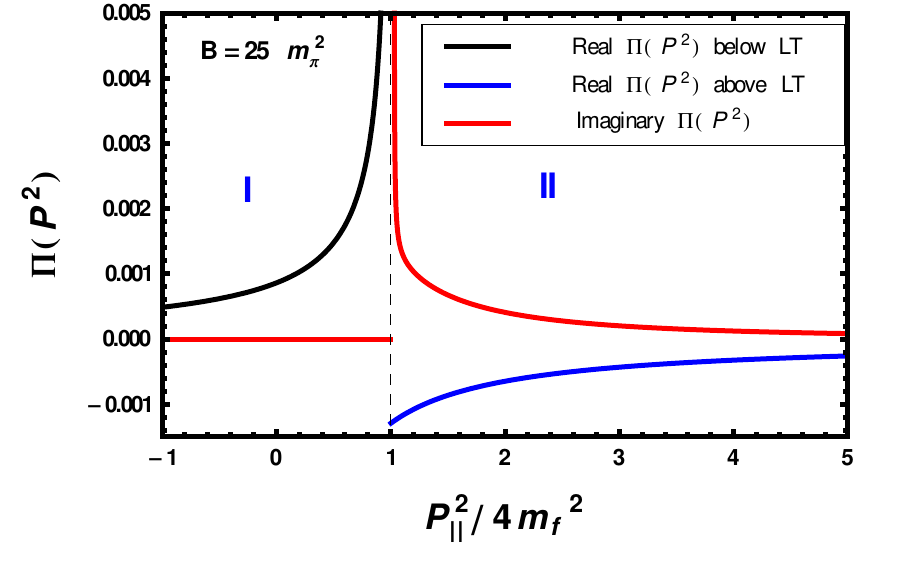}
\includegraphics[scale=0.8]{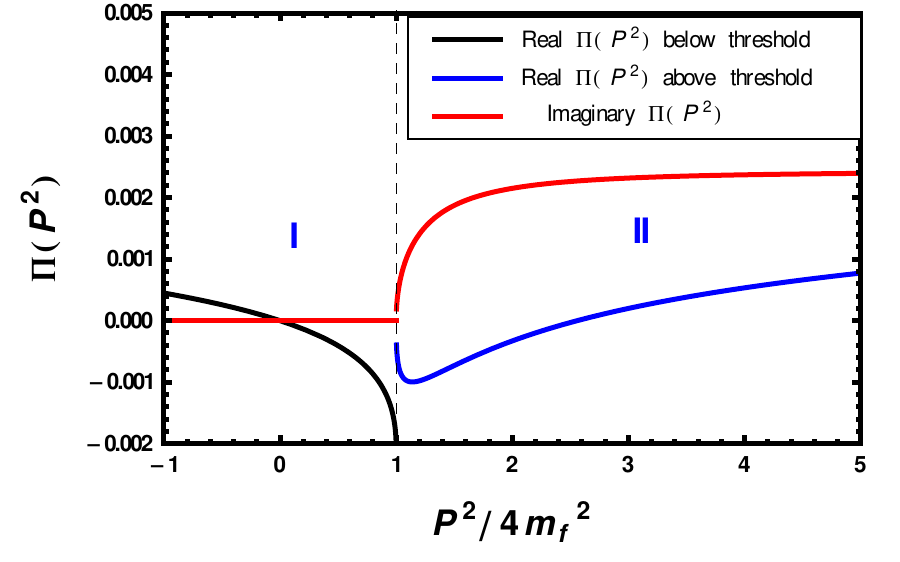}
\end{center}
\caption[Plot of real and imaginary parts of $\Pi(P^2)$ in presence and absence of strong magnetic field]{ Plot of real and imaginary parts of $\Pi(P^2)$ as a function scaled photon momentum square with respect to LT in various kinematic regions I and II as discussed in the text  in presence of a strong magnetic field (left panel) and in absence of a magnetic field (right panel).}
\label{pola_plot}
\end{figure}

We now extract the vacuum spectral function in presence of strong magnetic field 
following (\ref{spec_func}) as
\bea
\rho\Big\vert_{sfa}^{\textsf{vacuum}} &=& \frac{1}{\pi} 
\textrm{Im}~C^\mu_\mu(P)\Big\vert_{sfa}^{\textsf{vacuum}} \nn\\ &=& 
N_c\sum_{f}\frac{q_fBm_f^2}{\pi^2 P_\shortparallel^2}\ 
e^{-{P_\perp^2}/{2q_fB}}~\Theta\left(P_\shortparallel^2-4m_f^2\right)
\left(1-\frac{4m_f^2}{P_\shortparallel^2}\right)^{-{1}/{2}}. \label{spec_vac}
\eea
As seen above, the imaginary part is restricted by the LT, $P_\shortparallel^2=4m_f^2$. Below this threshold ($P^{2}_\shortparallel<4m_f^2$), $\Pi(P^2)$ is real and there is no electromagnetic spectral contribution in vacuum with strong magnetic field as can be seen from region I in the left panel of Fig.~\ref{pola_plot}. This implies that there is also no creation of particle and antiparticle in vacuum below LT because the width of the electromagnetic spectral function vanishes. Beyond LT there is also a continuous contribution (blue solid line in region II) in real part of $\Pi(P^2)$. As is seen the real part of $\Pi(P^2)$ is continuous both below and above the LT but has a discontinuity at the LT, $P^2_\shortparallel=4m_f^2$. Though we are interested in the imaginary part, we want to note that the real part can be associated with the dispersion property of vector boson~\footnote{This has been discussed in Refs.~\cite{Islam:2014sea,Greiner:2010zg} without  magnetic field and in  Ref.~\cite{Gusynin:1995nb} with magnetic field.}. 

On the  other hand  the imaginary part of the electromagnetic polarization tensor is associated with interesting spectral properties of the system. So, beyond the LT ($P^2_\shortparallel> 4m_f^2$) there is nonzero continuous contribution to the electromagnetic spectral function as given by (\ref{spec_vac}) and represented by a red solid line in region II in the left panel of Fig.~\ref{pola_plot}. The right panel of Fig.~\ref{pola_plot} displays the analytic structure of vacuum $\Pi(P^2)$ in absence of magnetic field~\cite{Peskin:1995}. In particular the comparison of the imaginary part of $\Pi(P^2)$ in absence of the magnetic field with that in presence of the strong  magnetic field reveals an opposite trend  around LT. This is due to the effect of dimensional reduction in presence of the strong magnetic field. As a consequence the imaginary part of $\Pi(P^2)$ in presence of strong magnetic field would provide a very strong width to the photon that decays into particle and antiparticle, vis-a-vis an enhancement of the dilepton production from the hot and dense medium produced in  heavy-ion collisions. 


\subsection{Photon polarization tensor in medium ($T\neq0$)}
\label{sfa_lll_pptm}

So far in \ref{sfa_lll_pptv} we have discussed some aspects of the electromagnetic polarization tensor with a strong background magnetic field in vacuum. Now we extend this to explore the spectral properties of a strongly magnetized medium created in non-central HICs. 

In presence of medium, without any loss of information we can contract the indices $\mu$ and $\nu$ in (\ref{pol_vacuum}), 
thus resulting in further simplification as
\bea
\Pi_\mu^\mu(P)\Big\vert_{sfa} &=&  -iN_c\sum_{f}~e^{{-P_\perp^2}/{2q_fB}}~~\frac{q_f^3B}{\pi}\int\frac{d^2K_\shortparallel}{(2\pi)^2} 
\frac{2m_f^2}{(K_\shortparallel^2-m_f^2)(Q_\shortparallel^2-m_f^2)}. 
\eea
At finite temperature this can be written by replacing the $p_0$ integral by Matsubara sum as 
\bea
\Pi_\mu^\mu(P)\Big\vert_{sfa} \!\!\!\!\!\!&=&\!\!\!\! -iN_c\sum_{f}e^{\frac{-P_\perp^2}{2q_fB}}~\frac{2q_f^3Bm_f^2}{\pi}\left(i T \sum_{k_0}\right)\!\!\int\!\!\frac{dk_3}{2\pi} 
\frac{1}{(K_\shortparallel^2-m_f^2)(Q_\shortparallel^2-m_f^2)}. 
\label{pola_ft}
\eea
We now perform the Matsubara sum using the mixed representation prescribed by Pisarski~\cite{Pisarski:1987wc}, where the trick is to dress 
the propagator in a way, such that it is spatial in momentum representation, but temporal in co-ordinate representation: 
\bea
\frac{1}{K_\shortparallel^2-m_f^2} \equiv \frac{1}{k_0^2-E_k^2} = \int\limits_0^\beta d\tau e^{k_0\tau} \Delta_M(\tau,k),
\label{mixed_representation}
\eea
 and 
\bea
\Delta_M(\tau,k) = \frac{1}{2E_k}\left[\left(1-n_F(E_k)\right)e^{-E_k\tau}-n_F(E_k)e^{E_k\tau}\right],
\eea
where $E_k=\sqrt{k_3^2+m_f^2}$ and $n_F(x)=(\exp(\beta x)+1)^{-1}$ is the Fermi-Dirac distribution function
with $\beta=1/T$. Using these, (\ref{pola_ft})
can be simplified as
\bea
\Pi_\mu^\mu(P)\Big\vert_{sfa} 
\!\!\!\!\!\!&=&\!\!\!\! N_c\sum_{f}e^{\frac{-P_\perp^2}{2q_fB}}~\frac{2q_f^3Bm_f^2}{\pi}~
T \sum_{k_0}\int\frac{dk_3}{2\pi}\nn\\
&&\times \int\limits_0^\beta~d\tau_1\int\limits_0^\beta~d\tau_2~ e^{k_0\tau_1}~e^{(k_0-p_0)\tau_2}\Delta_M(\tau_1,k)\Delta_M(\tau_2,q)\nn\\
&=&\!\!\!\! N_c\sum_{f}e^{\frac{-P_\perp^2}{2q_fB}}~\frac{2q_f^3Bm_f^2}{\pi}\int\frac{dk_3}{2\pi}\int\limits_0^\beta~d\tau~ e^{p_0\tau}~\Delta_M(\tau,k)\Delta_M(\tau,q).
\eea
Now the $\tau$ integral is trivially performed as
\bea
\Pi_\mu^\mu(P)\Big\vert_{sfa}\!\! \!\!\!\! &=&\!\!\!\! 
N_c\sum_{f}e^{\frac{-P_\perp^2}{2q_fB}}~\frac{2q_f^3Bm_f^2}{\pi}\int\frac{dk_3}{2\pi}
\sum_{l,r=\pm 1}\!\!
\frac{\left(1-n_F(rE_k)\right)\left(1-n_F(lE_q)\right)}{4(rl)E_kE_q(p_0-rE_k-lE_q)}\nn\\
&&\times\left[e^{-\beta(rE_k+lE_q)}-1\right] .
\label{Pi_sfa}
\eea
One can now easily  read off the discontinuity using
\bea
\textsf{Disc~}\left[\frac{1}{\omega +\sum_i E_i}\right]_\omega = - \pi\delta(\omega + \sum_i E_i),
\label{disc_delta}
\eea
which leads to
\bea
\mathcal{I}{ m}\, \Pi_\mu^\mu(P)\Big\vert_{sfa}\!\! \!\! \!\!&=& \!\!\! \! 
-N_c \pi 
\sum_{f}e^{\frac{-P_\perp^2}{2q_fB}}~~\frac{2q_f^3Bm_f^2}{\pi}\int\frac{dk_3}{2\pi}
\sum_{l,r=\pm 1}\!\!
\frac {\left(1-n_F(rE_k)\right)\left(1-n_F(lE_q)\right)}
{4(rl)E_kE_q} \nn \\
&& \times \left[ e^{-\beta(rE_k+lE_q)}-1\right]
\delta(\omega-rE_k-lE_q).
\label{Pi_sfa_gen}
\eea

The general form of the delta function in (\ref{Pi_sfa_gen})  corresponds to four processes for the choice of $r=\pm 1$ and $l=\pm 1$ as discussed below:
\begin{enumerate}
\item  $r=-1$ and $l=-1$ corresponds to a process with  $\omega <0$, which violates energy conservation as all the quasiparticles 
have positive energies.
\item (a)  $r=+1$ and $l=-1$ corresponds to a process, $q\rightarrow q\gamma$, where a quark with energy $E_k$ makes a transition to an energy $E_q$ after emitting a timelike photon of energy $\omega$. (b) $r=-1$ and $l=1$ corresponds to similar case as (a). It has explicitly been shown in Appendix~\ref{app_a} that both processes are not allowed by the phase space and the energy conservation. In other words, the production of a timelike photon from one loop photon polarization tensor  is forbidden by the phase space and the energy conservation. 
\item $r=1$ and $s=1$ corresponds to a process where a quark and an antiquark annihilate to a virtual photon, which is the only allowed process:
\end{enumerate}
So, considering the last case, one can write from (\ref{Pi_sfa_gen}) 
\bea
\textrm{Im}~\Pi_\mu^\mu(P)\Big\vert_{sfa} \!\!\!\!\!\!&=&\!\!\!\! 
N_c \!\!\sum_{f}
e^{\frac{-P_\perp^2}{2q_fB}}2q_f^3Bm_f^2\!\!\!
\int\!\!\!\frac{dk_3}{2\pi}~\delta(\omega\!-\!E_k\!-\!E_q)\frac{\left[1\!-\!n_F(E_k)\!-\!n_F(E_q)\right]}{
4E_kE_q}.
\label{Pi_im}
\eea
After performing the $k_3$ integral using (\ref{a3}) the spectral function  in strong 
field approximation  is finally obtained following (\ref{spec_func}) as
\bea
\rho \Big\vert_{sfa}\!\!\!\!\!\!&=&\frac{1}{\pi}\textrm{Im}~ 
C^\mu_\mu (P)\Big\vert_{sfa}\nn\\
&=&\!\!\!\!N_c\sum_{f}\frac{q_fBm_f^2}{\pi^2 P_\shortparallel^2}e^{-\frac{P_\perp^2}{2q_fB}}\Theta
\left(P_\shortparallel^2-4m_f^2\right)\!\!\left(1\!\!-\!\!\frac{4m_f^2}{P_\shortparallel^2}
\right)^{\!\!\!-\frac{1}{2}}\!\!\Bigl[1\!-\!n_F(p_+^s)\!-\!n_F(p_-^s)\Bigr],
\label{spec_sfa_general}
\eea
where 
\bea
p_\pm^s = \frac{\omega}{2}\pm \frac{p_3}{2}\sqrt{\left(1-\frac{4m_f^2}{P_\shortparallel^2}\right)}.
\eea

We note that the electromagnetic spectral function in strong field approximation obtained here in (\ref{spec_sfa_general}) using Schwinger method has a factor $[1-n_F(p_+^s)-n_F(p_-^s)]$. This thermal factor appears when a quark and an antiquark annihilate to a virtual photon in a thermal medium, which is the only process  allowed by the phase space as shown in our calculation. 

The vacuum part in presence of the strong magnetic field can be easily separated out from (\ref{spec_sfa_general}) as
\bea
\rho \Big\vert_{sfa}^{\textsf{vacuum}} &=& N_c\sum_{f}\frac{q_fBm_f^2}{\pi^2 
P_\shortparallel^2}~e^{-{P_\perp^2}/{2q_fB}}~\Theta
\left(P_\shortparallel^2-4m_f^2\right)\left(1-\frac{4m_f^2}{P_\shortparallel^2}\right)^{-{1}/{2}},
\eea
which agrees with that obtained in (\ref{spec_vac}).

We outline some of the important features of the spectral functions:

\begin{enumerate}
\item[(i)] In general the electromagnetic spectral function in (\ref{spec_sfa_general})  vanishes in the massless limit of  quarks. 
This particular feature arises because of the presence of magnetic field which  reduces the system  to ($1+1$) dimension. This can 
be further understood from the symmetry argument and is attributed to the CPT invariance of the theory~\cite{Das:2012pd}. Physically 
this observation further signifies that in ($1+1$) dimension an on-shell massless thermal fermion cannot scatter in the forward direction.
 
\item[(ii)] The threshold, $P_\shortparallel^2=4m_f^2$,  for LLL is  independent of the magnetic field 
strength. It is also independent of $T$ as $q_fB \gg T^2$ in the strong field approximation. 
Like vacuum case here also
the spectral function vanishes below the threshold and there is no pair creation of 
particle and antiparticle. 
This is because the polarization tensor is purely real below the threshold. This implies that the momentum of the external photon supplies energy and 
virtual pair in LLL becomes real via photon decay. 

\item[(iii)] When the photon longitudinal momentum square is equal to the LT,  $P_\shortparallel^2=4m_f^2 $, it strikes the LLL and the spectral strength diverges because of the factor $\left(1-{4m_f^2}/{P_\shortparallel^2}\right)^{-{1}/{2}}$ that appears due to the dimensional reduction. 
Since the LLL dynamics is (1+1) dimensional, there is a dynamical mass generation~\cite{Gusynin:1995nb,Ferrer:1998vw} of the fermions through mass operator (\textit{e.g.} chiral condensate), which causes the magnetic field induced chiral symmetry breaking in the system. This suggests that the strong fermion pairing takes place in LLL~\cite{Gusynin:1995nb} even at the weakest attractive interaction between fermions in (3+1) dimension. A (3+1) dimensional weakly interacting system in presence of strong magnetic field can be considered as a strongly correlated system 
in LLL dynamics which is (1+1) dimensional. In that case $m_f$ should be related to the dynamical mass provided by the condensates~\cite{ Gusynin:1995nb,Ferrer:1998vw}. One can incorporate it based on nonperturbative model calculations, then LT will change accordingly.

\item[(iv)] The spectral strength starts with a high value for the photon longitudinal 
momentum $P_\shortparallel > 2m_f$  due to the dimensional reduction  
or LLL dynamics and then falls off with increase of $\omega$ as there is nothing beyond 
the LLL in strong field approximation. 
\end{enumerate}

\begin{figure}[h]
\begin{center}
\includegraphics[scale=0.8]{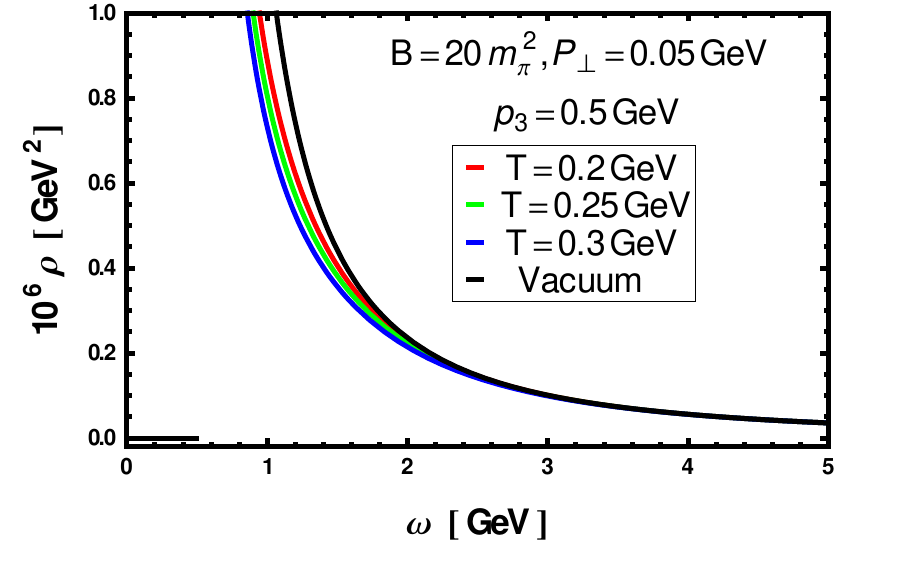}
\includegraphics[scale=0.8]{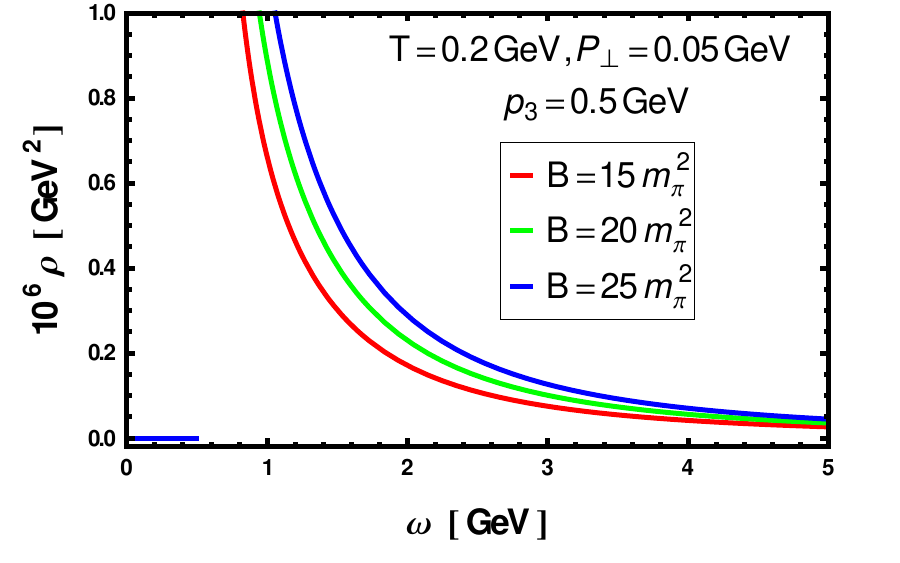}
\end{center}
\caption[Variation of spectral function with photon energy for different values of $T$ and $B$]{{\it Left panel:} Variation of the spectral function with photon energy for different values of $T$ at fixed $B$, $p_\perp$ and $p_3$.
{\it Right panel:} Same as left panel but for different values of magnetic field  at fixed $T$, $p_\perp$ and $p_3$.The value of the magnetic 
field is chosen in terms of the pion mass $m_\pi$.}
\label{spec_plot_1}
\end{figure}

In Fig. \ref{spec_plot_1} the variation of the spectral function with photon energy $\omega$ has been shown for different values of $T$ in the left panel and for
different values of magnetic field in the right panel. With increase in $T$  the spectral strength in the left panel gets depleted because of the presence 
of the thermal weight factor $[1-n_F(p_+^s)-n_F(p_-^s)]$ as the distribution functions $n_F(p_\pm^s)$ increase with $T$ that restricts the available 
phase space. Nevertheless the effect of temperature is small in the strong field 
approximation as $q_fB\gg T^2$.  On the other hand the spectral
strength in the right panel increases with the increase of the magnetic field $B$ as the 
spectral function is proportional to $B$. 

\begin{figure}[h]
\begin{center}
\includegraphics[scale=0.8]{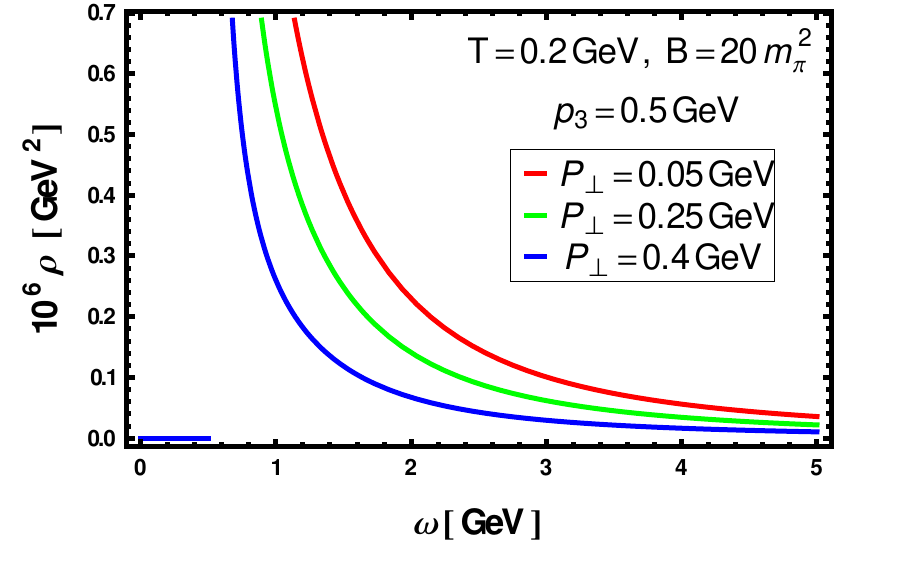}
\end{center}
\caption[Variation of spectral function with photon energy for different values of $P_\perp$]{Variation of the spectral function with photon energy $\omega$ for different values of transverse momentum at fixed $B$, $T$ and  $p_3$.}
\label{spec_plot_2}
\end{figure}

In Fig.~\ref{spec_plot_2} the variation of the spectral function with photon energy 
$\omega$  is shown for three different values of 
the transverse momentum $P_\perp$. The spectral function is found to get 
exponentially suppressed with 
the gradually increasing value of $P_\perp$.

We also consider a special case where the external three momentum ($\vec{p}$) of photon is 
taken to be zero and the simplified expression for the spectral function comes out to be,
\bea
\rho(\omega)\Big\vert_{sfa}\!\!\!\!&=&\!\!\!\!\frac{1}{\pi}\textrm{Im}~ C^\mu_\mu (\omega,\vec{p}=0)\Big\vert_{sfa}\nn\\ \!\!\!\!&=&\!\!\!\! N_c\sum_{f}\frac{q_fBm_f^2}{\pi^2 \omega^2}
\Theta\left(\omega^2-4m_f^2\right)
\left(1-\frac{4m_f^2}{\omega^2}\right)^{-{1}/{2}} \Bigl[1-2 n_F\left(\frac{\omega}{2}\right)\Bigr].
\label{spec_sfa}
\eea

\begin{figure}
\begin{center}
\includegraphics[scale=0.8]{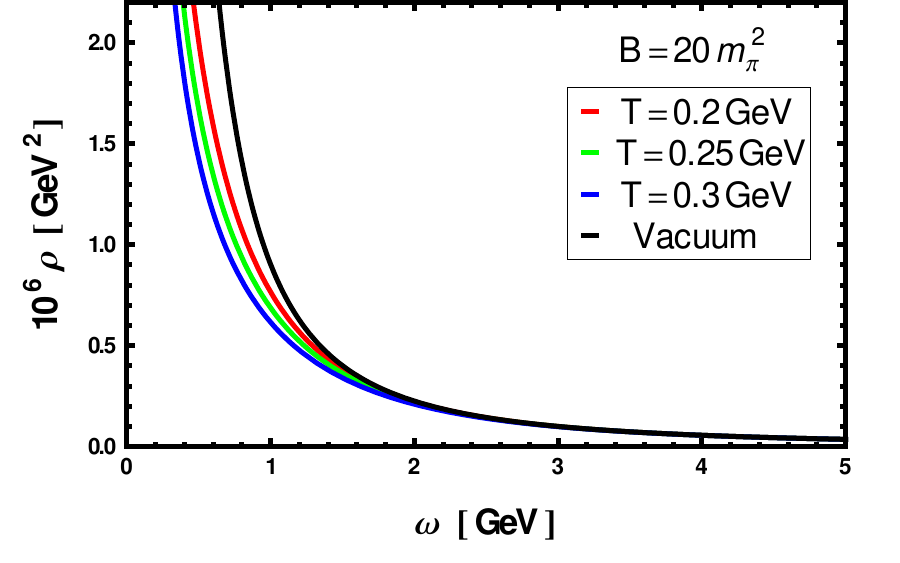}
\includegraphics[scale=0.8]{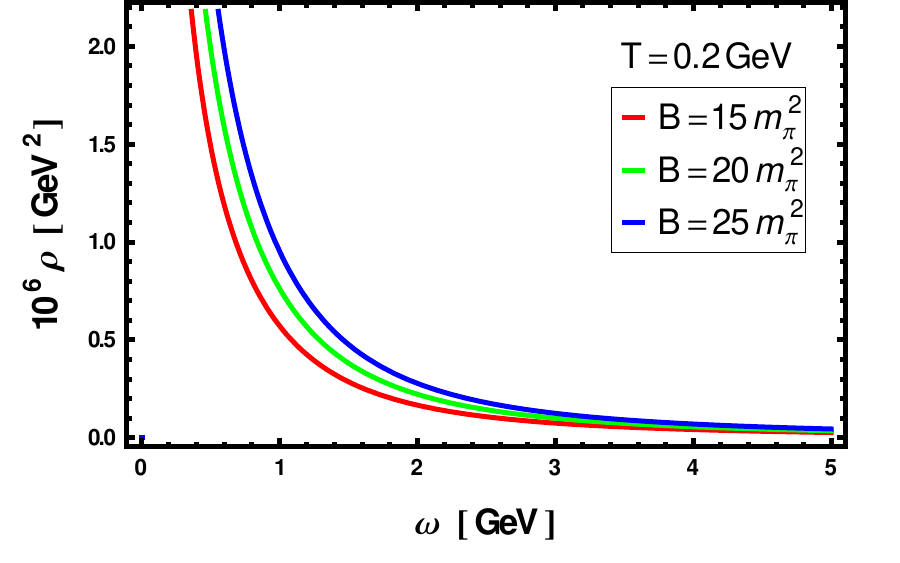}
\end{center}
\caption{Same as Fig.~\ref{spec_plot_1} but for zero external three momentum ($p$) of photon.}
\label{spec_plot_3}
\end{figure}

In Fig.~\ref{spec_plot_3} similar things are plotted as in Fig.~\ref{spec_plot_1} but for a 
simplified case of zero external three momentum of photon. As can be seen from 
(\ref{spec_sfa}), here the value of the threshold is shifted to photon energy as 
$\omega=2m_f$ and the shape of the plots are slightly modified. In the following 
subsection~\ref{sfa_lll_dil} as a spectral property we discuss the leading order thermal dilepton 
rate for a magnetized medium.


\subsection{Dilepton production rate}
\label{sfa_lll_dil}

In \ref{dil_mag}, the three types of scenarios for the DPR in a magnetized medium within LLL approximation is already discussed in details. In this subsection we present our results for the first two scenarios respectively. We also note that since the third scenario is a rare possibility, we skip that discussion here. Though it can easily be obtained by using the following results.

\begin{enumerate}
 \item {\bf Quarks move in a strong magnetized medium but not the final lepton pairs}

In this section the dilepton rate for 
massless ($m_l=0$) leptons  can be written following Eq.~(\ref{dlpm_case1}) as~\cite{Bandyopadhyay:2016fyd}
\bea
\frac{dN}{d^4Xd^4P}\Bigg\vert_1 &=& 
 \frac{5\alpha_{\textrm{em}}^2}{27\pi^2} 
\frac{n_B(p_0)}{P^2} \left [\rho\right]_{sfa} \nn \\
&&\!\!\!\!\!\!\!\!\!\!\!\!\!\!  =\frac{5N_c \alpha_{\textrm{em}}^2}{27\pi^4} n_B(\omega)
\sum_f \frac{q_fB m_f^2}{P^2P_\shortparallel^2}
~e^{-{P_\perp^2}/{2q_fB}}~\Theta
\left(P_\shortparallel^2-4m_f^2\right)\left(1-\frac{4m_f^2}{P_\shortparallel^2}
\right)^{-{1}/{2}} \nn \\
&& \!\!\!\!\!\!\!\!\!\!\times \Bigl[1-n_F(p_+^s)-n_F(p_-^s)\Bigr], 
\label{d9}
\eea
where the electromagnetic spectral function  $[\rho]_{sfa}$ 
in hot magnetized medium has been used from (\ref{spec_sfa_general}). The invariant mass 
of the lepton pair is $M^2 \equiv \omega^2-p_3^2-P_\perp^2=P_\shortparallel^2-P_\perp^2$.

In Fig.~\ref{dr_comparison} a ratio of the dilepton rate in the present scenario with 
strong field approximation to that of the PLO (Born) dilepton rate 
is displayed as a function of the invariant mass. The left panel is for finite external 
photon momentum  
whereas the right panel is for zero external photon momenta. 
The features of the spectral function as 
discussed above are reflected in these dilepton rates. The LLL dynamics in 
strong field approximation enhances the dilepton rate as compared to 
the Born rate for a very low invariant 
mass ($\le 100 $ MeV), whereas at high mass it falls off very 
fast similar to that of the spectral function since there is no higher LL in
strong field approximation as noted in 
point (iv) of \ref{sfa_lll_pptm}.

\begin{figure}
\begin{center}
\includegraphics[scale=0.8]{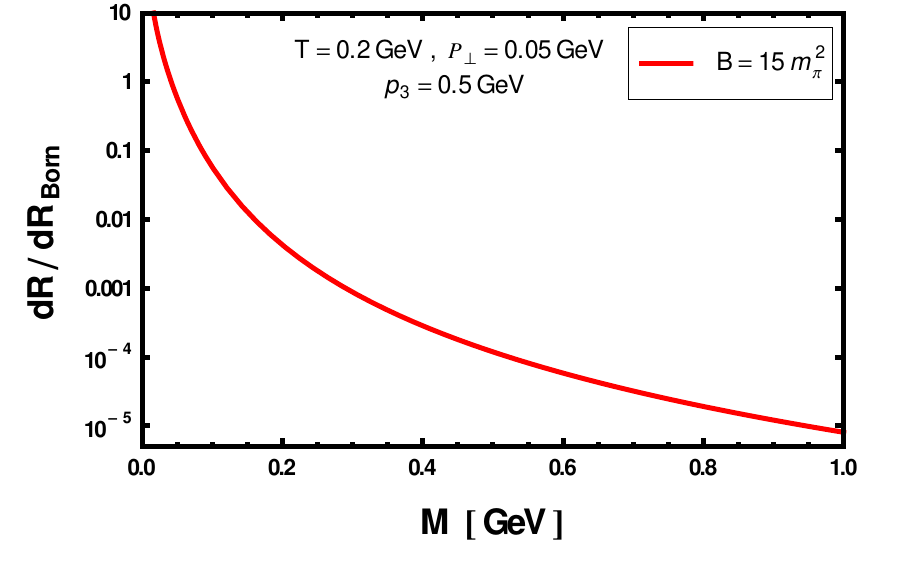}
\includegraphics[scale=0.8]{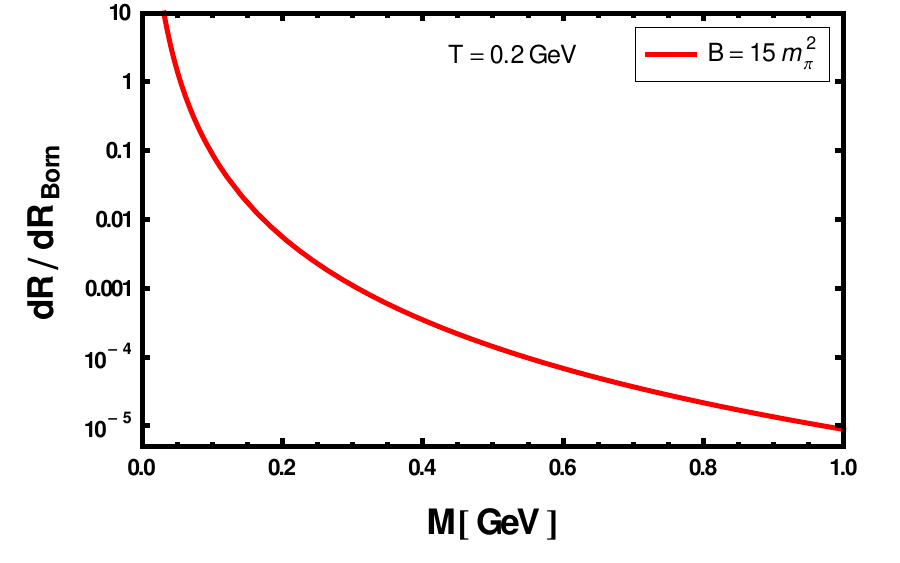}
\end{center}
\caption[Plot of ratio of the Dilepton rate in the strong magnetic field]{Plot of ratio of the Dilepton rate in the strong magnetic field 
approximation to the Born rate (PLO) for both finite (left panel)  
and zero (right panel) external three momentum of photon.}
\label{dr_comparison}
\end{figure}

\item {\bf Both quark and lepton move in magnetized medium in strong field approximation}

In the most general scenario the general expression of DPR is given by Eq.~(\ref{dlpm_case2}). For two-flavor case ($N_f=2$) it becomes~\cite{Bandyopadhyay:2016fyd}
\bea
\frac{dN^m}{d^4Xd^4P}\Bigg\vert_2 &=& 
\frac{10\alpha_{em}^2}{9\pi^2}\frac{n_B(p_0)}{P_\shortparallel^2P^4}
~|eB| m_l^2\left(1-\frac 
{4m_l^2}{P_\shortparallel^2}\right)^{-{1}/{2}}
\left[\rho\right]_{sfa} \nn \\
&&\!\!\!\!\!\!\!\!\!\!\!\!\!\!\!\!\!\!\!\! =\frac{10N_c\alpha_{\textrm{em}}^2}{9\pi^4} n_B(\omega) \sum_f  
\frac{|eB|\, q_fB m_f^2 m_l^2}{P_\shortparallel^4P^4} 
\Theta\left(P_\shortparallel^2-4m_l^2\right)
\left(1-\frac 
{4m_l^2}{P_\shortparallel^2}\right)^{-{1}/{2}} \times \nn\\
&&\!\!\!\!\!\!\!\!\!\!\!\!\!\!\!\!\!\!\!\Theta\left(P_\shortparallel^2-4m_f^2\right)\left(1-\frac{4m_f^2}{P_\shortparallel^2}
\right)^{-{ 1}/{2}} e^{-{P_\perp^2}/{2|q_fB|}} \Bigl[1-n_F(p_+^s)-n_F(p_-^s)\Bigr]\!\!.
\label{d15}
\eea

We now note that the dilepton production rate in (\ref{d15}) is of ${\cal O}[|eB|^2]$ in presence of magnetic field $B$ due to the effective dimensional reduction~\footnote{A factor $|eB|$ comes from leptonic part whereas $\sum_fq_fB \propto |eB|$ from electromagnetic spectral function involving quarks.}. This dimensional reduction also renders a factor $1/\sqrt{1-{4m_l^2}/{P_\shortparallel^2}}$ in the leptonic part $L^m_{\mu\nu}$ that provides another threshold $P_\shortparallel^2 \ge 4m_l^2$ in addition to that coming from electromagnetic part $ P_\shortparallel^2 \ge 4m_f^2$.  In general the mass of fermions in a magnetized hot medium will be affected by both temperature and magnetic field. The thermal effects~\cite{Kapusta:1989tk,Bellac:2011kqa} can be considered through thermal QCD and QED, respectively, for quark ($\sim g^2T^2$; $g$ is the QCD coupling)  and lepton ($\sim e^2T^2$) whereas the magnetic effect comes through the quantized LL ($2nq_fB$). However, in LLL ($n=0$), the magnetic effect to the mass correction vanishes in strong field approximation. Also in strong field approximation ($q_fB \gg T$), there could be dynamical mass generation through chiral condensates~\cite{Gusynin:1995nb} of quark and antiquark leading to magnetic field induced chiral symmetry breaking, which could play a dominant role. Nevertheless, the threshold will, finally, be determined by the effective mass $\tilde m =$ max(${ m}_l,{ m}_f$) as $\Theta\left(P_\shortparallel^2-4{\tilde m}^2\right)$ and the dilepton rate in LLL reads as
\bea
\frac{dN^m}{d^4Xd^4P}\Bigg\vert_2 \!\!\!\!&=&\!\!\!\!
\frac{10N_c\alpha_{em}^2}{9\pi^4}  \sum_f  
\frac{|eB|\, q_fB m_f^2 m_l^2}{P_\shortparallel^4P^4} 
\Theta\left(P_\shortparallel^2-4{\tilde m}^2\right)
\left(1-\frac 
{4m_l^2}{P_\shortparallel^2}\right)^{-\frac{1}{2}} 
 \times\nn \\
&& \left(1-\frac{4m_f^2}{P_\shortparallel^2}
\right)^{-\frac{ 1}{2}} e^{-\frac{P_\perp^2}{2q_fB}}\, n_B(\omega) \Bigl[1-n_F(p_+^s)-n_F(p_-^s)\Bigr],
\label{d16}
\eea
At this point we note that a comparison with the experimental results  or the results (dilepton spectra)  obtained by  Tuchin~\cite{Tuchin:2012mf} needs a space-time evolution of the dilepton rate in a hot magnetized medium produced in heavy-ion collision which is beyond the scope of this thesis.
\end{enumerate}


\section{Weak field approximation}
\label{wfa}

As the magnetic field created in noncentral HIC is a fast decreasing function of time so, it is expected that by the time the quarks and gluons thermalize in a QGP medium, the magnetic field strength becomes sufficiently weak. For this reason the temperature at that particular instant becomes the largest energy scale of the system. Exploiting this circumstance within the domain of weak magnetic field one uses suitable approximation for analytical simplicity. In weak field approximation the external magnetic field being weak is considered as a perturbation to the unmagnetized thermal tree level result. Eventually the CFs in the weak field approximation are represented as an expansion in powers of the magnetic field. In the regime of weak magnetic field, all the LLs are very closely placed (see Fig.~\ref{landau_levels}). This is because in the weak field approximation ($q_fB~\ll~T^2$), the energy spacing between consecutive LLs, $2(n+1)q_fB - 2nq_fB = 2q_fB$, gradually reduces. So, they can be considered as a continuous system instead of a discreet sum. This is why, weak field approximation is at the other end of the spectrum with respect to the strong field approximation.    

In the present section in contrast to the oft-used imaginary time formulation of thermal field theory~\cite{Matsubara:1955ws,Pisarski:1987wc,
Kapusta:1989tk,Bellac:2011kqa}, we shall use the real time formulation ~\cite{Niemi:1983nf,Kobes:1985kc, Mallik:2016}. The advantage of using RTF is that we do not have frequency sums for the propagators, but at the cost of dealing with $2\times 2$ matrices for them  in the intermediate stage of calculation as discussed briefly in section \ref{rtf}. The matrices admit spectral representations, just like the vacuum propagators, which we shall use in calculating the thermal correlation function. The procedure is briefly discussed in the next subsection.   


\subsection{Correlation function in real time formalism}
\label{wfa_cf}

In this subsection we briefly review how the current-current CF may be obtained in the real time thermal field theory \cite{Mallik:2016}. We start with the time contour of Fig. \ref{rtfc} and define the time-ordered two-point function $M_{\mn}(X,X')$ as
\bea
M_{\mn}(X,X') = \Theta_c(\tau-\tau') i \la J_\mu(X)J_\nu(X')\ra + 
\Theta_c(\tau'-\tau) i \la J_\nu(X')J_\mu(X)\ra.
\label{contourform}
\eea
where $X=(\tau,\vec{x}),~X'=(\tau',\vec{x}')$ with  the $`$times' $\tau$ and $\tau'$ on the contour shown in Fig. \ref{rtfc}. The subscript $c$ on the $\Theta$-functions refers to contour ordering. Beginning with the spatial Fourier transform, one can show that the vertical segments of the time contour does not contribute. Then the two-point function may be put in the form of a $2\times 2$ matrix~\cite{Mallik:2016}, which can be diagonalized with essentially one diagonal element
\bea
\ov{M}_{\mn}(Q) = \int\limits_{-\infty}^{+\infty} \frac{dq_0'}{2\pi}
\frac{\rho_{\mn}(q_0',\vec{q})}{q_0'-q_0-i\eta\epsilon(q_0)}
\label{diag_M}
\eea
where $\rho_{\mn} (Q)$ is the spectral function, expressed as 
\bea
\rho_{\mn}(Q) = 2~\rm{Im}\ov{M}_{\mn}(Q),
\label{rhotom}
\eea
and the dilepton rate is subsequently given by 
\bea
\frac{d^4N}{d^4Xd^4Q} = \frac{\alpha_{\textrm{em}}^2}{3\pi^3Q^2}\frac{W}{e^{\beta q_0}-1},
~~W=-g^{\mn} \textrm{Im} \ov{M}_{\mn}.
\label{dpr_rtf}
\eea
Using the matrix which diagonalizes the $2\times 2$ correlation matrix, we can relate the imaginary part of any one component, say the $11$, of the correlation function to that of its diagonal element,
\bea
\textrm{Im} \ov{M}_{\mn} = \epsilon(q_0) \tanh (\beta q_0)\textrm{Im}(M_{\mn})_{11}.
\label{parl_matrix_comp}
\eea

So far we have utilized general properties of two-point functions to relate the problem to 
$(M_{\mn})_{11}$. Taking $\tau$ and $\tau'$ on the real axis, the contour form 
(\ref{contourform}) gives it as 
\bea
M_{\mn}(X,X')_{11} = i\la TJ_\mu(\vec{x},t)J_\nu(\vec{x}',t')\ra,
\eea
where $T$ as usual time orders the operators. It is this quantity which we have to calculate. To leading order in strong interactions, it involves only the thermal quark propagator. The magnetic field enters into the problem through this propagator, which we find in the next section.


\subsection{Fermion propagator within weak magnetic field}
\label{wfa_prop}

Now, in the weak field approximation, generally one expands the exact propagator Eq.~(\ref{schwinger_propertime}), in powers of the magnetic field. Expanding the exponential and tangent functions, we immediately get $S(P)$ as a series in powers of $q_fB$. To order $(q_fB)^2$ it is

\bea
S(P)&=&\frac{-(\slashed{P}+m_f)}{P^2-m_f^2+i\eta}+q_fB\frac{i(\slashed{P}_\pl+m_f)\gm^1\gm^2}{(P^2-m_f^2)^2}
\nn\\&&-(q_fB)^2\left[\frac{2\slashed{P}_\pp}{(P^2-m_f^2)^3}-\frac{2P_\pp^2(\slashed{P}+m_f)}{(P^2-m_f^2)^4}\right].
\label{wfprop}
\eea
To put the propagator (\ref{wfprop}) in the form of a spectral representation, we 
introduce a variable mass $m_1$ to replace $1/(P^2-m_f^2)$ by $1/(P^2-m_1^2)$, 
keeping the physical mass $m_f$ unaltered at other places. The higher powers of the 
scalar propagator can then be expressed as derivatives of the propagator 
with respect to $m_1^2$. We thus get
\bea
S(P)=-F(P,m_f,m_1)\frac{1}{P^2-m_1^2}\Bigg|_{m_1=m_f},
\label{wfp_spec}
\eea
where 
\bea
F=(\slashed{P}+m_f)+a~i~\left(\slashed{P}_\pl + m_f\right)\gm^1\gm^2 + b\slashed{P}_\pp +cP_\pp^2 (\slashed{P}+m_f)
\eea
with coefficients $a, b$ and $c$ carrying the derivative operators,
\bea
a=-q_fB\frac{\partial}{\partial m_1^2};~~b=(q_fB)^2\frac{\partial^2}{\partial (m_1^2)^2};
~~c=-\frac{1}{3}(q_fB)^2\frac{\partial^3}{\partial (m_1^2)^3}.
\eea
From Eq.~(\ref{wfp_spec}) the spectral function for $S(P)$ will be recognized as 
\bea
\sg(P)=F(P,m_f,m_1)\rho(P,m_1)\big|_{m_1=m_f}
\label{spec_sg}
\eea
with $\rho$ being the spectral function for the scalar propagator of mass $m_1$ 
\bea
\rho(P,m_1)=2\pi\epsilon(p_0)\delta(P^2-m_1^2).
\eea
The desired spectral representation for the spinor propagator in vacuum 
(in presence of magnetic field) can be written in the form
\bea
S(P)=\int_{-\infty}^{+\infty}\frac{dp_0'}{2\pi}~\frac{\sg(p_0', \vec{p})}{p_0'-p_0-i\eta\eps(p_0)},
\eea
as can be readily verified by doing the $p_0'$ integral.


\subsection{Spectral properties and DPR}
\label{wfa_dpr}

\begin{center}
\begin{figure}[tbh]
\begin{center}
\includegraphics[scale=0.5]{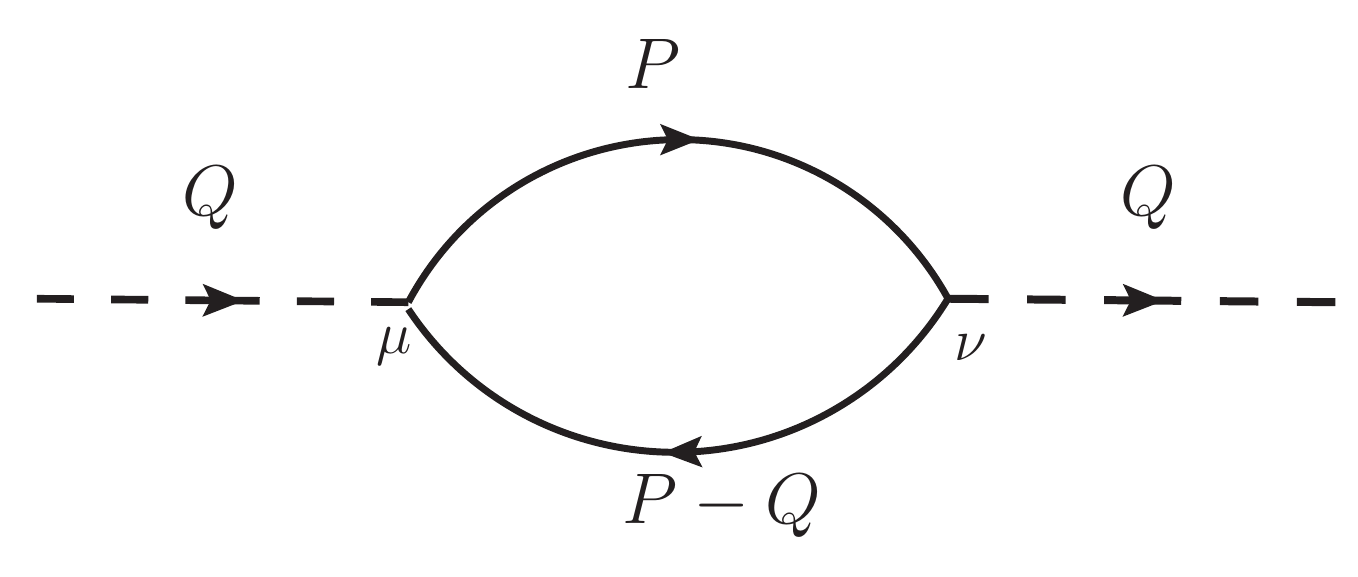} 
\caption[Current correlation function to one-loop]{Current correlation function to one-loop. The dashed and solid lines 
represent currents and quarks.}
\label{cc}
\end{center}
\end{figure}
\end{center}

Like the thermal correlation function of currents, the thermal quark
propagator can also be analyzed in the same way. In particular, its
$2\times 2$ matrix form can be diagonalized, again with essentially a
single diagonal element, which turns out to be the vacuum propagator (in
magnetic field) derived above. But in our calculation below, we need the
$11-$ element of the original matrix, which is conveniently written as
\cite{Mallik:2016},
\bea
S_{11}(P)=\int\limits_{-\infty}^{+\infty}
\frac{dp_0'}{2\pi}\sg(p_0', \vec{p})
\left\{\frac{1-\wt{f}(p_0')}{p_0'-p_0-i\eta}+\frac{\wt{f}(p_0')}{p_0'-p_0+i\eta}\right\},
~~\wt{f}(p_0) = \frac{1}{e^{\beta p_0}+1}
\label{prop_wfa_11}
\eea
where the spectral function $\sg$ is given by Eq.~(\ref{spec_sg}). 

For a two-flavor case ($N_f=2$), the graph of Fig.~\ref{cc} gives two terms involving $u$ and $d$ quark propagators.
Assuming these to be equal (which is true only for $q_fB \propto |eB|=0$)\footnote{For $q_fB 
\neq 0$, we include the necessary correction at the end of this section.}, we 
combine them to give
\bea
M_{\mn}(Q)_{11} = \frac{5i}{3}\int
\frac{d^4P}{(2\pi)^4}\textrm{Tr}\left[S_{11}(P)\gm_\nu S_{11}(P-Q) \gm_\mu\right]
\eea
where the prefactor includes a factor of 3 for color of the quarks.
Inserting the propagator Eq.~(\ref{prop_wfa_11}) in it, we want to work out the $p_0$
integral. For this purpose we write it as 
\bea
M_{\mn}(Q)_{11}=\int\frac{d^3p}{(2\pi)^3}\int \frac{dp_0'}{2\pi}\rho(p_0',\vec{p})\int
\frac{dp_0''}{2\pi}\rho(p_0'',\vec{p}-\vec{q})K_{\mn}(Q)
\label{p0integral}
\eea
where
\bea
K_{\mn}(Q) &=&
i\int_{-\infty}^{+\infty}\frac{dp_0}{2\pi}N_{\mn}(Q)\left(\frac{1-\wt{f}'}{p_0'-p_0-i\eta}
+\frac{\wt{f}'}{p_0'-p_0+i\eta}\right)\times\nn\\
&&
\left(\frac{1-\wt{f}''}{p_0''-(p_0-q_0)-i\eta}+\frac{\wt{f}''}{p_0''-(p_0-q_0)+i\eta}\right)
\eea
with $\wt{f}'=\wt{f}(p_0'), \wt{f}''=\wt{f}(p_0'')$ and
\bea
N_{\mn}(Q)=\frac{5}{3} \textrm{tr}\left\{\stackrel{\lrw}{F}(P,m_f,m_1) \gm_\nu
\stackrel{\lrw}{F}(P-Q,m_f,m_2)\gm_\mu\right\}.
\eea 
Here the masses $m_1$ and $m_2$ are variables on which the mass derivatives act in the 
two propagators. The left arrow on $F$ indicates the derivatives in it to be put 
farthest to the left (outside the integrals). As we are interested in the imaginary 
part of $K_{\mn}$, we can put $p_0=p_0',~ p_0-q_0=p_0''$ and bring $N_{\mn}$ outside 
the $p_0$ integral. Then it is simple to evaluate $K_{\mn}$, from which we get its 
imaginary part. Extracting a factor $coth(\beta q_0)$, it becomes linear in 
$\wt{f}'$ and $\wt{f}''$, 
\bea
\textrm{Im} K_{\mn}(Q) = N_{\mn}(p_0',p_0'')\pi (\wt{f}''-\wt{f}')\coth(\beta q_0)\de
\left(p_0''-p_0'+q_0\right)
\eea
(The hyperbolic function will cancel out in (II.14)). Next, the $p_0'$ and 
$p_0''$ integrals in Eq.~(\ref{p0integral}) can be removed, using the delta 
functions present in the spectral functions, namely $\de (p_0'\pm \om_1)$
and $\de (p_0''\pm \om_2)$ with $\om_1=\sqrt{p^2+m_1^2}$ and
$\om_2=\sqrt{(\vec{p}-\vec{q})^2+m_2^2}$. We need only the imaginary part in 
the physical region, $q_0 > (\omega_1 +\omega_2)$. From Eq.~(\ref{dpr_rtf}), (\ref{parl_matrix_comp}) and
(\ref{p0integral}), we then get
\bea
W=\pi\int\frac{d^3p}{(2\pi)^3}\frac{N_\mu^\mu(\om_1,-\om_2)}{4\om_1\om_2}
\left\{1-n_F(\omega_1)-n_F(\omega_2)\right\} \delta(q_0-\omega_1-\omega_2)
\eea
where we convert $\wt{f}$'s to distribution functions, $n_F(\om)=1/(e^{\beta\om}+1)$.

Working out the trace over $\gm$ matrices in $N_\mu^\mu$ we get
\bea
N_\mu^\mu \!\!&=&\!\! -\frac{40}{3}\bigl[(1-a_1a_2)~P\!\cdot\!(P-Q)\!-\!(b_1+b_2+a_1a_2)
[P\!\cdot\!(P-Q)]_\pp \nn\\
&& + P\!\cdot\!(P-Q)\left\{c_1P_\pp^2+c_2(P-Q)_\pp^2\right\}\bigr].\label{nmumu}
\eea
Let us now consider collision events in which the transverse components of momenta are 
small compared to the longitudinal ones, and we can omit the last two terms and 
calculate the dilepton rate analytically. Neglecting quark mass, we thus get
\bea
W=\frac{20\pi}{3} Q^2(1-a_1a_2)J
\label{wtoj}
\eea
where 
\bea
J=\int\frac{d^3p}{(2\pi)^34\om_1\om_2}\left\{1-n_F(\om_1)-n_F(\om_2)\right\}\delta(q_0-\om_1-\om_2).
\label{j}
\eea
After working out this integral analytically, we shall apply the mass derivatives 
contained in $a_1$ and $a_2$.
If $\th$ is the angle between $\vec{q}$ and $\vec{p}$, we can carry out the $\th$ integral 
by the delta function in Eq.~(\ref{j}). However a constraint remains to ensure that 
$\cos\th$ remains in the physical region, as we integrate over the angle. We get
\bea
J=\frac{1}{16\pi^2 q}\int d\om_1 \Theta(1-| \cos\th|)
\left\{1-n_F(\om_1)-n_F(\om_2)\right\}.
\eea
The $\Theta$-function constraint gives a quadratic expression in $\omega_1$,
\bea
(\omega_1-\omega_+)(\omega_1-\omega_-)\le 0
\eea
where
\bea
\omega_{\pm}=\frac{q_0 R \pm q\sqrt{R^2-4Q^2m_1^2}}{2Q^2},~~ R=Q^2+m_1^2-m_2^2,
\eea
With the corresponding limits on $\omega_1$, we get
\bea
J &=& \frac{1}{16\pi^2 q}\int\limits_{\om_-}^{\om_+} d\om_1 
\left(1-\frac{1}{e^{\beta\om_1}+1}-\frac{1}{e^{\beta(q_0-\om_1)}+1}\right)\nn\\
&=& \frac{1}{16\pi^2 q\beta} \left[\ln\left(\frac{\cosh(\beta\om_+/2)}
{\cosh(\beta\om_-/2)}\right)-\ln\left(\frac{\cosh(\beta(q_0-\om_+)/2)}
{\cosh(\beta(q_0-\om_-)/2)}\right)\right].
\eea

We now recall that the $u$ and $d$ quark charges were not included correctly in the propagators. The resulting correction will effect only $e^2$, contained in $a_1, a_2$ in the expression for $W$. We can readily find that we need to replace $q_f^2$ by 17/45 $e^2$ to restore the actual charges of the quarks in their propagators.
Carrying out the mass derivatives in Eq.~(\ref{wtoj}) and going to the limit of zero 
quark masses, we finally get 
\bea
W=\frac{5Q^2}{12\pi q\beta}\left[2\ln\left(\frac{\cosh \alpha_+}
{\cosh \alpha_-}\right)-\frac{17}{45}(|eB|)^2\mathcal{M}\right],
\eea
where $\mathcal{M}$ gives the effect of magnetic field to the leading order result,
\bea
\mathcal{M}=\frac{\beta^2}{8Q^2}\left(\sech^2\alpha_+-\sech^2\alpha_-\right)+\frac{\beta q}{q^4}\left(\tanh \alpha_+ +\tanh \alpha_-\right).
\label{mag_effect_final}
\eea
Here we use the abbreviation $\alpha_\pm=\beta(q_0\pm q)/4$. Note that 
$W$ is finite as $q \rightarrow 0$.

\begin{center}
\begin{figure}[tbh]
\begin{center}
\includegraphics[scale=0.8]{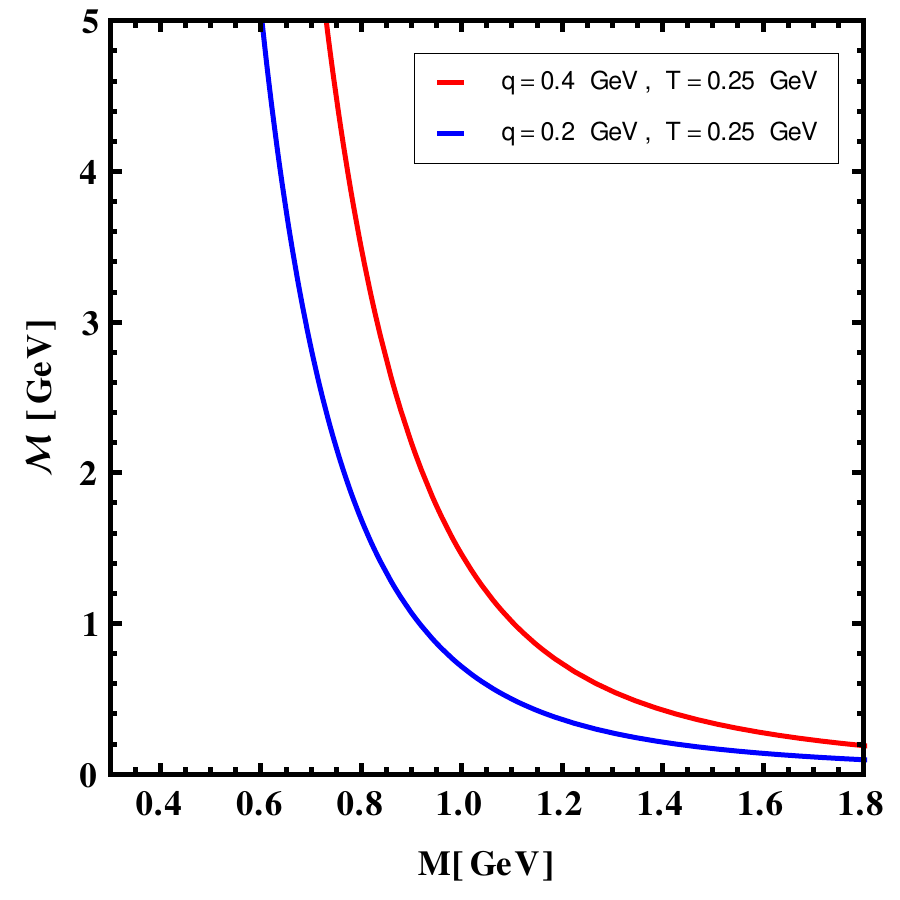}\includegraphics[scale=0.8]{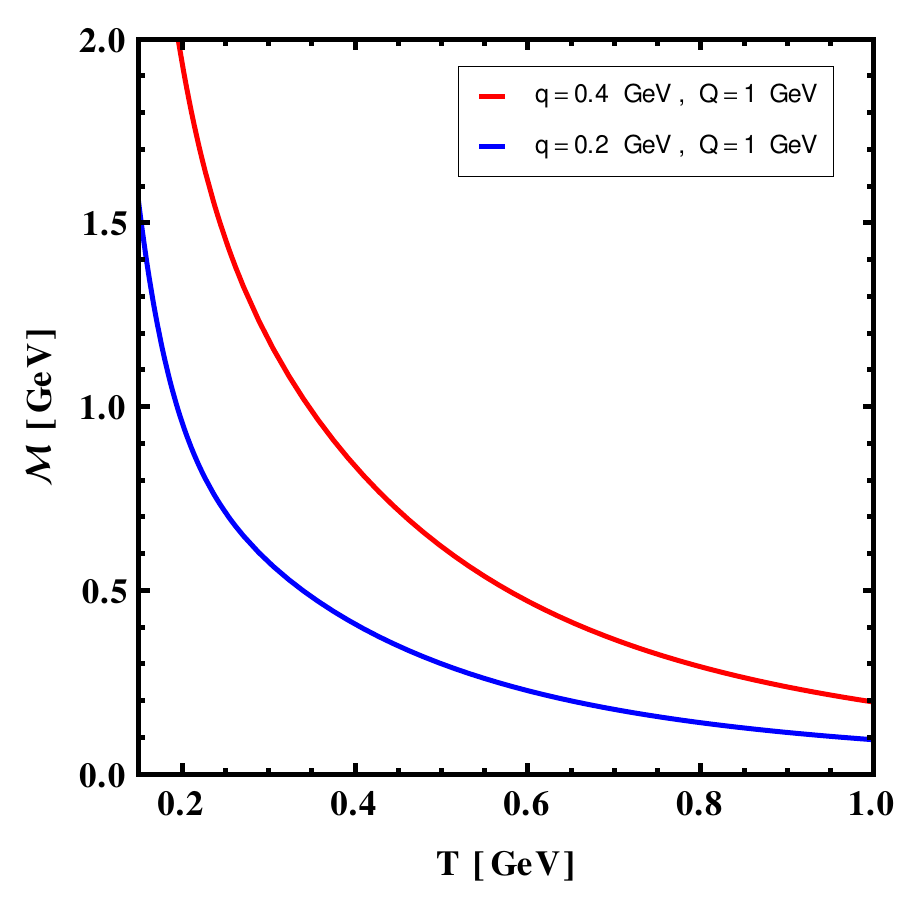}
\caption[Variation of $\mathcal{M}$ as a function of invariant mass and temperature]{Variation of $\mathcal{M}$ as a function of invariant mass
$M$ (left) and temperature $T$ (right).}
\label{mwq}
\end{center}
\end{figure}
\end{center}

\begin{center}
\begin{figure}[tbh]
\begin{center}
\includegraphics[scale=1.0]{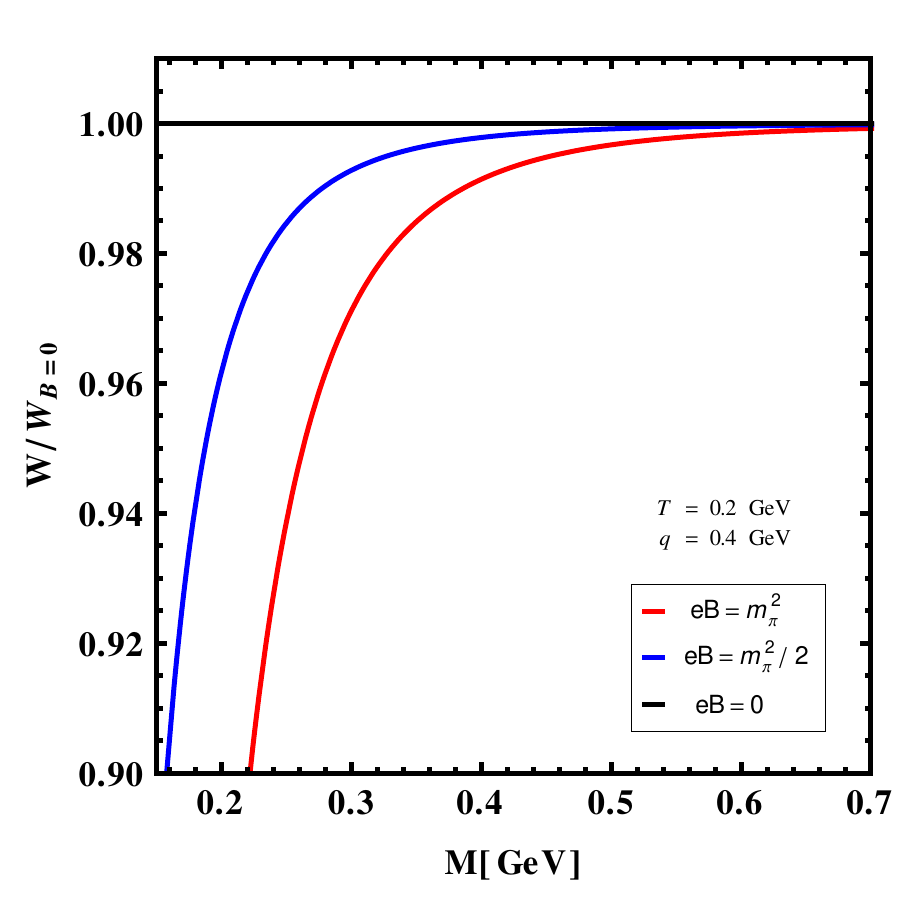}
\caption[Plot of ratio of the Dilepton rate in the weak magnetic field]{Plot of the ratio of the dilepton rate with and without the presence of the magnetic
 field as a function of $M$.}
\label{wr}
\end{center}
\end{figure}
\end{center}

The plots of $\mathcal{M}$, the coefficient of $(|eB|)^2$ in $W$, as functions 
of invariant dilepton mass $M=\sqrt{Q^2}$ and temperature $T$ are 
shown in Fig. \ref{mwq} for typical values of parameters. If the second order 
term in (IV.16) provides any indication of the behavior of the series, the
expansion parameters are $eB/Q^2$ and $eB/T^2$. In Fig. \ref{wr} we plot $W/W_{B=0}$~\cite{Cleymans:1986na}
as a function of $M$ for a few values of $eB$.

Fig. \ref{wr} shows that magnetic field changes the dilepton rate only at lower 
$Q^2$, reflecting the behavior of the first and second term of
 $\mathcal{M}~(Eq.~(\ref{mag_effect_final}))$ as $1/Q^2$ and $1/Q^4$ at low $Q^2$.
 We also note that earlier theoretical calculations without magnetic field 
disagree with experiment at low $Q^2$~\cite{Agakishiev:1998vt,Masera:1995ck,Abreu:1998wxa}. It would 
therefore be tempting to speculate if the effect of magnetic field can bring
the agreement, at least in part. However, to verify this speculation, we
have to improve our calculation in a number of ways. First, we should include
the terms in Eq.~(\ref{nmumu}) that are left out in our calculation. Then we need to
replace the constant magnetic field by one with its magnitude having
(adiabatic) time dependence, as realized in non-central collisions. One can also
include the first order $QCD$ correction to the current correlation function~\cite{Thoma:1997dk}.


\section{Conclusion and Outlook}
\label{mag_conclu}

In this chapter we have evaluated the in-medium electromagnetic spectral function by computing  the imaginary part of the photon polarization tensor, in presence of a magnetic field. We particularly dealt with the limiting cases, where either the magnetic field  is very strong  with  respect to the thermal scale ($q_fB \gg T^2$) of the system or it is weaker than the thermal scale ($q_fB \ll T^2$) where we can expand the quark propagator in terms of $q_fB$. 

In the strong field limit we have exploited the LLL dynamics that decouples the transverse and the longitudinal direction as a consequence of an effective dimensional reduction from (3+1)-dimension to (1+1)-dimension. The electromagnetic spectral function vanishes in the massless limit of quarks  which implies that in ($1+1$) dimension an on-shell massless thermal fermion cannot scatter in the forward direction. Since the LLL dynamics is (1+1) dimensional, the fermions are virtually paired up in LLL providing a strongly correlated system, which could possibly enhance the generation of fermionic mass through the chiral condensate. So, these massive quarks could provide a kinematical threshold to the electromagnetic spectral function at longitudinal photon momentum, $p_\shortparallel^2=4m_f^2$. Below the threshold the photon polarization tensor is purely real and the electromagnetic spectral function does not exist resulting in no pair creation of particle and antiparticle. This implies that the momentum of the external photon supplies energy to virtual fermionic pairs in LLL, which become real via photon decay. At threshold the photon strikes the LLL and the spectral strength diverges due to the dimensional reduction, since a factor of $\left(1-{4m_f^2}/{p_\shortparallel^2}\right)^{-{1}/{2}}$ appears in the spectral function, in strong field approximation. The spectral strength starts with a high value for the photon longitudinal momentum $p_\shortparallel > 2m_f$  due to the dimensional reduction or LLL dynamics and then falls off with increase of $\omega$ as there is nothing beyond the LLL in strong field approximation. 

After exploring the electromagnetic spectral function we analytically obtain the DPR for two scenarios: (i) the quarks and antiquarks are affected by the hot magnetized medium but not the final lepton pairs and (ii) when both  quark and lepton are affected by the magnetized medium. In the former case the dilepton rate is ${\cal O}[|q_fB|]$ and follows the properties of the electromagnetic spectral function along with a kinematical threshold provided by the quark mass. For the latter case the rate is found to be ${\cal O}[|eB|]^2$ with two kinematical thresholds provided by quark ($m_f$) and lepton ($m_l$) mass. Since the dynamics in LLL in strong field approximation is strongly correlated one, the threshold will finally be determined by $\tilde m= {\mbox{max}}(m_f,m_l)$. Overall, the DPR is found to be enhanced with respect to the leading perturbative result at very low invariant mass.  

In the weak field limit we have again exploited the perturbative expansion of the fermionic propagator in powers of the external weak magnetic field. The spectral representation for the spinor propagator is obtained in terms of the spectral function for the scalar propagator by introducing a variable mass $m_n$ and expressing the higher powers of the scalar propagator as derivatives with respect to $m_n^2$. Using the spectral representation of the fermionic propagator the CF, SF and the DPR is computed subsequently by the real time formalism. It is found that the effect of the weak magnetic field in the DPR is confined within the lower invariant mass region, just as in the case of strong magnetic field, though in contrast with the strong field result, this time the DPR decreases at lower $M^2$ with respect to the PLO result. 

The basis of the weak field approximation lies on the time dependence of the produced magnetic field in the non-central HIC. Generally, it is necessary to include the time dependence of the magnetic field in the space-time evolution of dilepton production, which is needed to determine its spectrum. Without going into the details of this evolution, we may estimate roughly the effect of the time dependence as follows. The magnetic field realized in the core in the thermalized QGP phase may be approximated as~\cite{Kharzeev:2007jp,McLerran:2013hla,Tuchin:2013bda}
\bea
eB(t) = \frac{8\alpha}{\gamma} \frac{Z}{t^2+(2R/\gamma)^2}
\label{mag_field_td}
\eea
where $\alpha$ is the fine structure constant ($=1/137$), $Z$ and $R$ are the atomic number and radius of the colliding nuclei and $\gamma$ is the Lorentz contraction factor. However for a different point of view, see~\cite{Tuchin:2012mf,Tuchin:2013bda,Tuchin:2013ie}, where 
the time dependence of magnetic field is shown to be adiabatic due to the high conductivity of the medium. As an example if we consider Au-Au collision at RHIC, for which $Z=79, R=6.5$ fm and $\gamma = 100$, giving $eB(t=0) = m_\pi^2/15$. However, considering Pb-Pb collision at LHC, where $Z=82, R=7.1$ fm and $\gamma = 2800$, we get $eB(t=0) = 1.4 m_\pi^2$. We also want to note that the expression in Eq.~(\ref{mag_field_td}) excludes large magnetic fields generated immediately after collisions. The strong field approximation could possibly be very appropriate for the initial stages of the noncentral HICs where the intensity of the produced magnetic field is expected to be very high. 

Concluding we note that the enhancement found in DPR within the strong field regime or the diminution found within the weak field approximation will contribute to the dilepton spectra at low invariant mass, which is however beyond the scope of the present detectors involved in heavy-ion collisions experiments.

%% file: text/ds.tex
\chapter{Debye Screening in a magnetized hot medium}
\label{th_ds}

In this chapter we will study another important spectral property of magnetized medium, i.e. the Debye screening or the screening of color charges in presence of quasi-free quarks and gluons. We will focus on both the limiting scenarios again and discuss how the behavior of different scales present in the hot magnetized medium affects the behavior of Debye mass (inverse of the Debye screening length). This chapter is based on  parts of \textit{Electromagnetic spectral properties and Debye screening of a strongly magnetized hot medium} by Aritra Bandyopadhyay, Chowdhury Aminul Islam and Munshi G Mustafa, {\bf Phys.Rev. D94 (2016) no.11, 114034} and \textit{The pressure of a weakly magnetized deconfined QCD matter within one-loop Hard-Thermal-Loop perturbation theory} by Aritra Bandyopadhyay, Najmul Haque and Munshi G Mustafa, communicated.


\section{Introduction}

We know that in a hot deconfined medium static electric fields are screened with a characteristic screening length. This screening is known as the Debye screening, the screening length as the Debye length and the inverse of the Debye length as the Debye mass. The Debye mass is related with the temporal part of the self energy tensor as 
\bea
m_D^2=\Pi_{00}(\omega=0,p\rightarrow 0),
\eea
where $\Pi_{00}$ is the temporal part of the self energy tensor $\Pi_{\mn}$ and the limit of zero external energy and vanishing external three momenta $(\omega=0,p\rightarrow 0)$ is known as the static limit. The QED Debye mass in a hot medium depends on the one loop polarization tensor, shown in Fig.~\ref{photon_se} and for electrons in the fermionic loop, it is simply expressed as 
\bea
m_D^2(B=0)\Big\vert_{QED} = \frac{e^2T^2}{3}.
\label{qed_dmass}
\eea
\begin{center}
\begin{figure}[hbt]
 \begin{center}
 \includegraphics[scale=0.6]{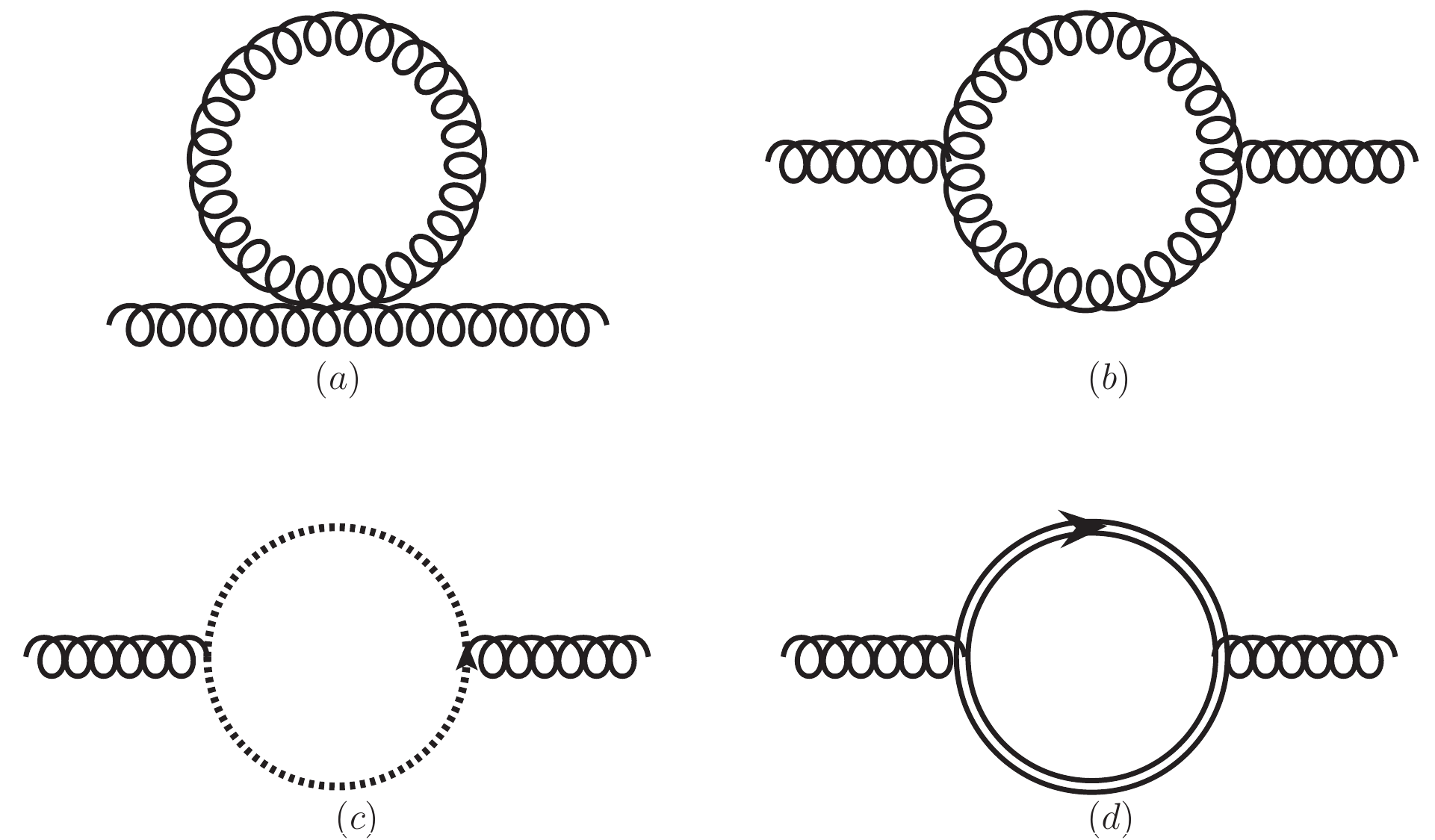} 
 \caption[Gluon self-energy diagrams in weak magnetic field approximation]{Gluon self-energy diagrams in weak magnetic field approximation. (a) 
Four gluon vertex, (b) Three gluon vertex, (c) Ghost loop and (d) Fermion loop. 
The double line in the fermionic loop indicates the modified fermionic 
propagators in presence of weak magnetic field.}
  \label{gse}
 \end{center}
\end{figure}
\end{center}
The QCD Debye mass on the other hand depends on the gluon polarization tensor or the gluon self-energy which can be evaluated from the four contributing diagrams as shown in Fig.~\ref{gse} and eventually in the absence of external magnetic field the Debye mass comes out to be~\cite{Haque:2012my}
\bea
\hat{m}_D^2(B=0)\Big\vert_{QCD} &=& \frac{g^2}{12\pi^2}\left(N_c+\frac{N_f}{2}\right), 
\label{qcd_dmass}
\eea
where $\hat{m}_D = m_D/2\pi T$. Now in presence of magnetic field only the fermionic diagram in Fig.~\ref{gse}(d) will pick up the correction because of the modified fermionic propagators. In the following two sections we will discuss the modification in the QED/QCD Debye mass due to the presence of an anisotropic external magnetic field, both in the regimes of strong field and weak field approximations. 


\section{Debye screening in a strongly magnetized medium}
\label{sfa_ds}

In this section we explore the Debye screening mass in strongly magnetized hot medium which is the extension of our previous work, i.e. evaluation of DPR discussed in chapter \ref{th_mag}. In a strongly magnetized hot medium, the self energy tensor is evaluated in Eq.~(\ref{pol_vacuum}). Following that, for the temporal part we get 
\bea
\Pi_{00}\Big\vert^{sfa}_{\omega=0,p\rightarrow 0} &=& 
N_c\sum_f\frac{q_f^3B}{\pi}\int\limits_0^{\infty}\frac{dk_3}{2\pi}~T\sum_{k_0}\frac{S_{00}
}{(K_\shortparallel^2-m_f^2)^2}\nn\\
&=& N_c\sum_f\frac{q_f^3B}{\pi}\int\limits_0^{\infty}\frac{dk_3}{2\pi}~\left 
[\frac{1}{4\pi i}\oint dk_0\frac{S_{00}\left[1-2n_F(k_0)\right]}{(k_0^2-E_k^2)^2}\right ],
\label{pi00sfa}
\eea
where, $E_k^2 = k_3^2+m_f^2$ and at the limit of vanishing external three momentum and 
zero external energy $S_{00}$ comes out to be
\bea
S_{00}&=&k_0q_0+k_3q_3+m_f^2\Big\vert_{\omega= 0,p\rightarrow0}\nn\\
&=&k_0^2+k_3^2+m_f^2,\nn\\
&=& (k_0^2-E_k^2)+2E_k^2. 
\eea
Now, the $k_0$ integration can be divided into two parts as
\bea
I_1 &=& \frac{1}{4\pi i}\oint dk_0\frac{\left[1-2n_F(k_0)\right]}{(k_0^2-E_k^2)}\nn\\
&=& \frac{1-2n_F(E_k)}{2E_k},
\eea
and
\bea
I_2 &=& \frac{1}{4\pi i}\oint 
dk_0\frac{2E_k^2\left[1-2n_F(k_0)\right]}{(k_0^2-E_k^2)^2}\nn\\
&=& 
2E_k^2 \frac{d}{dk_0}\left(\frac{1-2n_F(k_0)}{(k_0+E_k)^2}\right)\Big\vert_{k_0=E_k}\nn\\
&=& -\frac{1-2n_F(E_k)}{2E_k}+\beta n_F(E_k)\left[1-n_F(E_k)\right].\\
\therefore I_1+I_2 &=& \beta n_F(E_k)\left[1-n_F(E_k)\right].
\eea
From Eq.~(\ref{pi00sfa}) the temporal part of the polarization tensor in the limit of vanishing 
external three momentum (the long wavelength limit) and zero external energy comes 
out to be
\bea
\Pi_{00}\Big\vert^{sfa}_{\omega =0, p\rightarrow 0} &=& 
N_c\sum_f\frac{q_f^3B}{\pi T}\int\limits_0^{\infty}\frac{dk_3}{2\pi}~ 
n_F(E_k)\left[1-n_F(E_k)\right].
\label{dsmassive}
\eea
For massive case ($m_f \neq 0$) this expression cannot be reduced further, analytically, 
by performing the $k_3$ integration. We evaluate it numerically to extract the essence of 
Debye screening. On the other hand, for the massless case ($m_f = 0$) a simple analytical 
expression is obtained as
\bea
\Pi_{00}\Big\vert^{sfa}_{\omega,m_f=0;p\rightarrow 0} &=& 
N_c\sum_f\frac{q_f^3B}{\pi T}\int\limits_0^{\infty}\frac{dk_3}{2\pi}~ 
n_F(k_3)\left[1-n_F(k_3)\right],\nn\\
&=& N_c\sum_f\frac{q_f^3B}{\pi T} ~\frac{T}{4\pi}
= N_c\sum_f\frac{q_f^3B}{4\pi^2}.
\label{dsmassless}
\eea 

\begin{figure}
\begin{center}
\includegraphics[scale=0.7]{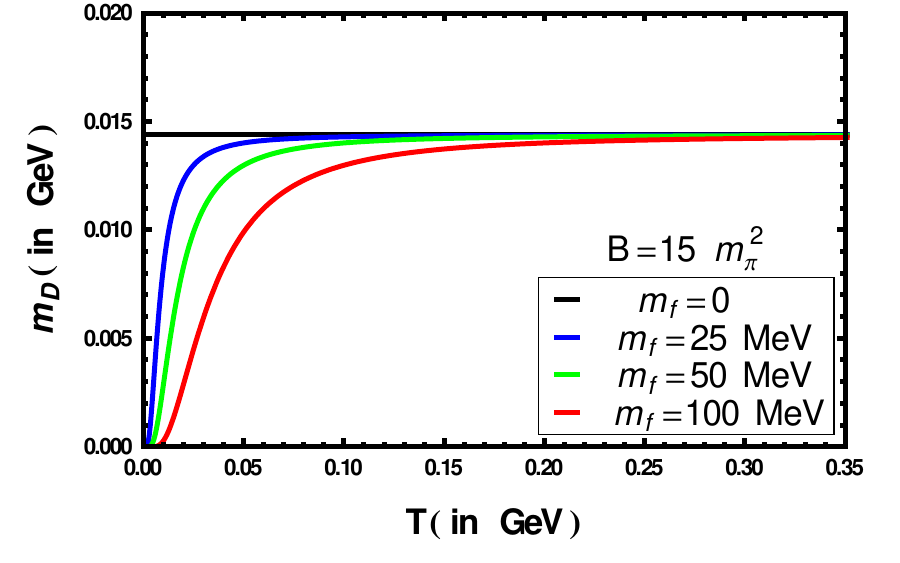}
\hspace{1cm}\includegraphics[scale=0.7]{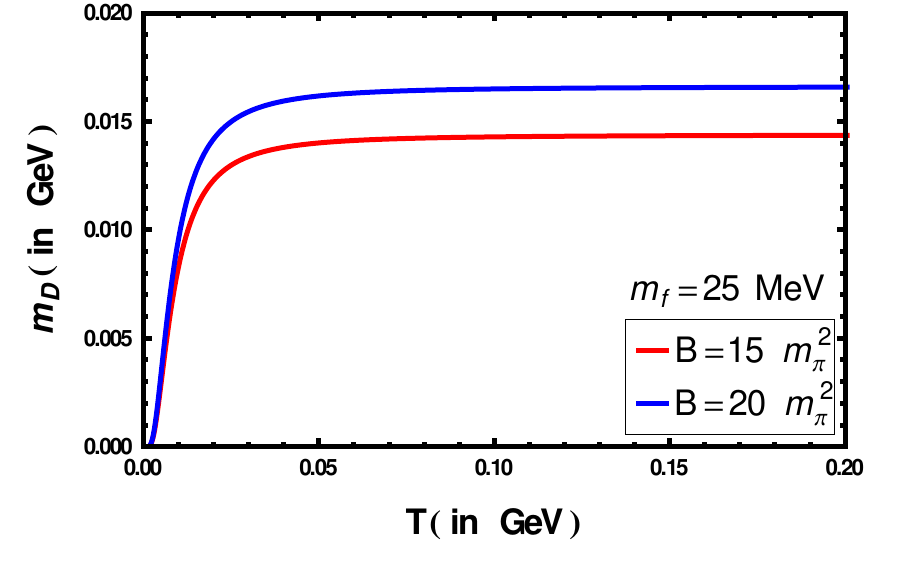}
\end{center}
\caption[Variation of the Debye screening mass with temperature at fixed values of strong magnetic field]{{\it Left panel:} Variation of the Debye screening mass with temperature for 
different quark masses massive at a fixed value of $B$.
{\it Right panel:} Comparison of the temperature Variation of the Debye screening mass for 
two values of $B~(=15m_{\pi}^2$ and $20m_{\pi}^2)$.}
\label{ds_plot_1}
\end{figure}

Before discussing the Debye screening we, first, note that the effective dimensional reduction in presence of strong magnetic field also plays an important role in catalyzing the spontaneous chiral symmetry breaking since the fermion pairing taking place in LLL strengthen the formation of spin-zero fermion-antifermion condensates. This enhances the generation of dynamical fermionic mass through the chiral condensate in strong field limit even at the weakest attractive interaction between fermions~\cite{Shovkovy:2012zn,Gusynin:1995nb} at $T=0$. The pairing dynamics is essentially (1+1) dimensional where the fermion pairs fluctuate in the direction of magnetic field. So, the zero temperature magnetized medium is associated 
with two scales: the dynamical mass $m_f$ and the magnetic field $B$ whereas a hot magnetized medium is associated with three scales: the dynamical mass $m_f$, temperature $T$ and the magnetic field $B$.

In the left panel of Fig.~\ref{ds_plot_1} the temperature variation of the Debye screening mass for quasiquarks in strongly magnetized medium with $B~=15m_{\pi}^2$ and for different quark masses is shown. When the quark mass, $m_f=0$, it is found to have a finite amount of Debye screening. This screening is independent of $T$ because the only scale in the system is  the magnetic field ($q_fB~\gg~T^2$), and the thermal scale gets canceled out exactly as found analytically  in Eq.~(\ref{dsmassless}) in contrast to Ref.~\cite{Alexandre:2000jc} where one needs to explicitly set the $T\rightarrow 0$ limit there. We would like to note that when $T$ drops below the phase transition temperature ($T_c$) the screening mass should, in principle, drop. However, it is found to remain constant in the region $0 \le T\le T_c$, because of the absence of any mass scale in the system.

For massive quarks, the three scales become very distinct and an interesting behavior of the Debye screening mass is observed in presence of strong magnetic field. For a given $m_f$, as the temperature is being lowered gradually than the value of the fermion mass $(T<m_f)$, the quasiquark mass brings the Debye screening down as shown in the left panel of Fig.~\ref{ds_plot_1}. Eventually the screening mass vanishes completely when $T=0$. When $T\sim m_f$, there is a shoulder in the Debye screening and as soon as the temperature becomes higher than the value of $m_f$ the screening becomes independent of other two scales ($m_f^2\le T^2\le q_f B$). So, in presence of strong magnetic field the Debye screening mass changes with temperature as long as $T<m_f$ and then saturates to a value determined by the strength of the magnetic field.
Further as the quasiquark mass is increased the shoulder and the saturation point are pushed towards the higher $T$. The point at which the saturation takes place  depends, particularly, on the strength of two scales, {\it viz.}, $m_f$ and $T$ associated with the hot magnetized system. In other words the dynamical mass generation catalyzes the spontaneous chiral symmetry breaking  indicating magnetic catalysis~\cite{ Shovkovy:2012zn,Gusynin:1995nb,Alexandre:2000jc} and in that case $T_c$ will be enhanced as a reflection of the dimensionally reduced system in presence of strong magnetic field. In the right panel of Fig.~\ref{ds_plot_1} a comparison of the Debye screening mass is being shown for massive quarks for two values of the magnetic field strength ($B=15m_\pi^2$ and $20m_\pi^2$ ) and the screening is enhanced for the latter as it is proportional to $B$.

Now in the next section we shall discuss the effect in Debye screening if the thermal scale is higher than the magnetic scale ($T^2~\gg~q_fB$). For this, one needs to employ a weak field approximation where higher LL contributions will lead to an almost continuous system. 


\section{Debye mass in a weakly magnetized medium}
\label{wfa_ds}

In this section we will study the modification of the Debye screening mass $m_D$ in presence of a weak external magnetic field. We noticed in chapter \ref{th_mag}, more specifically in Eq.~(\ref{wfprop}), that one can perturbatively expand the propagator in presence of an arbitrary magnetic field, to obtain the analytically simplified propagator in weak magnetic field. Similar procedure can be applied with other quantities also. The electromagnetic or QED Debye mass in presence of an arbitrary magnetic field was computed in~\cite{Alexandre:2000jc}. Generalizing  this to QCD we obtain the expression for the modified QCD Debye mass for QCD at finite chemical potential and an arbitrary magnetic field as 
\bea
\hat{m}_D^2 &=& \frac{g^2N_c}{12\pi^2}+\sum\limits_f 
\frac{g^2q_fB}{8\pi^4T^2}\int\limits_0^\infty e^{-x}dx\nn\\
&&\hspace{1cm}\times\sum\limits_{l=1}^\infty (-1)^{l+1}\coth\left(\frac{q_fBl^2}{4xT^2}\right)\exp\left(-\frac{m_f^2l^2}{4xT^2}\right)\!.
\label{md_full}
\eea
\begin{center}
\begin{figure}[h]
 \begin{center}
 \includegraphics[scale=0.8]{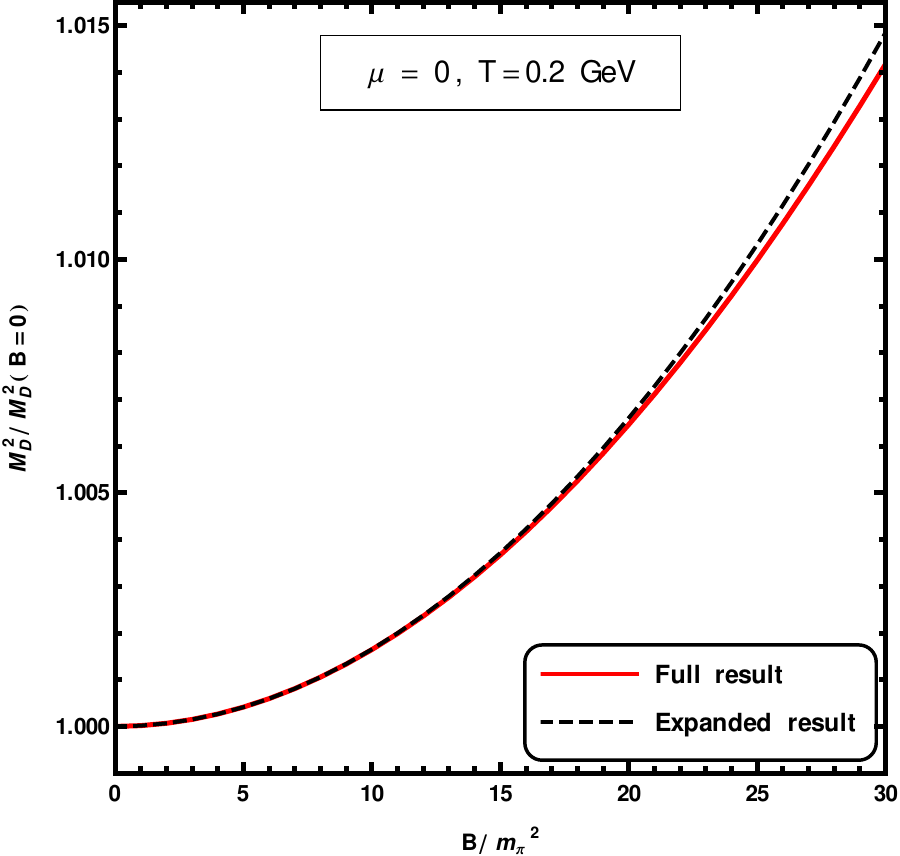} 
 \caption{Comparison of the scaled one-loop Debye masses in Eqs.~(\ref{md_full}) and 
(\ref{md_wfa}) varying with scaled 
magnetic field for $N_f=2$.}
  \label{md_full_vs_expanded}
 \end{center}
\end{figure}
\end{center}

In the weak field approximation ($T^2>q_fB$), the square of Debye mass can be obtained
from Eq.~(\ref{md_full}) by expanding $\coth\left({q_fBl^2}/{4xT^2}\right) $
as
\bea
\hat{m}_D^2 &\simeq& 
\frac{g^2}{12\pi^2}\left(N_c+\frac{N_f}{2}\right) \nn\\
&&+\sum\limits_f \frac{g^2(q_fB)^2}{48\pi^4T^4}
\sum\limits_{l=1}^\infty 
(-1)^{l+1}l^2  K_0\left(\frac{m_fl}{T}\right) + 
\mathcal{O}[(q_fB)^4],
\label{md_wfa}
\eea
where $K_n(z)$ represents the modified Bessel function of the second kind. In 
Eq.~(\ref{md_wfa}) the first term  is  the Debye mass contribution 
in the absence of the external magnetic field whereas the second term is the 
correction due to the presence of the weak external magnetic field. In 
Fig.~\ref{md_full_vs_expanded} the full expression in Eq.~(\ref{md_full}) and the weak 
field expression in Eq.~(\ref{md_wfa}) are displayed as a scaled magnetic field. In the 
strong field limit ($B/m_\pi^2\ge 20$)  the weak field result deviates slightly from that 
of the full result. However, there is no difference between the two in 
limit $B/m_\pi^2 < 20$. Hence it is a good approximation to work with 
Eq.~(\ref{md_wfa}) in the weak field limit.

\begin{center}
\begin{figure}[h]
 \begin{center}
 \includegraphics[scale=.8]{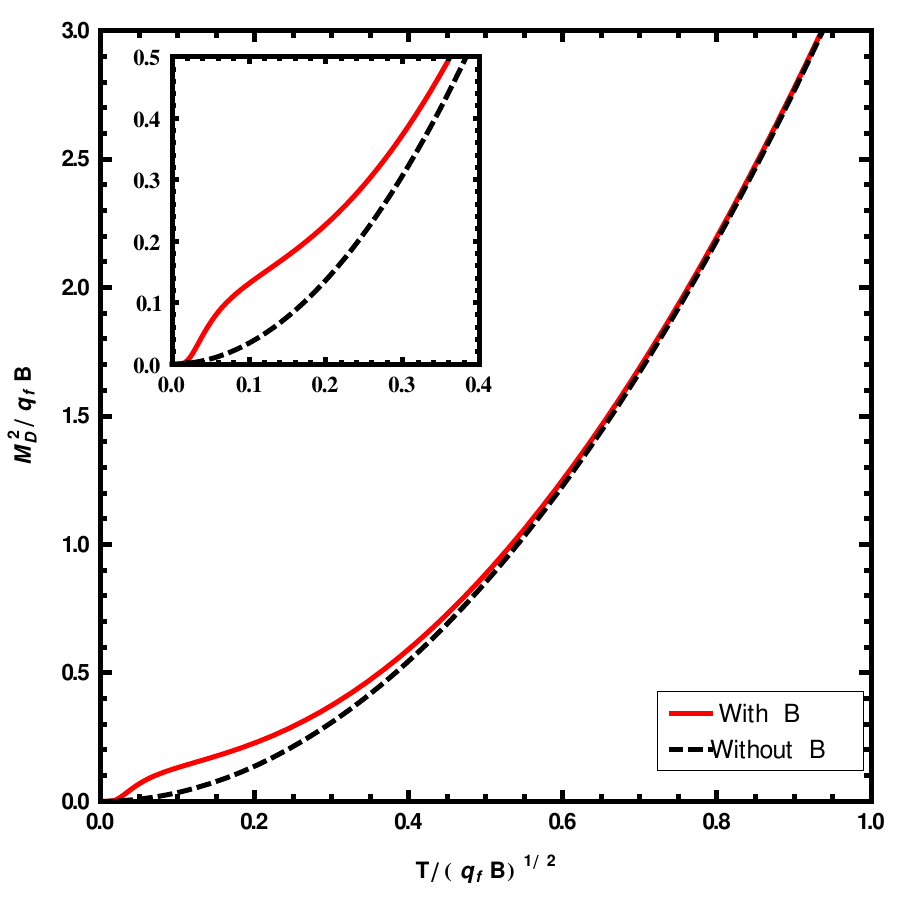} 
 \caption{Comparison of the scaled one-loop Debye masses 
varying with the scaled temperature for $N_f=2$.}
  \label{md_comparison}
 \end{center}
\end{figure}
\end{center}

In Fig.~\ref{md_comparison} we explicitly show some interesting features in the Debye mass due to the inclusion of the arbitrary external magnetic field by comparing it with the non-magnetized case. The $x$-axis is chosen to be $T/(q_fB)^{1/2}$, so that one can understand different limits of the magnetic fields. We note that Fig.~\ref{md_comparison} clearly reveals three different scales, $m_f$, $T$ and $B$, associated with the hot magnetized medium, as we discussed in the previous section. At $T=0$, $m_D=0$ and it gradually increases  with $T$ as long as $T < m_f <(q_fB)^{1/2}$. When $T\sim m_f<(q_fB)^{1/2}$ a shoulder appears in $m_D$ and then it increases a bit slowly\footnote {We further note that the shoulder is pushed towards the higher $T$ as quark mass increases (\ref{sfa_ds}). The appearance of the shoulder depends on the strength of two scales, viz., $m_f$ and $T$ associated with the hot magnetized system (\ref{sfa_ds}).} in the regime $m_f \le T \le (q_fB)^{1/2}$. This strong field behavior of $m_D$ is shown in the inset. Now we also note that if the thermal scale is higher than the magnetic scale ($T \gg (q_fB)^{1/2}$), then $m_D$ increases with $T$ like the usual hot but unmagnetized medium as shown in Fig.~\ref{md_comparison}.


\section{Conclusion}

In this chapter we have analyzed the electromagnetic screening effect through the Debye screening mass of the hot magnetized medium. This shows that there are three distinct scales in a hot magnetized medium, associated with the mass of the quasiquarks, temperature of the medium and the background magnetic field strength. When the mass of the quasiquarks are much higher than the temperature, the Debye screening is negligible. As 
the temperature increases, the screening mass starts increasing, a shoulder like structure appears when $T\sim m_f$,  and then it saturates to a fixed value when $q_fB \gg T^2\gg m_f^2$. Therefore in a strongly magnetized hot medium the Debye screening mass shows an interesting characteristic with temperature as long as $T\le m_f$ and then saturates to a value determined by the strength of the magnetic field. The point at which the saturation takes place  depends, especially, on the strength  of mass and temperature  scale associated with a hot magnetized system.

In strong field approximation  the fermion pairing takes place in LLL that could enhance the formation of quark-antiquark condensates, leading to a larger dynamical mass generation which catalyzes the spontaneous chiral symmetry breaking. This mass effect is reflected in the Debye screening as the shoulder and the saturation point are pushed towards a higher $T$  when the quasiquark mass increases. The effective dimensional reduction 
seems to play an important role in catalyzing the spontaneous chiral symmetry breaking, which indicates the occurrence of magnetic catalysis effect in presence of strong magnetic field. In the regime of weak field approximation, where the thermal scale is higher than the magnetic scale ($T^2 \gg q_fB$), the magnetic effect is negligible and Debye mass increases with $T$ like the usual hot unmagnetized medium.

%% file: text/qns.tex
\chapter{Quark Number Susceptibilities with the Gribov-Zwanziger action}
\label{th_qns}

In chapter \ref{th_gribov} we have computed the one-loop DPR with the GZ action within the HTL approximation. As an extension of that work we present in this chapter another experimentally relevant quantity, the one-loop QNS within the HTL approximation, evaluated with the non-perturbative GZ action. This chapter is based on a part of \textit{Dilepton rate and quark number susceptibility with the Gribov action} by Aritra Bandyopadhyay, Najmul Haque, Munshi G. Mustafa and Michael Strickland, {\bf Phys.Rev. D93 (2016) no.6, 065004}.  

\section{One-loop QNS - Computation}

Some generalities about QNS as well as its importance in characterizing QGP is already discussed in section \ref{qns_smgen}. In this section we present the computation of one-loop QNS with the GZ action within HTL approximation. In order to compute the QNS we need to calculate the imaginary part of the temporal component of the two one-loop diagrams given in Fig.~\ref{feyn_diag}. 
The contribution of the self energy diagram is
\bea
\Pi^s_{00}(Q)\!\!\!\!&=&\!\!\!\!N_fN_cT\sum_{p_0}\int\frac{d^3p}{(2\pi)^3} \textrm{Tr}\left[S(P)~\Gamma^0(K,Q,-P)~S(K)
~\Gamma^0(-K,-Q,P)\right]
\eea
where $K=P-Q$. After performing the traces of the self energy diagram, for zero external three momenta one obtains
\bea
\Pi^s_{00}(\vec{q}=0)&=& 2N_fN_cT\sum_{p_0}\int \!\!\!\frac{d^3p}{(2\pi)^3} 
\left[\frac{(a_G+b_G)^2}{D_+(\omega_1,p,\gamma_G)D_-(\omega_2,p,\gamma_G)}\right.\nonumber\\
&&\hspace{3.3cm}+\left.\frac{(a_G-b_G)^2}{D_-(\omega_1,p,\gamma_G)D_+(\omega_2,p,\gamma_G)}\right] , \label{se}
\eea
where $a_G$ and $b_G$ are given in Eq.~(\ref{gribov_vertex_coeff}) and
\bea
a_G + b_G &=& 1-\frac{2g^2C_F}{(2\pi)^2}\sum_\pm\int dk k \tilde{n}_\pm(k,\gamma_G)\frac{1}{\omega}\nn\\
&& \times \Bigl[Q_0(\tilde{\omega}_{11}^\pm,p) + Q_1(\tilde{\omega}_{11}^\pm,p)+Q_0(\tilde{\omega}_{21}^\pm,p)
- Q_1(\tilde{\omega}_{21}^\pm,p)\nn\\
&&+Q_0(\tilde{\omega}_{12}^\pm,p) + Q_1(\tilde{\omega}_{12}^\pm,p)+Q_0(\tilde{\omega}_{22}^\pm,p)- Q_1(\tilde{\omega}_{22}^\pm,p)\Bigr]\nn\\
&=& 1+\frac{1}{\omega}\left[D_+(\omega_1,p,\gamma_G)+D_-(\omega_2,p,\gamma_G)-\omega_1-\omega_2\right]\nn\\
&=& 1- \frac{\omega_1+\omega_2}{\omega} + \frac{D_+(\omega_1,p,\gamma_G)+D_-(\omega_2,p,\gamma_G)}{\omega}, \label{coeffsa}\\
a_G - b_G &=& 1-\frac{2g^2C_F}{(2\pi)^2}\sum_\pm\int dk k \tilde{n}_\pm(k,\gamma_G)\frac{1}{\omega}\nn\\
&& \times \Bigl[Q_0(\tilde{\omega}_{11}^\pm,p)- Q_1(\tilde{\omega}_{11}^\pm,p)+Q_0(\tilde{\omega}_{21}^\pm,p)
+ Q_1(\tilde{\omega}_{21}^\pm,p)\nn\\
&&+Q_0(\tilde{\omega}_{12}^\pm,p) - Q_1(\tilde{\omega}_{12}^\pm,p)+Q_0(\tilde{\omega}_{22}^\pm,p) + Q_1(\tilde{\omega}_{22}^\pm,p)\Bigr]\nn\\
&=& 1+\frac{1}{\omega}\left[D_-(\omega_1,p,\gamma_G)+D_+(\omega_2,p,\gamma_G)-\omega_1-\omega_2\right]\nn\\
&=& 1- \frac{\omega_1+\omega_2}{\omega} + \frac{D_-(\omega_1,p,\gamma_G)+D_+(\omega_2,p,\gamma_G)}{\omega}, \label{coeffsb}
\eea
where $D_\mp(\omega,p,\gamma_G)$ were defined in Eq.~(\ref{dpm}). We write only those terms of Eq.~(\ref{se}) which contain discontinuities
\bea
\frac{(a_G+b_G)^2}{D_+(\omega_1,p,\gamma_G)D_-(\omega_2,p,\gamma_G)} &=& \frac{(1- \frac{\omega_1+\omega_2}{\omega})^2}
{D_+(\omega_1,p,\gamma_G)D_-(\omega_2,p,\gamma_G)}\nonumber\\
&& \hspace{5mm} + \frac{1}{\omega^2}\left\{\frac{D_+(\omega_1,p,\gamma_G)}{D_-(\omega_2,p,\gamma_G)}+\frac{D_-(\omega_2,p,\gamma_G)}
{D_+(\omega_1,p,\gamma_G)}\right\}, \nn\\
~~~~~~~~~~&&~~~~~~~~~~~~~~~~~~~~~~~~~~~~~~~~~~~~~\nn\\
\frac{(a_G-b_G)^2}{D_-(\omega_1,p,\gamma_G)D_+(\omega_2,p,\gamma_G)} &=& \frac{(1- \frac{\omega_1+\omega_2}{\omega})^2}
{D_-(\omega_1,p,\gamma_G)D_+(\omega_2,p,\gamma_G)}\nonumber\\
&& \hspace{5mm} + \frac{1}{\omega^2}\left\{\frac{D_-(\omega_1,p,\gamma_G)}{D_+(\omega_2,p,\gamma_G)}+\frac{D_+(\omega_2,p,\gamma_G)}
{D_-(\omega_1,p,\gamma_G)}\right\}.\label{coeffs1}
\eea
Calculating the discontinuity using the BPY prescription given in Eq.~(\ref{bpy_pres}), one can write the imaginary part of Eq.~(\ref{se}) as
\bea
\textmd{Im}~\Pi^s_{00} &=& 4N_cN_f\pi(1-e^{\beta\omega})\int\frac{d^3p}{(2\pi)^3}\int d\omega_1 
\int d\omega_2 ~\delta(\omega-\omega_1-\omega_2)
n_F(\omega_1)n_F(\omega_2)\nn\\
&&\!\!\!\!\!\!\!\!\!\! \times \Bigl[\left(1- \frac{\omega_1+\omega_2}{\omega}\right)^2 \rho_+^G(\omega_1,p)\rho_-^G(\omega_2,p)+
\frac{C_1\rho_+^G(\omega_2,p)+C_2\rho_-^G(\omega_2,p)}{\omega^2}\Bigr], \label{im_se}
\eea
with
\bea
C_1 &=& \textmd{Im}~D_-(\omega_1,p)=0, \nn\\
C_2 &=& \textmd{Im}~D_+(\omega_1,p)=0. \label{c1c2}
\eea
The tadpole diagram of Fig.~\ref{feyn_diag} can now be written as
\bea
\Pi^t_{00}(Q)&=&N_fN_cT\sum_{p_0}\int\frac{d^3p}{(2\pi)^3} 
 \textrm{Tr}\biggl[S(P)~\Gamma_{00}(-P,P;-Q,Q)\biggr]. \label{qns_tad}
\eea
The four-point function $\Gamma_{00}$ at zero three-momentum can be obtained using Eq.~(\ref{wi_4pt}) giving
\bea
\Gamma^{00}&=& -(e_G\gamma^0+f_G~ \hat{p}\cdot\vec{\gamma}),\label{4pt}\\
e_G &=& \frac{2g^2c_F}{(2\pi)^2}\sum_\pm\int dk k \tilde{n}_\pm(k,\gamma_G)\frac{1}{(\omega_1-\omega_2)}\left[\delta Q_{01}^\pm+\delta Q_{02}^\pm+\delta Q_{01}^{\pm\prime}+\delta Q_{02}^{\pm\prime}\right],\nn\\
f_G &=& \frac{2g^2c_F}{(2\pi)^2}\sum_\pm\int dk k \tilde{n}_\pm(k,\gamma_G)\frac{1}{(\omega_1-\omega_2)}\left[\delta Q_{11}^\pm+\delta Q_{12}^\pm+\delta Q_{11}^{\pm\prime}+\delta Q_{12}^{\pm\prime}\right],\nn
\eea
where
\bea
\delta Q_{n1}^{\pm\prime} &=& Q_n(\tilde{\omega}_{11}^\pm,p)- Q_n(\tilde{\omega}_{21}^{\pm\prime},p){\rm{~for~}} n=0,1,2 \, \, , \nn\\
\tilde{\omega}_{21}^{\pm\prime}&=& E_\pm^0(\omega_2^\prime +k-E_\pm^0)/k  \, , \nn\\
\tilde{\omega}_{22}^{\pm\prime}&=& E_\pm^0(\omega_2^\prime -k+E_\pm^0)/k  \, , \nn\\
\omega_2^\prime &=& \omega_1 + \omega. \nn
\eea
Proceeding in a similar way as the self-energy diagram, the contribution from the tadpole diagram is
\bea
\textmd{Im}~\Pi_{00}^t &=& -4N_cN_f\pi(1-e^{\beta\omega})\int\frac{d^3p}{(2\pi)^3}\int d\omega_1 \int d\omega_2
~ \delta(\omega-\omega_1-\omega_2)\nn\\
&&\times \frac{n_F(\omega_1)n_F(\omega_2)}{\omega^2}\Bigl[C_1\rho_+^G(\omega_2,p)+C_2\rho_-^G(\omega_2,p)\Bigr]\nn\\
&=& 0. \label{tad_im}
\eea
Combining Eq.~(\ref{im_se}) and (\ref{tad_im}), the total imaginary contribution of the temporal part shown in Fig.~\ref{feyn_diag} can now be written as 

\bea
\textmd{Im}~\Pi_{00} &=&  \textmd{Im}~\Pi^s_{00} + \textmd{Im}~\Pi^t_{00} \nn \\
&=& 4N_cN_f\pi(1-e^{\beta\omega})\int\frac{d^3p}{(2\pi)^3}\int d\omega_1 \int d\omega_2 ~\delta(\omega-\omega_1-\omega_2)
n_F(\omega_1)n_F(\omega_2)\nn\\
&& \times \Bigl[\left(1- \frac{\omega_1+\omega_2}{\omega}\right)^2 \rho_+^G(\omega_1,p)\rho_-^G(\omega_2,p)\Bigr]. \label{tot_im}
\eea
It is clear from (\ref{im_se}) and (\ref{tad_im}) that the tadpole contribution 
in (\ref{tad_im}) exactly cancels with the second term of (\ref{im_se}) even if 
$C_1$ and $C_2$ are finite, e.g., for the HTL
case ($\gamma_G=0$) \cite{Chakraborty:2001kx, Chakraborty:2003uw}. At finite $\gamma_G$, the form 
of the sum of self-energy and tadpole diagrams remains the same, even though the 
individual contributions are modified.

\section{One-loop QNS - Results and discussion}

Putting Eq.~(\ref{tot_im}) in the expression for the QNS in Eq.~(\ref{defn_qns}), we obtain
\bea
\chi_q(T)&=&4N_cN_f\beta\int\frac{d^3p}{(2\pi)^3}\int\limits_{-\infty}^{\infty}d\omega\int d\omega_1 
\int d\omega_2 ~\delta(\omega-\omega_1-\omega_2)n_F(\omega_1)n_F(\omega_2) \nn \\
&&\hspace{1cm} \times \Big[\left(1- \frac{\omega_1+\omega_2}{\omega}\right)^2 
\rho_+^G(\omega_1,p)\rho_-^G(\omega_2,p)\Big] \nn \\
&=& 4N_cN_f\beta\int\frac{d^3p}{(2\pi)^3}\Bigl[
\left(\frac{\omega_+^2-p^2}{2m_q^2(\gamma_G)}\right)^2n_F(\omega_+)n_F(-\omega_+) \nn\\
&+&\left(\frac{\omega_-^2-p^2}{2m_q^2(\gamma_G)}\right)^2n_F(\omega_-)n_F(-\omega_-)+\left(\frac{\omega_G^2-p^2}{2m_q^2(\gamma_G)}\right)^2n_F(\omega_G)n_F(-\omega_G)
\Bigr] \nn \\
&&~~~~~~~~~~~~~~~~~~~~\nn\\
&=& \chi_q^{\rm{pp}}(T)\label{qns_pp}
\eea
where we represent the total $\chi_q(T)$ as $\chi^{\rm {pp}}_q(T)$ since there is only the pole-pole contribution for $\gamma_G\ne 0$. However for $\gamma_G=0$ there will be pole-cut ($\chi^{\rm {pc}}_q(T)$) and cut-cut ($\chi^{\rm{cc}}_q(T))$ contribution in addition to pole-pole contribution because the spectral function contains  pole part + Landau cut contribution of the quark propagator.

\begin{figure}[t]
\begin{center}
\includegraphics[width=0.48\linewidth]{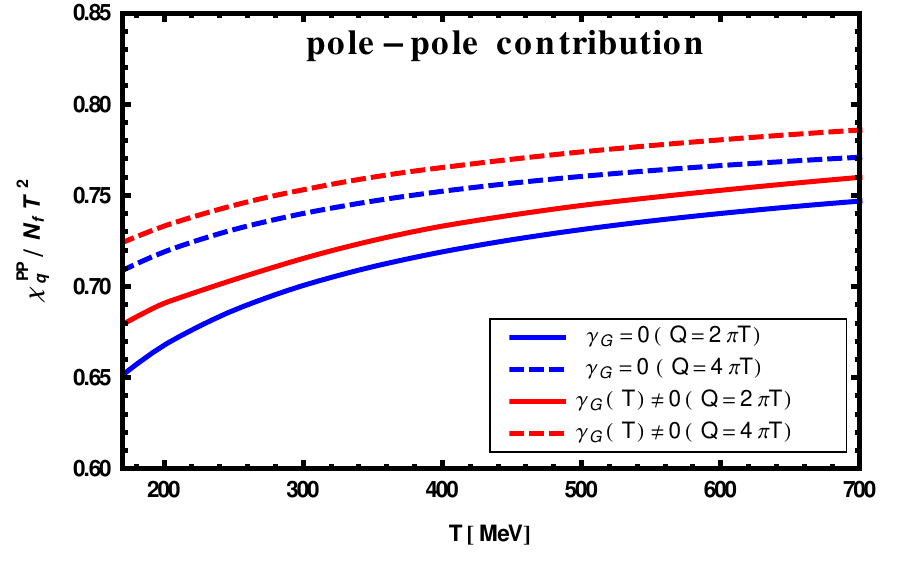}  
\hspace{2mm}
\includegraphics[width=0.48\linewidth]{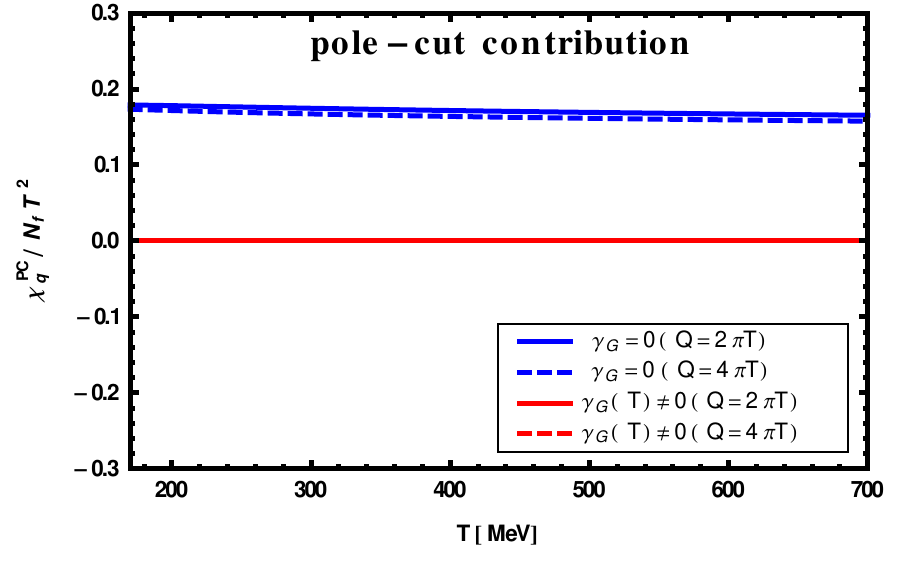}\\
\vspace{5mm}
\includegraphics[width=0.48\linewidth]{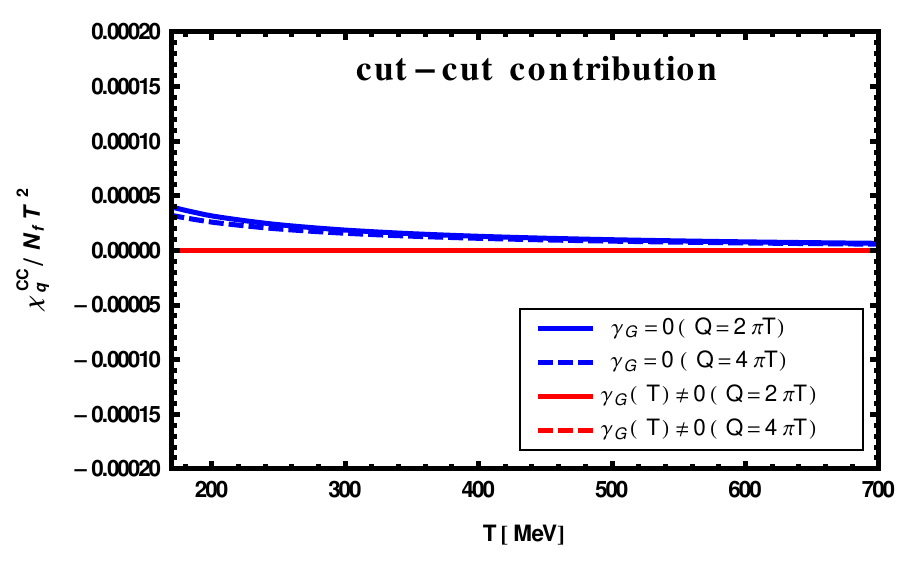}  
\hspace{2mm}
\includegraphics[width=0.48\linewidth]{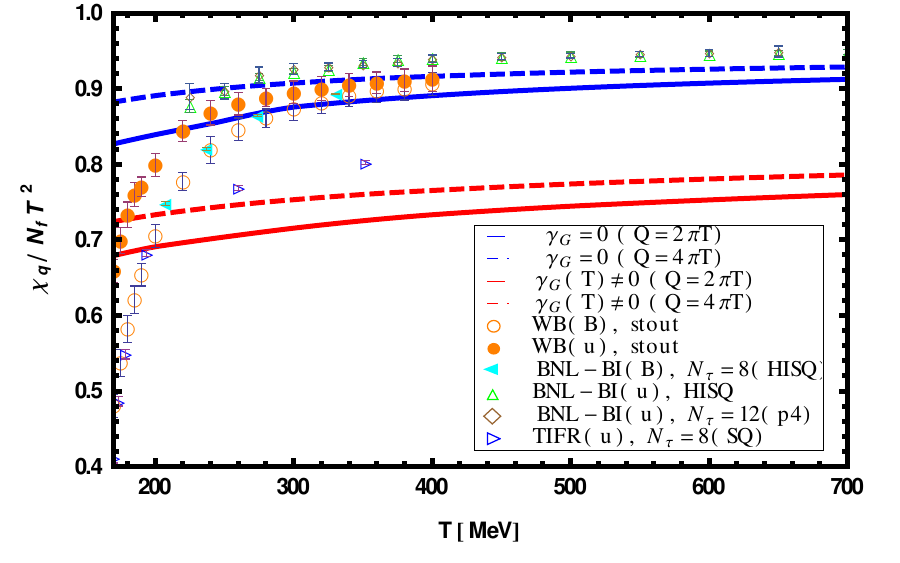}
\end{center}
\caption[QNS scaled with free values are compared with and without the 
inclusion of $\gamma_G$]{ QNS scaled with free values are compared with and without the 
inclusion of $\gamma_G$. In each case a band appears due to the choice of the 
two renormalization scales as $2\pi T$ and $4\pi T $.  The various symbols 
correspond to LQCD data from various groups labeled as  WB~\cite{Borsanyi:2012cr}, 
BNL-BI(B) and BNL-BI(u)~\cite{Bazavov:2013dta,Bazavov:2013uja}, and TIFR~\cite{Datta:2012pj}.}

\label{qns_Gribov}
\end{figure}

In Fig.~\ref{qns_Gribov} we have presented the different contribution of QNS scaled with the corresponding free values with and without the Gribov term. We, at first, note that the running coupling in (\ref{alpha_s})is a smooth function of $T$ around and beyond $T_c$ and our treatment is not valid below $T_c$. Now from the first panel of Fig.~\ref{qns_Gribov}, the pole-pole contribution to the QNS with the Gribov action is increased at low $T$, compared to that in absence of the Gribov term. This improvement at low $T$ is solely due to the presence of the non-perturbative Gribov mode in the collective excitations. However, at high $T$ both contributions become almost same as the Gribov mode disappears.  There are no pole-cut (pc) or cut-cut (cc) contribution for $\gamma_G(T)\ne 0$, compared to that for $\gamma_G = 0$. The pc and cc contributions in absence of magnetic scale are displayed in second and third panels. As a result, we find that the QNS in presence of magnetic scale contains 
only the pp-contribution due to collective excitations originating from the in-medium dispersion whereas, in absence of magnetic scale, the QNS is enhanced due to additional higher order Landau cut (i.e., pole-cut + cut-cut) contribution as shown in the fourth panel. When compared with LQCD data from various groups~\cite{Borsanyi:2012cr,Bazavov:2013dta,Bazavov:2013uja,Datta:2012pj}, the QNS in presence of magnetic scale lies around $(10-15)\%$ below the LQCD results whereas that in absence of magnetic scale is very close to LQCD data. This is expected due to the {\textit{additional higher-order Landau cut}} contribution  in absence of magnetic scale as discussed earlier. This also suggests that it is necessary to include higher loop orders in QNS in presence of the   magnetic scale, which is beyond the scope of this thesis. However, we hope to carry out this nontrivial task in near future.

\section{Conclusions}

After chapter \ref{th_gribov}, in this chapter also we considered the effect of inclusion of non-perturbative magnetic screening in another important observable, the quark number susceptibility. For the QNS, we found that like DPR, again due to the absence of Landau damping for $\gamma_G \neq 0$, the results do not agree well with available lattice data.  This can be contrasted with a standard HTLpt calculation, which seems to describe the lattice data quite well with no free parameters. It is possible that higher-order loop calculations could improve the agreement between the Gribov-scenario results and the lattice data; however, the success of HTLpt compared to lattice data as well as nonperturbative model calculations suggests that at \mbox{$T\gtrsim$ 200 MeV} the electric sector alone provides an accurate description of QGP thermodynamics. In conclusion, the present HTLpt results poses a serious challenge to the Gribov scenario simply for the inclusion of magnetic mass effects in the QGP.

%% file: text/conclusion.tex
\chapter{Summary and prospect}
\label{th_conclu}
The asymptotic freedom is a unique aspect of QCD, the guiding theory to explore the strong interaction.  By the virtue of asymptotic freedom, a deconfined state of matter (QGP) can be formed when the effective coupling strength of the strong interaction decreases, making the constituent quarks and gluons asymptotically free with respect to each other. Apart from its theoretical existence in the early universe or in the core of the neutron stars, QGP can be created transiently in the current HIC experiments. Characterization of QGP has great innuendos in understanding nature as it connects the dots in the particle era of the early universe and those in the underlying properties of the neutron star. In this dissertation we mainly studied the computation of some of the observables which are important for characterizing QGP. DPR is one such observable which is associated with the QGP signature dilepton spectra. After its production in the transient QGP phase, dileptons do not interact with the medium and carry least contaminated information to the detector. This is why DPR is considered to be a very important observable in the context of probing QGP phase. A large part of this thesis is dedicated to the computation of DPR in different circumstances employing different techniques. Another observable QNS is also computed and discussed in this dissertation which is associated with the charge fluctuation in the system. We have also evaluated Debye screening which is connected with another crucial signature of QGP, namely $J/\psi$ suppression.

Collating the experimental observations, theoretically these observables that are essential to characterize QGP can be evaluated both by means of perturbative and nonperturbative methods. Resummed perturbation theories like HTLpt or dimensional reduction techniques work well at a relatively higher temperature ($>2T_c$) to calculate various physical quantities associated with QGP. The ongoing HICs with their respective COM energies can produce QGP with a temperature closer to $T_c$.  But close to $T_c$, QGP could be completely non-perturbative in nature. First principle numerical technique LQCD is distinctly useful in this situation. Still shortcomings of LQCD suggests that we have some alternative analytic ways to include the non-perturbative effects in the theory. In this thesis we used some of such alternative analytic approaches to investigate the less-explored non-perturbative physics associated with the theory. In chapter \ref{th_dpr} we have discussed the ingredients of the tools or approaches that we used in the latter chapters to study various observables. All the observables we studied in this dissertation are related with the current-current CF and the corresponding spectral function of the system. Hence we shed light on the basics of CF and SF. We have also revisited the formulation of DPR and QNS in this chapter.  

Fluctuations of the quark and gluonic fields in the QCD vacuum generate local composite operators of quark and gluon fields, phenomenologically known as condensates. By considering the non-vanishing vacuum expectation values of such condensates the nonperturbative effect can be incorporated in physical observables. In chapter \ref{th_ope} the nonperturbative electromagnetic spectral function vis-a-vis the DPR in QCD plasma is evaluated in the intermediate mass range by taking those condensates into account within the framework of the OPE in $D=4$ dimension. The Wilson coefficients turns out to be free of infrared singularities as the mass singularity appearing in the self-energy diagram involving gluon condensate cancels out by judicious exploitation of the minimal subtraction through operator mixing. We consider the intermediate mass range because the formulation of OPE forbids it to be valid in the low mass regime ($ M \le 1$ GeV) whereas high mass regime ($ M \ge 2.5$ GeV) is well described by the perturbative approach. As for the DPR in the intermediate mass range, our result is found to be enhanced compared to other available non-perturbative results. So it could be of physical significance to describe the intermediate mass dilepton spectra from heavy-ion collisions incorporating the power corrections. As future prospects of this work we would like to note that the closure of the calculation of DPR using non-perturbative power corrections awaits computation of condensates at finite temperature via LQCD. Presently there is no estimate available in the literature for the quark and gluon condensates at finite $T$ which prompted us to exploit the Stefan-Boltzmann limits to have some limiting information of the nonperturbative effects associated with it. Also analytically this work can be extended by taking the temperature dependence in the Wilson coefficients along with the condensates. 

In chapter \ref{th_gribov} we considered a different way of inclusion of the non-perturbative effects in the theory through magnetic screening. The inclusion of magnetic scale is done using the Gribov-Zwanziger action, which describes the picture of confinement. The confinement effect is imposed by embedding a mass like Gribov parameter in the bare gluon propagator. Using this modified gluon propagator the HTL resummed quark propagator is obtained in a recent effort~\cite{Su:2014rma}. Some strikingly new key features were observed in this effective GZ-HTL quark propagator, like the formation of a new space like mode in the dispersion relation and the extermination of the discontinuity or the Landau damping part. Using this GZ-HTL quark propagator we then evaluated the DPR at vanishing three momenta by computing the spectral function.  It is observed that though it contains all the Van Hove singularities like that of the normal HTL result, but due to the absence of the Landau damping no other contribution is present. Interestingly, due to the new space like mode, all the energy gaps now disappear. It seems as if the Landau cut contribution in the DPR due to space like momenta is replaced by the filling of the energy gaps due to space like Gribov modes. This is why GZ-HTL result of DPR is in conflict with earlier perturbative and non-perturbative results. In chapter \ref{th_qns} again we exploited the GZ-HTL quark propagator to evaluate the QNS. Similarly like the DPR, the QNS results also effectively reduces in this approach due to the absence of the Landau cut contribution. In future it would be interesting to see if the novel space like excitation generated from the GZ action have important consequences for various other physical quantities relevant for the study of QGP. 

The effect of the anisotropic constant magnetic field generated in the non-central HIC is discussed in the chapter \ref{th_mag} and \ref{th_ds} of this thesis. Interestingly the produced magnetic field takes two different limiting values in two different situations along the timeline of the HIC. At the time of the collision the initial magnitude of this magnetic field is measured to be very high ($eB\approx m_\pi^2$ at RHIC and $eB\approx 10m_\pi^2$ at LHC) which can persist through the pre-equilibrium stage. Again as the generated magnetic field decreases very fast with time, it is assumed that after the thermalization the magnetic field becomes relatively weaker than the other relevant scale of the medium, i.e. temperature. Due to this limiting behavior of the magnetic field one can make suitable analytic approximations in both the regimes. In the first situation, at the very early stage of the collision, one can work with the strong field approximation. In the strong field approximation, all other Landau levels except the lowest lie very far away from the magnetic field affected fermion which ends up being confined in the LLL. In LLL approximation the fermion propagator and hence the one-loop self energy tensor shows some distinctive features because of the dimensional reduction from (3+1) to (1+1) and the corresponding electromagnetic spectral function diverges below a certain threshold, termed as LT. The DPR also reflects the nature of the spectral function and at very low invariant mass an enhancement from the leading perturbative result is found. The rate also gradually decreases and diminishes because of the assumption that in LLL approximation there is nothing beyond the LLL. In the weak field approximation things again become analytically simpler because of a perturbative expansion in orders of the magnetic field around the unmagnetized propagator, which is achieved by expanding the Schwinger's proper time propagator for arbitrary fields. Using the spectral representations of this weak field propagator the DPR is evaluated and again found to be decreasing only in the low invariant mass region. At higher invariant mass it merges with the leading perturbative unmagnetized result as expected. In chapter \ref{th_ds} we computed the Debye screening mass in both these regimes and found it to be heavily dependent on the relevant scales. In strong field approximation when the mass scale is higher than the temperature scale the Debye screening mass is negligible. Comparable values of the two scales yield a shoulder like structure, saturating to a fixed value when the magnetic field scale starts to dominate. In the regime of weak field approximation, the thermal scale dominates and then Debye screening mass increases with temperature like the usual hot but unmagnetized medium. As prospects of this currently young and active research area, several studies are going on to study the hot magnetized medium and to observe its consequences in various properties of QGP. Being part of this large group we would also like to continue the investigation of this fairly unexplored hot and magnetized systems.

%% file: text/app1.tex
\chapter{Evaluation of phase space integral $I_{\mu\nu}$}
\label{imunu}
\section{Massive case}
From Eq.~(\ref{imunu_defn}) we can straightway write down 
\bea
\!\!\!\!I_{\mu\nu}(Q) \!\!\!\!&=&\!\!\!\! \int\frac{d^3p_1}{E_1}\int\frac{d^3p_2}{E_2}~\delta^4(P_1+P_2-Q)\times\nn\\
&&\left[P_{1\mu}P_{2\nu}+P_{2\mu}P_{1\nu}-(P_1\cdot P_2+m_l^2)g_{\mu\nu}\right]
\label{structure_1}
\eea
Again, using the general structure of any second rank tensor, one can write down 
\bea
I_{\mu\nu}(Q) &=& AQ^2g_{\mu\nu}+BQ_\mu Q_\nu
\label{structure_2}
\eea
Now contracting both \ref{structure_1} and \ref{structure_2} by $g^{\mu\nu}$, and equating we get, 
\bea
(4A+B)Q^2 = \int\frac{d^3p_1}{E_1}\int\frac{d^3p_2}{E_2}~\delta^4(P_1+P_2-Q)\left[-2(P_1\cdot P_2)-4m_l^2\right]
\label{equate_1}
\eea
Again contracting both \ref{structure_1} and \ref{structure_2} by $Q^{\mu}Q^{\nu}$, and equating we get, 
\bea
(A+B)Q^4 &=& \int\frac{d^3p_1}{E_1}\int\frac{d^3p_2}{E_2}~\delta^4(P_1+P_2-Q)\times\nn\\
&&\left[2(Q\cdot P_1)(Q\cdot P_2)-Q^2(P_1\cdot P_2+m_l^2)\right]
\label{equate_2}
\eea
Both equations \ref{equate_1} and \ref{equate_2} have Lorentz invariant quantities on both their sides and so one can choose any reference frame to evaluate it. We shall choose the convenient center of mass frame, where $\vec{p}_1+\vec{p}_2=0$. Now, in COM frame, \ref{equate_1} yields
\bea
(P_1\cdot P_2)\Big\vert_{COM} \!\!\!\!&=&\!\!\!\! E_1E_2-\vec{p}_1\cdot\vec{p}_2 = E_1^2 + p_1^2 = 2E_1^2-m_l^2.\nn\\
\therefore(4A+B)Q^2\Big\vert_{COM} \!\!\!\!&=&\!\!\!\! \int\frac{d^3p_1}{E_1}\frac{(-4E_1^2-2m_l^2)}{E_1}\delta(2E_1-q_0)\nn\\
\!\!\!\!&=&\!\!\!\! -\int 16\pi~ p_1^2~dp_1\delta(2E_1-q_0)\left(1+\frac{m_l^2}{2E_1^2}\right)\nn\\
\!\!\!\!&=&\!\!\!\! -\int 8\pi~ E_1^2~dE_1\delta\left(E_1-\frac{q_0}{2}\right)\left(1+\frac{m_l^2}{2E_1^2}\right)\left(1-\frac{m_l^2}{E_1^2}\right)^{\frac{1}{2}}\nn\\
\!\!\!\!&=&\!\!\!\! -2\pi q_0^2\left(1+\frac{2m_l^2}{q_0^2}\right)\left(1-\frac{4m_l^2}{q_0^2}\right)^{\frac{1}{2}}\nn\\
\!\!\!\!&=&\!\!\!\! -2\pi Q^2\left(1+\frac{2m_l^2}{Q^2}\right)\left(1-\frac{4m_l^2}{Q^2}\right)^{\frac{1}{2}}\nn\\
\!\!\!\!&=&\!\!\!\! -F_1(m_l,Q^2)~2\pi~ Q^2\nn\\
\therefore (4A+B) \!\!\!\!&=&\!\!\!\! -2\pi~ F_1(m_l,Q^2) 
\label{AB_1}
\eea
Similarly, in COM frame, \ref{equate_2} gives us
\bea
2(Q\cdot P_1)(Q\cdot P_2)-Q^2(P_1\cdot P_2+m_l^2)\Big\vert_{COM} &=& 0 \nn\\
\therefore A+B &=& 0 
\label{AB_2}
\eea
Finally using \ref{AB_1} and \ref{AB_2} we obtain 
\bea
B=-A&=&\frac{2\pi}{3}F_1(m_l,Q^2)\nn\\
\therefore I_{\mu\nu}(Q) &=& AQ^2g_{\mu\nu}+BQ_\mu Q_\nu\nn\\
&=& \frac{2\pi}{3}F_1(m_l,Q^2)\left(Q_\mu Q_\nu-Q^2g_{\mu\nu}\right)
\label{unmagnetized_imn_massive}
\eea

\section{Massless case}

For the massless case, just putting $m_l=0$ in \ref{unmagnetized_imn_massive} we get, 
\bea
I_{\mu\nu}(Q)\Big\vert_{m_l=0}
&=& \frac{2\pi}{3}\left(Q_\mu Q_\nu-Q^2g_{\mu\nu}\right)
\label{unmagnetized_imn_massless}
\eea


\chapter{Evaluation of phase space integral $I^m_{\mu\nu}$}
\label{immunu}
Similarly as done in the unmagnetized case, here also from Eq.~(\ref{immunu_defn}) we get
\bea
\!\!\!\!I^m_{\mu\nu}(Q) \!\!\!\!&=&\!\!\!\! \int\frac{d^3p_1}{2E1}\int\frac{d^3p_2}{E2}~\delta^4(P_1+P_2-Q)\times\nn\\
\!\!\!\!&&\!\!\!\!\!\!\!\!\!\!\!\!\!\!\!\!\!\!\!\!\!\!\!\!\left[P^\shortparallel_{1\mu}P^\shortparallel_{2\nu}+P^\shortparallel_{1\nu}P^\shortparallel_{2\mu}-((P_1\cdot P_2)_\shortparallel + m_l^2)(g_{\mu\nu}^\shortparallel-g_{\mu\nu}^\perp-g_{1\mu}g_{1\nu}-g_{2\mu}g_{2\nu})\right].
\label{structure_1_mag}
\eea
Again using the general tensor structure in LLL approximation (see Eq.~(\ref{lll_genstruc})) we can write down
\bea
\!\!\!\!I^m_{\mu\nu}(Q) \!\!\!\!&=&\!\!\!\! AQ_\shortparallel^2g_{\mu\nu}^\shortparallel+BQ_\mu^\shortparallel Q_\nu^\shortparallel.
\label{structure_2_mag}
\eea
Now contracting both \ref{structure_1_mag} and \ref{structure_2_mag} by $g^{\mu\nu}_\shortparallel$ and equating we get 

\bea
(2A+B)Q_\shortparallel^2 = \int\frac{d^3p_1}{2E_1}\int\frac{d^3p_2}{E_2}~\delta^4(P_1+P_2-Q)\left[-2m_l^2\right].
\label{equate_1_mag}
\eea
Again contracting both \ref{structure_1_mag} and \ref{structure_2_mag} by $Q^{\mu}Q^{\nu}$ and equating we get
\bea
(A+B)Q_\shortparallel^4 &=& \int\frac{d^3p_1}{2E_1}\int\frac{d^3p_2}{E_2}\delta^4(P_1+P_2-Q)\times\nn\\
&& \left[2(Q\cdot P_1)_\shortparallel(Q\cdot P_2)_\shortparallel-Q_\shortparallel^2((P_1\cdot P_2)_\shortparallel+m_l^2)\right]
\label{equate_2_mag}
\eea
By the similar justification as in Appendix \ref{imunu}, in COM frame \ref{equate_1_mag} yields
\bea
\!\!\!\!\!\!\!\!\!&&\!\!\!\!\!\!\!\!\!\!(2A+B)Q_\shortparallel^2\Big\vert_{COM}\nn\\
\!\!\!\!\!\!\!\!\!&=& -m_l^2\int d^2P_1^\perp\int\frac{dp_1^z}{E_1}\int d^2P_2^\perp\int\frac{dp_2^z}{E_2}~\delta^2(P_1^\perp+P_2^\perp-Q^\perp)\delta^2(P_1^\shortparallel+P_2^\shortparallel-Q^\shortparallel)\nn\\
\!\!\!\!\!\!\!\!\!&=& -m_l^2\int d^2P_1^\perp\int\frac{dp_1^z}{E_1}\int\frac{dp_2^z}{E_2}~\delta^2(P_1^\shortparallel+P_2^\shortparallel-Q^\shortparallel)\nn\\
\!\!\!\!\!\!\!\!\!&=& -2\pi |eB|m_l^2\int\frac{dp_1^z}{E_1^2}~\delta(2E_1-q^0)\nn\\
\!\!\!\!\!\!\!\!\!&=& -\pi |eB|m_l^2\int\frac{dE_1}{E_1\sqrt{E_1^2-m_l^2}}~\delta(E_1-\frac{q^0}{2})\nn\\
\!\!\!\!\!\!\!\!\!&=& -\frac{4\pi |eB|m_l^2}{(Q_\shortparallel^2)^2}\left(1-\frac{4m_l^2}{Q_\shortparallel^2}\right)^{-\frac{1}{2}}Q_\shortparallel^2\nn\\
\!\!\!\!\!\!\!&&\!\!\!\!\!\!\therefore (2A+B) = -\frac{4\pi}{(Q_\shortparallel^2)^2} F_2(m_l,Q_\shortparallel^2) 
\label{AB_1_mag}
\eea
 by replacing $d^4P= d^2P^\perp d^2P^\shortparallel$ and $d^2P^\perp= 2\pi|eB|$ in the process. Again in COM frame, we obtain from \ref{equate_2_mag},
\bea
2(Q\cdot P_1)_\shortparallel(Q\cdot P_2)_\shortparallel-Q_\shortparallel^2((P_1\cdot P_2)_\shortparallel+m_l^2)\Big\vert_{COM} &=& 0 \nn\\
\therefore A+B &=& 0 
\label{AB_2_mag}
\eea
 Eventually from \ref{AB_1_mag} and \ref{AB_2_mag} we get the final value of the integral
 \bea
 B=-A&=& \frac{4\pi}{(Q_\shortparallel^2)^2} F_2(m_l,Q_\shortparallel^2) =\frac{4\pi |eB|m_l^2}{(Q_\shortparallel^2)^2}\left(1-\frac{4m_l^2}{Q_\shortparallel^2}\right)^{-\frac{1}{2}}\nn\\
 \therefore I^m_{\mu\nu}(Q) &=& AQ_\shortparallel^2g^\shortparallel_{\mu\nu}+BQ^\shortparallel_\mu Q^\shortparallel_\nu\nn\\
&=& \frac{4\pi}{(Q_\shortparallel^2)^2} F_2(m_l,Q_\shortparallel^2)\left(Q^\shortparallel_\mu Q^\shortparallel_\nu-Q_\shortparallel^2g^\shortparallel_{\mu\nu}\right)
\label{magnetized_imn_massive}
 \eea


\chapter{Massless Feynman Integrals}
\label{massless_feynman}
While computing the electromagnetic polarization tensor with gluon condensates, the 
following Feynman integrals for massless quarks have been used: 
\bea
\mathcal{I}_{mn}  &=& \int \frac{d^dK}{(2\pi)^d} \frac{1}{(K^2)^m~((K-P)^2)^n},\nn\\
\mathcal{I}_{mn}^\mu &=& \int \frac{d^dK}{(2\pi)^d} \frac{K^\mu}{(K^2)^m~((K-P)^2)^n},\nn\\
\mathcal{I}_{mn}^{\mu\nu}  &=& \int \frac{d^dK}{(2\pi)^d} \frac{K^\mu K^\nu}{(K^2)^m~((K-P)^2)^n}.\nn
\eea
The primary integrals can be represented as follows,
\bea
\mathcal{I}_{mn}\!\!\!\!&=&\!\!\!\! \frac{i}{(16\pi^2)^{\frac{d}{4}}}(-1)^{-m-n}(-P^2)^{-m-n+\frac{d}{2}}
\frac{\Gamma[m+n-\frac{d}{2}]}{\Gamma[m]~\Gamma[n]}B\left(\frac{d}{2}-n,
\frac{d}{2}-m\right),\\
\mathcal{I}_{mn}^\mu \!\!\!\!&=&\!\!\!\! \frac{i}{(16 
\pi^2)^{\frac{d}{4}}}(-1)^{-m-n}(-P^2)^{-m-n+\frac{d}{2}}\nn\\
&&~~~~~~~~~~~~~\Biggl\{\frac{\Gamma[m+n-\frac{d}{2}]~\Gamma[1+\frac{d}{2}-m]~\Gamma[\frac{d}
{2}-n]}{\Gamma[m]~\Gamma[n]~\Gamma[1+d-m-n]}\Biggr\},
\eea
\bea
\mathcal{I}_{mn}^{\mu\nu} \!\!\!\!&=&\!\!\!\! \frac{i}{(16 
\pi^2)^{\frac{d}{4}}}(-1)^{-m-n}(-P^2)^{-m-n+\frac{d}{2}}\nn\\
&&~~~~~~~~~~~~~\Biggl\{g^{\mu\nu}\frac{\Gamma[m+n+2-\frac{d}{2}]\Gamma[1+\frac{d}{2}-m]\Gamma[1+\frac{d}{2}-n]}
{2\Gamma[m]~\Gamma[n]~\Gamma[2+d-m-n]}\nn\\
&&~~~~~~~~~~~~~~+P^\mu P^\nu\frac{\Gamma[m+n-\frac{d}{2}]~\Gamma[2+\frac{d}{2}-m]~\Gamma[\frac{d}{2}-n]}{2\Gamma[m]~\Gamma[n]~\Gamma[2+d-m-n]}\Biggr\}.
\eea
Now putting $d = 4 - 2 \epsilon$, we obtain required results of $\mathcal{I}_{mn}, 
\mathcal{I}_{mn}^\mu$ and $\mathcal{I}_{mn}^{\mu\nu}$ 
for some given values of $m$ and $n$ needed for our purpose: 
\bea
\mathcal{I}_{12} &=& 
\mu^{-\epsilon}\frac{i}{16\pi^2}\frac{1}{P^2}\left(-\frac{1}{\tilde{\epsilon}}\right),
\nn\\
\mathcal{I}_{22} &=& 
\mu^{-\epsilon}\frac{i}{16\pi^2}\frac{2}{P^4}\left(-\frac{2}{\tilde{\epsilon}}-2\right),
\nn\\
\mathcal{I}_{22}^0 &=& 
\mu^{-\epsilon}\frac{i}{16\pi^2}\frac{p^0}{P^4}\left(-\frac{1}{\tilde{\epsilon}}-1\right),
\nn\\
\mathcal{I}_{31}^0 &=& 
\mu^{-\epsilon}\frac{i}{16\pi^2}\frac{p^0}{P^4}\left(-\frac{1}{2\tilde{\epsilon}}-\frac{1}
{2}\right),\nn\\
\mathcal{I}_{22}^{00} &=& 
\mu^{-\epsilon}\frac{i}{16\pi^2}\frac{1}{P^4}\left[\frac{P^2}{2}+\left(-\frac{1}{\tilde{
\epsilon}}-2\right)(p^0)^2\right],\nn\\
\mathcal{I}_{31}^{00} &=& \mu^{-\epsilon}\frac{i}{16\pi^2}\frac{1}{(P^2)^2}\left[P^2
\left(-\frac{1}{4\tilde{\epsilon}}-\frac{1}{4}\right)+\frac{(p^0)^2}{2}\right],\nn\\
\mathcal{I}_{41}^{00} &=& \mu^{-\epsilon}\frac{i}{16\pi^2}\frac{1}{(P^2)^3}\left[P^2
\left(\frac{1}{12\tilde{\epsilon}}-\frac{1}{12}\right)+(p^0)^2\left(-\frac{1}{3\tilde{
\epsilon}}-\frac{1}{2}\right)\right],\nn
\eea
where 
\bea
\mu &=& e^{\frac{\gamma_E}{2}}\frac{\Lambda^2}{4\pi}~;~~ \frac{1}{\tilde{\epsilon}} = 
\frac{1}{\epsilon}-\log\left(\frac{-P^2}{\Lambda^2}\right)\nn
\eea
with $\mu$ as the renormalization scale, $\Lambda$ as $\overline{\rm MS}$ renormalization 
scale and $\gamma_E$ as the Euler-Mascheroni constant.


\chapter{Forbidden processes in LLL with ($r=1,l=-1$)  and ($r=-1,l=1$) }
\label{app_a}
Here we demonstrate the $r=1,l=-1$ case. The $r=-1,l=1$ case can be done in an exact similar way. So, choosing $r=1,l=-1$ in (\ref{Pi_sfa_gen}) we obtain
\bea
\mathcal{I}m~\Pi_\mu^\mu(\omega,\vec{p})\Big\vert_{\abcom{r=1}{l=-1}}\!\! \!\! &=& \! \!\!\! 
N_c\pi \sum_{f}e^{\frac{-P_\perp^2}{2q_fB}}~\frac{2q_f^3Bm_f^2}{\pi}\int\!\!\frac{dk_3}{2\pi}
\frac{\left(1-n_F(E_k)\right)\left(1-n_F(-E_q)\right)}{4E_kE_q} \nn\\
&& \times \left[e^{ -\beta(E_k-E_q)}-1\right] \delta(p_0-E_k+E_q). \label{a1}
\eea
Now, using $1-n_F(-E_q)=n_F(E_q)$ one can write down
\bea
 \mathcal{I}m~\Pi_\mu^\mu(\omega,\vec{p})\Big\vert_{\abcom{r=1}{l=-1}} \!\! \!\! &=& \! \!\!\!
N_c\sum_{f}
e^{\frac{-P_\perp^2}{2q_fB}}~~\frac{2q_f^3Bm_f^2}{\pi}
\int\frac{dk_3}{2}~\delta(\omega-E_k+E_q)\nn\\
&&\times~\frac{\left[n_F(E_k)-n_F(E_q)\right]}{4E_kE_q}.
\label{a2}
\eea
The $k_3$ integral can now be performed  using the following property of the delta 
function 

\bea
\int\limits_{-\infty}^{\infty} dp_3~ f(p_3)~ \delta[g(p_3)] = \sum_{r} 
\frac{f(p_{zr})}{\vert g^\prime(p_{zr})\vert}, \label{a3}
\label{delta_property}
\eea
where the zeroes of the argument inside the delta function is called as $p_{zr}$.
Now $(\omega-E_k+E_q) = 0 $ yields
\bea
k_3^z = \frac{p_3}{2} \pm \frac{\omega}{2}\sqrt{1-\frac{4m_f^2}{(\omega^2-p_3^2)}},
\!\!\!\!&=&\!\!\!\! \frac{p_3}{2} \pm \frac{\omega R}{2}, \label{a4}\\
\vert g^\prime(p_{z})\vert = \Bigg\vert\frac{E_k(k_3-p_3)-E_qk_3}{E_kE_q}&&\!\!\!\!\!\!\!\!\!\!\!\!\!\Bigg\vert_{k_3=k_3^{z1},k_3^{z2}}~~, 
\label{a5}
\eea
with
\bea
&&E_k\Big\vert_{k_3=k_3^{z1}} = \frac{\omega}{2} + \frac{p_3 R}{2};~~ 
E_k\Big\vert_{k_3=k_3^{z2}} = \frac{\omega}{2} - \frac{p_3 R}{2},\label{a6} \\
&&E_q\Big\vert_{k_3=k_3^{z1}} = \frac{\omega}{2} - \frac{p_3 R}{2}; ~~ 
E_q\Big\vert_{k_3=k_3^{z2}}=\frac{\omega}{2} + \frac{p_3 R}{2},\label{a7} \\
\mbox{and}&& \, \, \Big\vert E_k(k_3-p_3)-E_qk_3\Big\vert_{k_3=k_3^{z1},k_3=k_3^{z2}} = 
\frac{\omega p_3}{2}(R^2-1).\label{a8}
\eea
Now using \ref{a6}, \ref{a7} and \ref{a8} one can write
\bea
 &&\mathcal{I}m~\Pi_\mu^\mu(\omega,\vec{p})\Big\vert_{\abcom{r=1}{l=-1}} \nn\\
 &=& N_c\sum_{f}
e^{\frac{-P_\perp^2}{2q_fB}}~~\frac{2q_f^3Bm_f^2}{\pi}\sum\limits_r\frac{\left[
n_F(E_k)-n_F(E_q)\right]}{8E_kE_q}\times 
\Big\vert\frac{E_kE_q}{E_k(k_3-p_3)-E_qk_3}\Big\vert\Bigg\vert_{k_3=k_3^{zr}}~~ \nn\\
&=& N_c\sum_{f}
e^{\frac{-P_\perp^2}{2q_fB}}~~\frac{2q_f^3Bm_f^2}{\pi}\sum\limits_r\frac{\left[
n_F(E_k)-n_F(E_q)\right]}{8\vert E_k(k_3-p_3)-E_qk_3\vert}\Bigg\vert_{k_3=k_3^{zr}}~~ 
\nn
\eea
\bea
&=& N_c\sum_{f}
e^{\frac{-P_\perp^2}{2q_fB}}~~\frac{2q_f^3Bm_f^2}{4\pi\omega p_3(R^2-1)}\times\nn\\
&&\left[n_F(E_k\Big\vert_{k_3=k_3^{z1}})-n_F(E_q\Big\vert_{k_3=k_3^{z1}}
)+n_F(E_k\Big\vert_{k_3=k_3^{z2}})-n_F(E_q\Big\vert_{k_3=k_3^{z2}})\right]\nn\\
&=& N_c\sum_{f}
e^{\frac{-P_\perp^2}{2q_fB}}~~\frac{2q_f^3Bm_f^2}{4\pi\omega p_3(R^2-1)}\times \nn\\
&&\left[\!n_F\left(\frac{\omega}{2} \!+\! \frac{p_3 R}{2}\right)\!-\!n_F\left(\frac{\omega}{2} \!-\! 
\frac{p_3 R}{2}\right)\!+\!n_F\left(\frac{\omega}{2} \!-\! \frac{p_3 
R}{2}\right)\!-\!n_F\left(\frac{\omega}{2} \!+\! \frac{p_3 R}{2}\right)\!\right] \nn\\
&=& 0. \label{a9}
\eea
This result shows that for the case $r=1,l=-1$ the phase space also does not 
allow the corresponding process. Same can be checked easily for $r=-1,l=1$.